\documentclass[twoside]{report}

\usepackage[cp1250]{inputenc}
\usepackage[T1]{fontenc}
\usepackage{graphicx}

\usepackage{array}
\usepackage{longtable}

\usepackage{amsmath}
\usepackage{amsthm}
\usepackage{amssymb}

\usepackage{bbm}

\usepackage{enumerate}
\usepackage{cite}

\usepackage{color}

\usepackage{MnSymbol}

\newcommand{\linspan}{\textnormal{span}}
\newcommand{\K}{\mathbbm{K}}

\def\setZ{\mathbbm{Z}}

\newtheorem{theorem}{Theorem}[chapter]
\newtheorem{lemma}[theorem]{Lemma}
\newtheorem{corollary}[theorem]{Corollary}
\newtheorem{remark}[theorem]{Remark}
\newtheorem{definition}[theorem]{Definition}
\newtheorem{example}[theorem]{Example}
\newtheorem{proposition}[theorem]{Proposition}

\newtheorem{conjecture}{Conjecture}

\newcommand{\fml}[1]{(\ref{fml.#1})}

\newcommand{\SEQ}[3]{$\left\{#1\right\}_{#2}^{#3} $}
\newcommand{\fnc}[2]{#1\left(#2\right)}

\newcommand{\id}{\mathbbm{1}}

\newcommand{\setC}{\mathbbm{C}}
\newcommand{\setR}{\mathbbm{R}}
\newcommand{\setN}{\mathbbm{N}}
\def\dV{\dim V}

\newcommand{\diag}[1]{\fnc{\textnormal{diag}}{#1}}
\def\dualsymb{\circ}

\def\conjsymb{\ast}

\newcommand{\bounded}[1]{\mathcal{B}\left(#1\right)}
\newcommand{\boundedplus}[1]{\bounded{#1}^+}
\newcommand{\bbounded}[2]{\mathcal{B}\left(#1,#2\right)}
\newcommand{\hilbertspaceone}{\mathcal{K}}
\newcommand{\hilbertspacetwo}{\mathcal{H}}
\newcommand{\kh}{\hilbertspaceone\otimes\hilbertspacetwo}

\newcommand{\bk}{\bounded{\hilbertspaceone}}
\newcommand{\bh}{\bounded{\hilbertspacetwo}}

\newcommand{\bkh}{\bounded{\kh}}
\newcommand{\bhplus}{\bh^+}
\newcommand{\bkplus}{\bk^+}
\newcommand{\mappingcone}{\mathcal{C}}
\newcommand{\innerpr}[2]{\left<#1,#2\right>}
\newcommand{\innerprtwo}[2]{\innerpr{#1}{#2}'}
\newcommand{\innerprthree}[2]{\innerpr{#1}{#2}''}
\newcommand{\seq}[3]{\left\{#1\right\}_{#2}^{#3}}
\newcommand{\Tr}{\mathop{\textnormal{Tr}}}
\newcommand{\Ad}{\mathop{\textnormal{Ad}}}

\newcommand{\conj}[1]{#1^{\conjsymb}}
\newcommand{\dual}[1]{#1^{\dualsymb}}
\newcommand{\ddual}[1]{#1^{\dualsymb\dualsymb}}
\newcommand{\convhull}{\mathop{\textnormal{convhull}}}
\newcommand{\Id}{\mathop{\textnormal{id}}}
\newcommand{\One}{\mathbbm{1}}
\newcommand{\diad}[1]{\left|#1\right>\left<#1\right|}
\newcommand{\Choimatr}[1]{C_{#1}}
\newcommand{\matrices}[2]{M_{#1}\left(#2\right)}

\newcommand{\HPmaps}{\mathcal{HP}}
\newcommand{\Pmaps}{\mathcal{P}}
\newcommand{\kPmaps}[1]{\Pmaps_{#1}}
\newcommand{\SPmaps}{\mathcal{SP}}
\newcommand{\kSPmaps}[1]{\SPmaps_{#1}}
\newcommand{\CPmaps}{\mathcal{CP}}

\newcommand{\Pmapsb}[1]{\Pmaps\left(#1\right)}
\newcommand{\kPmapsb}[2]{\kPmaps{#1}\left(#2\right)}

\newcommand{\kSPmapsb}[2]{\kSPmaps{#1}\left(#2\right)}
\newcommand{\CPmapsb}[1]{\CPmaps\left(#1\right)}
\newcommand{\HPmapsbb}[2]{\HPmaps\left(#1,#2\right)}
\newcommand{\Pmapsbb}[2]{\Pmaps\left(#1,#2\right)}
\newcommand{\kPmapsbb}[3]{\kPmaps{#1}\left(#2,#3\right)}
\newcommand{\SPmapsbb}[2]{\SPmaps\left(#1,#2\right)}
\newcommand{\kSPmapsbb}[3]{\kSPmaps{#1}\left(#2,#3\right)}
\newcommand{\CPmapsbb}[2]{\CPmaps\left(#1,#2\right)}
\newcommand{\wektor}[1]{#1}
\newcommand{\proj}[1]{\left|#1\right>\left<#1\right|}
\newcommand{\kernel}[1]{\textnormal{Ker}#1}
\newcommand{\range}[1]{R\left(#1\right)}
\newcommand{\rank}[1]{r\left(#1\right)}
\newcommand{\SL}{\textnormal{SL}}
\newcommand{\PSL}{\textnormal{PSL}}
\newcommand{\PSLt}{\PSL\left(3,\mathbbm{C}\right)}
\newcommand{\SLt}{\SL\left(3,\mathbbm{C}\right)}
\newcommand{\PSLtt}{\PSLt\times\PSLt}
\newcommand{\SLtt}{\SLt\otimes\SLt}

\input{Qcircuit}

\title{\textsc{Jagiellonian University}\\{\large M. Smoluchowski Institute of Physics}\vskip 3mm\includegraphics[scale=0.15]{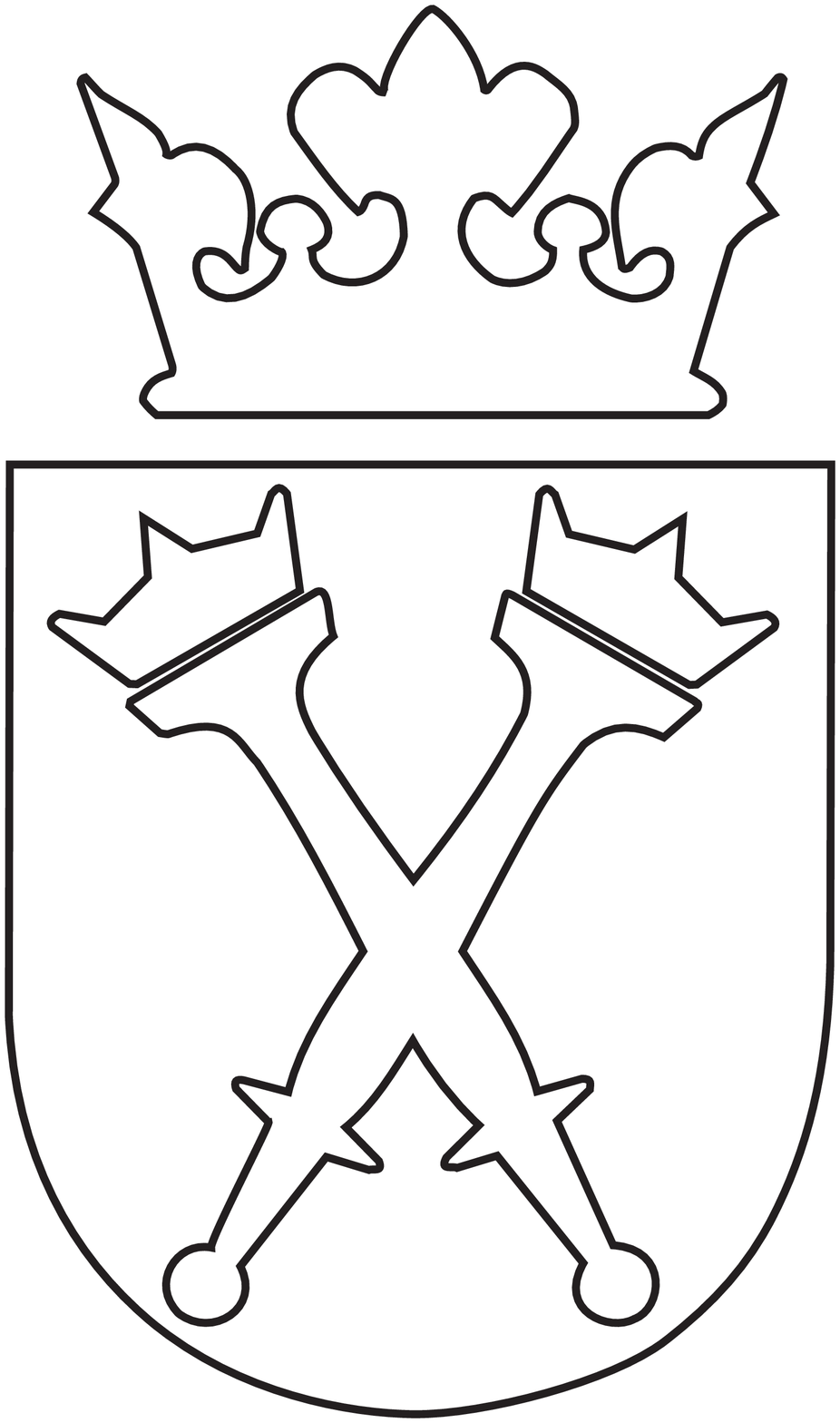}
\vskip 2 cm
A Few Algebraic Problems\\in the Theory of Quantum Entanglement\\
{\normalsize Thesis submitted for the fulfillment of the degree of Doctor of Philosophy}}
\author{Łukasz Skowronek}
\date{\today\vfill
\begin{figure}[b]\centering
\includegraphics[scale=0.5]{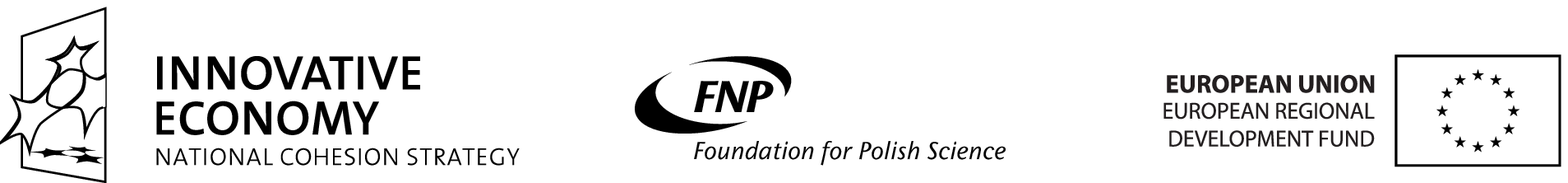}
\end{figure}
}

\begin{document}

\maketitle

\chapter*{Foreword}\addcontentsline{toc}{chapter}{Foreword}\,
Questions related to the practical use of quantum mechanics have grown extremely popular among physicists in the past two decades. The literature on the subject is extensive, but it seems not to make use of the advances of computational algebraic geometry, which is quite a natural framework when dealing with algebraic varieties like the set of product states. The lack of general interest can be partly attributed to the little popularity of algebraic geometry among the physicists working in the field, and partly to the fact that functions used as entanglement measures are not polynomials. Another reason may be the appearance of pairs of complex conjugate variables like $z$ and $\bar z$ in the polynomial equations that prevail in quantum information science, including the Knill-Laflamme equations, equations for Mutually Unbiased Bases and Symmetric Informationally Complete vectors, or for explicitly finding product vectors in the kernel of an entanglement witness. This makes the equations not truly polynomial, but functions of  both $z$ and $\bar z$ at the same time and thus apparently more difficult to solve. An important aim of the thesis is to present a number of specific questions that can nevertheless be solved using results from algebraic geometry, and in particular the technique of Groebner bases. The main result, on the other hand, which is a characterization of PPT bound entangled states of minimal rank, equal four, makes substantial use of Bezout's theorem, which can be described as a basic theorem in intersection theory. We also present a few problems solved by elementary algebra tricks. 

The structure of the thesis is the following. The first part, consisting of three chapters, discusses the basics of the theory of quantum entanglement, its practical uses, and the phenomenon of bound entanglement. In Chapter \ref{chfundamental}, the focus is on questions related to local realistic models of quantum mechanics. I familiarize the reader with separable quantum states and separability criteria. Later, we consider developments that go beyond the so-called \textit{separability paradigm}. In other words, we take a trip outside the reign of quantum entanglement. In Chapter \ref{chpractical}, I briefly describe the ideas behind several practical applications of quantum entanglement, such as quantum cryptography, quantum teleportation and dense coding, as well as quantum metrology. In Chapter \ref{chbound}, I included basic information about the distillation of quantum entanglement and about bound entangled states.

Chapter \ref{chVarIdGroeb} starts the second part of the thesis, which can be regarded as a standalone introduction to algebraic geometry for non-practitioners. I tried to make this part as rigorous as possible, however, in a number of places I had to refer to literature for proofs. The chapter begins from the definition of an affine variety and its ideal, and we proceed to the definition of a monomial ordering and a Groebner basis. I introduce the S-pair criterion and the Buchberger algorithm, which can be used to find Groebner bases of an ideal. In the end of Chapter \ref{chVarIdGroeb}, it should become clear why Groebner basis techniques can be useful for solving systems of polynomial equations. In Chapter \ref{chintersection}, the focus is on the basics of intersection theory. I try to explain how the dimension of an affine or projective variety relate to the number of monomials of certain total degree not in the corresponding ideal. I also introduce the important notion of the \textit{degree} of a projective variety. A theorem that two projective varieties of complementary dimension intersect appears as well. Finally, I give the Bezout's theorem in a simple form, which plays an important role later, in Chapter \ref{chPPT3x3}.
 
Part \ref{partIII} of the thesis, which starts with Chapter \ref{chmappingcones}, mostly contains the original results obtained and toy examples solved by the author. I first give a characterization theorem for a class of convex cones of maps from $n\times n$ to $m\times m$ matrices, which appear as $k$-positive and $k$-superpositive maps in the theory of entanglement. Next, in Chapter \ref{chhand}, we present three algebraic problems in the theory of quantum information, all of which can be solved by hand. They concern the following subjects:
\begin{itemize}
\item Product numerical range for a three-parameter family of operators,
\item Higher order numerical ranges (HONR) for three-by-three matrices,
\item Separable state of length three and Schmidt rank four.
\end{itemize}
In Chapter \ref{chGroebnerapplied}, we apply the Groebner basis approach to several types of equations that are of interest for the quantum information community. The problems we manage to solve relate to:
\begin{itemize}
\item Compression subspaces for Quantum Error Correction (QEC),
	\item Completely Entangled Subspaces (CES),
	\item Maximally entangled states in a linear subspace,
	\item Mutually Unbiased Bases (MUBs),
	\item Symmetric Informationally Complete vectors (SICs).
\end{itemize}
It should be kept in mind that the last two of the above subjects are presented here in a fully expository manner, because better solutions by other authors were available in the literature before I started my project. Finally, Chapter \ref{chPPT3x3}, which is the core element of the thesis, contains a proof of the above mentioned theorem relating positive-partial-transpose (PPT) states of minimal rank, equal four, to so-called Unextendible Product Bases (UPBs). I present a proof that the mentioned states can always be (stochastic) locally transformed to projections onto a subspace orthogonal to a UPB. On the way to prove the theorem, Bezout's theorem is applied, and some general observations concerning PPT states and so-called general Unextendible Product Bases are made. I conclude on page \pageref{chconclusion} and subsequently give a list of papers I published as a part of my PhD project. Most of them have strong relations to the results presented in this thesis. However, some of the contents has never been published.
\vskip 0.4 cm
There are a few people and organizations who helped me to succeed in my research project. Looking back in time, I can certainly say that my whole PhD studies were marked with a fair amount of good luck. Under different circumstances, it would have been much more difficult, if not impossible, to complete the thesis. Hence, I must first mention the support I received from the Foundation for Polish Science. Thanks to them, I was able to travel, meet other scientists, and to live a decent life for the most of the duration of my studies. Part of my contract with the foundation was a visit to Stockholm, where I got to know Jan Myrheim and Per {\O}yvind Sollid. Few months later, our interaction turned out to be very fruitful and resulted in a proof of Theorem \ref{maintheorem} of Chapter~\ref{chPPT3x3}, which is the backbone of the thesis. This could probably have never been possible, had I not received additional support from Stockholm University and the University of Oslo, all thanks to Ingemar Bengtsson and Erling St{\o}rmer. I wish to thank Ingemar for making a great discussion partner during my months in Sweden, and Erling for his grand hospitality during my two visits to Norway. It is Oslo where my best ideas were provisionally formed, including the results of Chapters \ref{chmappingcones} and \ref{chPPT3x3}. For the first visit there, I received additional funding from the Scholarship and Training Fund, operated by the Foundation for the Development of the Education System, which I am sincerely thankful for. It is also indisputable that the success of my research crucially depended on the constant support by my supervisor, Karol Życzkowski. His encouragement, wise judgment and great amount of understanding are difficult to overvalue. Besides the above, I owe special personal thanks to Per {\O}yvind Sollid for careful proofreading and detecting a flaw in a preliminary version of the manuscript on PPT states of rank four, included here as the crucial Chapter~\ref{chPPT3x3}. 
\vskip 0.4 cm
In the end, I wish to warmly thank my parents and my younger brother Michał, who were always there to help me when I needed it, especially during the sad days of my illness. Thank you! 



\part{Basics of quantum entanglement theory}\label{partI}\,

\chapter{Fundamental questions}\label{chfundamental}

\section{Local hidden variables}\label{secLHV}
Some strange consequences of quantum mechanics have bothered physicists from the very beginning of quantum theory. A classical example of this is the Einstein, Podolsky and Rosen paper \cite{EPR35}, where the authors argue that the quantum description of reality must be \emph{incomplete} if we accept two rather natural properties every physical theory should have. The first is the principle of \emph{physical reality}, which says that properties of physical systems such as spin direction or energy can be predicted with certainty before carrying out the corresponding measurement. They are \emph{elements of physical reality}. The second principle considered by Einstein, Podolsky and Rosen is that of \emph{locality}, which refers to the requirement that every system has its own properties, independently of any operations carried out on other, spatially separated systems. To see where the above two principles clash with the picture of reality given by quantum mechanics, let us consider a quantum system consisting of two two-level\footnote{we denote the levels by $0$ and $1$} subsystems $A$ and $B$, initially prepared in the so-called Bell state $\left|\Phi_+\right>=\left(\left|00\right>+\left|11\right>\right)/\sqrt{2}$. If the holder of the first subsystem measures it in the basis $\left\{\left|0\right>,\left|1\right>\right\}$, he or she obtains the result $0$ or $1$, both with probability $1/2$. This is not too surprising and may well happen in classical physics, however assuming that the state $\left|\Phi_+\right>$ does not contain a complete information about the degrees of freedom of the system. What is somewhat more interesting, is the prediction of quantum mechanics that \emph{after} $0$ or $1$ is measured in the subsystem $A$ using the $\left\{\left|0\right>,\left|1\right>\right\}$ basis, a corresponding measurement on the $B$ side yields identically the same result as the aforementioned measurement on the $A$ side. More generally, the holders of $A$ and $B$ never get two distinct results if they measure in the same basis. This is possible to reconcile with the principle of locality only if we accept that the outcomes of all possible measurements on the $A$ and $B$ sides are known \emph{beforehand}, i.e. before any measurements are done. It is possible to compare this to a macroscopic situation where a factory produces table tennis bats in two colors, say red and green, puts every single one into a box and groups these boxes into pairs with bats of the same color inside. It then sells these pairs without disclosing what colour the bats inside a particular pair of boxes are. The buyer of a table tennis set knows for sure what he or she has are two bats in the same colour, but does not know anything more. As soon as one of the boxes is opened, the colour of the bat inside the second box is revealed to the buyer. No matter how realistic the whole situation might seem in real life, it is clearly not excluded by classical physics, and it closely resembles the experiment with two two-level systems in the state $\left(\left|00\right>+\left|11\right>\right)/\sqrt{2}$, with $0$ corresponding to green and $1$ corresponding to red, or the other way round. Our aim in the following will be to shortly explain why a classical model similar to the table tennis set factory cannot nevertheless give us a proper description of the phenomena predicted by quantum mechanics. 
\begin{figure}\centering
\includegraphics[scale=0.3]{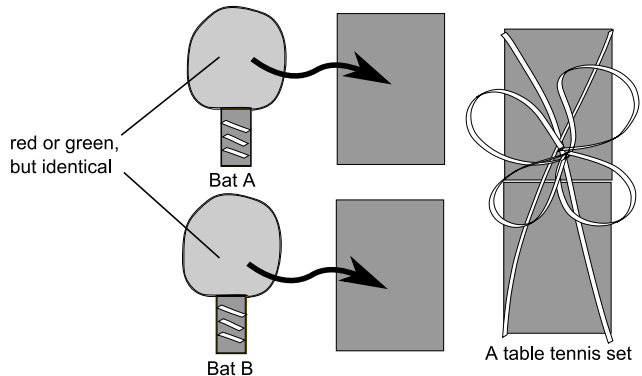}
\caption{A local realistic model: table tennis set factory at work.}%
\label{fig.pingpong}%
\end{figure}
In order for the discussion to stay general, let us introduce the following definition.
\begin{definition}
A \textbf{local hidden variable model} of an experiment on a bipartite system (consisting of parts $A$ and $B$) is a probability space $\left(\Omega,\Sigma,P\right)$ and a set of functions $S_A^x:\Omega\rightarrow\setR$ and $S_B^y:\Omega\rightarrow\setR$, where $x$ and $y$ refer to the possible measurement setups on subsystems $A$ and $B$, respectively and $S_A^x\left(\lambda\right)$, $S_B^y\left(\lambda\right)$ correspond to the measurements' outcomes. Here $\lambda$ represents the ``hidden variables'' or the true classical degrees of freedom of the system. Assuming that the measurement setup is fixed to $x$ for $A$ and to $y$ for $B$, the correlation coefficient between the measurement outcomes is given by the following formula
\begin{equation}\label{eqcorrcoeff}
\epsilon\left(x,y\right)=\int_{\Omega}S^x_A\left(\lambda\right)S^y_B\left(\lambda\right)dP\left(\lambda\right).
\end{equation}    
\end{definition}
To make a connection to the table tennis set factory model, let us mention that $\lambda$ in formula \eqref{eqcorrcoeff} corresponds to a ``mode'' of the factory, which is either the production of a pair of green bats or the production a red pair. The ``mode'' is hidden from the buyer a table tennis set, just as the additional degrees of freedom, represented by $\lambda$, are supposed to be hidden from the user of quantum mechanics. In the following, however, we show that it is possible, by a simple mathematical argument, to refute the idea of a local hidden variable model for quantum mechanics.

To this aim, let us consider a system consisting of two spin-$1/2$ particles, initially prepared in the state $\left|\Phi_+\right>=\left(\left|00\right>+\left|11\right>\right)/\sqrt{2}$, where $\left|0\right>$, $\left|1\right>$ represent the $\pm 1$ eigenstates of the operator $\sigma_z$. We measure the spin of the first particle in the direction $\vec a$ and the spin of the second particle in the direction $\vec b$. The corresponding observables are $\vec a\cdot\vec\sigma^A$ and $\vec b\cdot\vec\sigma^B$, where the subscripts $A$, $B$ refer to operators on the first and the second subsystem, respectively. The correlation coefficient between the two measurements, as predicted by quantum mechanics, is
\begin{equation}\label{eqcorrcoeff1}
\tilde\epsilon\left(\vec a,\vec b\right)=\left<\Phi_+\right|\left(\vec a\cdot\vec\sigma^A\right)\left(\vec b\cdot\vec\sigma^B\right)\left|\Phi_+\right>=a^1b^1-a^2b^2+a^3b^3,
\end{equation}
where the numbers $a^i$, $b^i$ for $i=1,2,3$ denote the coordinates of the vectors $\vec a$ and $\vec b$, resp.
For the specific choice of the vectors $\vec a=\left[\sin\alpha,0,\cos\alpha\right]$ and $\vec b=\left[\sin\beta,0,\cos\beta\right]$, we get $\tilde\epsilon\left(\vec a,\vec b\right)=\tilde\epsilon\left(\alpha,\beta\right)=\cos\left(\beta-\alpha\right)$. Let us now suppose that this form of correlation function can be reproduced by a local hidden variable model. Thus we need to have a probability space $\left(\Omega,\Sigma,P\right)$ and a set of functions $S^{\alpha}_A:\lambda\mapsto S^{\alpha}_A\left(\lambda\right)$ and $S^{\beta}_B:\lambda\mapsto S^{\beta}_B\left(\lambda\right)$ giving the measurement outcomes of the spin measurements for a fixed choice of the hidden variables $\lambda$. Since a spin measurement can only give $\pm 1$ as an answer, we have $S^{\alpha}_A\left(\lambda\right)\in\left\{-1,+1\right\}$, $S^{\beta}_B\left(\lambda\right)\in\left\{-1,+1\right\}$. Let us now consider the following combination of the functions $S^{\alpha}_A$ and $S^{\beta}_B$,
\begin{equation}\label{eqcombination}
S_A^{\alpha_2}\left(\lambda\right)\left[S_B^{\beta_1}\left(\lambda\right)+S_B^{\beta_2}\left(\lambda\right)\right]+S^{\alpha_1}_A\left(\lambda\right)\left[S^{\beta_1}_B\left(\lambda\right)-S_B^{\beta_2}\left(\lambda\right)\right].
\end{equation}
It is easy to see that for fixed $\lambda$, one of the expressions in squared brackets equals $0$, while the other one is equal to $\pm 2$. All in all, the whole expression in \eqref{eqcombination} equals $\pm 2$. Therefore we have, assuming that the hidden variable model we consider describes the quantum mechanical world, the following inequality for the previously considered correlation functions,
\begin{multline}\label{eqCHSH}
\left|\tilde\epsilon\left(\alpha_2,\beta_1\right)+\tilde\epsilon\left(\alpha_2,\beta_2\right)+\tilde\epsilon\left(\alpha_1,\beta_1\right)-\tilde\epsilon\left(\alpha_1,\beta_2\right)\right|\leqslant\\
\leqslant\int_{\Omega}\left|S_A^{\alpha_2}\left(\lambda\right)\left[S_B^{\beta_1}\left(\lambda\right)+S_B^{\beta_2}\left(\lambda\right)\right]+S^{\alpha_1}_A\left(\lambda\right)\left[S^{\beta_1}_B\left(\lambda\right)-S_B^{\beta_2}\left(\lambda\right)\right]\right|\leqslant 2.
\end{multline}
The above is the famous \textit{CHSH inequality}, named for J. F. Clauser, M. A. Horne, A. Shimony and R. A. Holt \cite{CHSH69}. For the choice $\alpha_1=45^{\circ}$, $\beta_1=90^{\circ}$, $\alpha_2=135^{\circ}$ and $\beta_2=180^{\circ}$, one can readily check that the correlation functions predicted by quantum mechanics do \emph{not} obey \eqref{eqCHSH}, since
\begin{equation}\label{eqTsirelson}
\left|\tilde\epsilon\left(\alpha_2,\beta_1\right)+\tilde\epsilon\left(\alpha_2,\beta_2\right)+\tilde\epsilon\left(\alpha_1,\beta_1\right)-\tilde\epsilon\left(\alpha_1,\beta_2\right)\right|=2\sqrt{2}.
\end{equation}
Moreover, the above violation of the CHSH inequality is the maximum allowed by quantum mechanics \cite{Cirelson80}. The value $2\sqrt{2}$ in \eqref{eqTsirelson}, called the \textit{Tsirelson bound}, is in clear contradiction with the assumption that quantum mechanics can be described as a local hidden variable theory. Thus, we are lead to the conclusion that there exists \emph{no local realistic model for quantum mechanics}. The question whether the quantum mechanical correlations are really observed in experiments, and how to close the possible experimental loopholes, is the subject of a separate field of research, with the first and most famous experiments done by the A. Aspect group \cite{Asp82}.  

\section{Separable states and separability criteria}\label{secsep}
Our next topic is closely related to hidden variable models, and was first studied by R. Werner in the late 80s \cite{Werner89}. He introduced a class of mixed states, which he called \textit{classically correlated}, but they are now generally referred to as \textit{separable}.
\begin{definition}\label{defseparablestate}
A state represented by a density matrix $\rho$ on a bipartite space $\hilbertspaceone\otimes\hilbertspacetwo$ is called \textbf{separable} if and only if it can be written as a convex combination of projections onto product states, i.e. a sum
\begin{equation}
\rho=\sum_{i=1}^n\lambda_i\proj{\phi_i\otimes\psi_i}
\label{eqseparablesum}
\end{equation}
with $n$ finite, $\lambda_i\geqslant 0$, $\sum_{i=1}^n\lambda_i=1$ and $\phi_i\in\hilbertspaceone$, $\psi_i\in\hilbertspacetwo$.
\end{definition}
Actually, in \cite{Werner89}, infinite sums of the type \eqref{eqseparablesum} were considered, but it follows from Carath{\'e}odory's theorem (cf. e.g. \cite[Chapter 13]{ref.Rockafellar}) that any such sum can be rewritten as a finite one. A generalization of Definition \ref{defseparablestate} to a multipartite setting is immediate.
\begin{definition}\label{defsepmulti}
A state represented by a density matrix $\rho$ on a multipartite space $\hilbertspaceone_1\otimes\ldots\otimes\hilbertspaceone_k$ is called \textbf{separable} if and only if it can be written as a sum
\begin{equation}
\rho=\sum_{i=1}^n\lambda_i\proj{\phi^1_i\otimes\ldots\otimes\phi^k_i}
\label{eqseparablesum2}
\end{equation}
with $n$ finite, $\lambda_i\geqslant 0$, $\sum_{i=1}^n\lambda_i=1$ and $\phi^l_i\in\hilbertspaceone_l\,\forall_{l=1,2,\ldots,k}$. 
\end{definition}
It is now also generally accepted that states which are \emph{not} of the form given in Definitions \ref{defseparablestate} and \ref{defsepmulti} are called \textit{entangled}.
\begin{definition}\label{defentstate}
A state represented by a density matrix $\rho$ on a multipartite space $\hilbertspaceone_1\otimes\ldots\otimes\hilbertspaceone_k$ is called \textbf{entangled} if and only if it cannot be written in the form \eqref{eqseparablesum2}.
\end{definition}
In case of pure states $\rho$, it can be shown \cite{G91} (cf. also \cite{GP92}) that the property of being entangled implies the lack of a local realistic model of the local measurements one can perform on $\rho$. More precisely, for a pure entangled state $\rho$, there always exists a Bell-type inequality\footnote{like the CHSH inequality we considered in Section \ref{secLHV}} that is not fulfilled by the correlation functions resulting from $\rho$. However, if  mixed states $\rho$ are taken into consideration, it was the main subject of the work \cite{Werner89} to show that \emph{there exist entangled states which do admit a local realistic description}. It should also be noted that in the paper \cite{Werner89}, the author never used the word ``entangled'' himself. It may thus be rather surprising to hear that what is now generally accepted as a synonym of something quantum-like, something \textit{entangled}, was born for the purpose to show that it can sometimes be described in a fully classical way. Fortunately, the apparent paradox was partially resolved by \cite{Popescu95}, where the author showed that sometimes hidden nonlocality in quantum states can be revealed by sequential measurements. A step in a similar direction was also taken by N. Gisin, who showed that local interaction can turn a state that does not violate any Bell-type inequality into one that is nonlocal \cite{Gisin96}. Additional justification for the importance of the notion of inseparability was provided by L. Masanes \cite{Mas06,Mas08}, who showed that entangled states are always  useful for certain tasks in quantum information processing.   Finally, the question about nonlocality of all bipartite entangled states was settled in the paper \cite{Mas08b}, by Masanes, Liang and Doherty. They managed to prove that bipartite entangled states $\rho$ are precisely those which \emph{do violate} some inequality of CHSH type, possibly after they are tensor multiplied by some state $\sigma$ that does \emph{not} violate any CHSH inequality itself. Being more precise,
\begin{equation}\label{eqMasanesEntangled}
\rho\textnormal{ is entangled }\Longleftrightarrow\rho\otimes\sigma\textnormal{ violates a CHSH type inequality}
\end{equation}
where $\sigma$ does \emph{not} violate any inequality of CHSH type, even after it undergoes arbitrary stochastic local operations with communication \cite{Mas08b}. Note that the tensor multiplication by $\sigma$ only plays a role of a catalyst in the process of discovering the nonlocality of $\rho$. Hence, it is legitimate to say that \textit{all bipartite entangled states have some kind of non-locally realistic properties}, and \textit{vice versa}.

Because of the result by Masanes, Liang and Doherty, we feel it is well-justified to accept the definition of entangled states as it is.  Hence we conform to the \textit{separability paradigm}. However, we shall go back to the question of separability versus local realism when we discuss distillation of entanglement in Section \ref{secdistill}. We should also give additional credit to the Werner's paper \cite{Werner89} and mention the famous family of states the author used to prove his result. They are now called \textit{Werner states} and are of the simple form
\begin{equation}\label{eqWernerstates}
W=\frac{1}{d^3-d}\left[\left(d-\Xi\right)\One+\left(d\,\Xi-1\right)V\right]
\end{equation}   
where $\Xi\in\left[-1,1\right]$, $d$ is the dimensionality of the Hilbert space $\hilbertspaceone$ such that $W$ is defined on $\hilbertspaceone\otimes\hilbertspaceone$, and $V:=\sum_{i,1=1}^d\left|i\right>\left<j\right|\otimes\left|j\right>\left<i\right|$. The choice of the specific parametrization in \eqref{eqWernerstates} is motivated by the equality $\Xi=\Tr\left(W V\right)$. A distinctive future of the Werner states is that they are invariant under the transformation $W\mapsto\left(U\otimes U\right)W\left(\conj{U}\otimes\conj{U}\right)$ for an arbitrary unitary $U$. In \cite{Werner89}, it was shown that the state $W$ is separable for $\Xi\geqslant 0$ and entangled otherwise. Moreover, for $\Xi=-1+\left(d+1\right)/d^2$ it \emph{admits a hidden variable description}. Since $-1+\left(d+1\right)/d^2\leqslant -1/4\leqslant 0$, the corresponding $W$ is entangled and at the same time it can be described in a local realistic manner.

Despite the above paradoxical property of some entangled states, it became widely accepted that the distinction between entanglement and separability plays a fundamental role in the theory of quantum information. Entanglement detection has become the subject of a separate research area, which we would very sparsely explore in the rest of this section. Much more information can be found in review articles like \cite{HHHH09,GT09}.

Probably the most famous separability test is the \textit{PPT criterion} by A. Peres \cite{Peres96}, where PPT stands for ``positive partial transpose''. The criterion was quickly proved by the Horodecki family to be a necessary and sufficient condition in the case of $2\times 2$ and $2\times 3$ systems\footnote{cf. \cite{LMO2006} for a nice explanation in the $2\times 2$ case} \cite{HHH96}. The criterion simply says that a density matrix $\rho$ on a bipartite space $\hilbertspaceone\otimes\hilbertspaceone$, if separable, must be positive under the following transformation
\begin{equation}\label{eqPPTcrit}
\rho\mapsto\left(\Id\otimes t\right)\rho,
\end{equation}  
where $t$ denotes the transposition map in $\bk$. Thus, if the \textit{partial transpose} $\rho^{T_2}:=\left(\Id\otimes t\right)\rho$ of a density matrix $\rho$ is found \emph{not} to be positive, we know that $\rho$ is entangled. Let us state this as a proposition.
\begin{proposition}[PPT criterion]\label{propPPTcrit}
If a state $\rho$ acting on a bipartite space $\hilbertspaceone\otimes\hilbertspaceone$ is separable, the partial transpose of $\rho$, given by the r.h.s. of \eqref{eqPPTcrit}, must be a positive operator.
\end{proposition}
States which do not satisfy the implication of Proposition \ref{propPPTcrit} are called \textit{NPT entangled}, where NPT stands for ``negative partial transpose''. It was a natural question to ask whether there exists entangled states with positive partial transpose (PPT). For $2\times 2$ and $2\times 3$ systems, this is impossible by \cite{HHH96}, but for $3\times 3$ systems, a \textit{PPT entangled} state was found by P. Horodecki \cite{Pawel97}. Different examples were earlier studied, in a slightly different context, by E. St{\o}rmer \cite{Erling82} and M.-D. Choi \cite{Choi82}. In order to prove his result, the author of \cite{Pawel97} needed a different separability test than the PPT criterion. What he used is now called the \textit{range criterion} for separability. 
\begin{proposition}[Range criterion]\label{proprange}
For a separable state $\rho$ on a bipartite space $\hilbertspaceone\otimes\hilbertspaceone$, there must exist a set of product vectors $\phi_i\otimes\psi_i$ that span the range of $\rho$, $\range{\rho}$. In addition to that, the partially conjugated vectors $\phi_i\otimes\conj{\psi}_i$ need to span the range of $\rho^{T_2}$, $\range{\rho^{T_2}}$.   
\end{proposition}
A whole family of separability criteria can be derived from the following result by the Horodecki family \cite{HHH96}, which generalises the PPT criterion.
\begin{proposition}[Positive maps criterion]
A separable state $\rho$ on a bipartite space $\hilbertspaceone\otimes\hilbertspaceone$ is separable if and only if
\begin{equation}\label{eqposmapscrit}
\left(\Id\otimes\Lambda\right)\rho\geqslant 0
\end{equation}
for all linear maps $\Lambda:\bk\rightarrow\bk$ that \textbf{preserve the positivity of operators}.
\end{proposition}
Maps that preserve positivity of operators are called \textit{positive maps}, and hence the name of the criterion. One of the main results of this thesis, presented in Chapter \ref{chmappingcones}, is a broad generalization of the positive maps criterion for different subclasses of the set of all density matrices, including states of Schmidt rank $k$ \cite{HT00}.

The problem with condition \eqref{eqposmapscrit} is that it needs to be checked for all positive maps, which is impossible as long as we do not know their full structure. However, for a fixed choice of the map $\Lambda$, the positive maps criterion always gives a necessary condition for separability. An example of this is when $\Lambda\left(\rho\right)=\One\Tr\rho-\rho$, so-called \textit{reduction map}. For such choice of the positive map, we get \cite{HH99}
\begin{proposition}[Reduction criterion]\label{propredcrit}
A separable state on a bipartite space has to fulfill the following condition
\begin{equation}
\left({\Tr}_B\rho\right)\otimes\One-\rho\geqslant 0,
\end{equation}
where ${\Tr}_B$ denotes the partial trace of $\rho$ with respect to the second subsystem, $\left({\Tr}_A\rho\right)_{ij}=\sum_k\rho_{ik,jk}$.
\end{proposition} 
Another possible choice of $\Lambda$ is $\Lambda:\rho\mapsto\One\Tr\rho-\rho-V\rho^t\conj{V}$, so-called Breuer-Hall map \cite{Breuer2006,Hall2006}. Here $\rho^t$ stands for the transposition of $\rho$ and $V$ is an \emph{antisymmetric}, \emph{unitary} matrix, $V^t=-V$. Such matrices $V$ only exist if the dimension of the space $\hilbertspaceone$ is even. 

Yet another, experimentally feasible approach to the discrimination of the set of separable states is by the use of so-called \textit{entanglement witnesses}. An entanglement witness\footnote{not to confuse with the Werner state introduced earlier} is an operator $W$ on a multipartite space $\hilbertspaceone_1\otimes\hilbertspaceone_2\otimes\ldots\otimes\hilbertspaceone_k$ with the property
\begin{equation}\label{eqwitness}
\left<\phi_1\otimes\phi_2\otimes\ldots\otimes\phi_k\right|W\left|\phi_1\otimes\phi_2\otimes\ldots\otimes\phi_k\right>\geqslant 0
\end{equation}
for all $\phi_1$, \ldots, $\phi_k$ in $\hilbertspaceone_1$, \ldots, $\hilbertspaceone_2$, resp. In terms of such operators, we have the following separability criterion
\begin{proposition}[Entanglement witness criterion]
A density matrix $\rho$ on a multipartite space $\hilbertspaceone_1\otimes\hilbertspaceone_2\otimes\ldots\otimes\hilbertspaceone_k$ is separable if and only if the following inequality,
\begin{equation}\label{witnesses}
\Tr\left(W\rho\right)\geqslant 0
\end{equation}
holds for all witnesses $W$ on $\hilbertspaceone_1\otimes\hilbertspaceone_2\otimes\ldots\otimes\hilbertspaceone_k$.
\end{proposition}
A big advantage of witnesses over positive maps is that the trace on the l.h.s. of \eqref{witnesses} can be measured in an experiment as an expectation value of an observable. Moreover, one can often find an optimal decomposition of the witness into locally measurable quantities \cite{Guhne02,Guhne03}, i.e. a decomposition of the form
\begin{equation}\label{optlocdec}
W=\sum_{l=1}^r\gamma_lX_1^l\otimes X_2^l\otimes\ldots\otimes X_k^l
\end{equation}
with $r$ minimal. One can also ask whether a witness $W$ is optimal in the sense that for no other witness $W'$ the inequality $\Tr\left(W\rho\right)<0$ implies $\Tr\left(W'\rho\right)<0$ \cite{LKCH00}.

Nevertheless, we should note that by the Jamiołkowski-Choi isomorphism \cite{ref.J72,ref.Choi75} (cf. also \cite{BZ2006}), every witness has a corresponding positive map $\Lambda_W$, and the corresponding positive map criterion $\left(\Id\otimes\Lambda_W\right)\rho\geqslant 0$ is much stronger than the criterion $\Tr\left(W\rho\right)\geqslant 0$. However, the first criterion is much more difficult to measure in an experiment \cite{HE02}. 

To finish, let us explain a relation of the CHSH inequality, introduced in Section \ref{secLHV}, to entanglement witnesses. It was first pointed out in \cite{Terhal00} that Bell-type inequalities can be perceived as vectors in the Farkas lemma \cite{ref.Rockafellar}, discriminating between the set of correlations with a local realistic description and the quantum correlations. The Farkas vectors can in turn be related to observables, which have the interpretation of witnesses. In the particular case of the CHSH inequality, the Terhal's theory boils down to the observation that the expression $\tilde\epsilon\left(\alpha_2,\beta_1\right)+\tilde\epsilon\left(\alpha_2,\beta_2\right)+\tilde\epsilon\left(\alpha_1,\beta_1\right)-\tilde\epsilon\left(\alpha_1,\beta_2\right)$ in equation \eqref{eqCHSH} can be written in the form $\Tr\left(\mathcal{B}\rho\right)$, where 
\begin{equation}\label{eqBelloperator}
\mathcal{B}:=\vec a_1\cdot\vec\sigma\otimes\left(\vec b_1+\vec b_2\right)\cdot\vec\sigma-\vec a_2\cdot\vec\sigma\otimes\left(\vec b_2-\vec b_1\right)\cdot\vec\sigma
\end{equation} 
is the CHSH operator, first introduced in \cite{BMR92}, and $\vec a_1$, $\vec a_2$, $\vec b_1$ and $\vec b_2$ are the spin direction vectors, corresponding to the previously used detector angles $\alpha_1$, $\alpha_2$, $\beta_1$ and $\beta_2$, respectively. Using $\mathcal{B}$, one can easily construct the operator $W=2\One-\mathcal{B}$, which is a witness according to the CHSH inequality and the fact that all separable states admit a hidden variable description. Moreover, the inequality
\begin{equation}\label{eqWentnonloc}
\Tr\left(W\rho\right)=2-\Tr\left(\mathcal{B}\rho\right)<0
\end{equation}
observed for some state $\rho$, does not only indicate that $\rho$ is entangled, but also that it is nonlocal. Thus $W$ plays a double role of an entanglement \emph{and} nonlocality witness.     
\section{Beyond quantum entanglement}\label{secbeyond}
Questions beyond the separability paradigm, or even beyond the frames of quantum mechanics, have been considered in the quantum information literature since the early days of the subject. A well-known example of this is the famous  paper \cite{PR94} by Popescu and Rohrlich, where nonlocality is considered as a possible axiom for quantum mechanics. More precisely, the authors consider nonlocal theories that do obey relativistic causality. It turns out that there can exist, at least in principle, theories of this type which are \emph{not} identical to quantum mechanics. To explain this in more detail, let us briefly repeat the simplified version of the argument in \cite{PR94}, as it was presented in a later paper \cite{PR96}. 

We consider a theory of a pair of spin-$\frac{1}{2}$ particles which yields, for some reason, identical probabilities for the measurement outcomes $\uparrow\uparrow$ and $\downarrow\downarrow$, as well as identical probabilities for the outcomes $\downarrow\uparrow$, $\uparrow\downarrow$, no matter what the measurement bases in the first and the second subsystem are. Such choice precludes the possibility of supraluminal communication using the two particles. We say that there are only \textit{non-signalling correlations} (cf. e.g. \cite{BLMPPR05}) between them. Another consequence is that the respective correlation function $\epsilon$ must depend only on the relative angle $\theta$ between the first and the second measuring device. Moreover, it has to fulfill $\epsilon\left(\pi-\theta\right)=-\epsilon\left(\theta\right)$. One possible choice of such a function is \cite{PR96},
\begin{equation}\label{eqdefcorrfunction}
\epsilon\left(\theta\right)=\begin{cases}
1&\textnormal{ for }\theta\in\left[0,\frac{\pi}{4}\right]\\
2\left(1-\frac{2x}{\pi}\right)&\textnormal{ for }\theta\in\left(\frac{\pi}{4},\frac{3\pi}{4}\right)\\
-1&\textnormal{ for }\theta\in\left[\frac{3\pi}{4},\pi\right]
\end{cases}
\end{equation}
By choosing the successive angles $\alpha_1=0$, $\beta_1=\frac{\pi}{4}$, $\alpha_2=\frac{\pi}{2}$ and $\beta_2=\frac{3\pi}{4}$ in an EPR experiment of the type discussed in Section \ref{secLHV}, we get
\begin{equation}\label{eqCHSHPopescuRohrlich}
\left|\epsilon\left(\alpha_2-\beta_1\right)+\epsilon\left(\alpha_2-\beta_2\right)+\epsilon\left(\alpha_1-\beta_1\right)-\epsilon\left(\alpha_1-\beta_2\right)\right|=4
\end{equation}
as an analogue of equation \eqref{eqTsirelson}. However, this time the violation of the classical bound \eqref{eqCHSH} is bigger than possible in quantum mechanics. Thus, a theory with a correlation function of the form \eqref{eqdefcorrfunction} obeys relativistic causality, yet it is not consistent with the quantum-mechanical description of the world.

\begin{figure}\centering
\includegraphics[scale=0.5]{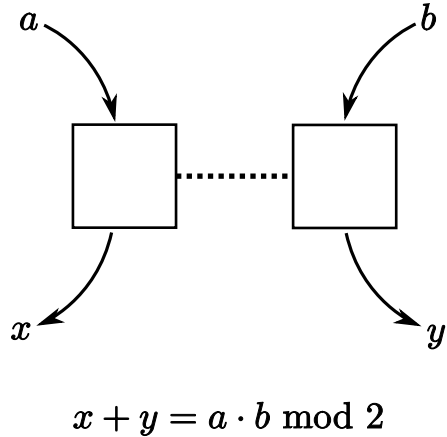}
\caption{A nonlocal box corresponding to the Popescu-Rohrlich thought experiment\label{figPRBox}}
\end{figure}

The above discussion shows that it is not possible to reproduce the laws of quantum mechanics just by using the principle of non-signalling. The Popescu-Rohrlich correlations constitute a toy model, useful for demonstrating this fact. However, after the seminal paper \cite{PR94}, a fair amount of work \cite{BLMPPR05,vD05,BGS05,BM06,BCUWW06,PHHH06,MRV07,FWW09} has been devoted to understanding the properties of Popescu-Rohrlich correlations and how they would affect communication complexity, had they been present in reality. Usually, such questions are formulated in the language of so-called \textit{nonlocal boxes}. In order to demystify this new notion, let us explain that a nonlocal box corresponding to the precise Popescu-Rohrlich setup discussed above, looks as in Figure \ref{figPRBox}. 
It is an imaginary device with two inputs $a$, $b$ and two (random) outputs $x$, $y$ that satisfy certain relation. The inputs, which take values $0$ or $1$, correspond to the measurement setups for the first and the second particle, respectively. For example, $a=0$ means that the spin of the first particle is measured in a basis rotated by $\alpha_1$. Similarly, $b=1$ indicates a measurement basis for the second particle is rotated by $\beta_2$. The outputs $x$ and $y$, on the other hand, correspond to the measurement results $\uparrow$ or $\downarrow$. For example, $y=1$ indicates that spin $\uparrow$ was measured for the second particle. A quick thought reveals that the above ``box'', called \textit{mod2NLB} in \cite{DGHMP07}, is just a more abstract way to express the properties of an imaginary EPR experiment with correlations given by the function \eqref{eqdefcorrfunction}. The only mathematical content of any such box, not necessarily related to the correlation function \eqref{eqdefcorrfunction}, is a conditional probability function $p\left(xy\,\vline\,ab\right)$ that fulfills so-called non-signalling conditions that guarantee the impossibility of supraluminal communication, cf. \cite{BLMPPR05}. Generalizations to a multipartite scenario are immediate. 

Notably, mod2NLB was postulated as a \textit{unit of nonlocality} \cite{BP05}, somewhat similar to the role played by the Bell singlet $1/\sqrt{2}\left(\left|\uparrow\downarrow\right>+\left|\downarrow\uparrow\right>\right)$ in entanglement theory. However, it was immediately realized \cite{BP05} that not all multipartite boxes can be simulated using a number of copies of mod2NLB. Moreover, in \cite{DGHMP07} the authors showed that in the bipartite scenario, there does \emph{not} exist a finite set of nonlocal boxes that could be used to simulate \emph{all} bipartite nonlocal boxes.
Interestingly, in the proof presented in \cite{DGHMP07} the Hilbert basis theorem was used, which also appears in Section \ref{secmonomialorders} of this thesis as Theorem \ref{thmHilbertbasis}.

As intellectually appealing as they are, general nonlocal boxes do not seem to have a counterpart in the real world. Still, most of the discussion by the quantum information community does stay within the framework of quantum mechanics, but not necessarily concentrates on entanglement. In particular, it was quickly recognized that there exist nonlocal phenomena in quantum mechanics which cannot be explained by the presence of entanglement. In the well-known paper \cite{BVFMRSSW99}, the authors show an example of a family of nine mutually orthogonal bipartite product states that cannot be distinguished using local measurements and classical communication by the two parties. They call this phenomenon ``nonlocality without entanglement'', hence pointing out to the difference between the two notions that tended to be taken as equivalent. However, it should be kept in mind that nonlocality in terms of the violation of Bell inequalities is very closely related, if not equivalent, to the property of being entangled. We briefly explained this in Section \ref{secLHV}, where we referred to a paper by L. Masanes, Y.-C. Liang and A. C. Doherty \cite{Mas08b}. Therefore, the notion of nonlocality in \cite{BVFMRSSW99} and in the research we describe in the rest of this section, significantly differs from what was traditionally perceived as the equivalent of being nonlocal, i.e. the violation of Bell inequalities and the lack of a local realistic description. 

In more recent days, the study of nonlocality largely revolves around its two quantitative measures, which are the \textit{quantum discord}, introduced by Żurek and Ollivier in \cite{OZ01}, and the \textit{quantum deficit}, studied by Oppenheim and the Horodecki family \cite{OHHH02}. For a recent review article on the subject, consult \cite{MBCPV11}. The basic idea behind the quantum discord is that two expressions for so-called \textit{mutual information} that are equivalent in the case of classical probability distributions, do not necessarily give the same answer when generalized to the quantum scenario. Indeed, let us define the entropy of a classical random variable $\mathcal{A}$ as
\begin{equation}\label{eqEntropy}
H\left(\mathcal{A}\right)=-\sum_ap\left(\mathcal{A}=a\right)\log p\left(\mathcal{A}=a\right)
\end{equation}
and the conditional entropy of $\mathcal{A}$ with respect to another classical variable $\mathcal{B}$ as
\begin{equation}\label{eqcondEntropy}
H\left(\mathcal{A}\,\vline\,\mathcal{B}\right)=\sum_bp\left(\mathcal{B}=b\right)H\left(\mathcal{A}\,\vline\,\mathcal{B}=b\right),
\end{equation}
where $H\left(\mathcal{A}\,\vline\,\mathcal{B}=b\right)$ is the entropy of the variable $\mathcal{A}$ conditioned on a particular value $b$ of the variable $\mathcal{B}$. We define the \textit{mutual information} of the variables $\mathcal{A}$ and $\mathcal{B}$ as
\begin{equation}\label{eqMutualInformation1}
J\left(\mathcal{A}:\mathcal{B}\right)=H\left(\mathcal{A}\right)-H\left(\mathcal{A}\,\vline\,\mathcal{B}\right)
\end{equation}
A little inspection shows that in the case of classical probability distributions, the above expression is equivalent to
\begin{equation}\label{eqMutualInformation2}
I\left(\mathcal{A}:\mathcal{B}\right)=H\left(\mathcal{A}\right)+H\left(\mathcal{B}\right)-H\left(\mathcal{A},\mathcal{B}\right),
\end{equation}
where $H\left(\mathcal{A},\mathcal{B}\right)$ stays for the entropy of the collective variable $\left(\mathcal{A},\mathcal{B}\right)$. Thus we have $I\left(\mathcal{A}:\mathcal{B}\right)=J\left(\mathcal{A}:\mathcal{B}\right)$ for arbitrary classical variables $\mathcal{A}$ and $\mathcal{B}$. However, as pointed out in \cite{OZ01}, the equality between the two expressions for mutual information does not generally hold in a quantum world. 

To show this, let us consider a bipartite quantum system described by a density matrix $\rho_{AB}$. The states of the subsystems are given by the partial traces of $\rho_{AB}$, $\rho_A=\Tr_B\rho_{AB}$ and $\rho_B=\Tr_A\rho_{AB}$. We immediately see that a quantum analogue of \eqref{eqMutualInformation2} is 
\begin{equation}\label{eqMutualInformation3}
I\left(\rho_{AB}\right)=H\left(\rho_{A}\right)+H\left(\rho_{B}\right)-H\left(\rho_{AB}\right)
\end{equation}
where $H\left(\rho\right):=-\Tr\left(\rho\log\rho\right)$. However, it is not obvious how to generalize $J\left(\mathcal{A}:\mathcal{B}\right)$ to the quantum case. The reason behind this is that the quantum subsystem $B$ can be measured in various bases, and one of them has to be selected before a sum similar to the $\sum_b$ in formula \eqref{eqcondEntropy} is calculated. Thus we have a whole family of conditional entropies $H\left(\rho_{AB}\,\vline\,\left\{\Pi_b\right\}\right)$, where $\left\{\Pi_b\right\}$ is an arbitrary complete set of one-dimensional projections on the subsystem $B$, satisfying $\sum_b\Pi_b=\One$. Explicitly, $H\left(\rho_{AB}\,\vline\,\left\{\Pi_b\right\}\right)$ is given by 
\begin{equation}\label{eqConditionalQuantum}
H\left(\rho_{AB}\,\vline\,\left\{\Pi_b\right\}\right)=\sum_bp_bH\left(\frac{\left(\One\otimes\Pi_b\right)\rho_{AB}\left(\One\otimes\Pi_b\right)}{p_b}\right)
\end{equation}
where $p_b=\Tr\left(\left(\One\otimes\Pi_b\right)\rho_{AB}\right)$ is the probability to obtain a result $b$ in a measurement corresponding to $\left\{\Pi_b\right\}$. Simply because the $H\left(\rho_{AB}\,\vline\,\left\{\Pi_b\right\}\right)$ are not all equal, there is no single quantum analogue of $J\left(\mathcal{A}:\mathcal{B}\right)$. Instead, we have a family of mutual information analogues, given by
\begin{equation}\label{eqMutualInformation4}
J\left(\rho_{AB}\,\vline\,\left\{\Pi_b\right\}\right)=H\left(\rho_A\right)-H\left(\rho_{AB}\,\vline\,\left\{\Pi_b\right\}\right)
\end{equation}
The supremum
\begin{equation}\label{eqClassicalCorrelations}
C_B\left(\rho_{AB}\right)=\sup_{\left\{\Pi_b\right\}}J\left(\rho_{AB}\,\vline\,\left\{\Pi_b\right\}\right)
\end{equation}
can be considered as a measure of classical correlations \cite{OZ01,HV01}. Note that there also exists a related quantity $C_A\left(\rho_{AB}\right)$ where the roles of $A$ and $B$ have been interchanged. The \textit{quantum discord} is now defined as the difference between $I\left(\rho_{AB}\right)$ and $C_B\left(\rho_{AB}\right)$,
\begin{equation}\label{eqDiscord1}
D_B\left(\rho_{AB}\right)=I\left(\rho_{AB}\right)-C_B\left(\rho_{AB}\right)
\end{equation}
Alternatively, the name ``discord'' may refer to
\begin{equation}\label{eqDiscord2}
D_A\left(\rho_{AB}\right)=I\left(\rho_{AB}\right)-C_A\left(\rho_{AB}\right)
\end{equation}
although the two quantities $D_A$ and $D_B$ do not generally coincide.

Due to the equality $I\left(\mathcal{A}:\mathcal{B}\right)=J\left(\mathcal{A}:\mathcal{B}\right)$ valid in the classical world, the non-vanishing of the discord for $\rho_{AB}$ is a sign of quantumness of the state. Unlike separability, the vanishing of the discord only occurs for a measure zero subset of the set of all states \cite{FACCA10}. In particular, $D_A$ and $D_B$ vanish simultaneously if and only if $\rho_{AB}$ has an eigenbasis consisting of product vectors, i.e.
\begin{equation}\label{eqClassicalState}
\rho_{AB}=\sum_{i,j}\lambda_{i,j}\proj{\phi_i}\otimes\proj{\psi_j}
\end{equation} 
where $\lambda_{ij}\geqslant 0$, while $\phi_i$ and $\psi_j$ constitute bases for the first and the second subsystem, respectively. Such states are called \textit{classically correlated} \cite{HHHOSSSR05}. They also play an important role in the alternative framework for correlation studies, developed by Oppenheim and the Horodecki family \cite{OHHH02,HHHOSSSR05}.

It is in general not easy to evaluate the quantum discord, but some results have been obtained e.g. for $2\times 2$ systems \cite{L08,ARA10}. Several conditions for zero and non-zero quantum discord are known as well \cite{FACCA10,BC10,DVB10}, and a missing operational interpretation of the quantity has been provided in \cite{CABMPW11} in terms of a quantum state merging protocol.  

Quantum deficit, on the other hand, has had a relatively clear physical interpretation from the very beginning when it was introduced in \cite{OHHH02}. The quantity is believed to be equal to the amount of work which can be extracted from a multipartite quantum state $\rho$ globally, minus the amount of work the parties can draw locally, possibly after transforming the state by an allowed family of transformations. This description may seem a little vague, but on the mathematical side, the discussion can easily be made more rigorous. For a quantum state $\rho$ in a $d$-dimensional space, we define
\begin{equation}\label{defI}
I\left(\rho\right)=\log_2 d-H\left(\rho\right)
\end{equation} 
as the information contained in $\rho$. For the allowed family of transformations, we take so-called \textit{closed local operations and classical communication} family, CLOCC for short \cite{HHHOSSSR05}. They can be decomposed into two basic types of operations
\begin{enumerate}[i)]
\item Local unitary transformations
\item Sending subsystems down a completely dephasing channel (i.e. a channel that destroys all non-diagonal elements of the transformed density matrix in some basis)
\end{enumerate}
Let us denote this family by $\mathcal{CL}$. In the bipartite scenario, the quantum deficit of a quantum state $\rho_{AB}$ is defined as
\begin{equation}\label{defDeficit}
\Delta\left(\rho_{AB}\right)=I\left(\rho_{AB}\right)-\sup_{\Phi\in\mathcal{CL}}\left(I\left({\Tr}_A\left(\Phi\left(\rho_{AB}\right)\right)\right)+I\left({\Tr}_B\left(\Phi\left(\rho_{AB}\right)\right)\right)\right)
\end{equation}
or equivalently
\begin{equation}\label{defDeficit2}
\Delta\left(\rho_{AB}\right)=\inf_{\Phi\in\mathcal{CL}}\left(H\left({\Tr}_A\left(\Phi\left(\rho_{AB}\right)\right)\right)+H\left({\Tr}_B\left(\Phi\left(\rho_{AB}\right)\right)\right)\right)-H\left(\rho_{AB}\right)
\end{equation}
Generalizations to multipartite cases are immediate. Similarly to the discord, the deficit vanishes for classically correlated states, i.e. states of the form \eqref{eqClassicalState}. Moreover, as explained in \cite{HHHOSSSR05}, reversible CLOCC transforms of classically correlated states play an important role in evaluation of $\Delta$ for a given state $\rho$.

On the physics side, the theoretical possibility to draw a maximal amount $kT\cdot I\left(\rho\right)$ of work from a heat bath in temperature $T$ using a state $\rho$ is a widely believed conjecture. It has been partly confirmed by papers like \cite{AHHH04} and \cite{LT11}. Hence, it seems plausible that the quantum deficit really has the physical interpretation we mentioned earlier, but one should remain cautions. The mathematical structure of the quantity, however, remains intact in either case.

Before we close this chapter, we should definitely mention that the principle of non-signalling, which appeared in the discussion by Popescu and Rohrlich, can be replaced by so-called \textit{information causality} principle, which is stronger than no-signalling and precludes correlations that are not allowed by quantum mechanics \cite{PPKSWZ09}. Hence, information causality may possibly be considered as an axiom for quantum theory \cite{PPKSWZ09, ABPS09}, unlike the non-signalling principle \cite{PR94}. However, this topic goes beyond the scope of this thesis. 

\chapter{Practical applications}\label{chpractical}

\section{Quantum cryptography}\label{seccrypto}
The idea of quantum cryptography or \textbf{quantum key distribution}, first put forward in the famous 1984 paper \cite{BB84} by Bennett and Brassard, has its origins in an early work by S. Wiesner \cite{Wiesner}. The main observation behind it was that two photon polarization bases, say $R$ and $D$ for rectilinear and diagonal, can be selected in such a way that photons fully polarized with respect to one of them give totally random results when measured in the other basis, and vice versa. Equally important was the fact that quantum measurements affect the measured systems in general. Bennett and Brassard used these quantum-mechanical features to construct a protocol, now called BB84, which allows two parties that do not initially share any secrets, to generate a random string of bits that is known to both of them, but not to anyone else. Such bits can subsequently be used as a shared secret key for perfectly secure classical data transmission. Let us call the two parties $A$ and $B$, or Alice and Bob. The protocol designed by Bennett and Brassard consists in the following steps:
\begin{enumerate}[1.]
\item Alice and Bob agree on two polarization bases, say $R$ and $D$, which are rotated by $45^{\circ}$ with respect to each other. Let us denote the corresponding pure polarization photon states by $\left|\leftrightarrow\right>$, $\left|\updownarrow\right>$ for the $R$ basis and $\left|\neswarrow\right>=1/\sqrt{2}\left(\left|\leftrightarrow\right>+\left|\updownarrow\right>\right)$, $\left|\nwsearrow\right>=1/\sqrt{2}\left(\left|\leftrightarrow\right>-\left|\updownarrow\right>\right)$ for the $D$ basis.
\item Alice generates random sequences of bits, $\left\{a_i\right\}_{i=1}^n$ and $\left\{b_j\right\}_{j=1}^n$, using a classical random number generator.
\item Bob generates a random sequence of bits $\left\{c_k\right\}_{k=1}^n$, also using a classical generator.
\item Alice then begins to send photons to Bob. The polarization state of the $i$-th photon is chosen according to the values of the random bits $a_i$ and $b_i$. The bit $a_i$ determines which polarization basis is used, with  $a_i=0$ standing for the $R$ and $a_i=1$ for the $D$ basis. The bit $b_i$ determines whether the first or the second pure polarization state with respect to the given basis is chosen. Table \ref{tabAbits} summarizes on Alice's choice of photon, depending on $\left(a_i,b_i\right)$.
\begin{table}[t]\centering
\begin{tabular}{|c|c|c|c|c|}
\hline
Random bits&$\left(0,0\right)$&$\left(0,1\right)$&$\left(1,0\right)$&$\left(1,1\right)$\\
\hline
Photon sent&$\left|\leftrightarrow\right>$&$\left|\updownarrow\right>$&$\left|\neswarrow\right>$&$\left|\nwsearrow\right>$\\
\hline
\end{tabular}
\caption{Photon polarization states choices corresponding to Alice's random bits $\left(a_i,b_i\right)$.\label{tabAbits}}
\end{table}
\item Bob measures the received $i$-th photon in the $R$ or $D$ basis, depending on the value of $c_i$. When $c_i=0$, Bob uses $R$. Otherwise, he uses $D$. The first vector in the selected basis ($\left|\leftrightarrow\right>$ or $\left|\neswarrow\right>$) is assigned the measurement result $0$, while the remaining vector ($\left|\updownarrow\right>$ or $\left|\nwsearrow\right>$) is assigned $1$. If Bob happens to choose the same basis as Alice did (i.e. $a_i=c_i$), his measurement result exactly matches $b_i$, assuming the photon transmission was not disrupted nor interfered with by an eavesdropper.
\item After measuring all the $n$ photons, Bob publicly discloses the bits $c_i$, and Alice does the same with $a_i$. Thus done, they know which measurement bases they used for individual photons and can single out the cases where their basis choices were identical. On average, they would have chosen the same basis in $n/2$ cases.
\item As their secret key, Alice and Bob choose the bits $b_i$ for which $a_i=c_i$. They both know these bits, as a result of using identical measurement bases.
\end{enumerate}
The power of the above protocol comes from the fact that any interference by an eavesdropper would very likely have been detected by Alice and Bob, provided that they perform an additional correctness check before they agree on the key. The required additional procedure can be summarized as follows:
\begin{enumerate}[1.']\setcounter{enumi}{6}
\item After performing Step 6., Alice and Bob select a random subset of the indices $i$ for which $a_i=c_i$. Assume the selected indices are $\left\{i_k\right\}_{k=1}^m$. Alice publicly discloses the bits $\left\{b_{i_k}\right\}_{k=1}^m$, and Bob discloses the corresponding measurement results he obtained. If both match, the transmission is assumed to be perfect and the remaining bits for which $a_i=c_i$ are used as a secret key. Otherwise, it is assumed that someone was eavesdropping, and the results of the whole secret key generation procedure are discarded. 
\end{enumerate}
An exemplary run of the procedure consisting of steps 1.-7., with 7.' included, is presented in Table \ref{tabABrun}. Note that in real life applications, it is impossible to avoid transmission errors, even if there is no one eavesdropping. Hence, a general strategy has to be developed to deal with transmission/eavesdropping errors, a strategy that would allow to produce a secret key, even if the transmission does not work perfectly. Suitable tools, borrowed from classical coding theory, were discovered some years after the advent of BB84 \cite{BBBSS92}. They are very generally described as \textbf{information reconciliation} and \textbf{privacy amplification}. For more details, cf. \cite{BBBSS92}. 

\begin{table}\centering
\begin{tabular}{|>{$}c<{$}|>{$}c<{$}|>{$}c<{$}|>{$}c<{$}|>{$}c<{$}|>{$}c<{$}|>{$}c<{$}|>{$}c<{$}|>{$}c<{$}|>{$}c<{$}|}
\hline
\left\{a_i\right\}&1&0&1&1&0&1&0&1&0\\
\left\{b_i\right\}&0&1&1&0&1&0&1&1&0\\
\left\{c_i\right\}&1&1&1&0&0&0&1&0&1\\
\textnormal{Alice's choice of basis}&D&R&D&D&R&D&R&D&R\\
\textnormal{Alice's photon state}&\neswarrow&\updownarrow&\nwsearrow&\neswarrow&\updownarrow&\neswarrow&\updownarrow&\nwsearrow&\leftrightarrow\\
\textnormal{Bob's choice of basis}&D&D&D&R&R&R&D&R&D\\
\textnormal{Bob's result}&0&\ast&1&\ast&1&\ast&\ast&\ast&\ast\\
\textnormal{The same basis?}&Y&N&Y&N&Y&N&N&N&N\\
\textnormal{Randomly selected bits}&&&&&1&&&&\\
\textnormal{Do they match?}&&&&&Y&&&&\\
\textnormal{Secure key}&0&&1&&&&&&\\
\hline
\end{tabular}
\caption{An exemplary run of the BB84 protocol\label{tabABrun}. The symbol $\ast$ denotes the fact that either $0$ or $1$ could have been obtained. The letters $Y$ and $N$ stand for ``Yes'' and ``No''.}
\end{table}

We need to point out that in the above procedures, no use of entanglement was made. However, in the early nineties, A. Ekert proposed the first entanglement-based quantum key distribution protocol, known as E91 \cite{E91}. Although the general idea behind E91 is the same as for BB84, there are several key differences: 
\begin{enumerate}[1)]
\item Instead of leaving the photon state preparation to Alice, both parties are assigned the identical task of measuring a subsystem in a two-partite maximally entangled photon state $\left(\left|00\right>+\left|11\right>\right)/\sqrt{2}$. The state is assumed to be externally given. Alice measures the first and Bob the second subsystem.
\item Three instead of two photon polarization bases are used at random by Alice and Bob. In case of Alice, the polarizer angles $\phi_1^A=0^{\circ}$, $\phi_2^A=45^{\circ}$ and $\phi_3^A=90^{\circ}$ are used. For Bob, it is $\phi_1^B=45^{\circ}$, $\phi_2^B=90^{\circ}$ and $\phi_3^B=135^{\circ}$.
\item Bob and Alice publicly disclose which bases they used in which measurement round. Then, they reveal the measurement results for which \emph{different} measurement setups were used. This permits them to calculate the CHSH quantity
\begin{equation}\label{eqCHSHquantity}
E\left(\phi^A_3,\phi^B_3\right)+E\left(\phi^A_3,\phi^B_1\right)+E\left(\phi^A_1,\phi^B_3\right)-E\left(\phi^A_1,\phi^B_1\right),
\end{equation}
where $E\left(\phi,\psi\right)$ is the correlation coefficient between the measurement results for Alice and Bob when their polarizer angles are $\phi$ and $\psi$, respectively. As in the example discussed in Section \ref{secLHV}, the value of the function \eqref{eqCHSHquantity} for a truly maximally entangled source state is $2\sqrt{2}$. By testing whether the equality between $2\sqrt{2}$ and \eqref{eqCHSHquantity} really occurs, Bob and Alice make sure that no eavesdropping takes place, nor that the source is corrupted. 
\item If there is (an approximate) equality between \eqref{eqCHSHquantity} and its theoretical value, the results which Bob and Alice obtained when they measured \emph{in the same bases}, should be perfectly correlated. They were not publicly disclosed so far, so they can be used as a secret key.
\end{enumerate}
Shortly after Ekert published his paper, Bennett, Brassard and Mermin \cite{BBM92} suggested another entanglement-based protocol, now called BBM92, which is basically a version of BB84 that exploits the properties of entangled quantum states. Thus, the difference from BB84 described by item $1)$ above still exists, but the other ones do not.

It is natural to ask how the above two-qubit key distribution methods generalize to higher dimensional quantum systems. The question was addressed by the authors of the paper \cite{CBKG2001}, who used so-called \textit{mutually unbiased bases} (MUBs) as a higher dimensional analogue of the pair of bases $\left\{\left|\leftrightarrow\right>,\left|\updownarrow\right>\right\}$ and $\left\{\left|\neswarrow\right>,\left|\nwsearrow\right>\right\}$. Let us explain that two orthonormal bases $\left\{\phi_i\right\}_{i=1}^d$ and $\left\{\psi_j\right\}_{j=1}^d$ of $\setC^d$ are called \textit{unbiased} if and only if the following equality 
\begin{equation}\label{eqMUBspre}
\left|\innerpr{\phi_i}{\psi_j}\right|^2=\frac{1}{d}
\end{equation}
 holds for all $i$ and $j$. The unbiasedness condition guarantees the desirable property that an element of one of the bases gives fully random results when measured in the other basis. 
 
There can exist at most $d+1$ mutually unbiased bases in $\setC^d$ \cite{WF89}. We shall discuss some of their further aspects in Section \ref{secMUBs}. Either a pair of them, or more can be used to design quantum key distribution protocols based on $d$-dimensional quantum systems \cite{CBKG2001}. These protocols do not differ significantly from the qubit ones. Let us also remark that in the qubit setting, there are three MUBs available, so that there exists an alternative to BB84 that uses six quantum states instead of four. This possibility was first studied in a paper by Bruss \cite{Bruss98}.

\section{Quantum teleportation and dense coding}\label{sectele}
As our next example of how the laws of quantum mechanics can be used for practical purposes, we shall discuss the two interconnected concepts of \textbf{dense coding} \cite{BW92} and \textbf{quantum state teleportation} \cite{BBCJPW93}.

In its most basic form, dense coding permits two parties, say Alice and Bob, to exchange \emph{two classical bits} of information by just transmitting \emph{one qubit}. The fundamental trick behind this feature is the use of one-sided Pauli transformations, acting on a maximally entangled state. We have
\begin{align}\label{eqfourPaulistates}
&\left(\One\otimes\One\right)\left|\Phi_+\right>=\left|\Phi_+\right>,&\left(\sigma_x\otimes\One\right)\left|\Phi_+\right>=\left|\Psi_+\right>,\\
&\left(\sigma_y\otimes\One\right)\left|\Phi_+\right>=-i\left|\Psi_-\right>,&\left(\sigma_z\otimes\One\right)\left|\Phi_+\right>=\left|\Phi_-\right>,\nonumber
\end{align}
so that the four states resulting from one-sided Pauli action on $\left|\Phi_+\right>$ are perfectly distinguishable. Hence, they can carry two bits of classical information. In the dense coding scheme proposed in \cite{BW92}, Alice and Bob initially share a maximally entangled state $\left|\Phi_+\right>$ of a two-partite system, and each of them has access to only one of the subsystems. Alice then performs one of the four Pauli transformations on her subsystem, and sends the subsystem to Bob. After this step, Bob is in possession of one of the two-partite maximally entangled states from the list \eqref{eqfourPaulistates}. Because these states can be perfectly distinguished by a quantum measurement, Bob can in principle tell which of the four Pauli operations Alice used. Consequently, two bits of classical information have been transmitted, even though only one qubit was exchanged between Alice and Bob.

The aim of quantum state teleportation is, on the other hand, to \textit{transmit an unknown quantum state} $\left|\psi\right>$ between the two parties. In the basic qubit teleportation model \cite{BBCJPW93}, the required resources are a maximally entangled state, i.e. $\left|\Psi_-\right>=\left(\left|01\right>-\left|10\right>\right)/\sqrt{2}$, which is shared between Alice an Bob, and the state to be teleported, initially held by Alice. Altogether, they have a tripartite system, initially in the state $\left|\psi\right>\left|\Psi_-\right>$. The first two subsystems are controlled by Alice, and the third one by Bob. In order to teleport $\left|\psi\right>$ to Bob, Alice performs a measurement on the first two qubits, using the measurement basis $\left\{\left|\Phi_+\right>,\left|\Phi_-\right>,\left|\Psi_+\right>,\left|\Psi_-\right>\right\}$. She then communicates the result to Bob. Provided this information, Bob can recover $\left|\psi\right>$ by performing a suitable unitary rotation on his subsystem. To see that this is actually the case, it suffices to notice the following identity
\begin{equation}
\left|\psi\right>\left|\Psi_-\right>=\frac{1}{2}\left(-\left|\Psi_-\right>\left|\psi\right>-\left|\Psi_+\right>\sigma_z\left|\psi\right>+\left|\Phi_-\right>\sigma_x\left|\psi\right>-i\left|\Phi_+\right>\sigma_y\left|\psi\right>\right)
\end{equation}
After the Alice's measurement on the first two qubits, Bob's subsystem is in one of the states $-\left|\psi\right>$, $-\sigma_z\left|\psi\right>$, $\sigma_x\left|\psi\right>$, $-i\sigma_y\left|\psi\right>$. Moreover, Alice can perfectly differentiate between these four cases, as she knows which of the states $\left|\Psi_-\right>$, $\left|\Psi_+\right>$, $\left|\Phi_-\right>$ and $\left|\Phi_+\right>$ she got in her measurement. If she is so kind to share this knowledge with Bob, he can then recover the state $\left|\psi\right>$ by simply undoing the suitable rotation $\sigma_x$, $\sigma_y$ or $\sigma_z$, if his state is not already a multiple of $\left|\psi\right>$.

Naturally, the above dense coding and teleportation schemes for qubits are expected to have generalizations to higher dimensional systems. Such generalizations do indeed exist and for the so-called \textit{tight} type, they have been completely characterized by Werner \cite{ref.Werner01}. Moreover, he showed that there is a one-to-one correspondence between tight dense coding and tight teleportation schemes. In order to fully understand his result, we first need to explain what a general dense coding and teleportation scheme is.
\begin{definition}Let $\mathcal{X}$ be a set of $d^2$ elements.
A tight \textbf{quantum teleportation scheme} consists of
\begin{itemize}
\item A density operator $\omega$ on $\setC^d\otimes\setC^d$
\item A collection of completely positive and trace preserving maps $T_x$, $x\in\mathcal{X}$, acting on operators on $\setC^d$
\item A collection of observables $F_x$ on $\setC^d\otimes\setC^d$, $x\in\mathcal{X}$, such that for all density operators $\rho$ on $\setC^d$ and all operators $A$ on $\setC^d$, the following equality holds
\begin{equation}\label{eqTeleport}
\sum_{x\in\mathcal{X}}\Tr\left(\left(\rho\otimes\omega\right)\left(F_x\otimes T_x\left(A\right)\right)\right)=\Tr\left(\rho A\right)
\end{equation}
\end{itemize}
\end{definition}
\begin{definition}Let $\mathcal{X}$ be a set of $d^2$ elements. 
A tight \textbf{dense coding scheme} consists of the same elements as a tight quantum teleportation scheme, however the condition \eqref{eqTeleport} is replaced by
\begin{equation}\label{eqDenseCoding}
\Tr\left(\omega\left(T_x\otimes\id\right)\left(F_y\right)\right)=\delta_{xy}
\end{equation}for all $x,y\in\mathcal{X}$
\end{definition}  
Note that in the above mentioned example of a dense coding scheme for qubits, we had  $\left\{F_x\right\}_{x\in\mathcal{X}}=\left\{\proj{\Phi_+},\proj{\Psi_+},\proj{\Psi_-},\proj{\Phi_-}\right\}$. We  used the maximally entangled state $\omega=\proj{\Phi_+}$ and the transformations $\left\{T_x\right\}_{x\in\mathcal{X}}=\left\{\id,\Ad_{\sigma_x},\Ad_{\sigma_y},\Ad_{\sigma_z}\right\}$, where $\Ad_{\sigma_x}:\rho\mapsto\sigma_x^{\ast}\rho\sigma_x$, and similarly for $\sigma_y$ and $\sigma_z$. In the qubit teleportation scheme, on the other hand, we had $\left\{F_x\right\}_{x\in\mathcal{X}}=\left\{\proj{\Psi_-},\proj{\Psi_+},\proj{\Phi_-},\proj{\Phi_+}\right\}$, $\omega=\proj{\Psi_-}$, as well as $\left\{T_x\right\}_{x\in\mathcal{X}}=\left\{\id,\Ad_{\sigma_z},\Ad_{\sigma_x},\Ad_{\sigma_y}\right\}$

Werner proves the following general result \cite{ref.Werner01}.
\begin{theorem}\label{thmWernertele}
All tight teleportation or dense coding schemes in $\setC^d$ are obtained by choosing $\omega=\proj{\Omega}$ for a maximally entangled state $\left|\Omega\right>\in\setC^d\otimes\setC^d$, $F_x=\proj{\Phi_x}$ for an orthonormal basis of maximally entangled states $\left\{\left|\Phi_x\right>\right\}_{x\in\mathcal{X}}\subset\setC^d\otimes\setC^d$ and $T_x=\Ad_{U_x}$, where $U_x$ is chosen such that $\left|\Phi_x\right>=\left(U_x\otimes\id\right)\left|\Omega\right>$.
\end{theorem}
In Particular, Theorem \ref{thmWernertele} applies that there is a one-to-one correspondence between tight teleportation and dense coding schemes. Every such scheme needs a basis of maximally entangled states. Let us remark that Werner proposed a construction of such bases, based on Latin squares and complex Hadamard matrices, which also appear in the context of mutually unbiased bases, to be discussed in more detail in Section \ref{secMUBs}.

\section{Quantum metrology}\label{secmetrology}
\begin{figure}[h!]\centering
\includegraphics[scale=0.75]{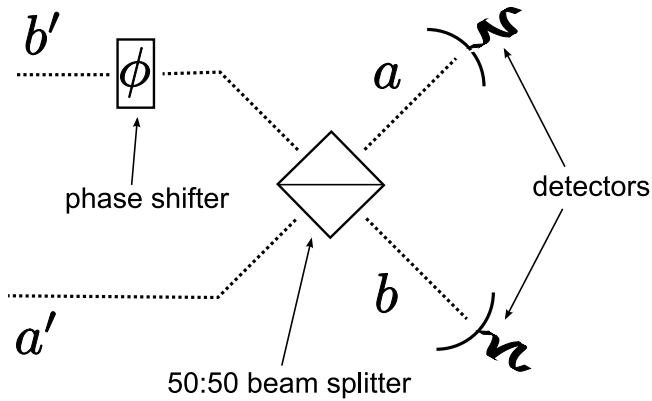}
\caption{A simple experimental setup for photon interferometry. The aim of the experiment is to estimate the phase $\phi$ using an appropriate input state and measurement}\label{figinterfere}
\end{figure}

In the last section concerning practical applications of quantum entanglement, we shall give an example of how entanglement can be used to increase phase sensitivity in a photon interferometry experiment. Our discussion is based on the paper \cite{GB02} by Gerry and Benmoussa, but we make a few remarks about related work by other authors. The very simple experimental setup we would like to discuss is depicted in Figure \ref{figinterfere}. It consists of two photodetectors, a beam splitter, and a phase shifter. Together, they make up a simple interferometer. An important part of the experiment is also the photonic quantum state which is fed into the arms of the interferometer, as well as the observable one calculates using the measurement results from the photodetectors. The aim is to estimate the phase $\phi$, induced by the phase shifter on single photons. Such phase may result e.g. from propagation through a thin layer of a medium that has an index of refraction greater than the environment. In the following, we argue that the estimation of $\phi$ can be made more precise if one does exploit entanglement between $N$ photons impinging on the beam splitter, instead of just repeating single-photon measurements $N$ times. 

We shall use the quantum-mechanical description of the optical experiment in Figure \ref{figinterfere}, the basics for which can be found in the textbook \cite[Chapter 6.]{GerryKnight}. In this formalism, the quantum state of the photons leaving the beam splitter is described as an element of a two-particle Fock space, with creation/annihilation operators $a/\conj{a}$ and $b/\conj{b}$ corresponding to the upper and the lower output arm of the interferometer, respectively. It should lead to no confusion if we call the upper and the lower arm itself $a$ and $b$ for convenience (cf. Figure \ref{figinterfere}). The corresponding creation/annihilation operators satisfy the commutation relations
\begin{equation}\label{eqcommutationrelations}
\left[a,\conj{a}\right]=\left[b,\conj{b}\right]=\One\quad\left[a,b\right]=\left[\conj{a},b\right]=\left[a,\conj{b}\right]=\left[\conj{a},\conj{b}\right]=0
\end{equation}
The vacuum state $\left|0,0\right>$ corresponds to no photons in arms $a$ and $b$, and it satisfies $a\left|0,0\right>=b\left|0,0\right>=0$. We assume that $\left|0,0\right>$ is normalized. Photon number states are subsequently defined as 
\begin{equation}\label{eqnumberstates}
\left|n,m\right>=\frac{\left(\conj{a}\right)^n\left(\conj{b}\right)^m}{\sqrt{n!\,m!}}\left|0,0\right>
\end{equation}
They have the clear interpretation of states with $n$ photons in arm $a$ and $m$ photons in arm $b$ of the interferometer. An analogous construction works for the upper and lower input arm of the interferometer, which we call $b'$ and $a'$, the same as the corresponding annihilation operators. Note that the upper arm is denoted with $b'$ and \emph{not} with $a'$, the same as in Figure \ref{figinterfere}. The corresponding photon number states are denoted with $\left|n,m\right>'$.

In accordance with \cite{GerryKnight}, if we have an input state $\left|\Phi\right>=f\left(\conj{a'},\conj{b'}\right)\left|0,0\right>'$ for some function $f$ of the creation operators $\conj{a'}$ and $\conj{b'}$, then the output state of the interferometer equals
\begin{equation}\label{eqoutput}
U_{BS}U\left(\phi\right)f\left(\conj{a},\conj{b}\right)\left|0,0\right>,
\end{equation} 
where $U_{BS}=\exp\left({i\pi\left(\conj{a}b+a\conj{b}\right)/4}\right)$ and $U\left(\phi\right)=\exp\left({i\phi\conj{b}b}\right)$. Note that we use the same function $f$, but we evaluate it for the creation operators $\conj{a}$ and $\conj{b}$, not for $\conj{a'}$, $\conj{b'}$. By a slight abuse of notation, we can therefore write \eqref{eqoutput} as $U_{BS}U\left(\phi\right)\left|\Phi\right>$ and consider the interferometer as a unitary transformation on the input state, which yields an output in the output Fock space. Let us denote $\left|\Psi\left(\phi\right)\right>:=U_{BS}U\left(\phi\right)\left|\Phi\right>$ The estimation of $\phi$ boils down to the calculation of the expectation value of an appropriately chosen observable $O$ on $\left|\Psi\left(\phi\right)\right>$, from which we recover $\phi$, i.e. we measure $\left<O\right>\left(\phi\right):=\left<\Psi\left(\phi\right)\right|O\left|\Psi\left(\phi\right)\right>$ and equate it to the theoretically predicted value of $\left<\Psi\left(\tilde\phi\right)\right|O\left|\Psi\left(\tilde\phi\right)\right>$ for some $\tilde\phi$. The number $\tilde\phi$ gives us an estimate of $\phi$. A widely applied formula for error propagation then provides us with an estimate of the error of $\tilde\phi$,
\begin{equation}\label{eqerrorphi}
\Delta\tilde\phi=\frac{\Delta O\left(\tilde\phi\right)}{\left|\frac{d\left<O\right>\left(\phi\right)}{d\phi}\left(\tilde\phi\right)\right|},
\end{equation}
where $\Delta O=\sqrt{\left<O^2\right>-\left<O\right>^2}$ is the standard deviation of $O$. As shown in \cite{GB02}, the choice
\begin{equation}\label{eqNOON}
\left|\Phi\right>=\frac{1}{\sqrt{2}}\left(\left|N,0\right>+\left|0,N\right>\right)
\end{equation}
allows for a significant improvement in the precision of the measurement of $\phi$ over a scenario where single-photon states of the type \eqref{eqNOON} are measured $N$ times. The states \eqref{eqNOON} are called \textit{NOON states} \cite{LKCD02} for obvious reasons. It is not easy to create them \cite{KLD02}, but significant progress has been made in that area in recent years, cf. e.g. \cite{AAS10}. In the following, we will briefly explain how the result of \cite{GB02} was obtained.

We already know which state $\left|\Phi\right>$ to use, but we have not yet specified the operator $O$ to measure. A suitable choice was suggested in \cite{BIWH96}, and it is 
\begin{equation}\label{eqchoiceO}
O=\exp\left(i\pi\conj{b}b\right)
\end{equation}
Note that $\conj{b}b$ is simply the photon number operator for the lower output arm, so the expectation value of $O$ can be estimated from experiment by measuring the number $n_b$ of clicks in the lower detector and calculating $\exp\left(i\pi n_b\right)=\left(-1\right)^{n_b}$. Of course, the experiment has to be repeated many times to get a reliable estimate, equal to the average of the expressions $\left(-1\right)^{n_b}$ over individual runs. Note that we assume that photodetectors are perfectly efficient, i.e. no photons are lost.

Once we know $\left|\Phi\right>$ and $O$, it is not very difficult to calculate $\left<\Psi\left(\phi\right)\right|O\left|\Psi\left(\phi\right)\right>$. In order to simplify the calculation, one can introduce
\begin{equation}\label{eqSchwinger}
J_0=\frac{\conj{a}a+\conj{b}b}{2},\quad J_1=\frac{\conj{a}b+a\conj{b}}{2},\quad J_2=\frac{\conj{a}b-a\conj{b}}{2i},\quad J_3=\frac{\conj{a}a-\conj{b}b}{2}
\end{equation} 
The operators $J_i$ with $i=1,2,3$ were introduced by Schwinger \cite{Schwinger} and they satisfy the angular momentum commutation relations, $\left[J_k,J_l\right]=i\sum_m\varepsilon_{klm}J_m$. The operator $J_0$ commutes with all of them and has the interpretation of the total photon number observable (divided by two).

From the very useful Hadamard lemma (cf. e.g. \cite{Miller})
\begin{equation}\label{eqexpcommutator}
e^XYe^{-X}=e^{\left[X,\cdot\right]}Y=Y+\left[X,Y\right]+\frac{1}{2!}\left[X,\left[X,Y\right]\right]+\ldots
\end{equation}
and the commutation relations \eqref{eqcommutationrelations}, one quickly obtains the following equalities
\begin{equation}\label{eqJtransforms}
e^{i\pi J_2}\conj{a}e^{-i\pi J_2}=-\conj{b},\quad e^{i\pi J_2}\conj{b}e^{i\pi J_2}=\conj{a},
\end{equation}
which give us
\begin{multline}\label{eqnmtransformed}
e^{i\pi J_2}\left|n,m\right>=e^{i\pi J_2}\frac{\left(\conj{a}\right)^n\left(\conj{b}\right)^m}{\sqrt{n!\,m!}}\left|0,0\right>=\\
=\frac{e^{i\pi J_2}\left(\conj{a}\right)^ne^{-i\pi J_2}e^{i\pi J_2}\left(\conj{b}\right)^me^{-i\pi J_2}}{\sqrt{n!\,m!}}e^{i\pi J_2}\left|0,0\right>=\\=\frac{\left(-\conj{b}\right)^n\left(\conj{a}\right)^m}{\sqrt{n!\,m!}}\left|0,0\right>=\left(-1\right)^n\left|m,n\right>,
\end{multline}
where we also used the equality $e^{i\pi J_2}\left|0,0\right>=\left|0,0\right>$. Another relation which follows from \eqref{eqexpcommutator} is
\begin{equation}\label{eqJtransforms2}
e^{-i\frac{\pi}{2}J_1}J_3e^{i\frac{\pi}{2}J_1}=J_2
\end{equation}
With \eqref{eqnmtransformed} and \eqref{eqJtransforms2} at hand, we can easily calculate $\left<\Psi\left(\phi\right)\right|O\left|\Psi\left(\phi\right)\right>$. Indeed, since $\conj{b}b=J_0-J_3$ and $J_0$ commutes with all $J_i$, we get
\begin{multline}\label{eqexpectedvalue1}
\left<\Psi\left(\phi\right)\right|O\left|\Psi\left(\phi\right)\right>=\\=\frac{1}{2}\left(\left<N,0\right|+\left<0,N\right|\right)\conj{U\left(\phi\right)}\conj{U_{BS}}e^{i\pi\left(J_0-J_3\right)}U_{BS}U\left(\phi\right)\left(\left|N,0\right>+\left|0,N\right>\right)=\\
=\frac{1}{2}\left(\left<N,0\right|+e^{-iN\phi}\left<0,N\right|\right)e^{-i\frac{\pi}{2}J_1}e^{i\pi\left(J_0-J_3\right)} e^{i\frac{\pi}{2}J_1}\left(\left|N,0\right>+e^{iN\phi}\left|0,N\right>\right)=\\
=\frac{1}{2}\left(\left<0,N\right|+e^{-iN\phi}\left<N,0\right|\right)e^{i\pi\left( J_0-J_2\right)}\left(\left|N,0\right>+e^{iN\phi}\left|0,N\right>\right)=\\
=\frac{1}{2}\left(\left<0,N\right|+e^{-iN\phi}\left<N,0\right|\right)e^{-i\pi J_2}e^{i\pi J_0}\left(\left|N,0\right>+e^{iN\phi}\left|0,N\right>\right)=\\
=\frac{1}{2}\left(\left(-1\right)^N\left<0,N\right|+e^{-iN\phi}\left<N,0\right|\right)e^{i\frac{N}{2}\pi}\left(\left|N,0\right>+e^{iN\phi}\left|0,N\right>\right)=\\=\frac{e^{iN\phi}+\left(-1\right)^N e^{-iN\phi}}{2i^N}
\end{multline}
Thus $\left<O\right>\left(\phi\right)=\left(-1\right)^{\frac{N-1}{2}}\sin\phi$ for $N$ odd and $\left<O\right>\left(\phi\right)=\left(-1\right)^{\frac{N}{2}}\cos\phi$ for $N$ even. These functions readily allow us to recover $\tilde\phi$ from $\left<O\right>\left(\tilde\phi\right)$, up to a multiple of $\pi/N$. Since $O^2=\One$, we have $\left<O^2\right>=1$ and formula \eqref{eqerrorphi} yields the following estimate for the error of $\tilde\phi$,
\begin{equation}\label{eqerrorfromparity}
\Delta\tilde\phi=\frac{1}{N}
\end{equation}
The above equality holds for both $N$ even and $N$ odd. The $\propto 1/N$ dependence in formula \eqref{eqerrorfromparity} corresponds to so-called \textit{Heisenberg limit}, which is widely accepted as the minimum phase estimation error allowed by quantum mechanics \cite{YMCSK86,O96,O97}. On the contrary, by simply repeating a single photon experiment $N$ times, one gets a precision $\Delta\tilde\phi\propto1/\sqrt{N}$, so-called \textit{shot-noise} or \textit{standard quantum limit}, which is significantly worse than \eqref{eqerrorfromparity} for large $N$. In this way, entanglement between the photons fed into the arms of the interferometer can increase the phase sensitivity in the experiment by a factor of $\sqrt{N}$. Compared to one single photon experiment, the sensitivity is increased $N$ times. A very practical use of this feature was proposed in \cite{BKABWD00}, where the authors suggest that NOON states could be used to imprint details of minimum resolution $N$ times better than usual in photolithography. In particular, diffraction patterns resulting from the use of NOON states would have the minimum resolution $N$ times greater than those obtained with unentangled photons. This was called \textit{quantum lithography} in \cite{BKABWD00}. However, the original argument of \cite{BKABWD00} has recently met with some criticism \cite{KBIB11}, and it is argued that in practice, the efficiency of quantum lithography would be rather low.

\chapter{Distillability and bound entanglement}\label{chbound}

\section{Distillation of quantum entanglement}\label{secdistill}
As we have seen above, a central role in the most popular quantum tasks, including quantum cryptography and teleportation, is played by maximally entangled states. However, states encountered in practice never match perfectly those used in the theory, due to experiment imperfections. In the early days of quantum information science therefore, it appeared to be crucial to answer the question whether a noisy entangled state can somehow be ``purified'' to yield one that is closer to being maximally entangled. A partially affirmative answer to this question was first provided in \cite{BBPSSW96} for the case of two qubits and refined by the authors of \cite{HHH97}. A method suitable for bipartite systems of arbitrary dimension, based on the reduction criterion for separability, was later presented in \cite{HH99}.

Let us briefly discuss a purification, or \textit{distillation} protocol developed by the authors of \cite{BBPSSW96}. The procedure starts with an arbitrary mixed state $\rho$ of two qubits. The following steps are designed to yield a state which is closer to $\left|\Phi_+\right>$ in a sense described below. However, it should be stressed that the method only works provided that $\left<\Phi_+\right|\rho\left|\Phi_+\right>>\frac{1}{2}$, i.e. $\rho$ is not too far from $\Phi_+$ at the outset. We call the parameter $\left<\Phi_+\right|\rho\left|\Phi_+\right>$   the \textit{fidelity} of $\rho$ with respect to the maximally entangled state $\left|\Phi_+\right>$.
\begin{enumerate}[1)]
\item First, we apply a local unitary rotation $\sigma_y$ to the second component of $\rho$. This yields $\rho'=\left(\One\otimes\sigma_y\right)\rho\left(\One\otimes\sigma_y\right)^{\ast}$, a state which is as close to $\left|\Psi_-\right>=\left(\left|01\right>-\left|10\right>\right)/\sqrt{2}$ as $\rho$ was to $\left|\Phi_+\right>$, in the sense that $\left<\Psi_-\right|\rho'\left|\Psi_-\right>=\left<\Phi_+\right|\rho\left|\Phi_+\right>$.
\item Second, we apply a \emph{random} bilateral $\textnormal{SU}\left(2\right)$ rotation to $\rho'$, which effectively yields
\begin{equation}\label{eqUUrotations}
\rho''=\int\left(U\otimes U\right)\rho'\left(U\otimes U\right)^{\ast}\textnormal{d}U,
\end{equation}
where $\textnormal{d}U$ refers to the Haar measure. In practice, the same goal can be achieved by randomly choosing the identity and bilateral $\sigma_x$, $\sigma_y$ and $\sigma_z$ rotations. The result of \eqref{eqUUrotations} is obviously $U\otimes U$-invariant, which implies that it must be one of the Werner states \eqref{eqWernerstates}. In the $2\times 2$ case considered here, the Werner states take the specific form
\begin{equation}\label{phiphipsipsi}
F\proj{\Psi_-}+\frac{1-F}{3}\left(\proj{\Psi_+}+\proj{\Phi_+}+\proj{\Phi_-}\right),
\end{equation}
where $\left|\Psi_+\right>=\left(\left|01\right>+\left|10\right>\right)/\sqrt{2}$ and $\Phi_-=\left(\left|00\right>-\left|11\right>\right)/\sqrt{2}$.
Therefore $\rho''$ is of the form given above, with $F=\left<\Psi_-\right|\rho''\left|\Psi_-\right>=\left<\Psi_-\right|\rho'\left|\Psi_-\right>$. The last equality follows from the fact that $\left|\Psi_-\right>$ is an $U\otimes U$-invariant state.
\item In the next step, a unilateral $\sigma_y$ rotation takes $\rho''$ to $\rho^{\left(3\right)}=\left(\One\otimes\sigma_y\right)\rho''\left(\One\otimes\sigma_y\right)^{\ast}$. In this way, the mostly $\left|\Phi_-\right>$ state is converted to a mostly $\left|\Phi_+\right>$ one.
\item Next, we take two copies of $\rho^{\left(3\right)}$, prepared in the way described above, and use one of them as a ``source'' and the second one as a ``target'' for a BXOR gate, depicted in Figure \ref{figBXOR}. A BXOR gate simply consists of two CNOT gates, applied to distinct pairs of source and target qubits. 
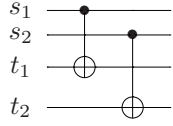
\begin{figure}\hfil
\Qcircuit @C=1em @R=.7em{
s_1&&\ctrl{2}&\qw&\qw\\
s_2&&\qw &\ctrl{2}&\qw\\
t_1&&\targ&\qw&\qw\\
t_2&&\qw &\targ&\qw\\
}\hfil
\caption{A BXOR gate applied to a pair of source ($s_1$, $s_2$) and a pair of target qubits ($t_1$, $t_2$).}\label{figBXOR}
\end{figure}
\item Next, the target pair of qubits is locally measured in the $\sigma_z$ basis, as depicted in Figure \ref{figparallel}, which also includes the BXOR operation described above.
\begin{figure}[h!]\hfil
\Qcircuit @C=1em @R=.7em{
&\meter&\targ&\measure{t_1 t_2}&\targ&\meter&\qw\\
&\qw&\ctrl{-1}&\measure{s_1 s_2}&\ctrl{-1}&\qw&\qw
}\hfil
\caption{BXOR operation followed by the measurement of the pair of target qubits.}\label{figparallel}
\end{figure}
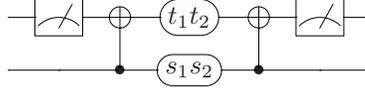
If the results are the same for the qubits $t_1$ and $t_2$, the remaining source pair $\left(s_1,s_2\right)$ is kept. Otherwise, it is discarded.
\item If in the previous step the source pair was kept, it is transformed to an almost $\left|\Psi_-\right>$ state by a unilateral $\sigma_y$ rotation. Next, it is made rotationally symmetric by applying random bilateral $\textnormal{SU}\left(2\right)$ rotations, as in equation \eqref{eqUUrotations}. Let us call the resulting state $\rho^{\left(4\right)}$. The corresponding parameter $F$ in formula \eqref{phiphipsipsi} is then equal to
\begin{equation}\label{eqFprime}
F^{\left(4\right)}=\frac{F^2+\frac{1}{9}\left(1-F\right)^2}{F^2+\frac{2}{3}F\left(1-F\right)+\frac{5}{9}\left(1-F\right)^2},
\end{equation}
which exceeds $F$ over the range $\left(1/2,1\right)$. Thus $\rho^{\left(4\right)}$ is closer to $\Psi_-$ than $\rho'$ in the sense that $\left<\Psi_-\right|\rho^{\left(4\right)}\left|\Psi_-\right>>\left<\Psi_-\right|\rho'\left|\Psi_-\right>$. 
\item In the last step, the almost $\left|\Psi_-\right>$ state $\rho^{\left(4\right)}$ is converted back to an almost $\left|\Phi_+\right>$ one by a unilateral $\sigma_y$ rotation. We call the resulting state $\rho^{\left(5\right)}$. The corresponding parameter $\left<\Phi_+\right|\rho^{\left(5\right)}\left|\Phi_+\right>$ is bigger than $\left<\Phi_+\right|\rho\left|\Phi_+\right>$. Thus, the resulting state is closer to $\left|\Phi_+\right>$ than $\rho$ was.
\end{enumerate}
As a result, by repeating the above procedure, states which are arbitrarily close to $\Phi_+$ can be obtained. Nevertheless, the number of copies of $\rho$ needed for the input grows very fast as the expected fidelity goes to $1$. Thus, for practical purposes, another procedure of distillation was designed by the authors of \cite{BBPSSW96}, which more efficiently uses the statistical properties of $\rho$. However, it needs a small input of $\left|\Phi_+\right>$ states, which may be obtained by the method described above. The mentioned procedure consists of two rounds of BXOR tests performed on suitably chosen \emph{subsets} of the whole supply of $\rho$ states, using the prepurified $\left|\Phi_+\right>$ states as targets. It also uses unilateral and bilateral $\sigma_y$ rotations, as well as unilateral $\sigma_z$ rotations to correct the discrepancies from $\left|\Phi_+\right>$ detected by the BXOR operations. More details of the procedure can be found in \cite{BBPSSW96}. All in all, from a theorist point of view, it is sufficient to say that all mixed states $\rho$ of two qubits with $\left<\Phi_+\right|\rho\left|\Phi_+\right>>1/2$ can be distilled to the maximally entangled state. This result was further extended, by using the technique of local filters \cite{Gisin96}, to \emph{arbitrary entangled states of two qubits} \cite{HHH97}. In this way, the authors of \cite{HHH97} showed that any entangled state of two qubits has some form of nonlocality, which is revealed by the distillation procedure. Note that a similar result for bipartite states of arbitrary dimension would have resolved the paradox of Werner's paper \cite{Werner89}, which we discussed in Section \ref{secsep}. However, it was quickly realized that the existence of PPT entangled states, first revealed to the physicist' community by the paper \cite{Pawel97}, immediately precludes the described strategy from working \cite{HHH98}. Let us briefly explain why this is the case.

In an ideal case, given a source characterized by a bipartite density matrix $\rho$, we have at our disposal the tensor product states $\rho^{\otimes n}$ for arbitrary $n$. The most general transformation one can perform on $\rho^{\otimes n}$ using only local operations and classical communication is of the form \cite{VPRK97}
\begin{equation}\label{eqgentransf}
\rho^{\otimes n}\mapsto\Theta\left(\rho^{\otimes n}\right):=\frac{1}{M}\sum_i\left(A_i\otimes B_i\right)\rho^{\otimes n}\left(A_i\otimes B_i\right)^{\ast},
\end{equation}
where $A_i$ and $B_i$ map into image space in the first and the second subsystem, respectively. In the case of entanglement distillation, both the image spaces are $\setC^2$, as we want to obtain the state $\left|\Phi_+\right>$, living in $\setC^2\otimes\setC^2$. Therefore, $A_i:\hilbertspaceone^{\otimes n}\rightarrow\setC^2$ and $B_i:\hilbertspacetwo^{\otimes n}\rightarrow\setC^2$, assuming that $\rho$ lives on $\hilbertspaceone\otimes\hilbertspacetwo$. Now assume that $\rho$ has a positive partial transpose. Thus, $\rho^{\otimes n}$ is a PPT state as well. One can also easily notice that the mapping $\rho^{\otimes n}\mapsto\sum_i\left(A_i\otimes B_i\right)\rho^{\otimes n}\left(A_i\otimes B_i\right)^{\ast}$ preserves the positivity of the partial transpose of $\rho^{\otimes n}$. Hence the state on the right-hand side of \eqref{eqgentransf} is a PPT state living on $\setC^2\otimes\setC^2$. Consequently, it is separable \cite{HHH96} and cannot be distilled (cf. also the discussion in Section \ref{secsep}). In this way we have proved the following \cite{HHH98}.
\begin{proposition}\label{propdistillPPT}
No PPT state living on a bipartite space $\hilbertspaceone\otimes\hilbertspacetwo$ can be distilled to $\left|\Phi_+\right>$.
\end{proposition}
As a result, all PPT entangled states, including those presented in \cite{Pawel97}, cannot be distilled to $\left|\Phi_+\right>$, even though they are not separable. Due to their undistillability, the states are called \textit{bound entangled}. For them, the paradox from the Werner's paper \cite{Werner89} cannot be resolved by using distillation protocols.  Let us mention, however, that the original Werner states are positive partial transpose if and only if they are separable. This still does \emph{not} allow us to conclude that all entangled Werner states can be distilled to $\left|\Phi_+\right>$, as there might exist \textit{NPT bound entangled} states, i.e. undistillable states which are not PPT. This has become a central, still unresolved problem in the theory of entanglement, so-called \textit{NPT bound entanglement existence problem}. Actually, it was demonstrated in \cite{DiVicenzo2000} (cf. also \cite{DCLB00}) that the question whether there exist NPT bound entangled states only needs to be answered for the Werner family of states, since all the other ones can be brought to the Werner form by transformations that do preserve the positivity of the partial transpose. However, it turns out that the question for Werner states becomes increasingly difficult to answer as the parameter $\Xi$ in the definition \eqref{eqWernerstates} tends to the boundary value $0$. Namely, it was proved in \cite{DiVicenzo2000} that for any $n\in\setN$, there exists $\varepsilon>0$ such that the state $W$ from eq. \eqref{eqWernerstates} with $\Xi\in\left(0,\varepsilon\right)$ cannot be distilled using operations of the form \eqref{eqgentransf} on $W^{\otimes n}$ (however, some of these states may be distillable using $n+1$ or more copies of $W$). Since then, considerable efforts have been made to prove or disprove the existence of NPT bound entangled states, none of which have lead to a conclusive answer \cite{Bandy03,Watrous04,Clarisse05,Clarisse06,VD06,Chatto06}. Moreover, two contradictory statements concerning the problem can be found in the preprints \cite{Simon06,SV09}, none of which is correct. One thing beyond any doubt is that the question of distillability intimately relates to the structure of $2$-positive maps, i.e. positive maps $\Lambda:\bk\rightarrow\bh$ with the property that the map
\begin{equation}
\mathcal{B}\left(\setC^2\otimes\hilbertspaceone\right)\ni\left[\begin{array}{cc}
A_{11}&A_{12}\\
A_{21}&A_{22}
\end{array}\right]\mapsto\left[\begin{array}{cc}
\Lambda\left(A_{11}\right)&\Lambda\left(A_{12}\right)\\
\Lambda\left(A_{21}\right)&\Lambda\left(A_{22}\right)
\end{array}\right]\in\mathcal{B}\left(\setC^2\otimes\hilbertspacetwo\right),
\end{equation}
denoted with $\Id_2\otimes\Lambda$, is also positive. To see the relation of $2$-positivity to distillability, let us first note  the following characterization of distillable states \cite{HHH98,DiVicenzo2000}.
\begin{proposition}\label{propdistill2Schmidt}
A state with a density matrix $\rho$ on $\hilbertspaceone\otimes\hilbertspacetwo$ is distillable if and only if there exists a finite $n$ and two-dimensional projections $P_1$, $P_2$ in $\hilbertspaceone^{\otimes n}$ and $\hilbertspacetwo^{\otimes n}$, resp. such that $\rho'=\left(P_1\otimes P_2\right)\rho\left(P_1\otimes P_2\right)^{\ast}$, supported on a $2\times 2$-dimensional space, is entangled. The last condition is equivalent to the statement that there exists a vector $\left|\psi\right>\in\hilbertspaceone\otimes\hilbertspacetwo$ of the form $\left|\xi_1\right>\left|\chi_1\right>+\left|\xi_2\right>\left|\chi_2\right>$ in $\hilbertspaceone^{\otimes n}\otimes\hilbertspacetwo^{\otimes n}$ such that 
\begin{equation}\label{eqSR2}
\left<\psi\right|\left(\rho^{\otimes n}\right)^{T_2}\left|\psi\right><0,
\end{equation}
where $T_2$ denotes the partial transpose with respect to the second subsystem, $\left(\id\otimes t\right)^{\otimes n}$.
\begin{proof}
As we mentioned above, the most general distillation operation one can perform on $\rho^{\otimes n}$ is of the form \eqref{eqgentransf}. In order for the transformed state $\Theta\left(\rho^{\otimes n}\right)$ transformed to be entangled, and thus distillable (remember that end up with states on $\setC^2\otimes\setC^2$), at least one of the terms $\rho_i:=\left(A_i\otimes B_i\right)\rho^{\otimes n}\left(A_i\otimes B_i\right)^{\ast}$, supported on a $2\times 2$-dimensional subspace, needs to be entangled. The operators $A_i$ and $B_i$ are of the form $\left|e_0\right>\left<\alpha_1\right|+\left|e_1\right>\left<\alpha_2\right|$ and $\left|e_0\right>\left<\beta_1\right|+\left|e_1\right>\left<\beta_2\right|$, where $\alpha_1,\alpha_2$ belong to $\hilbertspaceone$ and $\beta_1,\beta_2$ belong to $\hilbertspacetwo$. Let us denote with $P_1$ and $P_2$ the projections onto the subspaces $\linspan\left\{\alpha_1,\alpha_2\right\}$ and $\linspan\left\{\beta_1,\beta_2\right\}$, respectively. We have 
\begin{equation}\label{eqABPP}
\rho_i=\left(A_i\otimes B_i\right)\left(P_1\otimes P_2\right)\rho^{\otimes n}\left(P_1\otimes P_2\right)\left(A_i\otimes B_i\right)^{\ast}
\end{equation}
Since a product transformation cannot convert a separable state into an entangled one, we must have that $\rho'_i:=\left(P_1\otimes P_2\right)\rho^{\otimes n}\left(P_1\otimes P_2\right)$ is entangled. This proves the necessity in the first part of the proposition. In order to prove the sufficiency, it is enough to notice that the projected state $\left(P_1\otimes P_2\right)\rho^{\otimes n}\left(P_1\otimes P_2\right)$, if entangled, can be distilled, because it is supported on a $2\times 2$-dimensional subspace. 

To prove the second part of the proposition, we observe the following. Because $\rho'_i$ is supported on a $2\times 2$-dimensional subspace $\left(P_1\otimes P_2\right)\hilbertspaceone^{\otimes n}\otimes\hilbertspacetwo^{\otimes n}$, a necessary and sufficient condition for $\rho'_i$ to be entangled is that it does not have a positive partial transpose. The partial transpose equals
\begin{equation}\label{eqpartialofrho}
\left(\rho'_i\right)^{T_2}=\left(P_1\otimes \bar P_2\right)\left(\rho^{\otimes n}\right)^{T_2}\left(P_1\otimes \bar P_2\right),
\end{equation}
where $\bar P_2$ denotes an operator represented by the complex conjugated matrix of $P_2$. Thus $\bar P_2$ is also a two-dimensional projection.

The above operator is \emph{not} positive if and only if there exists a vector of the form $\left|\psi\right>=\left|\xi_1\right>\left|\chi_1\right>+\left|\xi_2\right>\left|\chi_2\right>$ in $\left(P_1\otimes\bar P_2\right)\hilbertspaceone^{\otimes n}\otimes\hilbertspacetwo^{\otimes n}$ that fulfills the inequality $\left<\psi\right|\left(\rho'_i\right)^{T_2}\left|\psi\right><0$. This simply follows because all the vectors in $\left(P_1\otimes\bar P_2\right)\hilbertspaceone^{\otimes n}\otimes\hilbertspacetwo^{\otimes n}$ are of the form  $\xi_1\otimes\chi_1+\xi_2\otimes\chi_2$, i.e. are of \textit{Schmidt rank} $2$. But $\left<\psi\right|\left(\rho'_i\right)^{T_2}\left|\psi\right>=\left<\psi\right|\left(\rho^{\otimes n}\right)^{T_2}\left|\psi\right>$ according to our choice of $\psi$, which finishes the proof of the second part of the proposition.
\end{proof}
\end{proposition}
Another way of phrasing the above result is that the operator $\left(\rho^{\otimes n}\right)^{T_2}$ is not \textit{$2$-block positive} for some $n$ (cf. e.g. \cite{ref.SSZ09}). By $k$-block positivity of an operator $X$ on a bipartite Hilbert space $\hilbertspaceone_1\otimes\hilbertspaceone_2$ we mean the property that $\left<\phi\right|X\left|\phi\right>\geqslant 0$ for all $\phi$ of the form $\sum_{i=1}^k\left|\xi_i\right>\left|\chi_i\right>$ (in particular, we can choose $\hilbertspaceone_1=\hilbertspaceone^{\otimes n}$ and $\hilbertspaceone_2=\hilbertspacetwo^{\otimes n}$). Thus, by Proposition \ref{propdistill2Schmidt}, a state $\rho$ on $\hilbertspaceone\otimes\hilbertspacetwo$ is distillable if for some $n$ the state $\rho^{\otimes n}$ is \emph{not} $2$-block positive. Since $2$-block positive operators are in a one-to-one Jamiołkowski-Choi correspondence to $2$-positive maps \cite{RA07,ref.SSZ09}, there is a direct link between distillability of entanglement and the property of \emph{not} being a $2$-positive map. For additional insights, consult \cite{Clarisse05}.

\begin{figure}\hfil
\Qcircuit @C=1em @R=.7em {
&\meter&\multigate{1}{U_{\textnormal{XOR}}} &\measure{\sigma_{\alpha}} &\multigate{1}{U_{\textnormal{XOR}}}&\meter&\qw\\
&\qw&\ghost{U_{\textnormal{XOR}}}& \measure{\varrho_{\textnormal{free}}}&\ghost{U_{\textnormal{XOR}}}&\qw&\qw\\
}\hfil
\caption{Bound entanglement activation procedure illustrated.}\label{figactivation}
\end{figure}
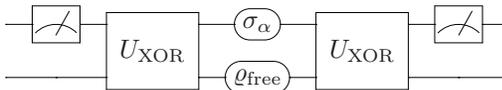

It was quickly realized \cite{HHH99} that bound entanglement, even though it is useless for entanglement distillation, can be used to improve fidelity of a given distillable ( $=$ free entangled) state $\varrho_{\textnormal{free}}$ in  a process very similar to the one depicted in Figure \ref{figparallel}. To this aim, a copy of the free entangled state $\varrho_{\textnormal{free}}$ together with a copy of a bound entangled state $\sigma_{\alpha}$ are passed as inputs to the circuit in Figure \ref{figactivation}, where $U_{\textnormal{XOR}}\left(e_i\otimes e_j\right)=e_i\otimes e_{i+j\textnormal{ mod
 }n}$, an analogue of the CNOT gate used in Fig \ref{figparallel}. Later, the target pair (the upper one in Fig. \ref{figactivation}) is measured in the basis $\left\{e_1,e_2,\ldots,e_n\right\}$. If both measurements agree, the source pair (initially in the state $\varrho_{\textnormal{free}}$) is kept and assumes a new state $\rho'_{\textnormal{free}}$ of higher fidelity. Otherwise, it is discarded an the whole procedure fails. If the run was successful, the described steps are repeated for $\varrho'_{\textnormal{free}}$ and another copy of $\sigma_{\alpha}$ as the source and the target pair, respectively. It can be shown that a sequence of successful runs of the above scheme leads, with a nonvanishing probability, to a state of an arbitrary high fidelity. This phenomenon is called \textit{bound entanglement activation} \cite{HHH99}. The precise form of the states $\varrho_{\textnormal{free}}$ and $\sigma_{\alpha}$ will be given in Section \ref{secexamplesbound}.
\section{Examples of bound entangled states}\label{secexamplesbound}
We already know from the previous  section that the question about the existence of undistillable states with negative partial transpose is still an unsolved problem in the theory of entanglement. Thus no example of an NPT bound entangled state is known. On the other hand, numerous successful efforts have been made to give explicit examples of bound entangled states that do obey the PPT criterion. Here, we give a list of references where the known examples can be found. For some of them, we provide the reader with the precise form of the state and briefly discuss how it was proved to be entangled.

Probably the most famous example in the physics literature is the $3\times 3$ Horodecki state, named after P. Horodecki work \cite{Pawel97}. The name refers to a one-parameter family of states, given in the canonical product basis of $\setC^3\otimes\setC^3$ by the matrices
\begin{equation}
\varrho_a={\frac{1}{8a + 1}}
\left[ \begin{array}{ccccccccc}
          a &0&0&0&a&0&0&0& a   \\
           0&a&0&0&0&0&0&0&0     \\
           0&0&a&0&0&0&0&0&0     \\
           0&0&0&a&0&0&0&0&0     \\
          a &0&0&0&a&0&0&0& a     \\
           0&0&0&0&0&a&0&0&0     \\
           0&0&0&0&0&0&{\frac{1+a}{2}}&0&{\frac{\sqrt{1-a^2}}{2}}\\
           0&0&0&0&0&0&0&a&0     \\
          a &0&0&0&a&0&{\frac{\sqrt{1-a^2}}{2}}&0&{\frac{1+a}{2}}\\
       \end{array}
      \right ],
\end{equation}
where $0<a<1$. As we already mentioned in Section \ref{secsep}, the state $\varrho_a$ can be proved to be entangled by using the range criterion, which is our Proposition~\ref{proprange}. Indeed, with some amount of algebra \cite{Pawel97}, one can show that the vectors in $\range{\varrho_a^{T_2}}$, the range of $\varrho_a^{T_2}$, belong to one of the following families 
\begin{eqnarray}
&A\left[\begin{array}{ccc}1&s&0\end{array}\right]\otimes\left[\begin{array}{ccc}1&s&0\end{array}\right],&A,s\in\setC, s\neq 0\label{eqprodvecfirst}\\
&F\left[\begin{array}{ccc}0&0&1\end{array}\right]\otimes\left[\begin{array}{ccc}1&0&x\end{array}\right],&F\in\setC\\
&D\left[\begin{array}{ccc}0&1&0\end{array}\right]\otimes\left[\begin{array}{ccc}0&1&0\end{array}\right],&D\in\setC\\
&A\left[\begin{array}{ccc}1&0&0\end{array}\right]\otimes\left[\begin{array}{ccc}1&0&0\end{array}\right],&A\in\setC\\
&C\left[\begin{array}{ccc}1&0&t\end{array}\right]\otimes\left[\begin{array}{ccc}\frac{1}{t}+\frac{1}{x}&0&1\end{array}\right],&C,t\in\setC, t\neq 0\label{eqprodveclast}
\end{eqnarray}
where $x:=\sqrt{\left(1+a\right)\left(1-a\right)}$. The partially conjugated vectors \eqref{eqprodvecfirst}-\eqref{eqprodveclast} do \emph{not} span the range of $\varrho_a$, as they cannot be linearly combined to yield $\left[\begin{array}{ccc}0&0&1\end{array}\right]\otimes\left[\begin{array}{ccc}0&1&0\end{array}\right]$, which is an element of $\range{\varrho_a}$. In this way, the author of \cite{Pawel97} arrived at a contradiction with the range criterion for the state $\varrho_a^{T_2}$. Hence $\varrho_a^{T_2}$ was proved to be entangled, so that $\varrho_a$ is entangled as well. A similar method was later used in the paper \cite{Clarisse06b}, which contains first examples of $3\times 3$ PPT entangled states of types $\left(5,5\right)$ and $\left(6,6\right)$. Here $\left(m,n\right)$ means that a PPT state $\rho$ has rank $m$, while the rank of $\rho^{T_2}$ equals $n$. The reduction criterion was also employed, in a very straightforward way, to prove  inseparability of a family of PPT chessboard states, introduced in \cite{BrussPeres}. They are states of the form $\frac{1}{N}\sum_{i=1}^4\proj{V_i}$, where
\begin{eqnarray}
V_1&=&\left[\begin{array}{ccccccccc}m&0&s&0&n&0&0&0&0\end{array}\right],\\
V_2&=&\left[\begin{array}{ccccccccc}0&a&0&b&0&c&0&0&0\end{array}\right],\\
V_3&=&\left[\begin{array}{ccccccccc}n^{\ast}&0&0&0&-m^{\ast}&0&t&0&0\end{array}\right],\\
V_4&=&\left[\begin{array}{ccccccccc}0&b^{\ast}&0&-a^{\ast}&0&0&0&d&0\end{array}\right].
\end{eqnarray}
According to the main result of the thesis, Theorem \ref{maintheorem}, the chessboard states are of the type $\left(4,4\right)$ and they are locally equivalent to states arising from the Unextendible Product Basis construction to be discussed below.

Shortly after the first example of a bound entangled state in the physics literature, C. H. Bennett and coworkers \cite{Bennett99} proposed a fully algorithmic way to construct more such examples. The method relies on the notion of an \textit{Unextendible Product Basis}, which is formally defined in the following way.
\begin{definition}\label{defUPB}
An Unextendible Product Basis, UBP for short, is a set of mutually orthogonal product vectors $\left\{\phi^i_1\otimes\ldots\otimes\phi^i_n\right\}_{i=1}^k$ in a multipartite Hilbert space $\hilbertspaceone_1\otimes\ldots\otimes\hilbertspaceone_n$ such that the orthogonal complement $\left(\linspan\left\{\phi^i_1\otimes\ldots\otimes\phi^i_n\right\}_{i=1}^k\right)^{\bot}$ \textbf{does not contain a product vector}.
\end{definition}
Given a UPB in a bipartite space, it is straightforward to give an example of a PPT entangled state.
\begin{proposition}\label{propUPBconstr}
Let $\left\{\phi^i_1\otimes\phi^i_2\right\}_{i=1}^k$ be an Unextendible Product Basis in $\hilbertspaceone\otimes\hilbertspacetwo$. The projection
\begin{equation}\label{eqprojectionUPB}
\pi_{\textnormal{UPB}}=
\One-\sum_{i=1}^k\proj{\phi^i_1\otimes\phi^i_2}
\end{equation}
defines a PPT bound entangled state $\rho_{\textnormal{UPB}}=\pi_{\textnormal{UPB}}/N$, where $N$ is a suitable normalization factor.
\end{proposition}
The proposition follows because the subspace on which $\rho_{\textnormal{UPB}}$ projects, contains no product vector. Hence, using the range criterion, the state proportional to \eqref{eqprojectionUPB} is entangled. The fact that it also has a positive partial transpose can be checked by a simple calculation. Indeed,
\begin{equation}
\pi_{\textnormal{UPB}}^{T_2}=\One-\sum_{i=1}^k\proj{\phi^i_1\otimes\left(\phi^i_2\right)^{\ast}},
\end{equation}
where $^{\ast}$ denotes componentwise conjugation, is another projection, hence positive definite. Generalizations to a multipartite setting are immediate.

In the main part of the thesis, we prove Theorem \ref{maintheorem}, which says that all PPT bound entangled states of rank $4$ in $3\times 3$ systems are locally equivalent to states of the form \eqref{eqprojectionUPB}. This means that any such state is proportional to $\left(A\otimes B\right)\pi_{\textnormal{UPB}}\left(A\otimes B\right)^{\ast}$ for some UPB and some $\textnormal{SL}\left(3,\setC\right)$ transformations $A$ and $B$. In this way we obtain a full characterization of simplest PPT entangled states, as all PPT states of ranks $\leqslant 3$ are separable \cite{HLVC2000}.

As far as the above examples are considered, the reduction criterion seems to be the only way to prove that a given PPT state is entangled. But in reality, it is not the only one known in literature. Another distinguished approach to the problem is by using so-called \textit{indecomposable positive maps}. By the positive maps criterion, the existence of a positive map $\Lambda$ such that $\left(\Id\otimes\Lambda\right)\rho\not\geqslant 0$ implies inseparability of a state $\rho$. It is precisely in this way that the earliest examples of PPT entangled states \cite{Choi82,Erling82} were obtained by mathematicians\footnote{Note however that the name ``bound entanglement'' was not used until \cite{HHH98}}. The exemplary PPT entangled state given in \cite{Erling82} is of the form
\begin{equation}\label{eqErling82}
x=\left[
\begin{array}{ccccccccc}
2\mu&0&0&0&2\mu&0&0&0&2\mu\\
0&4\mu^2&0&0&0&0&0&0&0\\
0&0&1&0&0&0&0&0&0\\
0&0&0&1&0&0&0&0&0\\
2\mu&0&0&0&2\mu&0&0&0&2\mu\\
0&0&0&0&0&4\mu^2&0&0&0\\
0&0&0&0&0&0&4\mu^2&0&0\\
0&0&0&0&0&0&0&1&0\\
2\mu&0&0&0&2\mu&0&0&0&2\mu
\end{array}
\right],
\end{equation}
which can be more concisely written as $x=2\mu\proj{\Phi_+}+4\mu^2\sigma_++\sigma_-$, where 
\begin{eqnarray}\label{eqsigmadef1}
\sigma_+=\frac{1}{3}\left(\proj{01}+\proj{12}+\proj{20}\right)\\
\sigma_-=\frac{1}{3}\left(\proj{10}+\proj{21}+\proj{02}\right)
\end{eqnarray}
and $\left|\Phi_+\right>$ stands for the maximally entangled vector, $\left|\Phi_+\right>=\left(\sum_{i=0}^2\left|ii\right>\right)/\sqrt{3}$. 

A slightly modified family of states $\sigma_{\alpha}=\frac{2}{7}\proj{\Phi_+}+\frac{\alpha}{7}\sigma_++\frac{5-\alpha}{7}\sigma_-$ was later used to demonstrate the phenomenon of bound entanglement activation \cite{HHH99}, which we briefly described in Section \ref{secdistill}. The authors of \cite{HHH99} also used a related family $\rho_{\textnormal{free}}=F\proj{\Phi_+}+\left(1-F\right)\sigma_+$ as their input free entangled states. 

Another notable example of a class of PPT entangled states revealed by indecomposable positive maps
was given in \cite{Piani06}. We should also mention a series of papers by K.-C. Ha and co-workers \cite{HaKyePark2003,HaKye2004,HaKye2005}, where the authors develop a possible general approach to constructing PPT entangled states from faces of the cone of all \emph{decomposable} positive maps. In particular, they consider a family of generalized Choi maps, introduced in \cite{CKL92} and use them to construct the corresponding bound states. For the definition of decomposability and related notions, check e.g. \cite{ref.SSZ09}.


\part{A brief introduction to algebraic geometry}\label{partII}\,

\chapter{Varieties, Ideals and Groebner bases}\label{chVarIdGroeb}

\section{Preliminaries}\label{secpreliminaries}
Just as we mentioned in previous parts of the thesis, problems encountered in the theory of quantum channels, measurement and entanglement are often of purely algebraic nature. More precisely, they pertain to the existence of solutions of certain algebraic equations or, for example, to positivity of a number of polynomials. In order to answer such questions in an effective way, one can uses techniques such as Groebner bases or resultans, which we briefly discuss in the following. By their effectiveness we mean the fact that a decisive answer to a question is obtained in a finite, though sometimes rather high, number of steps. We also include a proof of Bezout's theorem, which we later use to prove the main result of the thesis, concerning PPT bound entangled states of minimal rank. 

Before  we introduce the ideas of Groebner bases, let us begin with an introduction to basic notions of algebraic geometry. A more comprehensive treatment of the subject can be found in a book like \cite{IdealsVarieties}, which we recommend to everyone new to the subject. By $\mathbbm{K}\left[x_1,x_2,\ldots,x_n\right]$ we shall denote the set of $n$-variate polynomials in the variables $x_1,\ldots,x_n$ and coefficients in $\mathbbm{K}$. The two main cases considered in this thesis are $\mathbbm{K}=\mathbbm{C}$ and $\mathbbm{K}=\mathbbm{R}$. With this notation, let us define the basic object of algebraic geometry.
\begin{definition}\label{defaffine}
By an \textbf{affine variety} we mean a subset of $\mathbbm{K}^n$ defined by a set of equations
\begin{equation}\label{eqdefaffine}
f_1\left(x_1,\ldots,x_n\right)=0,\ldots, f_d\left(x_1,\ldots,x_n\right)=0
\end{equation}
We shall denote it by ${\bf V}\!\left(f_1,\ldots,f_d\right)$
\end{definition}
We can give a simple, although not an entirely trivial example of an affine variety in $\mathbbm{R}^n$ (or $\mathbbm{C}^n$), which reappears, in somewhat generalized form of a rational normal curve, in one of the papers related to the thesis \cite{S2011}.
\begin{example}[Twisted cubic]\label{extwisted}
The affine variety defined by the set of equations
\begin{equation}\label{eqtwisteddef}
y-x^2=0,\quad z-x^3=0
\end{equation}
is called the \textbf{twisted cubic} curve. 
\begin{figure}
\includegraphics[scale=0.65]{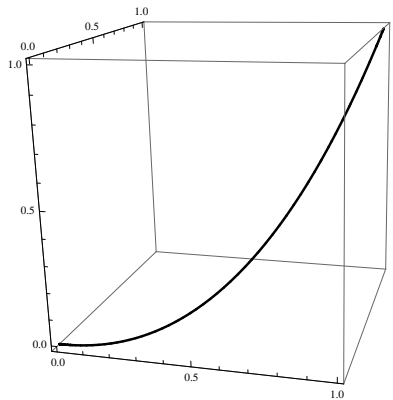}\centering
\caption{A twisted cubic curve in $\mathbbm{R}^3$.}\label{figtwisted}
\end{figure}
\end{example}
As we will explain, a concept intimately related to affine varieties is that of ideals. An ideal can be regarded as a generalization of a linear subspace, where the arbitrary scalar factors in linear combinations are replaced by arbitrary polynomials. More formally, we have the following definition.
\begin{definition}[Ideal]A subset $I\subset\mathbbm{K}\left[x_1,\ldots,x_n\right]$
\begin{enumerate}[(i)]
\item $0\in I$
\item $f,g\in I\Rightarrow f+g\in I$
\item $f\in I\land h\in\mathbbm{K}\left[x_1,\ldots,x_n\right]\Rightarrow hf\in I$
\end{enumerate}
is called an \textbf{ideal} in $\mathbbm{K}\left[x_1,\ldots,x_n\right]$.
\end{definition}
It turns out (cf. Theorem \ref{thmHilbertbasis}) that all ideals $I\subset\mathbbm{K}\left[x_1,\ldots,x_n\right]$ are finitely generated, which means that there always exists a finite set $f_1,\ldots,f_d\in\mathbbm{K}\left[x_1,\ldots,x_n\right]$ such that all elements of $I$ can be written in the form $\sum_{i=1}^dh_if_i$ with $h_i\in\mathbbm{K}\left[x_1,\ldots,x_n\right]$ and no element of $\mathbbm{K}\left[x_1,\ldots,x_n\right]\setminus I$ is of that form. Let us make it more formal.
\begin{definition}\label{deffinitegen}
For a subset $A\subset\mathbbm{K}\left[x_1,\ldots,x_n\right]$, we denote by $\left<A\right>$ the \textbf{ideal generated by $A$}, which is by definition the minimal ideal including $A$. If $A=\left\{f_1,\ldots,f_d\right\}$, we write $\left<A\right>=\left<f_1,\ldots, f_d\right>$ and say that the ideal $\left<f_1,\ldots, f_d\right>$ is finitely generated. Equivalently, $\left<f_1,\ldots, f_d\right>$ consists of all elements of the form $\sum_{i=1}^dh_if_i$, where $h_i\in\mathbbm{K}\left[x_1,\ldots,x_d\right]$ for all $i$.
\begin{proof}
Only the last statement needs a proof. First of all, let us denote by $\left<f_1,\ldots,f_d\right>'$ the set of all elements of the form $\sum_{i=1}^dh_if_i$. Clearly, by the definition of an ideal, we have $\left<f_1,\ldots,f_d\right>'\subset\left<f_1,\ldots,f_d\right>$. Let us also observe that $\left<f_1,\ldots,f_d\right>'$ is an ideal. Since $\left<f_1,\ldots,f_d\right>$ is by definition the smallest ideal containing $f_1,\ldots,f_d$, we must have $\left<f_1,\ldots,f_d\right>'\supset\left<f_1,\ldots,f_d\right>$, which gives us the equality $\left<f_1,\ldots,f_d\right>'=\left<f_1,\ldots,f_d\right>$.
\end{proof}
\end{definition}
A fixed ideal $I$ may have various sets of generators. One of the crucial observations of algebraic geometry is that the variety defined by a set of equations $f_1\left(x_1,\ldots,x_n\right)=0,\ldots,f_d\left(x_1,\ldots,x_d\right)=0$ depends only on the ideal $\left<f_1,\ldots,f_d\right>$ and not on the particular set of generators. 
\begin{proposition}\label{propidealsequal}
Let $\left<f_1,\ldots,f_d\right>=\left<g_1,\ldots, g_l\right>$. In such case ${\bf V}\!\left(f_1,\ldots,f_d\right)={\bf V}\!\left(g_1,\ldots,g_l\right)$
\begin{proof}
From the last part of Definition \ref{deffinitegen} we know that $g_i=\sum_{j=1}^dh^i_jf_j$ for some polynomials $h^i_j$. Thus $\left(x_1,\ldots,x_n\right)\in{\bf V}\!\left(f_1,\ldots,f_d\right)$ implies $\left(x_1,\ldots,x_n\right)\in{\bf V}\!\left(g_1,\ldots,g_l\right)$. Consequently, ${\bf V}\!\left(f_1,\ldots,f_d\right)\subset{\bf V}\!\left(g_1,\ldots,g_l\right)$. The inverse inclusion can be obtained in a similar way.
\end{proof}
\end{proposition}

Apart from  $\left<f_1,\ldots,f_d\right>$, there exists another ideal intimately related to ${\bf V}\!\left(f_1,\ldots,f_d\right)$, namely the ideal of polynomials that vanish on  ${\bf V}\!\left(f_1,\ldots,f_d\right)$.
\begin{definition}Let $ V={\bf V}\!\left(f_1,\ldots,f_d\right)$ be an affine variety in $\mathbbm{K}^n$. The \textbf{ideal of} $ V$ is by definition
\begin{equation}\label{eqidealof}
{\bf I}\left( V\right)=\left\{f\in\mathbbm{K}\left[x_1,\ldots,x_d\right]|f\left(x_1,\ldots,x_n\right)=0\,\,\forall_{\left(x_1,\ldots,x_n\right)\in V}\right\}
\end{equation}
\end{definition}
The above definition is easily generalized to arbitrary subsets in place of $V$.
\begin{definition}Let $S$ be a subset of $\mathbbm{K}^n$. The \textbf{ideal of} $\bf S$ is by definition
\begin{equation}\label{eqidealof2}
{\bf I}\left( S\right)=\left\{f\in\mathbbm{K}\left[x_1,\ldots,x_d\right]|f\left(x_1,\ldots,x_n\right)=0\,\,\forall_{\left(x_1,\ldots,x_n\right)\in S}\right\}
\end{equation}
\end{definition}
We leave it as an exercise for the reader to prove that ${\bf I}\left( S\right)$ is an ideal. Moreover, the maps ${\bf V}:I\mapsto{\bf V}\left(I\right)$ and ${\bf V}: S\mapsto{\bf I}\left( S\right)$ are inclusion reversing. We also have the following
\begin{proposition}\label{propVIV}
For any affine variety $V\subset\K\left[x_1,\ldots,x_n\right]$, we have
\begin{equation}
{\bf V}\left({\bf I}\left(V\right)\right)=V
\end{equation}
\begin{proof}
We know that $V={\bf{V}}\left(\left<f_1,\ldots,f_d\right>\right)$ and $\left<f_1,\ldots,f_d\right>\subset{\bf I}\left(V\right)$ because all the polynomials $f_i$ vanish on $V$. Consequently, ${\bf V}\left(\left<f_1,\ldots,f_d\right>\right)\supset{\bf V}\left({\bf I}\left(V\right)\right)$ since $I\mapsto{\bf V}\left(I\right)$ is inclusion-reversing. On the other hand, the inclusion $V\subset{\bf V}\left({\bf I}\left(V\right)\right)$ follows directly from the fact that every $f\in{\bf I}\left(V\right)$ vanishes on $V$.
\end{proof}
\end{proposition}
We can also characterize ${\bf V}\left({\bf I}\left(S\right)\right)$ for arbitrary subsets $S$ of $\K^n$.
\begin{proposition}\label{propVIS}
For $S\subset\K^n$, the affine variety ${\bf V}\left({\bf I}\left(S\right)\right)$ is the smallest variety that contains $S$.
\begin{proof}
Let $W$ be an affine variety such that $S\subset W$. Since ${\bf I}$ is inclusion-reversing, we have ${\bf I}\left(S\right)\supset{\bf I}\left(W\right)$. Moreover, ${\bf V}\left({\bf I}\left(S\right)\right)\subset{\bf V}\left({\bf I}\left(W\right)\right)$ because ${\bf I}$ is inclusion-reversing. Finally, ${\bf V}\left({\bf I}\left(W\right)\right)=W$, by Proposition \ref{propVIV} and the fact that $W$ is an affine variety. Thus ${\bf V}\left({\bf I}\left(S\right)\right)\subset W$ for any affine variety $W$ that contains $S$. 
\end{proof}
\end{proposition}

A natural question to ask is whether  ${\bf I}\left({\bf V}\!\left(f_1,\ldots,f_d\right)\right)=\left<f_1,\ldots,f_d\right>$. The answer in general is \emph{no}, however, under algebraically closed fields like $\mathbbm{C}$, there is a precise criterion, called {\it Nullstellensatz}, which allows to check whether the equality occurs. It can be found in Theorem \ref{thmNullstellen} of Section \ref{secelimination}.

A number of other questions come very naturally with the notions of an ideal and an affine variety. Let us give a list of three of them, which will be answered to in the following.  
\begin{enumerate}[1)]\label{lista}
\item Does every ideal in $\mathbbm{K}\left[x_1,\ldots,x_n\right]$ have a finite set of generators? In other words, can we always write $I=\left<f_1,\ldots,f_d\right>$ for some polynomials $f_1,\ldots,f_d$?
\item How can we check whether a given polynomial $f\in\mathbbm{K}\left[x_1,\ldots,x_n\right]$ belongs to an ideal $I$?
\item How can we solve a system of polynomial equations $f_1\left(x_1,\ldots,x_n\right)=0,$ $\ldots,$ $f_d\left(x_1,\ldots,x_n\right)=0$, i.e. find a parametric description of (a part of) the affine variety defined by the equations. Under which conditions solutions do exist at all?
\end{enumerate}
In order to better understand the above questions, it is useful to give a short summary of how they are answered in the univariate case, $\K\left[x\right]$. First of all, let us mention that the leading term of $f=\sum_{i=1}^d\alpha_ix^i\in\K\left[x\right]$ ($\alpha_i\in\K\setminus\left\{0\right\}$) is by definition equal to $\alpha_dx^d$, the leading coefficient is $\alpha_d$ and the leading monomial is $x^d$. Let us denote them by $\textnormal{LT}\left(f\right)$, $\textnormal{LC}\left(f\right)$ and $\textnormal{LM}\left(f\right)$, respectively. Let us also denote the degree of $F$ by $\deg f$. Given two univariate polynomials $f$, $g$, there is a unique way of writing $f$ as
\begin{equation}\label{eqdivision}
f=qg+r
\end{equation}
where $q,r\in\K\left[x\right]$ and either $r=0$ or $\deg r<\deg g$. The classical division algorithm in $\K\left[x\right]$ that produces $q$ and $r$ given $f$ and $g$ consists in the steps given in Figure \ref{figdivision}
\begin{figure}[h!]\centering
\begin{verbatim}
q=0
r=f
while r<>0 and LT(g) divides LT(r) do
    q=r+LT(r)/LT(g)
    r=r-(LT(r)/LT(g))g
\end{verbatim}
\caption{Polynomial division algorithm in $\K\left[x\right]$}\label{figdivision}
\end{figure}

We can now answer question one in the case of univariate polynomials.
\begin{proposition}\label{propfingenuni}
Every ideal in $\mathbbm{K}\left[x\right]$ is generated by a single polynomial $f$, which is the polynomial of lowest degree in $I$.
\begin{proof}
Clearly, there must exist a polynomial of lowest degree in $I$. Let us denote it by $g$. We shall prove that $I=\left<g\right>$. Clearly, $\left<g\right>\subset I$. If there existed a polynomial $f\in\K\left[x\right]\setminus\left<f\right>$, we could divide $f$ by $g$ and produce a polynomial $r$ as in formula \eqref{eqdivision}. Since $f\not\in\left<g\right>$, $r\neq 0$. It would satisfy $\deg r<\deg g$ and $r=f-qg\in I$, which is a contradiction, because we assumed that $g$ is the polynomial of minimal degree in $I$.
\end{proof}
\end{proposition}

Question two also has an immediate answer in the univariate case. Since every ideal in $\K\left[x\right]$ is of the form $\left<g\right>$ for some $g\in\K\left[x\right]$, it is sufficient to divide an arbitrary polynomial $f$ by $g$ to check whether $f$ belongs to the ideal or not. If $r=0$, it belongs to the ideal, and if $r\neq 0$, it does not. As it is well known from basic algebra courses, solutions to univariate polynomial equations of the form $f\left(x\right)=0$ always exist in case of $\K=\mathbbm{C}$ and other algebraically closed fields, but may fail to exist when the base field is not algebraically closed. Explicit general solutions in $\mathbbm{C}$ are only known for $f$ of degree up to $4$ as a consequence of the Abel-Ruffini theorem, cf. e.g. \cite{Fraleigh}. Note that the question, whether solutions exist or not, starts to be non-trivial if we pass to multiple polynomial equations or a multivariate setting, even if the base field is algebraically closed (e.g. when it equals $\mathbbm{C}$). In such case, the techniques of Groebner bases and resultants are of much help. We shall discuss both in subsequent sections of the thesis. 

\section{Monomial orders and Groebner bases}\label{secmonomialorders}

Let us now pass from one-variable polynomials, discussed at the end of the previous subsection, to the multivariate setting. We shall avoid excess notation by using the symbol $x^{\alpha}$ with multi-indices $\left(\alpha_1,\ldots,\alpha_n\right)$ in place of $x_1^{\alpha_1}\cdot\ldots\cdot x_n^{\alpha_n}$. In order to introduce an analogue of the division algorithm in $\K\left[x\right]$, we need to specify what is a leading term of a multivariate polynomial. Unlike for univariate polynomials, the notions of the leading term, leading coefficient or monomial are not uniquely defined. There are many possible choices and one needs to \emph{specify an ordering} of monomial terms in order to do multivariate polynomial division in a sensible way. The orderings also have to respect the multiplicative and additive structure of $\K\left[x_1,\ldots,x_n\right]$, so they fulfill a number of constraints. In such case we call them monomial orderings.
\begin{definition}[Monomial ordering]
A \textbf{monomial ordering} in $\K\left[x_1,\ldots,x_n\right]$ is any relation $>$ on the set of monomials in $\K\left[x_1,\ldots,x_n\right]$ which fulfills
\begin{enumerate}[(i)]
\item the ordering $>$ is linear, which means that for any monomials $x^{\alpha}$ and $x^{\beta}$, $\alpha\neq\beta$, either $x^{\alpha}<x^{\beta}$ or $x^{\alpha}>x^{\beta}$.
\item If $x^{\alpha}>x^{\beta}$ then $x^{\alpha+\gamma}=x^{\alpha}x^{\gamma}>x^{\beta}x^{\gamma}=x^{\beta+\gamma}$ for any multi-index $\gamma$.
\item The relation $>$ is w well-ordering, which means that for any set of monomials $\left\{x^{\alpha}\right\}_{\alpha\in A}$, there exists a smallest element under the ordering $>$.
\end{enumerate}
\end{definition}
In the following, we introduce three most common examples of monomial orderings.
\begin{example}[Lexicographic order]
Let $x^{\alpha}$ and $x^{\beta}$ be monomials in $\K\left[x_1,\ldots,x_n\right]$. We have $x^{\alpha}>_{lex}x^{\beta}$ if and only if $\alpha-\beta$ \textbf{has the left-most nonzero entry positive}.
\end{example}

\begin{example}[Graded lexicographic order]\label{exgradedlex}
Let $x^{\alpha}$ and $x^{\beta}$ be monomials in $\K\left[x_1,\ldots,x_n\right]$. We have $x^{\alpha}>_{grlex}x^{\beta}$ if and only if
\begin{equation}
\left|\alpha\right|=\sum_{i=1}^n\alpha_i>\sum_{j=1}^{n}\beta_j=\left|\beta\right|\textnormal{ or }\left|\alpha\right|=\left|\beta\right|\textnormal{ and }x^{\alpha}>_{lex}x^{\beta},
\end{equation}
where $\left|\alpha\right|$ denotes the total degree of $x^{\alpha}$. In other words $x^{\alpha}>_{grlex}x^{\beta}$ if and only if \textbf{$x^{\alpha}$
has a higher total degree than $x^{\beta}$ or has the same total degree and $x^{\alpha}>_{lex}x^{\beta}$}.
\end{example}

\begin{example}[Graded Reverse Lexicographic Order]\label{exgradedrevlex}
Let $x^{\alpha}$ and $x^{\beta}$ be monomials in $\K\left[x_1,\ldots,x_n\right]$. We have $x^{\alpha}>_{grevlex}x^{\beta}$ if and only if
\begin{equation}
\left|\alpha\right|=\sum_{i=1}^n\alpha_i>\sum_{j=1}^{n}\beta_j=\left|\beta\right|\textnormal{ or }\left|\alpha\right|=\left|\beta\right|,
\end{equation}
and \textbf{in $\alpha-\beta$ the right-most nonzero entry is negative}.
\end{example}

We can now introduce an analogue of the univariate division algorithm in Figure \ref{figdivision}. Let $f$ and $g_1,\ldots,g_d\in\K\left[x_1,\ldots,x_n\right]$ be arbitrary and fix a monomial ordering in $\K\left[x_1,\ldots,x_n\right]$. There exist $q_i\in\K\left[x_1,\ldots,x_n\right]$, $i=1,2,\ldots,d$ and $r\in\K\left[x_1,\ldots,x_n\right]$ such that
\begin{equation}\label{eqdivmulti}
f=\sum_{i=1}^dq_ig_i+r
\end{equation}
and no monomial of $r$ is divisible by any of the leading monomials $\textnormal{LM}\left(g_i\right)$. Moreover, $\textnormal{LM}\left(q_ig_i\right)\leqslant\textnormal{LM}\left(f\right)\,\forall_i$. Obviously, $r$ and $q_i$ in the above formula are analogues of $r$ and $q$ in equation \eqref{eqdivision}, while the condition on monomial terms of $r$ corresponds to $\deg r<\deg g$ in the univariate setting. An algorithm which gives a decomposition of the form \eqref{eqdivmulti} is shown in Figure \ref{figdivision2}. \begin{figure}[h!]
\begin{verbatim}
for i=1 to d do q_i=0
r=0
p=f
while p<>0 do {
      divisionocurred=0
      for j=2 to d do
          if LT(g_i) divides LT(p) do {
             divisionocurred=1
             q_i=g_i+LT(p)/LT(q_i)
             p=p-LT(p)/LT(q_i)
             }
      if not divisionocurred=1 do {
         r=r+LT(p)
         p=p-LT(p)
         }
      }
\end{verbatim}\label{figdivision2}
\caption{A division algorithm in $\K\left[x_1,\ldots,x_n\right]$}
\end{figure}

In short, the algorithm tries to \textbf{divide the leading term of $f$ by the leading terms of $g_i$}, $i=1,\ldots,d$. If this is not possible, the \textbf{leading term is added to the division remainder} and the whole procedure repeated from the beginning. Note that the ordering of the polynomials $g_1,\ldots,g_d$ has an influence on the result of division. In particular, the remainder $r$ \emph{may depend} on how the polynomials $g_1,\ldots,g_d$ are ordered and thus is \emph{not} uniquely defined. The last feature can be seen in the following example
\begin{example}\label{exremnotequal}
Let $g_1=xy+1$, $g_2=y^2-1$, $f=xy^2-x$ and the take the $>_{lex}$ order in $\K\left[x,y\right]$. The multivariate division algorithm gives us
\begin{equation}\label{eqexample51}
xy^2-x=y\cdot\left(xy+1\right)+0\cdot\left(y^2-1\right)+\left(-x-y\right).
\end{equation}
However, with the choice $g_2=xy+1$, $g_1=y^2-1$, $f=xy^2-x$, we get
\begin{equation}\label{eqexample52}
xy^2-x=x\cdot\left(y^2-1\right)+0\cdot\left(xy+1\right)+0
\end{equation}
instead.
\end{example}
We see from \eqref{eqexample51} and \eqref{eqexample52} that the condition $f\in\left<g_1,\ldots,g_d\right>$ is not equivalent to $r=0$. We shall see that with a proper choice of the ideal basis, a \textbf{Groebner basis}, both conditions can be made equivalent and the remainder $r$ ceases to be ordering dependent, though it still depends on the particular monomial order we choose in $\K\left[x_1,\ldots,x_n\right]$. 

First, we need to introduce the notion of monomial ideals and investigate their basic properties. A \textbf{monomial ideal} is simply the ideal generated by a set of monomials in $\K\left[x_1,\ldots,x_n\right]$. More formally, we have the following definition
\begin{definition}[Monomial ideal]
Let $\mathcal{A}$ be a subset of $\mathbbm{Z}^n$ consisting of componentwise nonnegative elements. A monomial ideal corresponding to $\mathcal{A}$ is the smallest  ideal in $\K\left[x_1,\ldots,x_d\right]$ containing $\left\{x^{\alpha}\right\}_{\alpha\in\mathcal{A}}$.
\end{definition}
We shall denote by $\left<x^{\alpha}\right>_{a\in\mathcal{A}}$ the monomial ideal generated by  $\left\{x^{\alpha}\right\}_{\alpha\in\mathcal{A}}$.
It turns out that all monomial ideals admit a finite set of generators. This is the contents of the following \textbf{Dickson's lemma}.
\begin{lemma}[Dickson's]\label{lemmaDicksons}
Let $I=\left<x^{\alpha}\right>_{\alpha\in\mathcal{A}}$ be a monomial ideal. There exists a finite set $\alpha_1,\ldots,\alpha_d\in\mathcal{A}$ such that $I=\left<x^{\alpha_1},\ldots,x^{\alpha_d}\right>$
\begin{proof}
Can be found in algebraic geometry textbooks like \cite{IdealsVarieties}.
\end{proof}
\end{lemma}
With the Dickson's lemma at hand, one can prove a key theorem about ideals in $\K\left[x_1,\ldots,x_n\right]$.
\begin{theorem}[Hilbert basis theorem]\label{thmHilbertbasis}
Every ideal $I\subset\K\left[x_1,\ldots,x_n\right]$ is finitely generated. Thus, there exist $g_1,\ldots,g_d\in I$ such that $I=\left<g_1,\ldots,g_d\right>$. In particular, every $g_1,\ldots,g_d$ with the property $\left<\textnormal{LT}\left(f\right)\right>_{f\in I}=\left<\textnormal{LT}\left(g_1\right),\ldots,\textnormal{LT}\left(g_d\right)\right>$ form an admissible set of generators of $I$.
\begin{proof}
Consider the monomial ideal  $J=\left<\textnormal{LT}\left(f\right)\right>_{f\in I}$. According to Dickson's lemma, there exist a finite set of generators of $J$, which are necessarily of the form $\textnormal{LT}\left(g_1\right),\ldots\textnormal{LT}\left(g_d\right)$. We shall prove that $g_1,\ldots,g_d$ generate $I$. If that was not the case, there would exist $f\in\K\left[x_1,\ldots,x_n\right]\setminus\left<g_1,\ldots g_d\right>$. Let us divide $f$ by $g_1,\ldots,g_d$ using the algorithm given in Figure \ref{figdivision2}. It necessarily gives us $f=\sum_iq_ig_i+r$ with $r\neq 0$ by our assumption that $f$ is not in $\left<g_1,\ldots g_d\right>$. However, $r=f-\sum_iq_ig_i$ is an element of $f$ and thus $\textnormal{LT}\left(f\right)$ an element of $J$. It must therefore be divisible by one of the generators of $J$. That is, it must be divisible by one of the $\textnormal{LT}\left(g_i\right)$, which is a contradiction because of the properties of the remainder $r$ on division by $g_1,\ldots,g_d$.
\end{proof} 
\end{theorem} 
We can also prove the following useful result
\begin{corollary}[Ascending chain condition]\label{corACC}
Let $I_1\subset I_2\subset I_3\subset\ldots$ be a sequence of ideals in $\K\left[x_1,\ldots,x_n\right]$. The sequence stabilizes for some finite $i$, i.e. $I_i=I_{i+n}$ for all $n\geqslant 0$.
\begin{proof}
It is easy to check that the set $I:=\bigcup_{j=1}^{+\infty}I_j$ is an ideal in $\K\left[x_1,\ldots,x_n\right]$. According to the above theorem, there exists a finite basis $g_1,\ldots,g_d$ of $I$. According to the definition of $I$, we must have $g_j\in I_{i_j}$ for some $i_j\geqslant 1$. Let us choose $i=\max\left\{i_1,\ldots,i_d\right\}$. Since $I=\left<g_1,\ldots,g_d\right>$ and $g_j\in I_i$ for all $i=1,\ldots,d$, we clearly see that $I=\bigcup_{j=1}^iI_j$. Thus, $I_k=I_i$ for all $k\geqslant i$.
\end{proof}
\end{corollary}

In the spirit of Theorem \ref{thmHilbertbasis}, a \textbf{Groebner basis} is defined as a finite subset $g_1,\ldots, g_d$ of an ideal $I$ with the property that $\left<\textnormal{LT}\left(g_1\right),\ldots,\textnormal{LT}\left(g_d\right)\right>=\left<\textnormal{LT}\left(f\right)\right>_{f\in I}$.
\begin{definition}[Groebner basis]
Let $I\subset\K\left[x_1,\ldots,x_n\right]$ be an ideal. A Groebner basis of $I$ is a finite subset $g_1,\ldots,g_d\in I$ such that $\left<\textnormal{LT}\left(g_1\right),\ldots,\textnormal{LT}\left(g_d\right)\right>=\left<\textnormal{LT}\left(f\right)\right>_{f\in I}$. In other words, a Groebner basis is a finite set of polynomials in $I$ with the property that their leading terms generate the ideal of leading terms of polynomials in $I$.
\end{definition}
Let us list a few properties of Groebner bases.
\begin{enumerate}[1)]
	\item A Groebner basis of an ideal $I$ generates $I$. In other words, it is a \textbf{basis of the ideal in the usual sense},
	\item There exists a Groebner basis an an arbitrary ideal $I\subset\K\left[x_1,\ldots,x_n\right]$,
	\item The remainder of $f\in\K\left[x_1,\ldots,x_n\right]$ on division by a Groebner basis $g_1,\ldots,g_d\in I$ is \textbf{uniquely defined}.
\end{enumerate}
Points one and two follow directly from the proof of Theorem \ref{thmHilbertbasis}. We shall give a more formal version of point three in the following proposition \cite{IdealsVarieties}.
\begin{proposition}\label{proprunique}
Let $G=\left\{g_1,\ldots,g_d\right\}$ be a Groebner basis for an ideal $I\subset\K\left[x_1,\ldots,x_n\right]$ and let $f\in\K\left[x_1,\ldots,x_n\right]$. Then there is a unique $r\in\K\left[x_1,\ldots,x_n\right]$ with the following two properties
\begin{enumerate}[(i)]
\item No term of $r$ is divisible by any of $\textnormal{LT}\left(g_1\right),\ldots,\textnormal{LT}\left(g_d\right)$,
\item There is $g\in I$ such that $f=g+r$.
\end{enumerate}
In particular, the polynomial $r$ is the remainder on division of $f$ by $G$, no matter how the elements of $G$ are listed when using the division algorithm.
\begin{proof}
An $r$ with the properties $(i)$ and $(ii)$ can be obtained using the division algorithm shown in Figure \ref{figdivision2}. Let us prove the uniqueness of $r$. Assume, on the contrary, that for some $f\in I$, $f=g+r=g'+r'$ where $r'\neq r$ and both $\left(g, r\right)$ and $\left(g',r'\right)$ satisfy $(i)$ and $(ii)$. Thus $r-r'=g'-g$ is an element of $I$ with $\textnormal{LT}\left(r-r'\right)\neq 0$. By the definition of a Groebner basis and $r-r'\in I$, the leading term must be divisible by some $\textnormal{LT}\left(g_i\right)$, $i=1,\ldots,d$, which is a contradiction, because by $(i)$, the monomials of $r$ and $r'$ are not divisible by any $\textnormal{LT}\left(g_i\right)$.   
\end{proof}
\end{proposition}
Note that by now, we already have an answer to the ideal membership question (number two) raised on page \pageref{lista}. Provided a Groebner basis $G$, we simply divide $f$ by $G$ using the division algorithm of Figure \ref{figdivision2} and check whether $r=0$ or not. Let us state this as a proposition.
\begin{proposition}\label{propmembership}
Let $G$ be a Groebner basis of an ideal $I\subset\K\left[x_1,\ldots,x_n\right]$. A polynomial $f\in\K\left[x_1,\ldots,x_n\right]$ belongs to $I$ if and only if the remainder of $f$ on division by $G$ equals $0$.
\begin{proof}
If the remainder is zero, we clearly have $f=\sum_iq_ig_i\in I$. On the other hand, assume that $f$ is an element of $I$ and $r\neq 0$. In such case, $\textnormal{LT}\left(r\right)\neq 0$ and $\textnormal{LT}\left(r\right)\neq\left<\textnormal{LT}\left(g_1\right),\ldots,\textnormal{LT}\left(g_d\right)\right>$
\end{proof}
\end{proposition}
We will see shortly that a Groebner basis of an ideal can be found by \textbf{Buchberger's algorithm} \cite{Buchberger} in a finite number of steps. Thus the ideal membership problem can also be solved in a finite number of steps by calculating the remainder of $f$ on division by a Groebner basis. For future convenience, let us denote such remainder by $f^{G}$. 

In the light of the above developments, it is important to know which bases of an ideal are Groebner bases, and how to find a Groebner basis of a given ideal, possibly of a nice form and unique in some sense. Fortunately, there exist simple answers to all these questions and we shall explain them in the following. First, we introduce the notion of so-called \textbf{$\bf S$-polynomial} \cite{Buchberger}.
\begin{definition}[$S$-polynomial]
Given two polynomials $f,g\in\K\left[x_1,\ldots,x_n\right]$ and some monomial order $>$, take $x^{\alpha}=\textnormal{LM}\left(f\right)$ and $x^{\beta}=\textnormal{LM}\left(g\right)$. The $S$-polynomial of $f$ and $g$ is defined to be
\begin{equation}\label{eqdefSpoly}
S\left(f,g\right):=\frac{x^{\gamma}}{\textnormal{LT}\left(f\right)}f-\frac{x^{\gamma}}{\textnormal{LT}\left(g\right)}g,
\end{equation}
where $\gamma$ is a multi-index $\left(\gamma_1,\ldots,\gamma_n\right)$ defined such that $\gamma_i=\max\left(\alpha_i,\beta_i\right)$ for $\alpha=\left(\alpha_1,\ldots,\alpha_n\right)$ and $\beta=\left(\beta_1,\ldots,\beta_n\right)$
\end{definition}
The $S$-polynomial is defined such that a cancellation of leading terms of $f$ and $g$ occurs, and some new leading terms can possibly be produced.

Let us now state without a proof a key result of Groebner basis theory, called Buchberger's \textbf{$\bf S$-pair criterion} \cite{Buchberger}.
\begin{theorem}[$S$-pair criterion]\label{thmSpaircrit}
A basis $F=\left(f_1,\ldots,f_d\right)$ of an ideal $I\subset\K\left[x_1,\ldots,x_n\right]$ is a Groebner basis of $I$ if and only if
\begin{equation}\label{eqSpaircrit}
S\left(f_i,f_j\right)^{F}=0\,\forall_{i,j\in\left\{1,\ldots,d\right\}}
\end{equation}
\end{theorem}
The above criterion suggests an algorithm how to find a Groebner basis of an ideal $I$, given a set of generators $f_1,\ldots,f_d$.  If we calculate all the possible remainders $S\left(f_i,f_j\right)^{\left(f_1,f_2,\ldots\right)}$ and some of them turn out to be nonzero, we add them to $\left(f_1,f_2,\ldots\right)$ and repeat the whole procedure for the extended set of generators. At some point, this extension procedure should terminate, and the $S$-pair criterion tells us that we have obtained a Groebner basis of the ideal $I$. A more precise description of the algorithm is shown in Figure \ref{figSalgorithm}. We also state its correctness as a separate theorem.
\begin{figure}[t]
\begin{verbatim}
l=d; m=d
for i=1 to l do g_i=f_i
repeat {
   l=m
    for i=1 to l do
        for j=1 to l do {
            if (r=Remainder(Spolynomial(g_i,g_j),{g_1,...,g_l}))<>0
               do {
               Append({g_1,g_2,...,g_l},r)
               m=l+1
               }
            }   
} until l=m   
\end{verbatim}
\caption{A rudimentary algorithm for calculation of a Groebner basis of an ideal $\left<f_1,\ldots,f_d\right>$\label{figSalgorithm}, given here according to \cite{IdealsVarieties}.}
\end{figure}

\begin{theorem}[Buchberger's algorithm]\label{thmBuchbergeralgorithm}
The algorithm given in Figure \ref{figSalgorithm} returns a Groebner basis $\left<g_1,\ldots,g_r\right>$ of the ideal $I=\left<f_1,\ldots,f_d\right>$ in a finite number of steps
\begin{proof}
The additional elements $g_i$, $i>d$, produced by the algorithm, belong to $I$. This follows inductively because at each step the $S$-polynomials $S\left(g_i,g_j\right)$ and their remainders $S\left(g_i,g_j\right)^{\left(g_1,g_2,\ldots\right)}$ belong to the same ideal as $\left(g_1,g_2,\ldots\right)$ do. Moreover, the algorithm terminates if and only if at some point all the remainders  $S\left(g_i,g_j\right)^{\left(g_1,g_2,\ldots\right)}$ vanish, which is equivalent to say, by Theorem \ref{thmSpaircrit}, that the set $\left(g_1,g_2,\ldots\right)$ is a Groebner basis of the ideal $I$. Thus, we only need to show that the algorithm terminates. This will be done with help of the ascending chain condition, Corollary \ref{corACC}.

Let us try to assume that a sequence $g_1,g_2,\ldots,g_d,g_{d+1},\ldots$ produced by the algorithm does not terminate. We have a corresponding sequence of ideals
\begin{eqnarray}
&\left<\textnormal{LT}\left(g_1\right),\textnormal{LT}\left(g_2\right),\ldots,\textnormal{LT}\left(g_d\right)\right>,\label{eqI1}\\
&\left<\textnormal{LT}\left(g_1\right),\textnormal{LT}\left(g_2\right),\ldots,\textnormal{LT}\left(g_d\right),\textnormal{LT}\left(g_{d+1}\right)\right>,\label{eqI2}\\
&\ldots\nonumber\\
&\left<\textnormal{LT}\left(g_1\right),\textnormal{LT}\left(g_2\right),\ldots,\textnormal{LT}\left(g_d\right),\textnormal{LT}\left(g_{d+1}\right),\textnormal{LT}\left(g_{d+2}\right)\right>,\label{eqI3}\\
&\ldots\nonumber
\end{eqnarray}
which must stabilize according to Corollary \ref{corACC}. However, the algorithm in Figure \ref{figSalgorithm} works is such a way that whenever an element $g_{l+1}$ is added to a sequence $g_1,g_2,\ldots,g_d,g_{d+1},\ldots,g_l$, its leading term $\textnormal{LT}\left(g_{l+1}\right)$ is not divisible by any of the leading terms $\textnormal{LT}\left(g_1\right),\textnormal{LT}\left(g_2\right),\ldots,\textnormal{LT}\left(g_d\right),\textnormal{LT}\left(g_{d+1}\right),\ldots,\textnormal{LT}\left(g_l\right)$. Thus 
\begin{multline}
\left<\textnormal{LT}\left(g_1\right),\textnormal{LT}\left(g_2\right),\ldots,\textnormal{LT}\left(g_d\right),\textnormal{LT}\left(g_{d+1}\right),\ldots,\textnormal{LT}\left(g_{l}\right)\right>\neq\\
\neq\left<\textnormal{LT}\left(g_1\right),\textnormal{LT}\left(g_2\right),\ldots,\textnormal{LT}\left(g_d\right),\textnormal{LT}\left(g_{d+1}\right),\ldots,\textnormal{LT}\left(g_{l}\right),\textnormal{LT}\left(g_{l+1}\right)\right>\label{ineqI}
\end{multline}
for all $l\geqslant d$. This shows that \eqref{eqI1}-\eqref{eqI3} forms a strictly increasing sequence of ideals in $\K\left[x_1,\ldots,x_n\right]$, which is impossible according to the ascending chain condition. The only possible solution is that the algorithm always terminates, so that it never produces an infinite sequence of polynomials $g_1,g_2,\ldots,g_d,g_{d+1},\ldots$
\end{proof}
\end{theorem}
The Groebner bases obtained by the algorithm in Figure \ref{figSalgorithm} are not optimal in many respects. First of all, different bases can be obtained, depending on the choice of the order of the inputs $f_1,\ldots,f_d$. Moreover, it may happen that a polynomial $g$ in the output sequence $G=\left\{g_1,\ldots,g_l\right\}$ has a leading term $\textnormal{LT}\left(g\right)$ which divisible by some of the leading terms of the polynomials in $G\setminus\left\{g\right\}$. In such case $G\setminus\left\{g\right\}$ is another Groebner basis of the ideal $\left<f_1,\ldots,f_d\right>$ with a smaller number of elements. A Groebner basis where no such reduction is possible and all the leading coefficients are equal to unity, is called a \textbf{minimal Groebner basis}.
\begin{definition}[Minimal Groebner basis]
A Groebner basis $G=\left\{g_1,\ldots,g_l\right\}$ of an ideal in $\K\left[x_1,\ldots,x_n\right]$ with $\textnormal{LC}\left(g_i\right)=1\,\forall_i$ is called \textbf{minimal} if and only if for any element $g\in G$, the leading term $\textnormal{LT}\left(g\right)$ is not divisible by any of the leading terms of the polynomials in $G\setminus\left\{g\right\}$.
\end{definition}
Clearly, a minimal Groebner basis of an ideal $I$ can be obtained from an arbitrary Groebner basis $G$ of $I$ by first normalizing the leading terms and then removing all the elements which have their leading term divisible by the leading term of some other polynomial in $G$. It can be proved \cite{IdealsVarieties} that the minimal Groebner bases of an ideal $I$ have identical sets of leading coefficients, however there usually exist multiple minimal Groebner bases of a given ideal $I$. This ambiguity can be entirely removed if we impose one further condition on the Groebner basis we are looking for.
\begin{definition}[Reduced Groebner basis]
A minimal Groebner basis $G$ of an ideal $I$ is called \textbf{reduced} if and only if for all $g\in G$, no monomial of $g$ is divisible by any of the leading terms of polynomials in $G\setminus\left\{g\right\}$.
\end{definition}
With this definition, we have
\begin{proposition}\label{thmreducedGB}
There exists a \textbf{unique reduced Groebner basis} of any ideal $I\subset\K\left[x_1,\ldots,x_n\right]$. Moreover, given a minimal Groebner basis $G'$ of $I$, the reduced Groebner basis $G$ can be found by the following procedure
\begin{verbatim}
for all g in G' do
      g=Remainder(g,G\{g})
\end{verbatim}
\begin{proof}
Can be found in \cite{IdealsVarieties}.
\end{proof}
\end{proposition}
Let us mention that Proposition \ref{thmreducedGB} provides one with an algorithmic way to solve the \textbf{ideal equality problem}. Given two ideals $I=\left<f_1,\ldots,f_d\right>$ and $J=\left<e_1,\ldots,e_c\right>$, one has the equality $I=J$ if and only if the corresponding reduced Groebner bases, which can be computed in a finite number of steps, are equal.

We see that Groebner bases allow us to answer a number of questions, including the ideal membership and ideal equality problems. Moreover, it turns out that they can be used to find solutions to sets of polynomial equations, which is very interesting from a practical perspective and it will turn out to be crucial in some parts of the thesis. A simplest way to see how Groebner bases can be used for this new task is to look into a concrete example. Consider the following set of polynomial equations:
\begin{eqnarray}\label{eqsetofpolyeqs1}
x y^2-z=0,\\
x z+y^2=0\label{eqsetofpolyeqs2},\\
x y-1=0\label{eqsetofpolyeqs3}.
\end{eqnarray}   
A Groebner basis calculation using the lexicographic order with $x>y>z$ for the ideal $\left<x y^2-z,x z+y^2,x y-1\right>$  provides us with $\left\{1+z^2,y-z,x+z\right\}$. Since these polynomials generate the same ideal as the polynomials in \eqref{eqsetofpolyeqs1}-\eqref{eqsetofpolyeqs3}, we have an equivalent set of equations:
 \begin{equation}
  z^2+1=0,\label{eqequiv1}\quad
 y-z=0,\quad 
 x+z=0.
 \end{equation}
 The equations \eqref{eqsetofpolyeqs1}-\eqref{eqsetofpolyeqs3} do not look much more complicated than those in \eqref{eqequiv1}, but at a first glance, it is not clear  how to solve them. On the other hand, the first equation in \eqref{eqequiv1} involves only the variable $z$ and it clearly has only two solutions, $z=\pm i$. The solutions hence obtained can later be substituted for $z$ in the latter two equations in \eqref{eqequiv1}. In this way, one can determine the corresponding values of $x$ and $y$ and find all solutions to the initial set of polynomial equations. Our aim in the following will be to explain that a similar phenomenon occurs in general when the lexicographical ordering of monomials is used for the calculation of Groebner bases.
 
\section{Elimination ideals}\label{secelimination} 
For a given ideal $I=\left<f_1,\ldots,f_d\right>$, we define the $k$-th \textbf{elimination ideal} $I_k$ as the intersection $I\cap\K\left[x_{k+1},\ldots,x_n\right]$. In other words, we pick up all polynomials in $I$ that involve only the variables $x_{k+1},\ldots,x_n$, or equivalently, they do not involve $x_1,\ldots,x_k$. In the simple example discussed above, we clearly had $z^2+1\in I_2$. The following theorem tells us that Groebner bases calculated with respect to a lexicographical order provide us with much information about elimination ideals.
\begin{theorem}\label{thmelimination}
Let $I\subset\K\left[x_1,\ldots,x_n\right]$ be an ideal with a Groebner basis $G$ with respect to the lexicographical order where $x_1>x_2>\ldots>x_n$. Then, for every $k=1,2,\ldots,n$ the set
\begin{equation}
G_k=G\cap\K\left[x_1,\ldots,x_n\right]
\end{equation}
is a Groebner basis of the $k$-th elimination ideal $I_k$.
\begin{proof}
By construction of $G_k$ and $I_k$, we have the inclusion $G_k\subset I_k$. It suffices to show that the monomial ideal $\left<\textnormal{LT}\left(f\right)\right>_{f\in I_k}$ of leading terms of $I_k$ is generated by $\left<\textnormal{LT}\left(g\right)\right>_{g\in G_k}$. For every $f\in I_k$, the leading term $\textnormal{LT}\left(f\right)$ is a polynomial in the variables $x_{k+1},\ldots,x_n$ only. Since $G$ is a Groebner basis of $I$, there must exist a $g$ in $G$ such that $\textnormal{LT}\left(g\right)$ divides $\textnormal{LT}\left(f\right)$, and the leading term $\textnormal{LT}\left(g\right)$ must necessarily be a monomial in $x_{k+1},\ldots,x_n$. Because we are using lexicographical order with $x_1>x_2>\ldots>x_n$, all the other monomials of $g$ do not involve the variables $x_1,\ldots,x_k$. Hence $g$ is a polynomial in $x_{k+1},\ldots,x_n$, $g\in G_k$. 
\end{proof}
\end{theorem} 
The importance of elimination ideals was obvious in the simple example we discussed above, where $I_2=\left<z^2+1\right>$, and it generally follows from their relation to projections of affine varieties in $\K^n$ onto ``axes'' in the high dimensional space. In terms of solving polynomial equations, we obtain partial solutions in a smaller number of variables and try to extend them to a full solution. More formally, we define the $k$-th projection map $\pi_k$ by the formula
\begin{equation}\label{eqdefpik}
\pi_k:\K^n\ni\left(x_1,\ldots,x_n\right)\mapsto\left(x_{k+1},\ldots,x_n\right)\in\K^{n-k}.
\end{equation}
We have the following
\begin{proposition}\label{propinclusion}
Let $I=\left<f_1,\ldots,f_d\right>$ be an ideal in $\K\left[x_1,\ldots,x_n\right]$. Let ${\bf V}\left(I\right)={\bf V}\left(f_1,\ldots,f_n\right)$ be the corresponding affine variety. We have
\begin{equation}
\pi_k\left({\bf V}\left(I\right)\right)\subset{\bf V}\left(I_k\right),
\end{equation}
where ${\bf V}\left(I_k\right)$ is the affine variety corresponding to the $k$-th elimination ideal $I_k$.
\begin{proof}
We want to show that $f\left(\pi_k\left(x_1,\ldots,x_k\right)\right)=0$ for all $\left(x_1,\ldots,x_k\right)\in{\bf V}\left(I\right)$ and $f\in I_k$. Since $f\in I$, we have $f\left(x_1,\ldots,x_n\right)=0$. But $f$ involves only the variables $x_{k+1},\ldots,x_n$, which gives us $f\left(x_1,\ldots,x_n\right)=f\left(x_{k+1},\ldots,x_n\right)=f\left(\pi_k\left(x_1,\ldots,x_n\right)\right)=0$. 
\end{proof}
\end{proposition}
The above proposition, although simple, tells us something important about ${\bf V}\left(I\right)$. A projection of ${\bf V}\left(I\right)$ onto $\K^{n-k}$ is contained in the affine variety  ${\bf V}\left(I_k\right)$, which is sometimes possible to determine explicitly, as in the case of $I_2=\left<z^2+1\right>$ discussed above. In this way, ${\bf V}\left(I_k\right)$ can be regarded as an easily computable approximation of $\pi_k\left({\bf V}\left(f_1,\ldots,f_d\right)\right)$. To make the statement more precise, we need some extra knowledge. Let us start with the following theorem.
\begin{theorem}[The Weak Nullstellensatz]\label{thmweakHilbertNull}
Let $\K$ be an algebraically closed field and let $I\subset\K\left[x_1,\ldots,x_n\right]$ be an ideal satisfying ${\bf V}\left(I\right)=\emptyset$. Then $I=\K\left[x_1,\ldots,x_n\right]$.
\begin{proof}
Can be found in algebraic geometry textbooks like \cite{IdealsVarieties} or \cite{Harris}.
\end{proof}
\end{theorem}
Intuitively speaking, the Weak Nullstellensatz asserts that the variety ${\bf V}\left(I\right)$ corresponding to an ideal $I\subset\K\left[x_1,\ldots,x_n\right]$ is an empty set if and only if $I$ contains all polynomials in $\K\left[x_1,\ldots,x_n\right]$. Thus, a set of polynomial equations $f_1=0,\ldots,f_d=0$ has no solutions in $\K^n$ if and only if the ideal generated by $f_1,\ldots,f_n$ is the whole $\K\left[x_1,\ldots,x_n\right]$.

Let us point out that the Weak Nullstellensatz allows us to answer the important question about the existence of solutions to systems of polynomial equations. We have the following
\begin{proposition}[Consistency condition]\label{propexistenceofsolutions} 
Let $f_1,\ldots,f_d$ be a set of polynomials in $\K\left[x_1,\ldots,x_n\right]$ over an \emph{algebraically closed} field $\K$. The system of equations
\begin{equation}\label{eqsetofeqs}
f_1\left(x_1,\ldots,x_n\right)=0,\quad f_2\left(x_1,\ldots,x_n\right)=0,\,\ldots\,,f_d\left(x_1,\ldots,x_n\right)=0
\end{equation}
has no solution in $\K^n$ if and only if the reduced Groebner basis of $\left<f_1,\ldots,f_d\right>$ with respect to some monomial order \textbf{equals $\bf\left\{1\right\}$}. In such case we say that the system \eqref{eqsetofeqs} is \textbf{inconsistent}.
\begin{proof}
If a Groebner basis of $\left<f_1,\ldots,f_d\right>$ equals $\left\{1\right\}$, then clearly the set of equations \eqref{eqsetofeqs} have no solutions in $\K$. Conversely, if ${\bf V}\left(f_1,\ldots,f_d\right)$ is the empty set, by the Weak Nullstellensatz we know that $\left<f_1,\ldots,f_d\right>=\K\left[x_1,\ldots,x_n\right]$. By Proposition \ref{thmreducedGB}, there is a unique reduced Groebner basis of $\left<f_1,\ldots,f_d\right>$. Since $\left\{1\right\}$ is the reduced Groebner basis of $\K\left[x_1,\ldots,x_n\right]$, it must be the reduced Groebner basis of $\left<f_1,\ldots,f_d\right>$.
\end{proof}
\end{proposition}
Note that the above proposition provides us with an algorithmic way to check consistency of a set of polynomial equations $f\left(x_1,\ldots,x_n\right)=0,\ldots,f\left(x_1,\ldots,x_n\right)=0$ over an algebraically closed field $\K$. We simply calculate the reduced Groebner basis of the ideal $\left<f_1,\ldots,f_d\right>$ and check whether it equals $\left\{1\right\}$ or not. If so, the system of equations is inconsistent. Otherwise, there exists a solution in $\K^n$.

By a clever trick, the Weak Nullstellensatz is equivalent to the following much celebrated result 
\begin{theorem}[Hilbert's Nullstellensatz]\label{thmHilbertNullstellen}
Let $\K$ be an algebraically closed field. Consider $f_1,\ldots,f_d\in\K\left[x_1,\ldots,x_d\right]$. If $f$ is a polynomial that vanishes on ${\bf V}\left(f_1,\ldots,f_d\right)$, then there exists $m\geqslant 1$ such that
\begin{equation}\label{eqfm}
f^m\in\left<f_1,\ldots,f_d\right>
\end{equation}
In other words, if $f\in{\bf I}\left({\bf V}\left(f_1,\ldots,f_d\right)\right)$, then the inclusion \eqref{eqfm} holds for some $m\geqslant 1$.
\begin{proof}
Consider the ideal 
\begin{equation}
\tilde I=\left<f_1,\ldots,f_d,1-yf\right>\subset\K\left[x_1,\ldots,x_n,y\right]
\end{equation}
where $f,f_1,\ldots,f_d$ are as above. It is not difficult to check that ${\bf V}\left(\tilde I\right)=\emptyset$. It is so because $f$ vanishes whenever $f_1=f_2=\ldots=f_d=0$, and hence $1-yf=1\neq 0$ in such case. By the Weak Nullstellensatz, we have $1\in\tilde I$. Therefore
\begin{equation}\label{equnitypq}
1=\sum_{i=1}^dp_i f_i+q\left(1-yf\right)
\end{equation}
for some polynomials $p_i,q\in\K\left[x_1,\ldots,x_n,y\right]$. Now set $y\rightarrow 1/f\left(x_1,\ldots,x_n\right)$. The relation \eqref{equnitypq} implies that
\begin{equation}\label{eqsumunity}
1=\sum_{i=1}^dp_i\left(x_1,\ldots,x_n,1/f\right)f_i.
\end{equation}
If we multiply both sides of \eqref{eqsumunity} by $f^m$, where $m$ is chosen sufficiently large to clear all the denominators, we get
\begin{equation}
f^m=\sum_{i=1}^sA_if_i
\end{equation}
for some polynomials $A_i\in\K\left[x_1,\ldots,x_n\right]$. Thus $f^m\in\left<f_1,\ldots,f_d\right>$.
\end{proof}
\end{theorem}
Another way to formulate the Hilbert's Nullstellensatz is by means of \textbf{radicals}.
\begin{definition}
Let $I\subset\K\left[x_1,\ldots,x_n\right]$ be an ideal. The \textbf{radical} of $I$, denoted by $\sqrt{I}$, is the set
\begin{equation}\label{eqdefradical}
\sqrt{I}=\left\{f\,\vline\,\exists_{m\geqslant 1}f^m\in I \right\}
\end{equation}
\end{definition}
We leave it as an exercise for the reader to prove that $\sqrt{I}$ is an ideal and $\sqrt{I}=\sqrt{\sqrt{I}}$. We call an ideal $J$ with the property $J=\sqrt{J}$ a \textbf{radical ideal}. Thus, $\sqrt{I}$ is a radical ideal. We can now formulate a version of Theorem \ref{thmHilbertNullstellen}, often simply called \textit{the Nullstellensatz}.
\begin{theorem}[The Nullstellensatz] \label{thmNullstellen}
Let $\K$ be an algebraically closed field. If $I$ is an ideal in $\K\left[x_1,\ldots,x_n\right]$, then
\begin{equation}\label{eqNullstellensatz}
{\bf I}\left({\bf V}\left(I\right)\right)=\sqrt{I}
\end{equation}
\begin{proof}
We certainly have $\sqrt{I}\subset{\bf I}\left({\bf V}\left(I\right)\right)$ because $f\in\sqrt{I}$ implies that $f^m\in I$. Therefore $f^m=0=f$ on ${\bf V}\left(I\right)$. Conversely, suppose that $f\in{\bf I}\left({\bf V}\left(I\right)\right)$. By Hilbert's Nullstellensatz, there exists an integer $m\geqslant 1$ such that $f^m\in I$. This means that $f\in\sqrt{I}$.
\end{proof}
\end{theorem}
With the help of Proposition \ref{propVIS} and the above results, we can now specify what we meant by saying that ${\bf V}\left(I_k\right)$ is an \emph{approximation} of the projection $\pi_k\left({\bf V}\left(I\right)\right)$.
\begin{theorem}\label{thmclosure}
Let $I=\left<f_1,\ldots,f_d\right>$ be an ideal in $\K\left[x_1,\ldots,x_n\right]$ and ${\bf V}\left(I\right)$ the corresponding affine variety. Let $I_k$ be the $k$-th elimination ideal of $I$. Then ${\bf V}\left(I_k\right)$ is the smallest affine variety containing $\pi_k\left({\bf V}\left(I\right)\right)$.
\begin{proof}
In view of Proposition \ref{propVIS}, we must show that ${\bf V}\left(I_k\right)={\bf V}\left({\bf I}\left(\pi_k\left(V\right)\right)\right)$. By Proposition \ref{propinclusion}, we have $\pi_k\left(V\right)\subset{\bf V}\left(I_k\right)$. Since ${\bf V}\left({\bf I}\left(\pi_k\left(V\right)\right)\right)$ is the smallest variety containing $\pi_k\left(V\right)$, it follows that ${\bf V}\left({\bf I}\left(\pi_k\left(V\right)\right)\right)\subset{\bf V}\left(I_k\right)$. 

On the other hand, let $f$ be an element of ${\bf I}\left(\pi_k\left(V\right)\right)$, thus a polynomial in $x_{k+1},\ldots,x_n$ that vanishes on $\pi_k\left(V\right)$. When considered as an element of $\K\left[x_1,\ldots,x_n\right]$, $f$ certainly vanishes on all of $V={\bf V}\left(f_1,\ldots,f_d\right)$. By the Nullstellensatz, $f^m\in\left<f_1,\ldots,f_d\right>$ for some $m\geqslant 0$. Since $f$ does not involve variables $x_1,\ldots,x_k$, $f^m$ does not either. As a consequence, $f^m$ is in the $k$-th elimination ideal $I_k$. This implies that $f\in\sqrt{I_k}$. The inclusion is true for any $f\in{\bf I}\left(\pi_k\left(V\right)\right)$, so ${\bf I}\left(\pi_k\left(V\right)\right)\subset\sqrt{I_k}$. Consequently ${\bf V}\left(I_k\right)={\bf V}\left(\sqrt{I_k}\right)\subset{\bf V}\left({\bf I}\left(\pi_k\left(V\right)\right)\right)$, where we used the fact that ${\bf V}:I\mapsto{\bf V}\left(I\right)$ is inclusion-reversing, as well as the equality ${\bf V}\left(I\right)={\bf V}\left(\sqrt{I}\right)$.
\end{proof}
\end{theorem}
The above theorem tells us that the variety ${\bf V}\left(I_k\right)$ corresponding to the $k$-th elimination ideal gives us the best approximation,  among all varieties in $\K^{n-k}$, of a projection of ${\bf V}\left(I\right)$ onto $\K^{n-k}$. Therefore elimination ideals should be expected to be helpful in solving systems of polynomial equations.

\chapter{A little intersection theory}\,\label{chintersection}
\section{Dimension and degree of a variety}\label{secdimdeg}
In the present section, we are going to introduce two basic properties of algebraic varieties, which are their \textbf{dimension} and \textbf{degree}. Before we do so, we need to introduce a distinction between projective and affine varieties, which has not yet appeared in our introduction to algebraic geometry. First, however, it is necessary to define the notion of a \textbf{projective space}. Note that we choose to work with the set of complex numbers, $\setC$, and polynomials with complex coefficients, $\setC\left[x_1,\ldots,x_n\right]$, but we could as well have chosen a different field of scalars.
\begin{definition}[Complex projective space]\label{defprojectivespace}
Let $n$ be a positive integer. The projective space $\mathbbm{P}^{n-1}$ equals the set of equivalence classes of $\setC^n\setminus\left\{0\right\}$ under the equivalence relation
\begin{equation}\label{eqprojective}
\left(x_1,\ldots,x_n\right)\sim\left(x'_1,\ldots,x'_n\right)\Leftrightarrow\exists_{z\in\setC\setminus\left\{0\right\}}\left(x_1,\ldots,x_n\right)=z\left(x'_1,\ldots,x'_n\right)
\end{equation}
\end{definition}
The elements of $\mathbbm{P}^{n-1}$ are often written simply as $\left[X_1,\ldots,X_n\right]$, where an element $\left(X_1,\ldots,X_n\right)\in\setC^n$ of an equivalence class is conveniently identified with the class itself, however the square brackets and capital letters indicate that we are dealing with the projective space. The variables $X_1,\ldots,X_n$ are called \textbf{homogeneous coordinates} in $\mathbbm{P}^{n-1}$. This is easy to understand if we notice that, given a set of homogeneous polynomials $h_1,\ldots,h_k\in\setC\left[x_1,\ldots,x_n\right]$, we may naturally identify the corresponding variety 
${\bf V}\left(h_1,\ldots,h_k\right)$
 with a subset of $\mathbbm{P}^{n-1}$ and write it as  $\left\{\left[X_1,\ldots,X_n\right]\vline\forall_ih\left(X_1,\ldots,X_n\right)=0\right\}$. We call such subsets projective varieties for obvious reasons, and we do not specify whether they belong to $\setC^n$ or $\mathbbm{P}^{n-1}$ as long as this is not necessary. More general varieties in $\setC^n$, not necessarily defined by the vanishing of a set of homogeneous polynomials, are called \textbf{affine varieties}, in accordance with Definition \ref{defaffine}.
\begin{definition}[Projective variety]\label{defprojvar}
Let $h_1,\ldots,h_k\in\setC\left[x_1,\ldots,x_n\right]$ be a set of homogeneous polynomials. The set of elements of $\mathbbm{P}^{n-1}$ corresponding to the points $\left(x_1,\ldots,x_n\right)$ with the property $h_1\left(x_1,\ldots,x_n\right)=0,\ldots,h_n\left(x_1,\ldots,x_n\right)=0$ is called a \textbf{projective variety}. One can write it as
\begin{equation}
\label{eqprojvar} 
\left\{\left[X_1,\ldots,X_n\right]\vline h_1\left(X_1,\ldots,X_n\right)=0,\ldots,h_k\left(X_1,\ldots,X_n\right)=0\right\}
\end{equation}
A shorter notation, ${\bf{V}}\left(h_1,\ldots,h_k\right)$, which does not explicitly refer to the property of being a projective variety, is also used.
\end{definition}
One typical example of a projective variety is the \textbf{Segre variety}.
\begin{example}[Segre variety]
Let $n$, $m$ be positive integers. The Segre variety in $\mathbbm{P}^{\left(m+1\right)\left(n+1\right)-1}$ is the image of $\mathbbm{P}^n\times\mathbbm{P}^m$ under the mapping
\begin{equation}\label{eqSegremapping}
S:\left[X_0,\ldots,X_n\right]\times\left[Y_0,\ldots,Y_m\right]\mapsto\left[X_0Y_0,X_1Y_0,\ldots,X_nY_0,X_0Y_1,\ldots,X_nY_m\right]
\end{equation}
Alternatively, it is the projective variety in $\mathbbm{P}^{\left(m+1\right)\left(n+1\right)-1}$, defined by the vanishing of the homogeneous polynomials
\begin{equation}\label{eqzizj}
Z_{i,j}Z_{k,l}-Z_{i,l}Z_{k,j}
\end{equation}
where $Z_{0,0},Z_{1,0},\ldots,Z_{n,0},Z_{0,1},\ldots,Z_{n,m}$ is the set of homogeneous coordinates in $\mathbbm{P}^{\left(m+1\right)\left(n+1\right)-1}$. We denote it by $\Sigma_{n,m}$
\end{example}
Note that in quantum entanglement theory, $\Sigma_{m,n}$ corresponds to the set of pure separable states in $\setC^{n+1}\otimes\setC^{m+1}$.

We can proceed to the definition of the dimension of an algebraic, i.e. projective or affine, variety. Definitions will be slightly different for affine and projective varieties, and it is somewhat more convenient to start from the affine case. Similar to the situation with the Dickson's lemma (Lemma \ref{lemmaDicksons}), it will also be useful to discuss varieties corresponding to monomial ideals first. As we know from Lemma \ref{lemmaDicksons}, monomial ideals are finitely generated by some monomials, hence for a monomial ideal $I$ in $\setC\left[x_1,\ldots,x_n\right]$, we can always assume that
\begin{equation}\label{eqmonomialideal}
I=\left<x^{\alpha^1},\ldots,x^{\alpha^l}\right>,
\end{equation}
where we used the multi-index notation introduced in Section \ref{secmonomialorders}, with $\alpha^i=\left(\alpha^i_1,\ldots,\alpha^i_n\right)$ for all $i$. It follows that ${\bf V}\left(I\right)=\bigcap_{i=1}^l{\bf V}\left(x^{\alpha^i}\right)$, where each ${\bf V}\left(x^{\alpha^i}\right)$ has a simple description as $\bigcup_{\alpha^i_j\neq 0}H_{j}$, $H_j=\left\{\left(x_1,\ldots,x_n\right)\in\setC^n\,\vline\, x_j=0\right\}$. Thus we have
\begin{equation}\label{eqintersectionVI}
{\bf V}\left(I\right)=\bigcap_{i=1}^l\bigcup_{\alpha^i_j\neq 0}H_{j}
\end{equation}
By intersecting $H_j$ for different $j$'s, we get
\begin{equation}\label{eqformofH}
H_{j_1}\cap\ldots\cap H_{j_p}=\left\{\left(x_1,\ldots,x_n\right)\in\setC^n\,\vline\, x_{j_1}=0,\ldots,x_{j_p}=0\right\}=:H_{j_1\ldots j_p},
\end{equation}
which is a linear subspace of dimension $n-p$. If some of the $j_i$'s were equal, the dimension of the subspace would have increased accordingly. From equations  \eqref{eqintersectionVI} and \eqref{eqformofH}, it follows that ${\bf V}\left(I\right)$ for a monomial ideal $I$ is a union of subspaces of the form $H_{j_1\ldots j_p}$. We identify the dimension of ${\bf V}\left(I\right)$ as the maximum dimension of a subspace $H_{j_1\ldots j_p}$ included in ${\bf V}\left(I\right)$. A little thought reveals that this number can be calculated explicitly, and it equals $n-\left|\mathcal{J}\right|$, where $\left|\mathcal{J}\right|$ denotes the minimum number of elements in a subset $\mathcal{J}\subset\left\{1,2,\ldots,n\right\}$ with the property $\forall_{i}\exists_{j\in\mathcal{J}}\alpha^i_j\neq 0$. Thus, for a monomial ideal $I$, we have
\begin{equation}\label{eqdimmonoexpl}
\dim{\bf V}\left(I\right)=n-\left|\mathcal{J}\right|,
\end{equation}   
and there is a simple way to calculate $\left|\mathcal{J}\right|$ from the generators of $I$. 

\begin{figure}\centering
\includegraphics[scale=0.7]{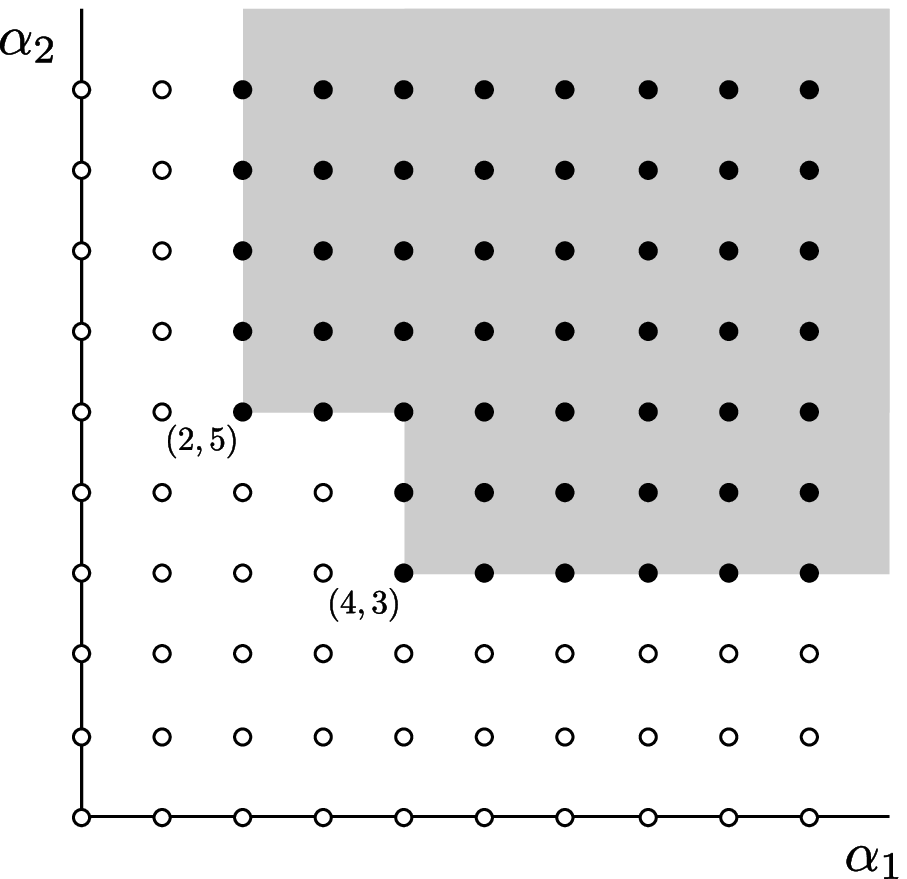}
\caption{A schematic picture of the monomial ideal $\left<x^2y^5,x^4y^3\right>$ in $\setC\left[x,y\right]$.}
\label{figMonomialIdeal}%
\end{figure}

A very important insight by Hilbert was that there exists an alternative way to obtain $\dim{\bf V}\left(I\right)$, which relates to the number of monomials of total degree lower or equal $s$ \emph{not} in $I$. To explain this in more detail, we need to introduce some extra notation. First of all, we define
\begin{equation}\label{eqCdef}
C\left(I\right)=\left\{\alpha\in\setN^n\,\vline\,x^{\alpha}\notin I\right\},
\end{equation}
i.e. the set of multi-indices corresponding to the monomials not in $I$. We will also be using a basis of multi-indices, $e_i:=\left[0,\ldots,1,\ldots,0\right]$, with $1$ on the $i$-th position and zeros elsewhere, and the notation 
\begin{equation}\label{eqcoordinatespace}
\left[e_{i_1},\ldots,e_{i_r}\right]:=\left\{a_1e_{i_1}+\ldots+a_re_{i_r}\,\vline\,a_j\in\setN\forall_{j=1,\ldots,r}\right\}
\end{equation}
for so-called \textit{coordinate subspaces}. Their translates by $\alpha=\left(\alpha_1,\ldots,\alpha_n\right)\in\setN^n$ will be denoted, in a natural way, by $\alpha+\left[e_{i_1},\ldots,e_{i_r}\right]$. When using this notation, it is assumed that $\alpha_{i_j}=0$ for all $j=1,\ldots,r$, so that $\alpha$ is perpendicular to the coordinate subspace. We have the following.
\begin{proposition}\label{propcoordinatesubspaces}
Let $I\subset\setC\left[x_1,\ldots,x_n\right]$
be a monomial ideal.
\begin{enumerate}[i)]
\item The set $\left\{\left(x_1,\ldots,x_n\right)\in\setC^n\,\vline\, x_{j}=0\forall_{j\notin\left\{i_1,\ldots,i_r\right\}}\right\}$, which can also be denoted as $H_{l_1\ldots l_{n-r}}$ with $\left\{l_1,\ldots,l_{n-r}\right\}=\left\{1,2,\ldots,n\right\}\setminus\left\{i_1,\ldots,i_r\right\}$, is contained in ${\bf V}\left(I\right)$ if and only if $\left[e_{i_1},\ldots,e_{i_r}\right]\subset C\left(I\right)$
\item The dimension ${\bf V}\left(I\right)$ is the dimension of the largest coordinate subspace in $C\left(I\right)$
\end{enumerate}
\begin{proof}
We first prove $i)$. Let us assume that $H_{l_1\ldots l_{n-r}}$ with $\left\{l_1,\ldots,l_{n-r}\right\}=\left\{1,2,\ldots,n\right\}\setminus\left\{i_1,\ldots,i_r\right\}$ is in ${\bf V}\left(I\right)$. In particular, the point $\left(x_1,\ldots,x_n\right)$ with coordinates
\begin{equation}\label{eqdefx10}
x_i=\begin{cases}
1,\textnormal{ if }i\in\left\{i_1,\ldots,i_r\right\}\\
0,\textnormal{ if }i\notin\left\{i_1,\ldots,i_r\right\}
\end{cases}
\end{equation}
belongs to ${\bf V}\left(I\right)$. Assume $\left[e_{i_1},\ldots,e_{i_r}\right]\notin C\left(I\right)$. If so, there must exist a monomial $x^{\alpha}\in I$ such that $\alpha$ belongs to $\left[e_{i_1},\ldots,e_{i_r}\right]$. However, all such monomials give $1$ when evaluated on $\left(x_1,\ldots,x_n\right)$ from equation \eqref{eqdefx10}, which leads to a contradiction with $\left(x_1,\ldots,x_n\right)\in{\bf V}\left(I\right)$. Thus we have proved the $\Rightarrow$ implication in $i)$. Conversely, if $\left[e_{i_1},\ldots,e_{i_r}\right]\in C\left(I\right)$, it means that every monomial in $I$ is of nonzero degree in some of the variables $x_{l_1},\ldots,x_{l_{n-r}}$, $\left\{{l_1},\ldots,{l_{n-r}}\right\}=\left\{1,2,\ldots,n\right\}\setminus\left\{i_1,\ldots,i_r\right\}$. Therefore, the monomials in $I$ give $0$ when evaluated on elements of $H_{l_1\ldots l_{n-r}}$. In other words, $H_{l_1\ldots l_{n-r}}\subset{\bf V}\left(I\right)$, which proves the $\Leftarrow$ implication in part $i)$ of the theorem. Part $ii)$ follows immediately from $i)$, since $\dim{\bf V}\left(I\right)$ is defined as the maximum dimension of a subspace $H_{l_1\ldots l_{n-r}}$ included in ${\bf V}\left(I\right)$. If $\left\{l_1,\ldots,l_{n-r}\right\}=\left\{1,2,\ldots,n\right\}\setminus\left\{i_1,\ldots,i_r\right\}$, the dimension of $H_{l_1\ldots l_{n-r}}$ equals $r$, which is precisely the dimension of the coordinate subspace $\left[e_{i_1},\ldots,e_{i_r}\right]$.
\end{proof}
\end{proposition} 
An illustrative picture of a monomial ideal $I=\left<x^2y^5,x^4y^3\right>$ in $\setC\left[x,y\right]$ is presented in Figure \ref{figMonomialIdeal}. Empty dots denote the monomials with multi-indices in $C\left(I\right)$, and black dots correspond to monomials in $I$. Generalizing from this example, it is easy to believe in the following proposition, which we give without a proof \cite{IdealsVarieties}.

\begin{proposition}\label{propcoordinatetranslates}
For any monomial ideal $I\subset\setC\left[x_1,\ldots,x_n\right]$, the set $C\left(I\right)$ can be written as a finite (not necessarily disjoint) union of translates $T^i=\alpha^i+\left[e_{j^i_1},\ldots,e_{j^i_{r_i}}\right]$ of some coordinate subspaces $\left[e_{j^i_1},\ldots,e_{j^i_{r_i}}\right]$.
\end{proposition}

We claim that the number of elements $\left(\alpha_1,\ldots,\alpha_n\right)\in C\left(I\right)$ with the property $\left|\alpha\right|:=\sum_{i=1}^n\alpha_i\leqslant s$ can be expressed, for $s$ sufficently large, as a polynomial $a_0s^d+a_1s^{d-1}+\ldots+a_d$ of degree $\dim{\bf V}\left(I\right)$, with $a_0>0$. Equivalently, the number of monomials of total degree no larger than $s$, not in $I$, is given by such polynomial for $s$ sufficiently large. 
\begin{proposition}\label{propnumberinCI}
Let $I\subset\setC\left[x_1,\ldots,x_n\right]$ be a monomial ideal. Denote by ${^a}HF_I\left(s\right)$ the number of multi-indices $\alpha=\left(\alpha_1,\ldots,\alpha_n\right)\in\setN^n$  in $C\left(I\right)$ with the property $\left|\alpha\right|\leqslant s$. For $s$ sufficiently large, ${^a}HF_I\left(s\right)$ can be written as a polynomial
\begin{equation}\label{eqHilbertpolynomial1}
a_0s^d+a_1s^{d-1}+\ldots+a_d,
\end{equation}
where $a_0>0$ and $d$ equals the dimension of ${\bf V}\left(I\right)$. The function ${^a}HF_I\left(s\right)$ and the polynomial \eqref{eqHilbertpolynomial1} are called the \textbf{(affine) Hilbert function} and the \textbf{(affine) Hilbert polynomial} of $I$, respectively. The latter will be denoted by ${^a}HP_I$.
\begin{proof}
To prove the statement, we first notice that the number of multi-indices $\left(\alpha_1,\ldots,\alpha_n\right)$ with the property $\left|\alpha\right|\leqslant s$ is equal to $n +s \choose s$. From this, it is easy to conclude that the number of multi-indices $\alpha$ such that $\left|\alpha\right|\leqslant s$ and $\alpha\in\alpha^i+\left[e_{j^i_1},\ldots,e_{j^i_{r_i}}\right]$ equals
\begin{equation}\label{eqchoosersalpha}
\binom{r_i+s-\left|\alpha^i\right|}{s-\left|\alpha^i\right|}=\frac{1}{r_i!}\left(r_i+s-\left|\alpha^i\right|\right)\left(r_i+s-\left|\alpha^i\right|-1\right)\cdot\ldots\cdot\left(s-\left|\alpha^i\right|+1\right)
\end{equation}
for $s$ sufficiently large. Thus, the above formula gives precisely an expression for the number of multi-indices $\left|\alpha\right|\leqslant s$ in the translates $T^i$ from Proposition \ref{propcoordinatetranslates}. Of course, it can be applied to other translates as well. Note that \eqref{eqchoosersalpha} is a polynomial in $s$ of degree $r_i$, which is precisely the dimension of the coordinate subspace $\left[e_{j^i_1},\ldots,e_{j^i_{r_i}}\right]$. 

For convenience, let us denote the set of multi-indices $\left|\alpha\right|\leqslant s$ in $T^i$ by $T^i_{\leqslant s}$. By the well-known \textbf{inclusion-exclusion principle} from combinatorics, we get
\begin{equation}\label{eqinclusionexclusion}
{^a}HF_I\left(s\right)=\sum_i\left|T_{\leqslant s}^i\right|+\sum_{i<j}\left|T_{\leqslant s}^i\cap T_{\leqslant s}^j\right|+\sum_{i<j<k}\left|T_{\leqslant s}^i\cap T_{\leqslant s}^j\cap T_{\leqslant s}^k\right|+\ldots
\end{equation}
A key point is now that $T^i_{\leqslant s}\cap T^j_{\leqslant s}$ as well as $T_{\leqslant s}^i\cap T_{\leqslant s}^j\cap T_{\leqslant s}^k$ and higher-order intersections are either empty, or equal to $T_{\leqslant s}$ for some translated coordinate space $T$ of dimension $<r_i$, simply because $T^i_{\leqslant s}\cap T^j_{\leqslant s}$ and higher-order intersections are either empty or equal to some coordinate space $T$ of the mentioned property. By \eqref{eqchoosersalpha}, the second and further terms in the sum on the right-hand side of \eqref{eqinclusionexclusion} are equal to some polynomials of degrees $<\max\left(\left\{r_i\right\}\right)$ for $s$ sufficiently large. Hence, for $s$ sufficiently large, they cannot cancel the leading term of $\sum_i\left|T^i_{\leqslant s}\right|$, which sum is also a polynomial, of degree $\max\left(\left\{r_i\right\}\right)$ and a positive leading term. The last statement is again a consequence of formula \eqref{eqchoosersalpha}. All in all, for $s$ sufficiently large, the sum in \eqref{eqinclusionexclusion} is given by a polynomial of degree $\max\left(\left\{r_i\right\}\right)$ with a nonnegative leading coefficient. 
\end{proof}
\end{proposition}
The degree of the Hilbert polynomial, which we obtained in the above proof, is equal to the maximum dimension of a coordinate subspace in $C\left(I\right)$. By Proposition \ref{propcoordinatesubspaces}, this is equal to $\dim{\bf V}\left(I\right)$. Thus we have obtained an alternative characterization of the dimension of a variety corresponding to a monomial ideal, which can be rather conveniently generalized to all affine varieties. Before we discuss the general affine case however, it is important to notice that the varieties corresponding to monomial ideals in $\setC\left[x_1,\ldots,x_n\right]$ can be regarded as projective varieties in $\mathbbm{P}^{n-1}$ as well. If we look at them in this way, the definition of their dimension needs to be slightly modified. First of all, we call $n-1$ the \textbf{projective dimension} of $\mathbbm{P}^{n-1}$. It is therefore natural to call $d-1$ the projective dimension of a $d$-dimensional linear subspace of $\setC^n$, when we regard it as a subset of $\mathbbm{P}^{n-1}$. Consequently, the projective dimension of ${\bf V}\left(I\right)$ for a monomial ideal $I$ in $\setC\left[x_1,\ldots,x_n\right]$ is defined as $d-1$, where $d$ is the maximum dimension of a linear subspace contained in ${\bf V}\left(I\right)$. Following \eqref{eqdimmonoexpl}, the projective dimension can be calculated as $n-\left|\mathcal{J}\right|-1$. On the other hand, using the Hilbert approach, we can calculate the projective dimension of $I$ as the degree of the polynomial 
\begin{equation}
HP_I\left(s\right):={^{a}}HP_I\left(s\right)-{^{a}}HP_I\left(s-1\right),
\end{equation}
which is called simply the \textbf{Hilbert polynomial} of $I$. For $s$ sufficiently large, it equals the number of monomials not in $I$ and of total degree \emph{equal} $s$. The last definition of projective dimension of a variety corresponding to a monomial ideal is the one which conveniently generalizes to all projective varieties.

Let us also note that the Hilbert polynomial and affine Hilbert polynomial are customarily written in the form
\begin{equation}\label{eqformHP}
HP_I\left(s\right)=\sum_{i=0}^{d-1}b_i\binom{s}{d-1-i}\quad\textnormal{and}\quad{^{a}}HP_I\left(s\right)=\sum_{i=0}^{d}a_i\binom{s}{d-i},
\end{equation}
where $b_i,a_i\in\mathbbm{Z}$, $b_0>0$, $a_0>0$ and $d=\dim{\bf V}\left(I\right)$. The possibility to write the Hilbert polynomials in the above form is a direct consequence of the fact that a general polynomial $p\left(s\right)$ of degree $d$ that takes integer values for integer $s$ can be written as ${^{a}}HP_I\left(s\right)$ in \eqref{eqformHP} \cite{IdealsVarieties}. 

After the above lengthy discussion of monomial ideals, we can smoothly define the dimension of arbitrary projective or affine varieties. Given an ideal $I\subset\setC\left[x_1,\ldots,x_n\right]$, we define its \textbf{affine Hilbert function} as
\begin{equation}\label{eqdefaffHF}
{^{a}}HF_I\left(s\right)=\dim\setC\left[x_1,\ldots,x_n\right]_{\leqslant s}-\dim I_{\leqslant s},
\end{equation}
where $\setC\left[x_1,\ldots,x_n\right]_{\leqslant s}$ is the set of polynomials of degree $\leqslant s$, $I_{\leqslant s}$ equals $I\cap\setC\left[x_1,\ldots,x_n\right]_{\leqslant s}$, and $\dim$ refers to the dimensionality of these sets when regarded as $\setC$-linear subspaces of $\setC\left[x_1,\ldots,x_n\right]$. For monomial ideals $I$, it is easy to see that the above definition of ${^{a}}HF_I$ coincides with the one we gave earlier. A key observation is that for general $I$, the Hilbert function of $I$ can be computed from a suitably chosen monomial ideal. Similar to the situation we encountered in the proof of the Hilbert basis theorem (Theorem \ref{thmHilbertbasis}), the monomial of leading terms $\left<\textnormal{LT}\left(f\right)\right>_{f\in I}$ with respect to some monomial ordering $>$ turns out to be of great importance. However, in the affine case, we additionally need to assume that $>$ is a \textbf{graded order}, i.e. $x^{\alpha}>x^{\beta}$ whenever $\left|\alpha\right|>\left|\beta\right|$. We then have the following result.
\begin{proposition}\label{propLTidealHF}
Let $I\subset\setC\left[x_1,\ldots,x_n\right]$ be an ideal and let $>$ be a graded order on $\setC\left[x_1,\ldots,x_n\right]$. The monomial ideal $\left<\textnormal{LT}\left(I\right)\right>_{f\in I}$ has the same affine Hilbert function as $I$.
\begin{proof}
Can be found in \cite[Chapter 9, {\S}3]{IdealsVarieties}.
\end{proof}
\end{proposition} 
From the above proposition and the earlier discussion about monomial ideals, we conclude that for $s$ sufficiently large, ${^{a}}HF_I\left(s\right)$ equals ${^{a}}HP_{\left<\textnormal{LT}\left(f\right)\right>_{f\in I}}\left(s\right)$, the Hilbert polynomial of $\left<\textnormal{LT}\left(f\right)\right>_{f\in I}$. We call the same function the \textbf{affine Hilbert polynomial} of $I$ and denote it by ${^{a}}HP_I$. The same as in equation \eqref{eqformHP}, ${^{a}}HP_I$ can be written as a sum of terms $a_i\binom{s}{d-i}$ with $a_i\in\mathbbm{Z}$, $a_0>0$. For closed scalar fields like $\setC$, the dimension of the affine variety ${\bf V}\left(I\right)$ is now simply defined as the degree of ${^{a}}HP_I$, cf. Theorem 8 in \cite[Chapter 9, {\S}3]{IdealsVarieties}.  
\begin{definition}[Dimension of an affine variety]\label{defaffinedimensiongeneral}
Let $I\subset\setC\left[x_1,\ldots,x_n\right]$ be an ideal in $\setC\left[x_1,\ldots,x_n\right]$. Let ${^{a}}HP_I$ be the polynomial which equals ${^{a}}HF_I(s)$ for large $s$. The \textbf{dimension} of ${\bf V}\left(I\right)$ is defined to be equal to the degree of ${^a}HF_I$.
\end{definition}
Such defined dimension can be calculated from the generators of $I$. A suitable procedure consists of two elementary steps:
\begin{enumerate}[1.]
\item Choose a graded monomial order in $\setC\left[x_1,\ldots,x_n\right]$ such as the graded lexicographic order of Example \ref{exgradedlex} or graded reverse lexicographic order of Example \ref{exgradedrevlex}. Compute a Groebner basis $\left\{g_1,\ldots,g_t\right\}$ of $I$ using the selected ordering.
\item Compute the maximal dimension of a subspace $H_{i_1,\ldots,i_r}$ contained in the variety ${\bf V}\left(\left<\textnormal{LT}\left(g_1\right),\ldots,\textnormal{LT}\left(g_t\right)\right>\right)$, using the approach outlined above formula \eqref{eqdimmonoexpl}.
\end{enumerate}
 
To define the dimension of a general projective variety, we can proceed similar as above. First, we denote by $\setC\left[x_1,\ldots,x_n\right]_s$ the set of all \emph{homogeneous} polynomials of total degree $s$, together with the zero polynomial. We also set $I_s=I\cap\setC\left[x_1,\ldots,x_n\right]_s$ for an ideal $I$, generated by homogeneous polynomials. The \textbf{Hilbert function} of $I$ is defined as
\begin{equation}\label{eqdefHilbertfunction}
HF_I\left(s\right)=\dim\setC\left[x_1,\ldots,x_n\right]_s-\dim I_s,
\end{equation} 
where $\dim$ refers to the dimension as a $\setC$-linear subspace of $\setC\left[x_1,\ldots,x_n\right]$. In full analogy to Proposition \ref{propLTidealHF}, we have \cite[Chapter 9, {\S}3]{IdealsVarieties}
\begin{proposition}\label{propLTidealHFprojective}
Let $I\in\setC\left[x_1,\ldots,x_n\right]$ be an ideal generated by homogeneous polynomials. Consider \textbf{any monomial order} $>$ in $\setC\left[x_1,\ldots,x_n\right]$. The monomial ideal $\left<\textnormal{LT}\left(f\right)\right>_{f\in I}$ has the same Hilbert function as $I$.
\end{proposition}
Note that this time, unlike in the affine case, it is possible to use \emph{any} monomial ordering to obtain the desired monomial ideal.

For monomial ideals like $\left<\textnormal{LT}\left(f\right)\right>_{f\in I}$, the above definition of Hilbert function coincides with the one we gave previously. It immediately follows that for large $s$, $HF_I\left(s\right)$ equals $HP_{\left<\textnormal{LT}\left(f\right)\right>_{f\in I}}\left(s\right)$, where $HP$ refers to the Hilbert polynomial, which we have already defined for monomial ideals. To no surprise, we call the latter function the \textbf{Hilbert polynomial} of $I$ and denote it with $HP_I$. By the formula on the left-hand side of \eqref{eqformHP}, we can write the Hilbert polynomial of an arbitrary ideal $I$ generated by homogeneous polynomials as
\begin{equation}\label{eqformHP2}
HP_I\left(s\right)=\sum_{i=0}^db_i\binom{s}{d-i-1}
\end{equation}
for some $d\in\setN$, $b_i\in\mathbbm{Z}$ and $b_0>0$. For algebraically closed scalar fields like $\setC$, we define the projective dimension of ${\bf V}\left(I\right)$ simply as the degree of $HP_I$, i.e. $d-1$ in the above formula.
\begin{definition}[Projective dimension]\label{defprojdim}
Let $I\subset\setC\left[x_1,\ldots,x_n\right]$ be a monomial generated by homogeneous polynomials. Let $HP_I\left(s\right)$ be the polynomial which equals $HF_I\left(s\right)$ for large $s$ (i.e. the Hilbert polynomial of $I$). The \textbf{projective dimension} of the projective variety ${\bf V}\left(I\right)$ is defined to be equal to the degree of $HP_I$.
\end{definition}
Again, the dimension of a projective variety ${\bf V}\left(I\right)$ can be calculated by a procedure completely analogous to the one we outlined for affine varieties. The projective dimension is well behaved under many operations, cf. \cite[Chapter 9, {\S}4]{IdealsVarieties} and it plays a key role in the following elegant result (cf. Theorem 7.2 in \cite{Hartshorne}).
\begin{theorem}Let $\mathcal{V}$ and $\mathcal{U}$ be two projective varieties in $\mathbbm{P}^n$.  Let $r$ and $s$ be the projective dimensions of $\mathcal{V}$ and $\mathcal{U}$, respectively. If $r+s\geqslant n$, the intersection $\mathcal{V}\cap\mathcal{U}$ is nonempty and of dimension $\geqslant r+s-n$. 
\end{theorem}
Another important characteristic of a projective variety, which can be read off its Hilbert polynomial, is the \textit{degree}. 
\begin{definition}[Degree]\label{defprojdegree}
Let $I\subset\setC\left[x_1,\ldots,x_n\right]$ be a monomial generated by homogeneous polynomials. Let $HP_I$ be the Hilbert polynomial of $I$. Write $HP_I\left(s\right)$ as in \eqref{eqformHP2},
\begin{equation}\label{eqformHP3}
HP_I\left(s\right)=\sum_{i=0}^db_i\binom{s}{d-i-1},
\end{equation}
where $b_i\in\mathbbm{Z}$, $b_0>0$. The \textbf{degree} of the projective variety ${\bf V}\left(I\right)$ is defined to be equal to $b_0$ -- the \textbf{leading term} of $HP_I$.
\end{definition}
As we shall learn from Section \ref{secBezout}, the degree of a projective variety $\mathcal{V}\subset\mathbbm{P}^n$ of dimension $d$ equals, under certain assumptions, the number of intersection points of $\mathcal{V}$ with a projective variety $\mathcal{U}$ of complementary dimension $n-d$.

 In the thesis, we are particularly interested in Segre varieties. The following remark tells us about their dimension and degree.
\begin{remark}[Dimension and degree of a Segre variety]
Let $\Sigma_{n,m}$ denote the Segre variety in $\mathbbm{P}^n\times\mathbbm{P}^m\cong\mathbbm{P}^{\left(m+1\right)\left(n+1\right)-1}$. The projective dimension of $\Sigma_{n,m}$ is $n+m$ whereas its degree equals $\binom{m+n}{n}$.

\end{remark}
A short discussion of the above facts can be found in the classical textbook by J. Harris \cite[Lectures 12 and 18]{Harris}. 
\section{Tangent spaces. Smoothness}\label{sectangent}
The notion of the tangent space to a curve or a surface in $\setR^3$ is something intuitively well understood. As we will see, it can be easily generalized to affine and projective varieties. We choose to work with $\setC$ as the field of scalars, but definitions can as well be formulated for general fields $\mathbbm{K}$ in place of complex numbers.

Let us start with an affine variety $V\subset\setC^n$ and consider the ideal $I={\bf I}\left(V\right)$, i.e. the set of polynomials $f\in\setC\left[x_1,\ldots,x_n\right]$ that vanish on $V$. 
We know from the Hilbert basis theorem that $I$ is finitely generated, so we can write it as $\left<f_1,\ldots,f_l\right>$ for some polynomials $f_i$.
\begin{definition}\label{deftangentspaceaffine}
Let $p$ be a point in an affine variety $V\subset\setC\left[x_1,\ldots,x_n\right]$. The \textbf{Zariski tangent space} to $V$ at $p$ is defined as
\begin{equation}\label{eqZariskitspace}
T_pV:=\left\{v\in\setC^n\,\vline\,\left(df\right)\left(v\right)=0\,\forall_{f\in{\bf I}\left(V\right)}\right\},
\end{equation}
where $df$ denotes the derivative of a polynomial $f$. Equivalently,
\begin{equation}\label{eqZariskitspace2}
T_pV:=\left\{v\in\setC^n\,\vline\,\left(df_i\right)\left(v\right)=0\forall_{i=1,2,\ldots,l}\right\},
\end{equation}
where $f_1,\ldots,f_l$ is a set of generators of ${\bf I}\left(V\right)$.
\end{definition}
Note that the calculation of $df$ or $df_i$ can be done in a purely formal manner, since we are dealing with polynomials.

\begin{definition}\label{defaffinetangent}
Let $V$ and $p$ be as in Definition \ref{deftangentspaceaffine}. We call $p+T_pV$ the \textbf{affine tangent space} to $V$ at $p$. More explicitly, the affine tangent space is defined as
\begin{equation}\label{eqaffinetangent}
{^a}T_pV:=\left\{q\in\setC^n\,\vline\,\left(df\right)\left(q-p\right)=0\forall_{f\in{\bf I}\left(V\right)}\right\}
=\left\{q\in\setC^n\,\vline\,\left(df_i\right)\left(q-p\right)=0\forall_{i}\right\}
\end{equation}
\end{definition}
Using the Zariski tangent space to $V$ at $p$, we can define what it means for $p$ to be \textbf{smooth}.
\begin{definition}
Let $V\subset\setC^n$ be an affine variety of (affine) dimension $\dim V$ and such that $p\in V$. We call $p$ a \textbf{smooth point} of $V$ if and only if $\dim\left(T_pV\right)=\dim V$.
\end{definition}

Given a set of generators of the ideal ${\bf I}\left(V\right)$, smoothness of a $p\in V$ can readily be checked by the following Jacobi criterion, cf. e.g. \cite{Farkas}
\begin{proposition}[Jacobi criterion for smoothness]\label{propJacobi}Let $V\subset\setC^n$ be an affine variety of (affine) dimension $\dim V$, such that ${\bf I}\left(V\right)=\left\{f_1,\ldots,f_l\right\}$. A point $p\in V$ is smooth if and only if the rank of the matrix
\begin{equation}\label{eqJacobimatrix}
\left[\frac{\partial f_i}{\partial x_j}\right]_{i=1,2,\ldots,l\atop j=1,2,\ldots,n}
\end{equation}
is equal to $n-\dim V$.
\end{proposition}

For projective varieties, definitions of the tangent space and smoothness are very similar to the ones presented above. To define the projective tangent space to a projective variety $\mathcal{V}\subset\mathbbm{P}^n$, consider first a dehomogenized version of the polynomials in ${\bf I}\left(\mathcal{V}\right)$. Namely, for a homogeneous polynomial $h\in{\bf I}\left(V\right)$ taking points $\left[X_0,X_1,\ldots,X_n\right]\in\mathbbm{P}^n$ as input and and giving $h\left(X_0,X_1,\ldots,X_n\right)$ as output, let us define $\tilde h\subset\setC\left[x_1,\ldots,x_n\right]$ by the formula
\begin{equation}\label{eqdehomo1}
\tilde h\left(x_1,\ldots,x_n\right):=h\left(1,x_1,\ldots,x_n\right).
\end{equation}
Consider the affine variety $\tilde{\mathcal{V}}\subset\setC^n$ consisting of the common zeros of the polynomials $\tilde h$, $h\in{\bf I}\left(\mathcal{V}\right)$. Its affine tangent space at a point $z=\left(z_1,\ldots,z_n\right)$ equals
\begin{equation}\label{eqaffinedehomo1}
T_z\tilde{\mathcal{V}}=\left\{\left(y_1,\ldots,y_n\right)\,\vline\,\sum_{i=1}^n\frac{\partial\tilde h}{\partial x_i}\left(z\right)\cdot\left(y_i-z_i\right)=0\forall_{h\in{\bf I}\left(\mathcal{V}\right)}\right\}.
\end{equation}
To get a projectivized version of $T_x\tilde{\mathcal{V}}$, we can homogenize the defining polynomial equations in \eqref{eqaffinedehomo1}, i.e. consider 
\begin{equation}\label{eqprojanalogue1}
\left\{\left[Y_0,Y_1,\ldots,Y_n\right]\,\vline\,\sum_{i=1}^n\frac{\partial\tilde h}{\partial x_i}\left(z\right)\cdot\left(Y_i-z_iY_0\right)=0\forall_{h\in{\bf I}\left(\mathcal{V}\right)}\right\}.
\end{equation}
as a projective analogue of $T_z\tilde{\mathcal{V}}$. A key observation is now that partial derivatives of a homogeneous polynomial $h$ of degree $d$ satisfy the following \textbf{Euler relations}
\begin{equation}\label{eqEulerrelations}
\sum_{i=0}^n\frac{\partial h}{\partial X_i}\left(Z_0,Z_1,\ldots,Z_n\right)Z_i=d\cdot F\left(Z_0,Z_1,\ldots,Z_n\right).
\end{equation}
In particular, the above relation can be applied to $\left[Z_0,Z_1,\ldots,Z_n\right]=\left[1,z_1,\ldots,z_n\right]$ to yield
\begin{equation}\label{eqEulerused1}
\sum_{i=1}^n\frac{\partial\tilde h}{\partial x_i}\left(z\right)z_i=-\frac{\partial h}{\partial X_0}\left(z\right),
\end{equation}
where we used the fact that $\tilde h$ vanishes at $\left(1,z_1,\ldots,z_n\right)$. We can use \eqref{eqEulerused1} and the identity $\partial\tilde h/\partial x_i\left(z\right)=\partial h/\partial X_i\left(Z\right)$, where $Z=\left[1,z_1,\ldots,z_n\right]$  to rewrite \eqref{eqprojanalogue1} as
\begin{equation}\label{eqprojanalogue2}
\left\{\left[Y_0,Y_1,\ldots,Y_n\right]\,\vline\,\sum_{i=0}^n\frac{\partial h}{\partial X_i}\left(Z\right)Y_i=0\forall_{h\in{\bf I}\left(\mathcal{V}\right)}\right\}.
\end{equation}
The tangent space to a projective variety $\mathcal{V}$ at a point $Z=\left[Z_0,\ldots,Z_n\right]$ is now simply defined by formula \eqref{eqprojanalogue2} with the requirement $Z=\left[1,z_1,\ldots,z_n\right]$ dropped. Thus we have the following definition
\begin{definition}\label{defprojtangentspace}
Let $\mathcal{V}\subset\mathbbm{P}^n$ be a projective variety and let $Z$ be an element of $\mathcal{V}$. Let us write the elements of $\mathbbm{P}^n$ as $\left[X_0,X_1,\ldots,X_n\right]$. The projective tangent space to $\mathcal{V}$ at a point $Z\in\mathcal{V}$ is defined as the following subspace of $\mathbbm{P}^n$,
\begin{equation}\label{eqprojtangentspace2}
\mathbbm{T}_Z\mathcal{V}:=\left\{\left[Y_0,Y_1,\ldots,Y_n\right]\,\vline\,\sum_{i=0}^n\frac{\partial h}{\partial X_i}\left(Z\right)Y_i=0\forall_{h\in{\bf I}\left(V\right)}\right\}.
\end{equation}
Alternatively, given a set of generators $h_1,\ldots,h_l$ of ${\bf I}\left(V\right)$, we can restate the definition \eqref{eqprojtangentspace2} as
\begin{equation}\label{eqprojtangentspace3}
\mathbbm{T}_Z\mathcal{V}:=\left\{\left[Y_0,Y_1,\ldots,Y_n\right]\,\vline\,\sum_{i=0}^n\frac{\partial h_j}{\partial X_i}\left(Z\right)Y_i=0\forall_{j=1,2,\ldots,l}\right\}.
\end{equation} 
\end{definition}
Similar as in the affine case, the smoothness of a point $Z\in\mathcal{V}$ is defined by a suitable condition for the dimension of $\mathbbm{T}_Z\mathcal{V}$. 
\begin{definition}
Let $\mathcal{V}\subset\mathbbm{P}^n$ be a projective variety of projective dimension $\dim\mathcal{V}$. A point $Z\in\mathcal{V}$ is called a \textbf{smooth point} of $\mathcal{V}$ if and only if the projective dimension of $\mathbbm{T}_Z\mathcal{V}$ equals $\dim\mathcal{V}$.
\end{definition}
Clearly, there exists a projective analogue of the Jacobi criterion for smoothness \cite{Farkas}. We state it as the following proposition.
\begin{proposition}[Projective Jacobi criterion]\label{propprojectiveJacobi}Let $\mathcal{V}\subset\mathbbm{P}^n$ be a projective variety of projective dimension $\dim\mathcal{V}$, such that ${\bf I}\left(V\right)=\left\{h_1,\ldots,h_l\right\}$. A point $Z\in\mathcal{V}$ is smooth if and only if the rank of the matrix
\begin{equation}\label{eqJacobimatrix2}
\left[\frac{\partial h_i}{\partial X_j}\right]_{i=1,2,\ldots,l\atop j=0,1,\ldots,n}
\end{equation}
is equal to $n-\dim\mathcal{V}$.
\end{proposition}

Let us discuss the above notions in the example of Segre varieties, which is crucial for the main result of the thesis.
\begin{example}[Segre varieties]\label{exSegretangent}
The tangent space to the Segre variety $\Sigma_{n,m}\subset\mathbbm{P}^{\left(n+1\right)\left(m+1\right)-1}$ at a point $S\left(\left[X_0,\ldots,X_n\right],\left[Y_0,\ldots,Y_m\right]\right):=S\left(X,Y\right)$ is spanned by the points $S\left(X,Y'\right)$
and $S\left(X',Y\right)$ with $X'\in\mathbbm{P}^n$ and $Y'\in\mathbbm{P}^m$ arbitrary.
In particular, it follows that $\Sigma_{n,m}$ is smooth at every point $Z\in\Sigma_{n,m}$, for all $m,n\in\setN$.
\begin{proof}
A linear transformation $X\times Y\mapsto AX\times BY$, with $A$ and $B$ nonsingular linear maps, brings $X\times Y$ to $\left[0,\ldots,0,1\right]\times\left[0,\ldots,0,1\right]$. At the same time, it transforms all the pairs of the form $X'\times Y$ and $X\times Y'$ to
$X''\times\left[0,\ldots,0,1\right]$ and $\left[0,\ldots,0,1\right]\times Y''$ with $X''=AX'$ and $Y''=BY'$. These $AX'$ and $BY'$ still run over all elements of $\mathbbm{P}^n$ and $\mathbbm{P}^m$ if $X'$ and $Y'$ can be taken as arbitrary. As a result, we see that it is sufficient to prove our assertions about $\Sigma_{n,m}$ for the single point $Z_0=S\left(\left[0,\ldots,0,1\right]\times\left[0,\ldots,0,1\right]\right)$, and the rest will follow. Recall that $\Sigma_{n,m}$ is defined as the common zero of the polynomials 
\begin{equation}\label{eqpolysSegre2}
h_{ijkl}:=Z_{ij}Z_{kl}-Z_{il}Z_{kj},
\end{equation}
where $0\leqslant i<k\leqslant n$, $0\leqslant j<l\leqslant m$ and $Z_{00},Z_{01},\ldots,Z_{0m},Z_{10},\ldots,Z_{nm}$ denote the homogeneous coordinates in $\mathbbm{P}^{\left(n+1\right)\left(m+1\right)-1}$. The calculation of the derivative of $h_{ijkl}$ at the point $S\left(\left[0,\ldots,0,1\right]\times\left[0,\ldots,0,1\right]\right)$ is very simple. We have
\begin{equation}\label{eqSegretangentX}
\frac{\partial h_{ijkl}}{\partial Z_{ab}}\left(Z_0\right)=\delta_{ai}\delta_{bj}\delta_{kn}\delta_{lm},
\end{equation}
where $i<n$, $j<m$ and $\delta$ denotes the Kronecker delta. As it is not difficult to see, points in $\mathbbm{P}^{\left(n+1\right)\left(m+1\right)-1}$ with coordinates $Z_{00},Z_{01},\ldots,Z_{0m},Z_{10},\ldots,Z_{nm}$ do \emph{not} satisfy
\begin{equation}\label{eqSegretangentY}
\sum_{a=1}^n\sum_{b=1}^m\frac{\partial h_{ijkl}}{\partial Z_{ab}}\left(Z_0\right)Z_{ab}=0
\end{equation}
if $Z_{ab}\neq 0$ for some $a<n$ and $b<m$. All other points in $\mathbbm{P}^{\left(n+1\right)\left(m+1\right)-1}$, with vanishing $Z_{ab}$ whenever $a<n$ and $b<m$, \emph{do satisfy} \eqref{eqSegretangentY}. As it is not difficult to check, all such points can be written as linear combinations of $S\left(\left[0,\ldots,0,1\right]\times Y''\right)$ and $S\left(X''\times\left[0,\ldots,0,1\right]\right)$ for some $X''\in\mathbbm{P}^n$ or $Y''\in\mathbbm{P}^m$, and all points of the latter form do satisfy \eqref{eqSegretangentY}. Hence, they are good candidates for a basis of $\mathbbm{T}_{Z_0}\Sigma_{n,m}$. However, to remain in compliance with the above definition of projective tangent space, we should prove that points of the form  $S\left(\left[0,\ldots,0,1\right]\times Y''\right)$ and $S\left(X''\times\left[0,\ldots,0,1\right]\right)$ satisfy an analogue of \eqref{eqSegretangentY},
\begin{equation}
\label{eqSegretangentZ}
\sum_{a=1}^n\sum_{b=1}^m\frac{\partial h}{\partial Z_{ab}}\left(Z_0\right)Z_{ab}=0
\end{equation}
for \emph{all} elements $h$ of ${\bf I}\left(\Sigma_{n,m}\right)$. However, this easily follows because the points $Z_0+\lambda S\left(\left[0,\ldots,0,1\right]\times Y''\right)$ and $Z_0+\lambda S\left(X''\times\left[0,\ldots,0,1\right]\right)$ are again elements of $\Sigma_{n,m}$, for all $\lambda\in\setC$. 
In conclusion, the tangent space to $\Sigma_{n,m}$ at $Z_0$ is spanned by elements of $\mathbbm{P}^{\left(n+1\right)\left(m+1\right)-1}$ of the form $ S\left(\left[0,\ldots,0,1\right]\times Y''\right)$ and $S\left(X''\times\left[0,\ldots,0,1\right]\right)$. From them, we can choose a basis, consisting of $m+n+1$ elements, so the projective dimension of $\mathbbm{T}_{Z_0}\Sigma_{n,m}$ is $m+n$. Thus, $\Sigma_{n,m}$ is smooth at $Z_0$. 
By our earlier comments, the same applies to any point $Z$ of $\Sigma_{n,m}$. Moreover, the tangent spaces $\mathbbm{T}_Z\Sigma_{n,m}$ have the asserted form for all $Z\in\Sigma_{n,m}$.
\end{proof}
\end{example}
In Section \ref{secprodvecPPT}, we are going to use the above characterization of the tangent space of $\Sigma_{n,m}$ to make a key step in the proof of the strongest result of the thesis, which is Theorem \ref{maintheorem}. 

\section{Bezout's theorem}\label{secBezout} 
In the last part of our basic introduction to intersection theory, we will discuss a powerful theorem that allows, among others, to calculate the number of intersection points between two projective varieties of complementary dimension. The theorem works under certain assumptions. To explain them, we need to introduce the notion of \textbf{transverse intersection} of two projective varieties.
\begin{figure}\centering
\includegraphics[scale=0.5]{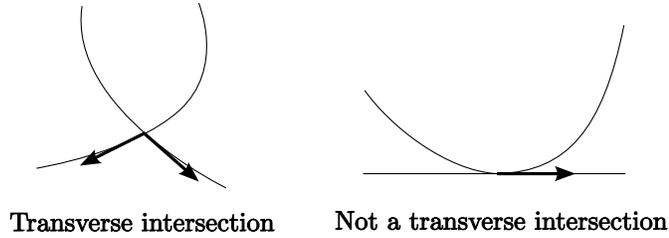}%
\caption{A schematic picture showing the difference between a transverse and not a transverse intersection.}%
\label{figtangentintersection}%
\end{figure} 
\begin{definition}[Transverse intersection]\label{deftransverseintersection}
Let $\mathcal{V}$ and $\mathcal{U}$ be two projective varieties in $\mathbbm{P}^n$ of complementary dimension, i.e. $\dim\mathcal{V}+\dim\mathcal{U}=n$ where $\dim$ refers to the projective dimension of a variety. We say that $\mathcal{V}$ and $\mathcal{U}$ \textbf{intersect transversely} if and only if for any $Z\in\mathcal{V}\cap\mathcal{U}$, the tangent spaces $\mathbbm{T}_Z\mathcal{V}$ and $\mathbbm{T}_Z\mathcal{U}$ span $\mathbbm{P}^n$. 
\end{definition}
Figure \ref{figtangentintersection} in the previous page shows, in a schematic way, the difference between a transverse intersection of two varieties and a one which is not transverse. There also exists the notion of generic transverse intersection \cite[Chapter 18]{Harris}. It plays a role in the formulation of Bezout's theorem, which is the result mentioned at the beginning of this section. However, we think for the purpose of this thesis, it is sufficient to state Bezout's theorem in its very basic form, which we do in the following. For more general formulations, consult the classical book by J. Harris \cite[Chapter 18]{Harris}. 
\begin{theorem}[Bezout]\label{thmBezout}
Let $\mathcal{V}$ and $\mathcal{U}$ be two projective varieties in $\mathbbm{P}^n$ of complementary dimension, i.e. $\dim\mathcal{V}+\dim\mathcal{U}=n$ where $\dim$ refers to the projective dimension of a variety. Let the degrees of $\mathcal{V}$ and $\mathcal{U}$ be $c$ and $d$. Assume that $\mathcal{V}$ and $\mathcal{U}$ intersect transversely. In such case, $\mathcal{V}\cap\mathcal{U}$ consists of \textbf{precisely $\mathbf{cd}$ points}. 
\end{theorem} 
We also have the immediate
\begin{corollary}\label{corBezout}
Let $\mathcal{V}$ be a projective variety in $\mathbbm{P}^n$ of projective dimension $\dim\mathcal{V}$ and let $\mathcal{P}$ be a projective plane of complementary dimension, i.e. $\dim\mathcal{V}+\dim\mathcal{P}=n$, where $\dim$ refers to the projective dimension of a variety. Let the degree of $\mathcal{V}$ be $d$. Assume that $\mathcal{V}$ and $\mathcal{P}$ intersect transversely. In such case, $\mathcal{V}\cap\mathcal{P}$ consists of \textbf{precisely $\mathbf{d}$ points}. 
\end{corollary} 
The above corollary of Bezout's theorem proves to be a key ingredient in the proof of the main result of the thesis, which we present in Chapter \ref{chPPT3x3}. Note, once again, that there exists a very general version of Bezout's theorem, which refers to so-called \textit{intersection multiplicities} \cite{Harris} and does not require the two projective varieties to be of complementary dimension. However, this topic is beyond the focus of the thesis.

\part{Results obtained  and examples solved}\label{partIII}\,
\chapter{A structure theorem for a class of cones of positive maps}\label{chmappingcones}\,
In Sections \ref{secsep}, \ref{secdistill} and \ref{secexamplesbound} of the introductory Part \ref{partI} of the thesis, we refered to the notion of positive maps, i.e. maps that preserve the set of positive-definite matrices. It may seem that positive maps are perfectly suited for the description of physical processes, as they map density matrices into density matrices, or positive definite matrices at least. However, a more careful analysis, which can be found e.g. in \cite{BZ2006}, shows that the first impression is wrong. It turns out that a physical process that can be described as a map $\Phi:\rho\mapsto\Phi\left(\rho\right)$ must necessarily have $\Phi$ not only positive, but also \textit{completely positive}. By complete positive positivity of a $\Phi$ we mean the property that the map
\begin{equation}\label{eqCPdef}
\left[
\begin{array}{cccc}
A_{11}&A_{12}&\ldots&A_{1n}\\
A_{21}&A_{22}&\ldots&A_{2n}\\
\vdots&\vdots&\ddots&\vdots\\
A_{n1}&A_{n2}&\ldots&A_{nn}
\end{array}
\right]\mapsto
\left[
\begin{array}{cccc}
\Phi\left(A_{11}\right)&\Phi\left(A_{12}\right)&\ldots&\Phi\left(A_{1n}\right)\\
\Phi\left(A_{21}\right)&\Phi\left(A_{22}\right)&\ldots&\Phi\left(A_{2n}\right)\\
\vdots&\vdots&\ddots&\vdots\\
\Phi\left(A_{n1}\right)&\Phi\left(A_{n2}\right)&\ldots&\Phi\left(A_{nn}\right)
\end{array}
\right],
\end{equation}
mapping operators on $\setC^n\otimes\hilbertspaceone$ into operators on the same space, is positive for arbitrary $n$. Here $\hilbertspaceone$ denotes the space in which $\rho$ lives. To see that for $\Phi$ corresponding to a physical process the map \eqref{eqCPdef} must indeed be positive for all $n$, one can imagine two very distant quantum systems, which do not interact at the present moment. However, they may have interacted in the past. Let one of them be described by states on $\hilbertspaceone$, and let the other one be an $n$-dimensional system with states on $\setC^n$. The initial state of the composite system can in principle be an arbitrary state on $\setC^n\otimes\hilbertspaceone$. The map acting on the composite system when the first subsystem undergoes the process $\Phi$ and the second subsystem remains untouch, is given by $\One\otimes\Phi$. Here $1$ denotes identity on $n\times n$ matrices. This is precisely the map \eqref{eqCPdef}, and it must be positive since, as we mentioned, the initial state of the composite system can be arbitrary.   

Nevertheless, we have already seen that maps which are positive, but not completely positive are not useless in the theory of quantum information. In Section \ref{secsep} we explained the role of entanglement witnesses, which correspond to positive but not completely positive maps, for entanglement detection. On the other hand, in Section \ref{secdistill} we showed a direct connection of distillability of quantum states to the property of being $2$-positive. In the following, we introduce a unifying framework for completely positive, $2$-positive and several other natural classes of positive maps. The idea comes from an early work by St{\o}rmer \cite{Stormer86} and consists in distinguishing the class of cones with certain symmetry property.
They are called \textit{mapping cones}, and in the context discussed here, \textit{cones with a mapping cone symmetry} or \textit{mcs-cones}, due to a minor difference from the original definition by St{\o}rmer.

Let us describe the setup for our discussion. Let $\hilbertspaceone$ and $\hilbertspacetwo$ be two Hilbert spaces. We denote with $\innerpr{.}{.}$ the inner product in $\hilbertspaceone$ or $\hilbertspacetwo$. In the following, we shall assume that $\hilbertspaceone$ and $\hilbertspacetwo$ are finite-dimensional and thus equivalent to $\setC^m$ and $\setC^n$ for some $m,n\in\setN$, $\dim\hilbertspaceone=m$, $\dim\hilbertspacetwo=n$. We also fix orthonormal bases $\seq{f_j}{j=1}{m}$ and $\seq{e_i}{i=1}{n}$ of $\hilbertspaceone$ and $\hilbertspacetwo$, respectively. Thus we have a very specific setting for our discussion, but we shall keep the abstract notation of Hilbert spaces, hoping to bring the attention of the reader to possible generalizations to the infinite-dimensional case. Let us denote with $\bk$ and $\bh$ the spaces of bounded operators on $\hilbertspaceone$ and $\hilbertspacetwo$ respectively, and choose their canonical bases $\seq{f_{kl}}{k,l=1}{m}$, $\seq{e_{ij}}{i,j=1}{n}$. That is, $f_{kl}\left(e_j\right)=\delta_{lj}f_k$ and similarly for the $e_{ij}$.  Positive elements of $\bk$ are operators $A\in\bk$ such that $\innerpr{v}{A\left(v\right)}\geqslant 0\,\forall_{v\in\hilbertspacetwo}$. Similarly for elements of $\bh$. The sets of positive elements of $\bk$ and $\bh$ will be denoted by $\bkplus$ and $\bhplus$. In the finite-dimensional case, there exists a natural inner product in $\bk$, given by the formula
\begin{equation}\label{HSProd}
 \innerprtwo{A}{B}:=\Tr\left(A\conj{B}\right)
\end{equation}
for $A,B\in\bk$. An identical definition works for $A,B\in\bh$ and we do not distinguish notationally between the inner products in $\bh$ and $\bk$. Note that the bases $\seq{f_{kl}}{k,l=1}{m}$ and $\seq{e_{ij}}{i,j=1}{n}$ are orthonormal with respect to $\innerprtwo{.}{.}$.

In the following, we will be mostly dealing with linear maps from $\bk$ to $\bh$. Because of the finite-dimensionality assumption, they are all elements of $\bbounded{\bk}{\bh}$, the space of bounded operators from $\bk$ to $\bh$. Given a map $\Phi\in\bbounded{\bk}{\bh}$, we define its conjugate $\conj{\Phi}$ as a map from $\bh$ into $\bk$ satisfying $\innerprtwo{A}{\Phi\left(B\right)}=\innerprtwo{\conj{\Phi}\left(A\right)}{B}$ for all $A\in\bh$ and $B\in\bk$. In our setting, there also exists a natural inner product in $\bbounded{\bk}{\bh}$, given by the formula
\begin{equation}\label{HSProdtwo}
 \innerprthree{\Phi}{\Psi}:=\sum_{k,l=1}^{m}\innerprtwo{\Phi\left(f_{kl}\right)}{\Psi\left(f_{kl}\right)}.
\end{equation}
Note that the spaces $\bbounded{\bh}{\bk}$, $\bounded{\bk}$ and $\bounded{\bh}$ can be endowed with analogous inner products and we shall not notationally distinguish between them. The following proposition summarizes a few elementary facts about $\innerprthree{.}{.}$ that will be useful for our later discussion.

\begin{proposition}\label{propinnerpr}
 For all $\Phi,\Psi\in\bbounded{\bk}{\bh}$ and $\alpha\in\bounded{\bh}$, $\beta\in\bounded{\bk}$, and $\circ$ denoting the composition of maps, one has the following equalities
\begin{enumerate}
\item $\innerprthree{\Phi\circ\beta}{\Psi}=\innerprthree{\beta}{\conj{\Phi}\circ\Psi}=\innerprthree{\conj{\Psi}\circ\Phi}{\conj{\beta}}$,
\item $\innerprthree{\alpha\circ\Phi}{\Psi}=\innerprthree{\alpha}{\Psi\circ\conj{\Phi}}=\innerprthree{\Phi\circ\conj{\Psi}}{\conj{\alpha}}$,
\item $\innerprthree{\alpha\circ\Phi\circ\beta}{\Psi}=\innerprthree{\Phi}{\conj{\alpha}\circ\Psi\circ\conj{\beta}}$.
\end{enumerate}

\begin{proof} The first equality in point one follows  directly from $\innerprtwo{\Phi\circ\beta\left(f_{kl}\right)}{\Psi\left(f_{kl}\right)}=\innerprtwo{\beta\left(f_{kl}\right)}{\conj{\Phi}\circ\Psi\left(f_{kl}\right)}$ and the definition of $\innerprthree{.}{.}$, eq. \eqref{HSProdtwo}. To prove the other equalities, we can use a simple lemma.
\begin{lemma}\label{lemmapoema}
 For any finite-dimensional Hilbert spaces $\hilbertspaceone$, $\hilbertspacetwo$ and maps $\Phi,\Psi\in\bbounded{\bk}{\bh}$, we have
\begin{equation}
\innerprthree{\Phi}{\Psi}=\innerprthree{\conj{\Psi}}{\conj{\Phi}}. 
\end{equation}
\begin{proof} Starting from the definition of $\innerprthree{.}{.}$, we get
\begin{multline}\label{proofLemma1}
\innerprthree{\Phi}{\Psi}=\sum_{k,l=1}^m\innerprtwo{\Phi\left(f_{kl}\right)}{\Psi\left(f_{kl}\right)}=\sum_{i,j=1}^n\sum_{m,n=1}^n\sum_{k,l=1}^m\Phi_{ij,kl}\overline{\Psi_{mn,kl}}\innerprtwo{e_{ij}}{e_{mn}}=\\=\sum_{i,j=1}^n\sum_{k,l=1}^m\Phi_{ij,kl}\overline{\Psi_{ij,kl}}
=\sum_{i,j=1}^n\sum_{k,l=1}^m\sum_{r,s=1}^m\Phi_{ij,rs}\overline{\Psi_{ij,kl}}\innerprtwo{f_{rs}}{f_{kl}}=\\=\sum_{i,j=1}^n\sum_{k,l=1}^m\sum_{r,s=1}^m\innerprtwo{\overline{\Psi_{ij,rs}}f_{r,s}}{\overline{\Phi_{ij,kl}}f_{kl}}=\sum_{i,j=1}^n\innerprtwo{\conj{\Psi}\left(e_{ij}\right)}{\conj{\Phi}\left(e_{ij}\right)},
\end{multline}
where the last equality follows because $\conj{\Phi}\left(e_{ij}\right)=\sum_{k,l=1}^m\overline{\Phi_{ij,kl}}f_{kl}$ as a consequence of $\innerprtwo{f_{kl}}{\conj{\Phi}\left(e_{ij}\right)}=\innerprtwo{\Phi\left(f_{kl}\right)}{e_{ij}}=\sum_{r,s=1}^m\overline{\Phi_{rs,kl}}\innerprtwo{e_{rs}}{e_{ij}}=\overline{\Phi_{ij,kl}}$. Similarly, $\conj{\Psi}\left(e_{ij}\right)=\sum_{r,s=1}^m\overline{\Phi_{ij,rs}}f_{rs}$ holds.
The final expression in \eqref{proofLemma1} clearly equals $\innerprthree{\conj{\Psi}}{\conj{\Phi}}$.
\end{proof} 
\end{lemma}
Note that the assertion of Lemma \ref{lemmapoema} holds for any choice of $\hilbertspaceone$ and  $\hilbertspacetwo$, and thus also when the two finite-dimensional Hilbert spaces are different from the $\hilbertspaceone$ and  $\hilbertspacetwo$ referred to in the statement of the proposition. Using the lemma, we get $\innerprthree{\beta}{\conj{\Phi}\circ\Psi}=\innerprthree{\conj{\Psi}\circ\Phi}{\conj{\beta}}$, which proves the second equality in point one. Furthermore,
\begin{multline}\label{eqpointtwo}
\innerprthree{\alpha\circ\Phi}{\Psi}=\innerprthree{\conj{\Psi}}{\conj{\Phi}\circ\conj{\alpha}}=\overline{\innerprthree{\conj{\Phi}\circ\conj{\alpha}}{\conj{\Psi}}}=\\=\overline{\innerprthree{\conj{\alpha}}{\Phi\circ\conj{\Psi}}}=\innerprthree{\Phi\circ\conj{\Psi}}{\conj{\alpha}}=\innerprthree{\alpha}{\Psi\circ\conj{\Phi}},
\end{multline}
where we successively used Lemma \ref{lemmapoema}, the conjugate symmetry of $\innerprthree{.}{.}$, the first equation in point one, the conjugate symmetry again, and finally Lemma \ref{lemmapoema} for the second time. Obviously, the first, the fifth and the sixth term in equation \eqref{eqpointtwo} are the same as in point two of the proposition. Hence the only remaining thing to prove is point three. We have
\begin{multline}\label{eqpointthree} 
\innerprthree{\alpha\circ\Phi\circ\beta}{\Psi}=\innerprthree{\alpha}{\Psi\circ\conj{\beta}\circ\conj{\Phi}}=\innerprthree{\beta\circ\conj{\Psi}\circ\alpha}{\conj{\Phi}}=\innerprthree{\Phi}{\conj{\alpha}\circ\Psi\circ\conj{\beta}},
\end{multline}
where we used the two properties $\innerprthree{\alpha\circ\Phi}{\Psi}=\innerprthree{\alpha}{\Psi\circ\conj{\Phi}}$ with $\Phi\rightarrow\Phi\circ\beta$, $\innerprthree{\beta}{\conj{\Phi}\circ\Psi}=\innerprthree{\Phi\circ\beta}{\Psi}$ with $\beta\rightarrow\alpha$, $\Phi\rightarrow\beta\circ\conj{\Psi}$ and $\Psi\rightarrow\conj{\Phi}$, and finally Lemma \ref{lemmapoema}.
 \end{proof}
\end{proposition}

Consider the tensor product $\kh$. This space has a natural inner product, inherited from $\hilbertspaceone$ and $\hilbertspacetwo$, and an orthonormal basis $\seq{f_{kl}\otimes e_{ij}}{i,j=1;k,l=1}{n;m}$. Similarly to $\bk$ and $\bh$, the space $\bkh$ of bounded operators on $\kh$ is endowed with a natural Hilbert-Schmidt product, defined by formula \eqref{HSProd} with $A,B\in\bkh$. We shall again denote the inner product with $\innerprtwo{.}{.}$ to avoid excess notation. As we explained in previous sections, there exists a one-to-one correspondence between linear maps $\Phi$ of $\bk$ into $\bh$ and elements of $\bkh$, given by
\begin{equation}\label{Jamisodef}
\Phi\mapsto\Choimatr{\Phi}:=\sum_{k,l=1}^m f_{kl}\otimes\Phi\left(f_{kl}\right).
\end{equation}
The symbol $\Choimatr{\Phi}$ denotes the {\it Choi matrix} of $\Phi$ \cite{ref.Choi75} and the mapping $J:\Phi\mapsto\Choimatr{\Phi}$ is sometimes called the {\it Jamio\l kowski-Choi isomorphism} \cite{ref.J72}. In fact, $J$ is not only an isomorphism, but also an \emph{isometry} between $\bbounded{\bk}{\bh}$ and $\bkh$ in the sense of Hilbert-Schmidt type inner products. One has the following
\begin{lemma}\label{propertythree}The Jamio\l kowski-Choi isomorphism is an isometry. One has
\begin{equation}
 \innerprthree{\Phi}{\Psi}=\innerprtwo{\Choimatr{\Phi}}{\Choimatr{\Psi}}
\end{equation}
for all $\Phi,\Psi\in\bbounded{\bk}{\bh}$ (with $\Choimatr{\Phi},\Choimatr{\Psi}\in\bkh$).
\begin{proof}By the definition of $\Choimatr{\Phi}$ and $\Choimatr{\Psi}$,
\begin{equation}\label{proofJamisoiso1}
 \innerprtwo{\Choimatr{\Phi}}{\Choimatr{\Psi}}=\innerprtwo{\sum_{k,l=1}^mf_{kl}\otimes\Phi\left(f_{kl}\right)}{\sum_{r,s=1}^mf_{rs}\otimes\Psi\left(f_{rs}\right)}=\ldots
\end{equation}
Since $\Tr\left(\left(A\otimes A'\right)\conj{\left(B\otimes B'\right)}\right)=\Tr\left(A\conj{B}\right)\Tr\left(A'\conj{B'}\right)$ for arbitrary $A,B\in\bk$ and $A',B'\in\bh$, by formula \eqref{HSProd} we have
\begin{equation}\label{proofJamisoiso2}
 \ldots=\sum_{k,l=1}^m\sum_{r,s=1}^m\innerprtwo{f_{kl}}{f_{rs}}\innerprtwo{\Phi\left(f_{kl}\right)}{\Psi\left(f_{rs}\right)}=\sum_{k,l=1}^m\innerprtwo{\Phi\left(f_{kl}\right)}{\Psi\left(f_{kl}\right)},
\end{equation}
where we used orthonormality of $\seq{f_{kl}}{k,l=1}{m}$.
The last expression equals $\innerprthree{\Phi}{\Psi}$ by definition \eqref{HSProdtwo}.
 \end{proof}
\end{lemma}

Let us recall that a linear map $\Phi$ from $\bk$ to $\bh$ is called {\it positive} if it preserves positivity of operators, which means $\Phi\left(\bkplus\right)\subset\bhplus$. Moreover, $\Phi$ is called {\it$k$-positive} if $\Phi\otimes\Id_{\matrices{k}{\setC}}$ is positive as a map from $\bk\otimes\matrices{k}{\setC}$ into $\bh\otimes\matrices{k}{\setC}$, where $\matrices{k}{\setC}$ denotes the space of $k\times k$ matrices with complex entries and $\Id$ refers to the identity map. A map $\Phi$ is called {\it completely positive} if it is $k$-positive for all $k\in\setN$. From the Choi's theorem on completely positive maps \cite{ref.Choi75} (cf. also Lemma \ref{lemmaCofAdV}) it follows that every such map has a representation $\Phi=\sum_i\Ad_{V_i}$ as a sum of {\it conjugation maps}, $\Ad_{V_i}:\rho\mapsto V_i\rho\conj{V_i}$ with $V_i\in\bbounded{\hilbertspaceone}{\hilbertspacetwo}$. Conversely, every map $\Phi$ of the form $\sum_i\Ad_{V_i}$ is completely positive. If all the $V_i$'s can be chosen of rank $\leqslant k$ for some $k\in\setN$, $\Phi$ is said to be {\it$k$-superpositive} \cite{ref.SSZ09}. One-superpositive maps are simply called {\it superpositive} \cite{ref.Ando04}. The sets of positive, $k$-positive, completely positive, $k$-superpositive and superpositive maps from $\bk$ to $\bh$ will be denoted with $\Pmapsbb{\bk}{\bh}$, $\kPmapsbb{k}{\bk}{\bh}$, $\CPmapsbb{\bk}{\bh}$, $\kSPmapsbb{k}{\bk}{\bh}$, $\SPmapsbb{\bk}{\bh}$ or $\Pmaps$, $\kPmaps{k}$, $\CPmaps$, $\kSPmaps{k}$, $\SPmaps$ for short. It is clear that all of them are closed convex cones contained in $\Pmapsbb{\bk}{\bh}$. They also share a more special property that the product $\Upsilon\circ\Phi\circ\Omega$ of $\Phi\in\mappingcone$, $\Upsilon\in\CPmapsb{\bh}$ and $\Omega\in\CPmapsb{\bk}$ is an element of $\mappingcone$ again, where $\mappingcone$ stands for one of the sets $\Pmaps$, $\kPmaps{k}$, $\CPmaps$, $\kSPmaps{k}$ and $\SPmaps$ (cf. e.g. \cite{ref.SSZ09}). Thus, following rather closely the original definition by St{\o}rmer \cite{Stormer86}, we make
\begin{definition} A \textbf{cone with a mapping cone symmetry}, or an \textbf{mcs-cone} for short, is defined as a closed convex cone $\mappingcone$ in $\Pmapsbb{\bk}{\bh}$, different from $\left\{0\right\}$, such that 
\begin{equation}
\Upsilon\circ\Phi\circ\Omega\in\mappingcone 
\end{equation}
for all $\Phi\in\mappingcone$, 
$\Upsilon\in\CPmapsb{\bh}$ and $\Omega\in\CPmapsb{\bk}$. 
\end{definition}
In the following, the convexity assumption could sometimes be skept, and we do include appropriate comments. 

Note that the set of positive maps from $\bk$ into $\bh$ is contained in the real-linear subspace $\HPmapsbb{\bk}{\bh}\subset\bbounded{\bk}{\bh}$ ($\HPmaps$ for short) consisting of all Hermiticity-preserving maps, i.e. $\Phi$ such that $\Phi\left(\conj{X}\right)=\conj{\Phi\left(X\right)}$. Moreover, the image of $\HPmapsbb{\bk}{\bh}$ by $J:\Phi\mapsto\Choimatr{\Phi}$ equals the set of self-adjoint elements of $\bkh$ \cite{ref.Pillis}. Therefore $\innerprthree{.}{.}$ induces a \emph{symmetric}
 inner product on $\HPmapsbb{\bk}{\bh}$ (cf. Property \ref{propertythree}). By definition, all mapping cones are subsets of $\Pmaps$ and thus of $\HPmaps$. Since $\HPmaps$ is a finite-dimensional space over $\setR$ with a symmetric inner product $\innerprthree{.}{.}$, one can easily apply to it tools of convex analysis. In particular, given any cone $\mappingcone\subset\HPmaps$, one defines its {\it dual} $\dual{\mappingcone}$ as the cone of elements $\Psi\in\HPmaps$ such that $\innerprthree{\Psi}{\Phi}\geqslant 0$ for all $\Phi\in\mappingcone$,
\begin{equation}\label{dualconedef}
 \dual{\mappingcone}:=\left\{\Psi\in\HPmapsbb{\bk}{\bh}\,\vline\,\innerprthree{\Psi}{\Phi}\geqslant 0\,\forall_{\Phi\in\mappingcone}\right\}.
\end{equation}
 Obviously, $\dual{\mappingcone}$ is closed and convex. It has a clear geometrical interpretation as the convex cone spanned by the normals to the supporting hyperplanes for $\mappingcone$. The dual cone has a well-known counterpart in convex analysis \cite{ref.Rockafellar}, $\mappingcone^{\star}=-\dual{\mappingcone}$, which is called the {\it polar} of $\mappingcone$. We have the following
\begin{lemma}\label{propertyfour}
 Let $\mappingcone$ be a closed convex cone. Then
$\mappingcone=\ddual{\mappingcone}$.
\begin{proof}
 The formula $\ddual\mappingcone=\mappingcone$ is equivalent to $\mappingcone^{\star\star}=\mappingcone$ for a closed convex cone $\mappingcone$. The latter equality is a known fact in convex analysis. A proof can be found e.g. in \cite{ref.Rockafellar} (Theorem 14.1). 
\end{proof}
\end{lemma}
It can be shown (cf. e.g. \cite{ref.SSZ09}) that a duality relation $\dual{\kPmaps{k}}=\kSPmaps{k}$ holds for all $k\in\setN$. The converse relation $\dual{\kSPmaps{k}}=\kPmaps{k}$ is also true, as a consequence of Property \ref{propertyfour}. In particular, for $k=1$ we get $\dual{\SPmaps}=\Pmaps$ and $\dual{\Pmaps}=\SPmaps$. Taking $k=\min\left\{m,n\right\}$, one obtains $\dual{\CPmaps}=\CPmaps$, which is in accordance with Choi's theorem on completely positive maps \cite{ref.Choi75} and with Property \ref{propertythree}. 

In the following, we shall be interested in duality relations between mcs-cones. This is in general a well-posed problem, because the operation $\mappingcone\rightarrow\dual{\mappingcone}$ acts within the ``mcs'' class. We have
\begin{proposition}\label{dualisamappingcone}
Let $\mappingcone\subset\Pmapsbb{\bk}{\bh}$ be an arbitrary mcs-cone. Then $\dual{\mappingcone}$, defined as in \eqref{dualconedef}, is an mcs-cone as well.
\begin{proof}
 Let $\Psi$ be an element of $\dual{\mappingcone}$. First we prove that $\Upsilon\circ\Psi\circ\Omega\in\dual{\mappingcone}$ for all $\Upsilon\in\CPmapsb{\bh}$ and $\Omega\in\CPmapsb{\bk}$. We have $\conj{\Upsilon}\in\CPmapsb{\bh}$ and $\conj{\Omega}\in\CPmapsb{\bk}$ because the sets of completely positive maps are $\conjsymb$-invariant. Therefore $\conj{\Upsilon}\circ\Phi\circ\conj{\Omega}\in\mappingcone$ for an arbitrary element $\Phi$ of the cone $\mappingcone$. By the definition \eqref{dualconedef} of $\dual{\mappingcone}$, we have
$ \innerprthree{\Psi}{\conj{\Upsilon}\circ\Phi\circ\conj{\Omega}}\geqslant 0\,\forall_{\Phi\in\mappingcone}$. Using Proposition \ref{propinnerpr}, point three, we can rewrite this as
\begin{equation}\label{dualineq2}
 \innerprthree{\Upsilon\circ\Psi\circ\Omega}{\Phi}\geqslant 0\,\forall_{\Phi\in\mappingcone}.
\end{equation}
According to definition \eqref{dualconedef}, condition \eqref{dualineq2} means that $\Upsilon\circ\Psi\circ\Omega\in\dual{\mappingcone}$. This holds for arbitrary $\Upsilon\in\CPmapsb{\bh}$ and $\Omega\in\CPmapsb{\bk}$. The only thing which is left to prove is $\dual{\mappingcone}\subset\Pmapsbb{\bk}{\bh}$. The inclusion holds because every mcs-cone $\mappingcone$ contains all the conjugation maps $\Ad_V$ with $\textnormal{rank}\,V=1$. Consequently, $\dual{\mappingcone}\subset\dual{\convhull\left\{\Ad_V\,\vline\,\textnormal{rank}\,V=1\right\}}=\dual{\SPmaps}=\Pmaps$. To show that indeed $\left\{\Ad_V\vline\textnormal{rank}\,V=1\right\}\subset\mappingcone$ for any mcs-cone $\mappingcone$, take an arbitrary nonzero $\Phi\in\mappingcone$. There must exist normalized vectors $\upsilon\in\hilbertspaceone$ and $\omega\in\hilbertspacetwo$ such that $\innerprtwo{\proj{\omega}}{\Phi\left(\proj{\upsilon}\right)}\geqslant 0$, where $\proj{\upsilon}$ and $\proj{\omega}$ are orthogonal projections onto the one-dimensional subspaces spanned by $\upsilon$ and $\omega$. Denote $\chi:=\innerprtwo{\proj{\omega}}{\Phi\left(\proj{\upsilon}\right)}$. Consider a pair of maps, $U:\hilbertspaceone\ni a\mapsto\innerpr{a}{\upsilon'}\upsilon\in\hilbertspaceone$ and $W:\hilbertspacetwo\ni b\mapsto\innerpr{b}{\omega}\omega'\in\hilbertspacetwo$, where $\upsilon'$ and $\omega'$ are arbitrary normalized vectors in $\hilbertspaceone$ and $\hilbertspacetwo$. A map $\Phi'$, defined as $\lambda/\chi\left(\Ad_W\circ\,\Phi\circ\Ad_U\right)$ acts in the following
 way, $\Phi':\rho\mapsto\lambda\innerprtwo{\proj{\upsilon'}}{\rho}\proj{\omega'}$ or $\Phi'=\Ad_V$ with $V:\hilbertspaceone\ni c\mapsto\lambda\innerpr{\upsilon'}{c}\omega'$. Any rank one operator $V$ can be written in the latter form for some $\upsilon'$ and $\omega'$. But $\Phi'$ is an element of $\mappingcone$ because of the assumption that $\mappingcone$ is an mcs-cone. Thus indeed $\Ad_V\in\mappingcone$ for all $V\in\bbounded{\hilbertspaceone}{\hilbertspacetwo}$ such that $\textnormal{rank}\,V=1$. In the case of $\hilbertspaceone=\hilbertspacetwo$ and mapping cones $\mappingcone$ as in the original definition by St{\o}rmer,  the inclusion $\Ad_V\in\mappingcone$ follows from Lemma 2.4 in \cite{Stormer86}. Note that we never used convexity of $\mappingcone$ in the proof.
\end{proof}
\end{proposition}

Using the lemmas introduced above, we can almost immediately prove a surprising characterization theorem for mcs-cones, which was strongly suggested by earlier results on the subject \cite{ref.St09dual,ref.St09mappingcones,ref.SSZ09}. It holds without any additional assumptions about the cone, and is noteworthy as it links the condition that two maps $\Phi$, $\Psi$ lay in a pair of dual mcs-cones to the fact that the product $\conj{\Psi}\circ\Phi$ is a $\CPmaps$ map. Thus it reveals a connection between convex geometry and a fact which is more likely to be called algebraic than geometrical. Before we proceed with the proof, let us show a simple lemma, which is a version of \cite[Lemma 1$(i)$]{ref.SS10} for $\hilbertspaceone\neq\hilbertspacetwo$.
\begin{lemma}\label{lemmaCofAdV}Let $V:\hilbertspaceone\ni a\mapsto\sum_{i=1}^n\sum_{j=1}^mV_{ij}\innerpr{a}{f_j}e_i\in\hilbertspacetwo$ be an arbitrary operator in $\bbounded{\bk}{\bh}$ and consider the map $\Ad_V:\rho\mapsto V\rho\conj{V}$. Then
\begin{equation}\label{CAdV1}
\Choimatr{\Ad_V}=\diad{\upsilon}, 
\end{equation}
where $\upsilon=\sum_{i=1}^n\sum_{j=1}^mV_{ij}f_j\otimes e_i$ is a vector in $\hilbertspaceone\otimes\hilbertspacetwo$ and $\diad{\upsilon}:w\mapsto\innerpr{w}{\upsilon}\upsilon$ is proportional to an orthogonal projection onto the subspace spanned by $\upsilon$.
\begin{proof}Obviously, the map $\conj{V}$ acts in the following way,
\begin{equation}\label{eqVstaracts}
 \conj{V}:\hilbertspacetwo\ni b\mapsto\sum_{i=1}^n\sum_{j=1}^m\overline{V_{ij}}\innerpr{b}{e_i}f_j\in\hilbertspaceone.
\end{equation}
 Thus 
\begin{equation}\label{VfklV}
Vf_{kl}\conj{V}:\hilbertspacetwo\ni b\mapsto\sum_{i,r=1}^n\sum_{j,s=1}^mV_{rs}\innerpr{f_{kl}\left(f_j\right)}{f_{s}}\overline{V_{ij}}\innerpr{b}{e_i}e_{r}\in\hilbertspacetwo,
\end{equation}
where the last expression is easily verified to be equal to $\sum_{i,r=1}^nV_{rk}\overline{V_{il}}\innerpr{b}{e_i}e_{r}$. Thus we have $Vf_{kl}\conj{V}=\sum_{i,r=1}^nV_{rk}\overline{V_{il}}e_{ri}$ and by the definition \eqref{Jamisodef} of the Choi matrix,
\begin{equation}\label{CAdV2}
 \Choimatr{\Ad_V}=\sum_{k,l=1}^m\sum_{i,r=1}^nV_{rk}\overline{V_{il}}f_{kl}\otimes e_{ri}=\diad{\upsilon},
\end{equation}
with $\upsilon=\sum_{i=1}^n\sum_{j=1}^mV_{ij}f_j\otimes e_i$. A proof of the last equality in \eqref{CAdV2} is left as an elementary exercise for the reader. 
\end{proof}
\end{lemma}

We are ready to prove the following result, which is an extension of Theorem~1 in \cite{ref.St09dual}.

\begin{theorem}\label{MainTheorem}
 Let $\mappingcone\subset\Pmapsbb{\bk}{\bh}$ be an mcs-cone. The following conditions are equivalent,
\begin{enumerate}
 \item $\Phi\in\mappingcone$,
\item $\conj{\Psi}\circ\Phi\in\CPmapsb{\bk}$ for all $\Psi\in\dual{\mappingcone}$,
\item $\Phi\circ\conj{\Psi}\in\CPmapsb{\bh}$ for all $\Psi\in\dual{\mappingcone}$.
\end{enumerate}
\begin{proof}
 We first show $1\Leftrightarrow 2$. Let us start with $2\Rightarrow 1$. Since $\conj{\Psi}\circ\Phi\in\CPmaps\,\forall_{\Psi\in\dual{\mappingcone}}$, we can use the facts that $\dual{\CPmaps}=\CPmaps$ and $\Id\in\CPmaps$ to get 
\begin{equation}\label{conjPsiPhiid}
 \innerprthree{\conj{\Psi}\circ\Phi}{\Id}\geqslant 0\,\forall_{\Psi\in\dual{\mappingcone}}.
\end{equation}
By using point one of Proposition \ref{propinnerpr} with the identity map $\Id$ substituted for $\beta$, we get $\innerprthree{\Phi}{\Psi}\geqslant 0\,\forall_{\Psi\in\dual{\mappingcone}}$, which means that $\Phi\in\ddual{\mappingcone}$. But $\ddual{\mappingcone}=\mappingcone$ because $\mappingcone$ is a closed convex cone and Property \ref{propertyfour} holds. Hence $\Phi\in\mappingcone$. The proof of $1\Rightarrow 2$ strongly builds on the assumption that $\mappingcone$ has the mapping cone symmetry. By Proposition \ref{dualisamappingcone}, we know that $\dual{\mappingcone}$ is an mcs-cone as well. Therefore $\Psi\circ\Ad_V\in\dual{\mappingcone}$ for an arbitrary $\Psi\in\dual{\mappingcone}$ and $V\in\bk$. We have $\innerprthree{\Psi\circ\Ad_V}{\Phi}\geqslant 0\,\forall_{V\in\bk}\forall_{\Psi\in\dual{\mappingcone}}$. By Proposition \ref{propinnerpr}, point one, we get $\innerprthree{\Psi\circ{\Ad}_V}{\Phi}=\innerprthree{{\Ad}_V}{\conj{\Psi}\circ\Phi}$. Using Property \ref{propertythree} and Lemma \ref{lemmaCofAdV} with $\hilbertspacetwo=\hilbertspaceone$, the last term can be rewritten as
\begin{equation}\label{fourequalitiesforinnerpr}
 \innerprthree{{\Ad}_V}{\conj{\Psi}\circ\Phi}=\innerprtwo{\Choimatr{\Ad_V}}{\Choimatr{\conj{\Psi}\circ\Phi}}=\innerprtwo{\diad{v}}{\Choimatr{\conj{\Psi}\circ\Phi}}=\innerpr{\upsilon}{\Choimatr{\conj{\Psi}\circ\Phi}\left(\upsilon\right)},
\end{equation}
where $\upsilon=\sum_{i,j=1}^m V_{ij}f_j\otimes f_i$ for $V:\hilbertspaceone\ni a\mapsto\sum_{i,j=1}^mV_{ij}\innerpr{a}{f_j}f_i\in\hilbertspaceone$. The vector $\upsilon\in\hilbertspaceone\otimes\hilbertspaceone$ can be arbitrary, since we do not assume anything about the operator $V$. Consequently, the condition $\innerprthree{\Psi\circ\Ad_V}{\Phi}\geqslant 0\,\forall_{V\in\bk}\forall_{\Psi\in\dual{\mappingcone}}$ is equivalent to
\begin{equation}\label{posChoimatr}
 \innerpr{\upsilon}{\Choimatr{\conj{\Psi}\circ\Phi}\left(\upsilon\right)}\geqslant 0\,\forall_{\upsilon\in\hilbertspaceone\otimes\hilbertspaceone}\,\forall_{\Psi\in\dual{\mappingcone}},
\end{equation}
which means that $\Choimatr{\conj{\Psi}\circ\Phi}\in\boundedplus{\hilbertspaceone\otimes\hilbertspaceone}$ for all $\Psi\in\dual{\mappingcone}$. By the Choi theorem on completely positive maps \cite{ref.Choi75}, $\conj{\Psi}\circ\Phi\in\CPmapsb{\bk}$ for all $\Psi\in\dual{\mappingcone}$. Thus we have finished proving that $1\Leftrightarrow 2$. The proof of the equivalence $1\Leftrightarrow 3$ only needs a minor modification of the above argument. Instead of using point one of Proposition \ref{propinnerpr}, point two of the same proposition has to be used. Other details are practically the same as above and we shall not give them explicitly.
\end{proof}
\end{theorem}
In case of $\hilbertspacetwo=\hilbertspaceone$ and a $\conjsymb$-invariant mcs-cone $\mappingcone\in\Pmapsb{\bk}$, Theorem~\ref{MainTheorem} can be further simplified. 
\begin{theorem}\label{MainTheoremconjugatesymmetric}
Let $\mappingcone\subset\Pmapsb{\bk}$ be a $\conjsymb$-invariant mcs-cone. Then the following conditions are equivalent,
\begin{enumerate}
 \item $\Phi\in\mappingcone$,
\item $\Psi\circ\Phi\in\CPmapsb{\bk}$ for all $\Psi\in\dual{\mappingcone}$,
\item $\Phi\circ\Psi\in\CPmapsb{\bk}$ for all $\Psi\in\dual{\mappingcone}$.
\end{enumerate}
\begin{proof}
 Obvious from Theorem \ref{MainTheorem}.
\end{proof}
\end{theorem}
This result was earlier known for $\kPmapsb{k}{\bk}$ and $\kSPmapsb{k}{\bk}$ \cite{ref.SSZ09}, and inexplicitly for all so-called symmetric (and convex) mapping cones \cite{ref.St09mappingcones}. As it was pointed to the author by Erling St{\o}rmer, in the case of $k$-positive maps, not necessarily from $\bk$  into itself, an even stronger characterization of the type of Theorems \ref{MainTheorem} and \ref{MainTheoremconjugatesymmetric} is valid. First, we have the simple
\begin{theorem}\label{MainTheoremkpositivemaps1}The following conditions are equivalent
 \begin{enumerate}
  \item $\Phi\in\kPmapsbb{k}{\bk}{\bh}$,
\item $\Ad_{\conj{V}}\circ\,\Phi\in\CPmapsb{\bk}$ for all $V\in\bbounded{\hilbertspaceone}{\hilbertspacetwo}$ such that $\textnormal{rank}\,V\leqslant k$,
\item $\Phi\circ\Ad_{\conj{V}}\in\CPmapsb{\bh}$ for all $V\in\bbounded{\hilbertspaceone}{\hilbertspacetwo}$ such that $\textnormal{rank}\,V\leqslant k$.
 \end{enumerate}
\begin{proof}
 Obvious from Theorem \ref{MainTheorem}. The duality relation 
\begin{multline}
 \dual{\kPmapsbb{k}{\bk}{\bh}}=\kSPmapsbb{k}{\bk}{\bh}=\\=\convhull\left\{{\Ad}_V|V\in\bbounded{\hilbertspaceone}{\hilbertspacetwo},\textnormal{rank}\,V\leqslant k\right\}
\end{multline}
holds (cf. \cite{ref.SSZ09}) and we can substitute $\Psi$ in Theorem \ref{MainTheorem} with $\Ad_V$, $\textnormal{rank}\,V\leqslant k$. We also use the elementary fact that $\conj{\Ad_V}=\Ad_{\conj{V}}$.
\end{proof}
\end{theorem}
The next result on $k$-positive maps seems to be less obvious.
\begin{theorem}\label{MainTheoremkpositivemaps2}Denote with $\Pi_k\left(\hilbertspaceone\right)$ and $\Pi_k\left(\hilbertspacetwo\right)$ the sets of $k$-dimensional projections in $\hilbertspaceone$ and $\hilbertspacetwo$, resp.
 The following conditions are equivalent
 \begin{enumerate}
  \item $\Phi\in\kPmapsbb{k}{\bk}{\bh}$,
\item $\Ad_E\circ\,\Phi\in\CPmapsbb{\bk}{\bh}$ for all $E\in\Pi_k\left(\hilbertspacetwo\right)$,
\item $\Phi\circ\Ad_F\in\CPmapsbb{\bk}{\bh}$ for all $F\in\Pi_k\left(\hilbertspaceone\right)$,
\item $\Ad_E\circ\,\Phi\circ\Ad_F\in\CPmapsbb{\bk}{\bh}$ for all  $E\in\Pi_k\left(\hilbertspacetwo\right)$, $F\in\Pi_k\left(\hilbertspaceone\right)$.
 \end{enumerate}
\begin{proof}
 We shall prove the equivalence $1\Leftrightarrow 4$. The other ones follow analogously. Since $\dual{\CPmaps}=\CPmaps$ and any $\CPmaps$ map can be written as $\sum_i\Ad_{V_i}$ with $V_i$ arbitrary, the condition $\Ad_E\circ\,\Phi\circ\Ad_F\in\CPmapsbb{\bk}{\bh}$ is equivalent to
\begin{equation}\label{equivalentoffour}
 \innerprthree{{\Ad}_E\circ\,\Phi\circ{\Ad}_F}{{\Ad}_V}\geqslant 0\,\forall_{E\in\Pi_k\left(\hilbertspacetwo\right),F\in\Pi_k\left(\hilbertspaceone\right)}\forall_{V\in\bbounded{\hilbertspaceone}{\hilbertspacetwo}}.
\end{equation}
By Proposition \ref{propinnerpr}, point three, equation \eqref{equivalentoffour} can be rewritten as
\begin{equation}\label{equivalentoffour2}
 \innerprthree{\Phi}{{\Ad}_{EVF}}\geqslant 0\,\forall_{E\in\Pi_k\left(\hilbertspacetwo\right),F\in\Pi_k\left(\hilbertspaceone\right)}\forall_{V\in\bbounded{\hilbertspaceone}{\hilbertspacetwo}},
\end{equation}
where we used the fact that $\Ad_E\circ\Ad_V\circ\Ad_F=\Ad_{EVF}$ and the self-adjointness of $E$ and $F$. Note that $U=EVF$ is an element of $\bbounded{\hilbertspaceone}{\hilbertspacetwo}$ of rank $\leqslant k$. Conversely, every map in $U\in\bbounded{\hilbertspaceone}{\hilbertspacetwo}$ of rank $\leqslant k$ can be written in the form $EVF$ for some $V\in\bbounded{\hilbertspaceone}{\hilbertspacetwo}$, $E\in\Pi_k\left(\hilbertspacetwo\right)$ and $F\in\Pi_k\left(\hilbertspaceone\right)$. It is sufficient to take $V=U$ and $E$, $F$ as the range and rank projections for $U$, resp. Therefore the condition \eqref{equivalentoffour2} is equivalent to $\innerprthree{\Phi}{\Ad_U}\geqslant 0$ for all $U\in\bbounded{\hilbertspaceone}{\hilbertspacetwo}$ s.t. $\textnormal{rank}\,U\leqslant 0$. But this is the same as $\innerprthree{\Phi}{\Psi}\geqslant 0$ for all $\Psi\in\kSPmapsbb{k}{\bk}{\bh}$, or $\Phi\in\dual{\kSPmapsbb{k}{\bk}{\bh}}=\kPmapsbb{k}{\bk}{\bh}$. Thus $1\Leftrightarrow 4$.
\end{proof}
\end{theorem}

Let us note that Theorem \ref{MainTheorem} can be perceived as a very broad generalization  of the so-called \textbf{positive maps entanglement criterion} by the Horodecki family \cite{HHH96}. To see this, we prove the following general
\begin{proposition}[Generalized positive maps criterion]\label{propgenposmaps}
Let $\mappingcone$ be an mcs-cone in $\Pmapsbb{\bk}{\bh}$. An operator $\rho\in\bkh$ belongs to the image $J\left(\mappingcone\right)$ if and only if the following condition
\begin{equation}\label{eqpositivecrit}
\left(\Psi^{\ast}\otimes\Id\right)\rho\in\mathcal{B}^+\left(\mathcal{K}\otimes\mathcal{K}\right)
\end{equation}
holds for all $\Psi\in\dual{\mappingcone}$.
\begin{proof}The proof relies on Theorem \ref{MainTheorem} and the formula \eqref{Jamisodef} for the isomorphism $J$. Let us note that 
\begin{equation}\label{eqequivPIR}
\left(\Psi^{\ast}\otimes\Id\right)\rho=\left(\Psi^{\ast}\otimes\Id\right)\left(J^{-1}\left(\rho\right)\otimes\Id\right)\sum_{k,l=1}^mf_{kl}\otimes f_{kl}=J\left(\Psi^{\ast}\circ J^{-1}\left(\rho\right)\right)
\end{equation}
where $m$ denotes the dimension of the space $\mathcal{K}$. Thus the condition $\left(\Psi^{\ast}\otimes\Id\right)\rho\in\mathcal{B}^+\left(\mathcal{K}\otimes\mathcal{K}\right)$ is the same as $J\left(\Psi^{\ast}\circ J^{-1}\left(\rho\right)\right)\in\mathcal{B}^+\left(\mathcal{K}\otimes\mathcal{K}\right)$, which is equivalent, by the Choi theorem on completely positive maps \cite{ref.Choi75}, to $\Psi^{\ast}\circ J^{-1}\left(\rho\right)\in\CPmapsb{\bk}$. If the last inclusion holds for all $\Psi\in\dual{\mappingcone}$, we know by Theorem \ref{MainTheorem} that $J^{-1}\left(\rho\right)$ is in $\mappingcone$, or $\rho\in J\left(\mappingcone\right)$. Conversely, if $\rho$ is in $J\left(\mappingcone\right)$, then $J^{-1}\left(\rho\right)$ belongs to $\mappingcone$. By Theorem \ref{MainTheorem}, $\Psi^{\ast}\otimes J^{-1}\left(\rho\right)$ belongs to $\CPmapsb{\bk}$ for all $\Psi\in\dual{\mappingcone}$, which is equivalent to $J\left(\Psi^{\ast}\otimes J^{-1}\left(\rho\right)\right)\in\mathcal{B}^+\left(\mathcal{K}\otimes\mathcal{K}\right)$ according to the Choi theorem on completely positive maps. By formula \eqref{eqequivPIR} the last expression is equivalent to $\left(\Psi^{\ast}\otimes\Id\right)\rho\in\mathcal{B}^+\left(\mathcal{K}\otimes\mathcal{K}\right)$ for all $\Psi\in\dual{\mappingcone}$. 
\end{proof}
\end{proposition}
\begin{remark}
For the choice $\mappingcone=\mathcal{SP}\left(\bk\right)$, the above theorem reduces to the positive maps criterion by Horodeccy \cite{HHH96}. We have the following equivalence
\begin{equation}
\rho\textnormal{ is separable }\Leftrightarrow\,\left(\Psi\otimes\id\right)\rho\in\mathcal{B}^+\left(\mathcal{K}\right)\forall_{\Psi\in\mathcal{P}\left(\bk\right)}
\end{equation}
\begin{proof}
Follows from Proposition \ref{propgenposmaps} if we recall that the set of separable operators equals $J\left(\mathcal{SP}\left(\bk\right)\right)$ and the dual of $\mathcal{SP}\left(\bk\right)$ is $\mathcal{P}\left(\bk\right)$ \cite{ref.SSZ09}.
\end{proof}
\end{remark}

\chapter{Algebraic problems solved by hand}\,\label{chhand}

\section{Product numerical range for a three-parameter family of operators}\label{secabcrange}\,

\newcommand{\matrfour}[1]{\left[\begin{array}{cccc}#1_{00,00} & #1_{00,01} & #1_{00,10} & #1_{00,11}\\#1_{01,00} & #1_{01,01} & #1_{01,10} & #1_{01,11}\\#1_{10,00} & #1_{10,01} & #1_{10,10} & #1_{10,11}\\#1_{11,00} & #1_{11,01} & #1_{11,10} & #1_{11,11}\end{array}\right]}
\def\dV{\dim V}

Product numerical range is a concept derived from the well-known numerical range (cf. e.g. \cite{HornJohnson}). For an operator $A$ on a Hilbert space $\mathcal{H}$, the numerical range of $A$ is by definition the set of numbers which can be obtained as $\left<v,A\left(v\right)\right>$ for some vector $v\in\mathcal{H}$ of unit norm. Accordingly, for an operator $A$ on a \emph{bipartite} space $\mathcal{H}_1\otimes\mathcal{H}_2$ the \textbf{product numerical range} is defined as
\begin{equation}
\Lambda^{\otimes}\left(A\right)=\left\{\left<v\otimes u,A\left(v\otimes u\right)\right>\vline v\in\mathcal{H}_1,u\in\mathcal{H}_2,\left|v\right|=\left|u\right|=1\right\}
\end{equation}
A generalization to a multipartite setting is possible and very straightforward. The definition was introduced in \cite{ref.Product11} and demonstrated to have various links to problems in the quantum information science \cite{ref.Restricted11}, including the evaluation of minimum output entropy \cite{Petz}, checking whether two unitary operations are locally distinguishable \cite{WSHV00,DuanFengYing08} or the identification of local dark spaces and error correcting codes \cite{KL97,MMZ2010}.
In the present section we analytically calculate the product numerical range for a three-parameter family of $4\times 4$ matrices introduced in \cite{SZ09}. In order to obtain explicit formulas, some additional constraints need to put on the parameters of the matrices. We take
\begin{equation}\label{abcfamilydef}
 F=\matrfour{F}=\left[\begin{array}{cccc}
               	\frac{1}{2}&a&0&0\\
		\bar a&\frac{1}{2}&b&0\\
		0&\bar b&\frac{1}{2}&c\\
		0&0&\bar c&\frac{1}{2}\\
              \end{array}\right],
\end{equation}
which represent operators on $\mathcal{H}_1\otimes\mathcal{H}_2=\setC^2\otimes\setC^2$. In order to find $\Lambda^{\otimes}\left(F\right)$, we first calculate the quantities $\left(F^{\left(2\right)}_u\right)_{\alpha\gamma}:=F_{\alpha\beta\gamma\delta}\bar u^{\beta}u^{\delta}$. The result is
\begin{equation}
 \label{blocksabc}
  F^{\left(2\right)}_u\left(a,b,c\right)=\left[
  \begin{array}{cc}
    \frac{1}{2}\left(\left|u_1\right|^2+\left|u_2\right|^2\right)&a\left|u_1\right|^2+c\left|u_2\right|^2+\bar bu_1\bar u_2\\
    \bar a\left|u_1\right|^2+\bar c\left|u_2\right|^2+b\bar u_1 u_2&\frac{1}{2}\left(\left|u_1\right|^2+\left|u_2\right|^2\right)\\
  \end{array}
 \right].
\end{equation}
To calculate the product numerical range of $F$, we only need to find the maximum and the minimum of $\innerpr{v\otimes u}{ F\left(u\otimes v\right)}=\innerpr{v}{F^{\left(2\right)}_u\left(v\right)}$, where $u,v\in\setC^2$ and $\left|u\right|=\left|v\right|=1$. Obviously, $\Tr F^{\left(2\right)}_u=\left|u_1\right|^2+\left|u_2\right|^2=\left|u\right|^2=1$ for all $u$ that meet the constraint $\left|u\right|=1$. The characteristic polynomial of $F^{\left(2\right)}_u$ is $\lambda^2-\Tr F^{\left(2\right)}_u\lambda+\det F^{\left(2\right)}_u=\lambda^2-\lambda+\det F^{\left(2\right)}_u$, which has the roots 
\begin{equation}\label{eigenvalues}
\lambda_{\pm}=\frac{1\pm\sqrt{1-4\det F^{\left(2\right)}_u}}{2}=\frac{1\pm\left|a\left|u_1\right|^2+c\left|u_2\right|^2+\bar b u_1 \bar u_2\right|}{2}.
\end{equation}
The last equality follows from a direct calculation of the determinant of $F^{\left(2\right)}_u$, $\det F^{\left(2\right)}_u=\frac{1}{4}-\left|a\left|u_1\right|^2+c\left|u_2\right|^2+\bar b u_1 \bar u_2\right|^2$. We see that the product numerical range of $F^{\left(2\right)}_u$ is
\begin{equation}\label{localrange}
\left[\frac{1-M}{2},\frac{1+M}{2}\right],
\end{equation}
where $M=\max_{\left|u\right|=1}\left|a\left|u_1\right|^2+c\left|u_2\right|^2+\bar b u_1 \bar u_2\right|$. Hence to determine the product numerical range of $F^{\left(2\right)}_u$, it is enough to calculate the maximum of the expression $\left|a\left|u_1\right|^2+c\left|u_2\right|^2+\bar b u_1 \bar u_2\right|$ over the elements $\left(u_1,u_2\right)\in\setC^2$ with unit norm. First we observe that for $x:=\left|u_1\right|$, $y:=\left|u_2\right|$ fixed, the function $\left|a\left|u_1\right|^2+c\left|u_2\right|^2+\bar b u_1 \bar u_2\right|$ attains the maximum value $\left|ax^2+cy^2\right|+\left|b\right|xy$. Thus the calculation of $M$ reduces to finding the maximum of $\left|ax^2+cy^2\right|+\left|b\right|xy$
over $x,y\in\setR$ nonnegative and such that $x^2+y^2=1$. Equivalently, we may skip the nonnegativity condition on $x$ and $y$, substitute $x\rightarrow\cos\phi,y\rightarrow\sin\phi$ and maximize $\left|a\cos^2\phi+c\sin^2\phi\right|+\left|b\right|\sin\phi\cos\phi$ over real $\phi$. Using simple algebra, it is easy to show that $\left|a\cos^2\phi+c\sin^2\phi\right|+\left|b\right|\sin\phi\cos\phi$ is equal to $\frac{1}{2}\left(\left|\left(a+c\right)+\left(a-c\right)\cos\psi\right|+\left|b\right|\sin\psi\right)$ for $\psi=2\phi$. The maximum of this expression over $\psi\in\setR$ can be easily found if $a$ and $c$ satisfy one of the following conditions,
\begin{itemize}
\item[a)] $\left|a\right|=\left|c\right|$ or
\item[b)] $a=rc$ for real $r$.
\end{itemize}
In the case a), we get $\left|\left(a+c\right)+\left(a-c\right)\cos\psi\right|=\sqrt{\left|a+c\right|^2+\left|a-c\right|^2\cos^2\psi}$, and thus we are left with the problem of maximizing 
\begin{equation}\label{ffunction}
 f\left(\psi\right):=\frac{1}{2}\left(\sqrt{\left|a+c\right|^2+\left|a-c\right|^2\cos^2\psi}+\left|b\right|\sin\psi\right)
\end{equation}
over real $\psi$. The maximum can be calculated explicitly. The result reads
\begin{equation}\label{Mcasea}
 M=\max_{\psi\in\setR}f\left(\psi\right)=\begin{cases}
                                          \left|b\right|+\left|a+c\right|,\,\left|b\right|\left|a+c\right|>\left|a-c\right|^2\\
					\sqrt{\left|b\right|^2+\left|a-c\right|^2}\sqrt{1+\frac{\left|a+c\right|^2}{\left|a-c\right|^2}},\,\left|b\right|\left|a+c\right|\leqslant\left|a-c\right|^2
                                         \end{cases}.
\end{equation}
Here we only outline how \eqref{Mcasea} was obtained. The first derivative of $f$ is 
\begin{equation}\label{exprderf}
f'\left(\psi\right)=\frac{1}{2}\left(\frac{\left|a-c\right|^2\sin\psi\cos\psi}{\sqrt{\left|a+c\right|^2+\left|a-c\right|^2\cos^2\psi}}+\left|b\right|\cos\psi\right),
\end{equation}
and there are either two or four solutions to the equation $f'\left(\psi\right)=0$ in $\left[0,2\pi\right)$, depending on the sign of the expression $\left|b\right|\left|a+c\right|-\left|a-c\right|^2$. If inequality $\left|b\right|\left|a+c\right|-\left|a-c\right|^2>0$ holds, we get a single maximum, equal to $\left|b\right|+\left|a+c\right|$, at $\psi=\pi/2$. Let us define $\psi_0=\arccos\left(\frac{\sqrt{\left|a-c\right|^4-\left|b\right|^2\left|a+c\right|^2}}{\left|a-c\right|\sqrt{\left|a-c\right|^2+\left|b\right|^2}}\right)$. When $\left|b\right|\left|a+c\right|-\left|a-c\right|^2<0$, the maximum at $\psi=\pi/2$ turns into a minimum, but two new maxima of $f$ appear at $\psi=\psi_0$ and $\psi=\pi-\psi_0$. The value of $f$ in both of these maxima is the same and equals $\sqrt{\left|b\right|^2+\left|a-c\right|^2}\sqrt{1+\frac{\left|a+c\right|^2}{\left|a-c\right|^2}}$. Thus we have explained formula \eqref{Mcasea} but for the case $\left|b\right|\left|a+c\right|=\left|a-c\right|^2$. With little additional effort, it can be shown that \eqref{Mcasea} also works in that special case. Therefore \eqref{Mcasea} is true whenever $\left|a\right|=\left|c\right|$ and we have found the product numerical range \eqref{localrange} of $F$ in the case a).

When $a=rc$ for real $r$, it is even simpler to calculate $M$ than in the situation considered above. Since then we have the equality $\left|\left(a+c\right)+\left(a-c\right)\cos\psi\right|=\left|\left|a+c\right|+\left|a-c\right|\cos\psi\right|$, we can first maximize the expression \begin{equation}
\frac{1}{2}\left(\left|\left|a+c\right|+\left|a-c\right|\cos\psi\right|+\left|b\right|\sin\psi\right)
\end{equation}
while keeping $s:=\left|\sin\psi\right|$ and $c:=\left|\cos\psi\right|$ constant. This yields
\begin{equation} 
\frac{1}{2}\left(\left|a+c\right|+\left|a-c\right|c+\left|b\right|s\right)
\end{equation}
and we are left with the task of maximizing this expression over all nonnegative $s,c$ such that $s^2+c^2=1$ holds. The calculation of the maximum is elementary, so we only give the final result,
\begin{equation}\label{Mcaseb}
 M=\frac{1}{2}\left(\left|a+c\right|+\sqrt{\left|a-c\right|^2+\left|b\right|^2}\right).
\end{equation}
Hence we have obtained the product numerical range \eqref{localrange} of $F$ in the case b). In the case of general $a,b,c\in\setC$, it does not seem easy to calculate the product numerical range of $F$.

For the cases where the calculation of the product numerical range of $F$ turned out to be possible, the results obtained can be used to find a part of the boundary of the set of entanglement witnesses. Namely, one can consider the minimal $\lambda\in\setR$ such that $W\left(\lambda\right)=\left(1-\lambda\right)F+\lambda\One$ is positive on product vectors. From \eqref{localrange}, it is not difficult to see that the appropriate $\lambda$ equals $\frac{M-1}{M+1}$, which we can explicitly calculate under certain assumptions on $a$, $b$ and $c$.
With a little more effort, the above argument also shows how to explicitly find the specific product vectors $v\otimes u$ that satisfy $\innerpr{v\otimes u}{W\left(\lambda\right)\left(v\otimes u\right)}=0$. The set of product vectors $v\otimes u$ that satisfy $\innerpr{v\otimes u}{W\left(v\otimes u\right)}=0$ for an entanglement witness $W$ often turns out to be important when considering the optimality of $W$ \cite{LKCH00}.

\section{Higher order numerical ranges and code carriers for the qutrit case}\label{sec3x3matrix}\,
We already know from the introduction to Chapter \ref{chmappingcones} that physical processes in quantum systems are best described by completely positive maps. Every such map, if not simply a unitary transformation, can be understood as some kind of noisy evolution induced upon the system by an environment. More precisely, two initially orthogonal pure states of the system are often no longer orthogonal after the evolution, which is an analogue of a spontaneous bit flip in classical computing. A way to deal with the noise in a classical setting is by representing the logical $0$ and $1$ by multiple physical bits, for example $000$ and $111$, resp. Even if one of those is physically flipped, there is sufficient information in the remaining ones to recover the initial value $0$ or $1$. An identical solution encounters severe difficulties in the quantum setting, since by the no-cloning theorem \cite{WZ82}, there exists no transformation that could transform an arbitrary quantum state $\rho$ into $\rho\otimes\rho$, let alone $\rho\otimes\rho\otimes\rho$.

However, nothing prevents us from encoding, in the qubit case, an arbitrary pure state $a\left|0\right>+b\left|1\right>$ of a qubit as $a\left|000\right>+b\left|111\right>$. In this way, a similar resistance to single bit flips as in the classical case is achieved, since the set of bit-flipped states $\left|000\right>$ is orthogonal to the bit-flipped $\left|111\right>$. This is the basic idea behind \textit{quantum error correction} \cite{Shor95,Gottesman2010}, but more details need to be accounted for before it really works. For a fixed completely positive transformation $\Phi$, describing the noise affecting a quantum system, a general criterion for quantum error correction was provided in the paper \cite{KL97} by E. Knill and R. Laflamme.  Note that by the Choi theorem on completely positive maps \cite{ref.Choi75}, the map $\Phi$ can be written in the form $\Phi:\rho\mapsto\sum_iA^{\ast}_i\rho A_i$ for some operators $A_i$ on the space in which $\rho$ lives. The Knill-Laflamme criterion now says that we can encode a $d$-dimensional quantum states and send them through the ``quantum channel'' described by $\Phi$ if and only if the conditions  
\begin{equation}\label{eqKLcond}
P_kA_i^{\ast}A_jP_k=\lambda_{ij}P_k\,\forall_{i,j}
\end{equation}
hold for some $k$-dimensional projection $P_k$ and a set of numbers $\lambda_{ij}\in\mathbbm{C}$. The equations \eqref{eqKLcond} are called \textit{Knill-Laflamme equations} accordingly. All of them are of the form $P_k M P_k=\lambda P_k$, where $M$ is some matrix and $\lambda$ a constant. This problem is a generalization of the eigenvalue problem and, more generally, of the question about the so-called {\it numerical range} of an operator, $\Lambda_1\left(M\right):=\left\{\left<\psi\right|M\left|\psi\right>\vline\psi\in\mathcal{H}\right\}$, where $\psi$ runs over all vectors of unit norm in the respective Hilbert space $\mathcal{H}$. We already mentioned numerical ranges in Section \ref{secabcrange}. Because of the form of Knill-Laflamme conditions, it is natural to introduce so-called {\it higher order numerical ranges} \cite{ref.CKZ06} (HONR),
\begin{equation}\label{defHONR}
\Lambda_k\left(M\right):=\left\{\lambda|\exists_{P_k}P_kMP_k=\lambda P_k\right\}.
\end{equation}
where $P_k$ is a $k$-dimensional projection.
It is also important to know the description of the set of all projections which give rise to some $\lambda$ in the above formula. We denote the set of such projections by $\Pi_k\left(M\right)$ and call it a {\it code carrier}, because it relates to the set of all possible error correcting subspaces.
\begin{equation}
\Pi_k\left(M\right):=\left\{P_k|\exists_{P_k}P_kMP_k=\lambda P_k\right\}.
\end{equation}

 In the following, we are going to show how to find $\Lambda_2\left(M\right)$ and $\Pi_{2}\left(M\right)$ for an arbitrary matrix $M$ of order three. By solving the problem for $k=2$, we shall give a full description of higher order numerical ranges and code carriers for $3\times 3$ matrices. This is so because the other cases, $k=1,3$, are trivial.
 
 Let us first observe that for a general matrix $M$, not necessarily of order three, the equation $P_kMP_k=\lambda P_k$ is equivalent to
 \begin{equation}\label{eqsymasym}
 \begin{cases}
 P_k\frac{M+M^{\ast}}{2}P_k=\xi P_k\\
 P_k\frac{M-M^{\ast}}{2i}P_k=\zeta P_k
 \end{cases},
 \end{equation} 
 where $\zeta$ and $\xi$ are real numbers. In this way, the compression equation $P_kMP_k=\lambda P_k$ is transformed into a pair of compression equations for Hermitian matrices $\frac{M+M^{\ast}}{2}$ and $\frac{M-M^{\ast}}{2i}$. Thus by solving the compression equations for a general Hermitian matrix $H$ of respective dimension and finding $\Pi_k\left(H\right)$ and $\Lambda_k\left(H\right)$, we may hope to be able to find $\Pi_k\left(M\right)$ and $\Lambda_k\left(M\right)$ for a general matrix $M$ just by intersecting $\Pi_k\left(\frac{M+M^{\ast}}{2}\right)$ and $\Pi_k\left(\frac{M-M^{\ast}}{2i}\right)$ and reading off the $\zeta$'s and $\xi$'s corresponding the elements in the intersection. Note that an almost complete description of code carriers and numerical ranges for Hermitian matrices of arbitrary dimension was obtained in \cite{ref.CKZ06}. In the present section, however, we shall give an alternative proof in the case of dimension $3$, which is mainly justified by the fact that we solve algebraic equations. 
 
 Let us first consider a Hermitian $3\times 3$ matrix $H$ with three distinct eigenvalues $\lambda_1<\lambda_2<\lambda_3$ and the corresponding eigenvectors $\left|x_1\right>,\left|x_2\right>,\left|x_3\right>$. By adding a factor proportional to identity to $H$, we may assume that all the $\lambda_i$'s are nonzero. We know from \cite{ref.CKZ06} that $\Lambda_2\left(H\right)=\left\{\lambda_2\right\}$. We shall find $\Pi_2\left(H\right)$.
 
Note that the condition $P_2 H P_2=\lambda_2 P_2$ is equivalent to the existence of vectors $\left|v_1\right>, \left|v_2\right>$ such that
\begin{equation}\label{conddelta}
\left<v_i, v_j\right>=\delta_{ij}\textnormal{ and }\left<v_i,H\left(v_j\right)\right>=\lambda_2\delta_{ij}.
\end{equation}
If we denote with $v_i^n$ the $n$-th coordinate of $v_i$ with respect to the basis $\left(\left|x_1\right>,\left|x_2\right>\right)$, conditions \eqref{conddelta} can be rewritten as 
\begin{eqnarray}
\sum_{k=1}^3\left|v^k_i\right|^2=1,&\ \sum_{k=1}^3\lambda_k\left|v_i^k\right|^2=\lambda_2,\label{algeq1}\\
\sum_{k=1}^3\bar v_1^kv_2^k=0,&\ \sum_{k=1}^3\lambda_k\bar v_1^kv_2^k=0,\label{algeq2}
\end{eqnarray}
with $i=1,2$.
By appropriately transforming a solution of \eqref{algeq1} and \eqref{algeq2} according to the following prescription: $v_1^k\rightarrow v_1^ke^{i\phi_k}$ $v_2^k\rightarrow v_2^ke^{i\phi_k}$,  we can get another solution, where $v_1$ has real numbers as coefficients. Indeed, the transformations of the form given above do not affect equalities \eqref{algeq1} and \eqref{algeq2}, and the phases $e^{iphi_k}$ can be chosen as $\bar v_1^k/\left|v_1^k\right|$ to make all the coordinates $v_1^k$ real. Therefore in the following, we assume that all the coordinates of $v_1$ are real.

From equations \eqref{algeq2} it follows that
\begin{equation}
\left(\lambda_1-\lambda_2\right) v_i^1 v_j^1+\left(\lambda_3-\lambda_2\right) v_i^2v_i^2=0,
\label{22minus}
\end{equation}
where we removed the bars over $v_1^k$ using the reality assumption explained above. Equation \eqref{22minus} implies the existence of a phase $e^{i\psi}$ such that the numbers  $e^{i\psi}v_2^1$ and $e^{i\psi}v_2^2$ are real. Moreover, the first equation in \eqref{algeq2} now implies that also $e^{i\psi}v_j^2$ has to be a real number. We can now transform $v_2$ according to the following prescription $v_2\rightarrow e^{-i\psi}v_2$ and obtain another solution to equations \eqref{algeq1} and \eqref{algeq2}, where both $v_1$ and $v_2$ have real coefficients. Consequently, it is possible first to find all \emph{real} solutions to the following set of equations,
\begin{eqnarray}
\sum_{k=1}^3\left(v^k_i\right)^2=1,&\ \sum_{k=1}^3\lambda_k\left(v_i^k\right)^2=\lambda_2\label{algeq3},\\
\sum_{k=1}^3v_1^kv_2^k=0,&\ \sum_{k=1}^3\lambda_k v_1^kv_2^k=0\label{algeq4}
\end{eqnarray}
and later recover all the solutions to \eqref{algeq1} and \eqref{algeq2} by transforming the variables according to the prescription $v_1^k\rightarrow e^{i\phi_k}v_1^k$ and $v_2^k\rightarrow e^{i\left(\phi_k-\psi\right)}v_2^k$ with \emph{arbitrary} angles $\phi_k$ and $\psi$. This follows because the transformations of the type just described do not affect equations \eqref{algeq1} and \eqref{algeq2} and on the other hand, they allow us to bring any solution of \eqref{algeq1} and \eqref{algeq2} to a real solution of equations \eqref{algeq3} and \eqref{algeq4}. Thus, let us look for real solutions of equations \eqref{algeq3} and \eqref{algeq4}. By multiplying the first equation in \eqref{algeq4} by $\lambda_2$ and subtracting the result from the second equation in the same line, one easily gets
\begin{equation}\label{fourlambdas1}
v_1^1v_2^1=\frac{\lambda_3-\lambda_2}{\lambda_2-\lambda_1}v_i^3v_2^3.
\end{equation}
Substitution of this equality back to the first equation in \eqref{algeq4} yields
\begin{equation}\label{eq27}
v_1^1v_2^1=-\frac{\lambda_3-\lambda_2}{\lambda_3-\lambda_1}v_1^2v_2^2\quad\textnormal{and}\quad v^3_1v^3_2=-\frac{\lambda_2-\lambda_1}{\lambda_3-\lambda_1}v_1^2v_2^2.
\end{equation}
In a similar fashion, equations \eqref{algeq3} give us
\begin{equation}\label{eq28}
\left(v_i^1\right)^2=\frac{\lambda_3-\lambda_2}{\lambda_3-\lambda_1}\left(1-\left(v_i^2\right)^2\right)\quad\textnormal{and}\quad\left(v_i^3\right)^2=\frac{\lambda_2-\lambda_1}{\lambda_3-\lambda_1}\left(1-\left(v_i^2\right)^2\right).
\end{equation}
If we multiply the first equation in \eqref{eq28} for $i=1$ by the same equation but for $i=2$, we obtain
\begin{equation}\label{eq29}
\left(v_1^1v_2^1\right)^2=\left(\frac{\lambda_3-\lambda_2}{\lambda_3-\lambda_1}\right)^2\left(1-\left(v_1^2\right)^2\right)\left(1-\left(v_2^2\right)^2\right).
\end{equation}
In a similar way
\begin{equation}\label{eq30}
\left(v^3_1v^3_2\right)^2=\left(\frac{\lambda_2-\lambda_1}{\lambda_3-\lambda_1}\right)^2\left(1-\left(v^2_1\right)^2\right)\left(1-\left(v^2_2\right)^2\right).
\end{equation}
On the other hand, we may square the equations in \eqref{eq27} to obtain
\begin{equation}\label{eq31}
\left(v_1^1v_2^1\right)^2=\left(\frac{\lambda_3-\lambda_2}{\lambda_3-\lambda_1}\right)^2\left(v_1^2v_2^2\right)^2\textnormal{ and 
}\left(v_1^3v_2^3\right)^2=\left(\frac{\lambda_2-\lambda_1}{\lambda_3-\lambda_1}\right)^2\left(v_1^2v_2^2\right)^2
\end{equation}
Now we can subtract the first equation in \eqref{eq31} from \eqref{eq29} and the second equation in \eqref{eq31} from \eqref{eq30} to get
\begin{eqnarray}\label{eq32}
\left(\frac{\lambda_3-\lambda_2}{\lambda_3-\lambda_1}\right)^2\left( \left(1-\left(v_1^2\right)^2\right)\left(1-\left(v_2^2\right)^2\right)
- \left(v_1^2v_2^2\right)^2 \right)=0,\\
\left(\frac{\lambda_2-\lambda_1}{\lambda_3-\lambda_1}\right)^2\left( \left(1-\left(v^2_1\right)^2\right)\left(1-\left(v^2_2\right)^2\right) -
\left(v_1^2v_2^2\right)^2  \right)=0.
\label{eq33}
\end{eqnarray}
According to our assumption $\lambda_1<\lambda_2<\lambda_3$, the factors $\frac{\lambda_3-\lambda_2}{\lambda_3-\lambda_1}$ and $\frac{\lambda_2-\lambda_1}{\lambda_3-\lambda_1}$ are non-zero. Therefore the equations \eqref{eq32} and \eqref{eq33} are equivalent to
\begin{equation}\label{eq34}
\left(1-\left(v^2_1\right)^2\right)\left(1-\left(v^2_2\right)^2\right) -
\left(v_1^2v_2^2\right)^2  =1-\left(v^2_1\right)^2-\left(v_2^2\right)^2=0.
\end{equation}
The solution of \eqref{eq34} is of the form $v_1^2=\cos\gamma$, $v_2^2=\sin\gamma$ for an arbitrary $\gamma$. We can substitute this in \eqref{eq28} to obtain a general solution to equations \eqref{algeq3} and \eqref{algeq4} in the following form,
\begin{equation}
\left[\begin{array}{c}
v_1^1\\
v_1^2\\
v_1^3
\end{array}\right]=
\left[
\begin{array}{c}
s^1\kappa\sin\gamma\\
s^2\cos\gamma\\
s^3\eta\sin\gamma
\end{array}
\right]\textnormal{ and }
\left[
\begin{array}{c}
s^1s\kappa\cos\gamma\\
-s^2s\sin\gamma\\
s^3s\eta\cos\gamma
\end{array}
\right],\label{eq35}
\end{equation}
where $s^i=\pm 1$ and $s=\pm 1$ are arbitrary and we have introduced the notation $\kappa:=\sqrt{\frac{\lambda_3-\lambda_2}{\lambda_3-\lambda_1}}$ and $\eta:=\sqrt{\frac{\lambda_2-\lambda_1}{\lambda_3-\lambda_1}}$. Note that we have used equations \eqref{eq28} to establish sign relations between the coordinates of $v_1$ and $v_2$.

Now, if we recall the discussion preceding equations \eqref{algeq3} and \eqref{algeq4}, we can recover a general solution to \eqref{conddelta} by introducing complex phases back into \eqref{eq35}
\begin{equation}
\left[\begin{array}{c}
v_1^1\\
v_1^2\\
v_1^3
\end{array}\right]=
\left[
\begin{array}{c}
e^{i\phi_1}\kappa\sin\gamma\\
e^{i\phi_2}\cos\gamma\\
e^{i\phi_3}\eta\sin\gamma
\end{array}
\right]\textnormal{ and }
\left[
\begin{array}{c}
e^{i\left(\phi_1-\psi\right)}\kappa\cos\gamma\\
-e^{i\left(\phi_2-\psi\right)}\sin\gamma\\
e^{i\left(\phi_3-\psi\right)}\eta\cos\gamma
\end{array}
\right],\label{eq36}
\end{equation}
where the phases $\phi_i$ and $\psi$ are arbitrary. 

Since we are interested in $\Pi_2\left(H\right)$ rather than the vectors $v_i$, it is sufficient for us to know that  $v_1$ and $v_2$ given in \eqref{eq36} span the two-dimensional subspace
\begin{equation}
\linspan\left\{\left|x_1\right>,\kappa e^{i\phi}\left|x_1\right>+\eta\left|x_3\right>\right\}.
\end{equation}
with $\phi$ arbitrary.

In this way we obtain the following description of $\Pi_2$.
\begin{proposition}\label{propPi2}
Let $H$ be a Hermitian operator on $\mathbbm{C}^3$ with eigenvalues $\lambda_1<\lambda_2<\lambda_3$ and the corresponding eigenvectors $\left|x_1\right>,\left|x_2\right>,\left|x_3\right>$. The rank $2$ code carrier of $H$ is given as
\begin{equation}\label{Pi2formula}
\Pi_2\left(H\right)=\left\{P_2\,\vline\,\exists_{\phi\in\mathbbm{R}}P_2 \textnormal{ projects onto }\linspan\left\{\left|x_2\right>,\kappa e^{i\phi}\left|x_1\right>+\eta\left|x_3\right>\right\}\right\}.
\end{equation}
\qed
\end{proposition}

Obviously, two orthogonal projections are equal iff they project onto the same subspace. Furthermore, two linear subspaces are identical if and only if all the vectors spanning one of the subspaces are linearly dependent of the vectors spanning the second subspace. Using Proposition \ref{propPi2}, we can now find the intersection of rank $2$ code carriers of two distinct Hermitian operators on $\mathbbm{C}^3$. We have
\begin{proposition}
\label{propintersect}
Let $H$, $H'$ be Hermitian operators on $\mathbbm{C}^3$ with eigenvalues $\lambda_1<\lambda_2<\lambda_3$, $\lambda_1'<\lambda'_2<\lambda'_3$, respectively. Let the corresponding eigenvectors be $\left|x_1\right>,\left|x_2\right>,\left|x_3\right>$ ($\left|x'_1\right>,\left|x'_2\right>,\left|x'_3\right>$). Let $\kappa:=\sqrt{\frac{\lambda_3-\lambda_2}{\lambda_3-\lambda_1}}$, $\eta:=\sqrt{\frac{\lambda_2-\lambda_1}{\lambda_3-\lambda_1}}$,
$\kappa':=\sqrt{\frac{\lambda'_3-\lambda'_2}{\lambda'_3-\lambda'_1}}$, $\eta':=\sqrt{\frac{\lambda'_2-\lambda'_1}{\lambda'_3-\lambda'_1}}$. The intersection $\Pi_2\left(H\right)\cap\Pi_2\left(H'\right)$ is nonempty if and only if there exist $\phi,\phi'\in\mathbbm{R}$ such that the family of vectors 
\begin{equation}
\left\{\left|x_2\right>,\kappa e^{i\phi}\left|x_1\right>+\eta\left|x_3\right>,\left|x'_2\right>\right\},
\end{equation}
as well as the family of vectors
\begin{equation}
\left\{\left|x_2\right>,\kappa e^{i\phi}\left|x_1\right>+\eta\left|x_3\right>,\kappa e^{i\phi'}\left|x'_1\right>+\eta'\left|x'_3\right>\right\}
\end{equation}
are linearly dependent. If this is the case,
\begin{equation}\label{eqintersect}
\Pi_2\left(H\right)\cap\Pi_2\left(H'\right)=\left\{\left|x_2\right>\left<x_2\right|+\left|\chi\right>\left<\chi\right|\,\vline\,\left|\chi\right>=\kappa e^{i\phi}\left|x_1\right>+\eta\left|x_3\right>,\phi\in\Xi\right\}
\end{equation}
where $\Xi$ is the set of all $\phi\in\mathbbm{R}$ such that there exists $\psi$ for which the families of vectors 
\begin{equation}
\left\{\left|x_2\right>,\kappa e^{i\phi}\left|x_1\right>+\eta\left|x_3\right>,\left|x'_2\right>\right\}
\end{equation} and 
\begin{equation}
\left\{\left|x_2\right>,\kappa e^{i\phi}\left|x_1\right>+\eta\left|x_3\right>,\kappa e^{i\phi'}\left|x'_1\right>+\eta'\left|x'_3\right>\right\}
\end{equation}
are both linearly dependent.
\begin{proof}
Obvious from Proposition \ref{propPi2}
\end{proof}
\end{proposition}

Note that the above results have been derived using the assumption $\lambda_1<\lambda_2<\lambda_3$ ($\lambda'_1<\lambda'_2<\lambda'_3$) about the eigenvalues of $H$ ($H'$, resp.). However, in case that these assumptions do not hold, we can still easily give a description of $\Pi_2\left(H\right)$ ($\Pi_2\left(H'\right)$) and find $\Pi_2\left(H\right)\cap\Pi_2\left(H'\right)$. First of all, if $H$ or $H'$ is proportional to identity, the corresponding code carrier equals the set of all two-dimensional projections in $\mathbbm{C}^3$. It is also easy to prove the following proposition.
\begin{proposition}\label{proplambdaeq}
Let $H$ be a Hermitian operator on $\mathbbm{C}^3$ with eigenvalues $\lambda_1=\lambda_2<\lambda_3$ or $\lambda_1<\lambda_2=\lambda_3$. Then $\Pi_2\left(H\right)$ consists of an orthogonal projection onto the eigenspace corresponding to $\lambda_2$.
\end{proposition}
We leave the proof of the proposition as an exercise for the reader (it is enough to check what conditions \eqref{algeq1} imply when exactly two of the eigenvalues are equal). We should notice that formulas \eqref{eq36} still apply, so we can easily generalize Proposition \ref{propintersect} to a situation where the eigenvalues of $H$ (or $H'$) are not all distinct.
\begin{proposition}\label{propnotalldistinct}
Let $H$, $H'$ be Hermitian operators on $\mathbbm{C}^3$ with eigenvalues $\lambda_1\leqslant\lambda_2\leqslant\lambda_3$, $\lambda_1'\leqslant\lambda'_2\leqslant\lambda'_3$, respectively. Assume that neither $H$ nor $H'$ is proportional to identity. Let the corresponding eigenvectors be $\left|x_1\right>,\left|x_2\right>,\left|x_3\right>$ ($\left|x'_1\right>,\left|x'_2\right>,\left|x'_3\right>$). Let $\kappa:=\sqrt{\frac{\lambda_3-\lambda_2}{\lambda_3-\lambda_1}}$, $\eta:=\sqrt{\frac{\lambda_2-\lambda_1}{\lambda_3-\lambda_1}}$,
$\kappa':=\sqrt{\frac{\lambda'_3-\lambda'_2}{\lambda'_3-\lambda'_1}}$, $\eta':=\sqrt{\frac{\lambda'_2-\lambda'_1}{\lambda'_3-\lambda'_1}}$. The intersection $\Pi_2\left(H\right)\cap\Pi_2\left(H'\right)$ is nonempty if and only if there exist $\phi,\phi'\in\mathbbm{R}$ such that the family of vectors
\begin{equation}
\left\{\left|x_2\right>,\kappa e^{i\phi}\left|x_1\right>+\eta\left|x_3\right>,\left|x'_2\right>\right\},
\end{equation} 
as well as the family of vectors 
\begin{equation}
\left\{\left|x_2\right>,\kappa e^{i\phi}\left|x_1\right>+\eta\left|x_3\right>,\kappa e^{i\phi'}\left|x'_1\right>+\eta'\left|x'_3\right>\right\}
\end{equation}
are linearly dependent. If this is the case,
\begin{equation}\label{eqintersect1}
\Pi_2\left(H\right)\cap\left(H'\right)=\left\{\left|x_2\right>\left<x_2\right|+\left|\chi\right>\left<\chi\right|\vline\left|\chi\right>=\kappa e^{i\phi}\left|x_1\right>+\eta\left|x_3\right>,\phi\in\Xi\right\}
\end{equation}
where $\Xi$ is the set of all $\phi\in\mathbbm{R}$ such that there exists $\psi$ for which the families of vectors 
\begin{equation}
\left\{\left|x_2\right>,\kappa e^{i\phi}\left|x_1\right>+\eta\left|x_3\right>,\left|x'_2\right>\right\}
\end{equation} 
and 
\begin{equation}
\left\{\left|x_2\right>,\kappa e^{i\phi}\left|x_1\right>+\eta\left|x_3\right>,\kappa e^{i\phi'}\left|x'_1\right>+\eta'\left|x'_3\right>\right\}
\end{equation}
are both linearly dependent.
\end{proposition}
The case of $H$ or $H'$ proportional to identity, which we excluded in the above proposition, can be handled in an obvious way. Thus we have fully characterized the intersection $\Pi_2\left(H\right)\cap\Pi_2\left(X'\right)$ for a pair of Hermitian operators on $\mathbbm{C}^3$. Following the discussion after equations \eqref{eqsymasym}, we can now use Proposition \ref{propnotalldistinct} to obtain the numerical range $\Lambda_2\left(M\right)$ for an arbitrary (not necessarily Hermitian or normal) matrix of dimension three. Let us discuss this in an example.
\begin{example}\label{exJordan}
Consider the Jordan matrix
\begin{equation}\label{eqJordan}
J=\left[\begin{array}{ccc}
0&0&0\\
1&0&0\\
0&1&0
\end{array}\right].
\end{equation}
The second order numerical range $\Lambda_2\left(J\right)$ equals $\left\{0\right\}$ and the corresponding code carrier consists of a single element, $\Pi_2\left(J\right)=\left\{\left|1\right>\left<1\right|+\left|2\right>\left<2\right|\right\}$,
where $\left|1\right>=\frac{1}{\sqrt{2}}\left[1,0,1\right]$, $\left|2\right>=\frac{1}{\sqrt{2}}\left[1,0,-1\right]$.
\begin{proof}
We have
\begin{equation}
\frac{J+J^{\ast}}{2}=\frac{1}{2}\left[\begin{array}{ccc}
0&1&0\\
1&0&1\\
0&1&0
\end{array}\right]=:H\quad\textnormal{and}\quad
\frac{J-J^{\ast}}{2i}=\frac{1}{2}\left[\begin{array}{ccc}
0&i&0\\
-i&0&i\\
0&-i&0
\end{array}\right]=:H'.
\end{equation}
Let us denote the eigenvalues of $H$ with $\lambda_1\leqslant\lambda_2\leqslant\lambda_3$ and the corresponding eigenvectors with $\wektor{v_1},\wektor{v_2},\wektor{v_3}$. For $H'$, similarly define $\lambda'_1\leqslant\lambda'_2\leqslant\lambda'_3$ and the eigenvectors $\wektor{v'_1},\wektor{v'_2},\wektor{v'_3}$. One can easily check that
\begin{equation}\label{eqlambdasJ}
\lambda_1=\lambda_1'=-\sqrt{2},\quad\lambda_2=\lambda'_2=0,\quad\lambda_3=\lambda'_3=\sqrt{2}.
\end{equation}
Thus $\kappa=\eta=\kappa'=\eta'=1/\sqrt{2}$. The eigenvectors of $H$ and $H'$ are
\begin{eqnarray}
\wektor{v_1}=\frac{1}{2}\left[1,\sqrt{2},1\right],&\wektor{v_2}=\frac{1}{2}\left[-1,0,1\right],&\wektor{v_3}=\frac{1}{2}\left[1,-\sqrt{2},1\right],\\
\wektor{v'_1}=\frac{1}{2}\left[1,\sqrt{2}i,-1\right],&\wektor{v'_2}=\frac{1}{2}\left[1,0,1\right],&\wektor{v'_3}=\frac{1}{2}\left[-1,\sqrt{2}i,1\right],
\end{eqnarray}
We can now easily check that the equations
\begin{eqnarray}
\det\left[\begin{array}{ccc}
-\frac{1}{\sqrt{2}}&0&\frac{1}{\sqrt{2}}\\
\frac{1}{2}\left(e^{i\phi}+1\right)&\frac{1}{\sqrt{2}}\left(e^{i\phi}-1\right)&\frac{1}{2}\left(e^{i\phi}+1\right)\\
\frac{1}{\sqrt{2}}&0&\frac{1}{\sqrt{2}}
\end{array}
\right]&=0,\\
\det\left[\begin{array}{ccc}
-\frac{1}{\sqrt{2}}&0&\frac{1}{\sqrt{2}}\\
\frac{1}{2}\left(e^{i\phi}+1\right)&\frac{1}{\sqrt{2}}\left(e^{i\phi}-1\right)&\frac{1}{2}\left(e^{i\phi}+1\right)\\
\frac{1}{2}\left(e^{i\phi'}-1\right)&\frac{i}{\sqrt{2}}\left(e^{i\phi'}+1\right)&-\frac{1}{2}\left(e^{i\phi'}-1\right)
\end{array}
\right]&=0.
\end{eqnarray}
have a single solution in $\left[0;2\pi\right)\times\left[0;2\pi\right)$, which is $\left(\phi,\phi'\right)=\left(0,\pi\right)$. This corresponds to the subspace in the assertion of the theorem.
\end{proof}
\end{example}

Note that with the methods described in this section, it is possible to find solutions to Knill-Laflamme equations or prove their non-existence for \emph{all} qutrit quantum channels $\Phi$. As proud as it sounds, the solutions will however almost never exist. It should also be kept in mind that solving the problems for all qutrit channels simply means finding all qutrit channels that allow for encoding of a single qubit. This is quite a simple setup. 
\section[A separable state of length four and Schmidt rank three]{A separable state of length four\\ and Schmidt rank three}\label{seclength4rank3}\,
In Section \ref{secsep}, we introduced the concept of separability and discussed some fundamental subtleties about the distinction between separable and entangled quantum states. Here, we consider so-called \textit{length of a separable state}, which is the minimum number of terms in its separable decomposition. More precisely, we have the following.
\begin{definition}[Length of a separable state]\label{deflength}
Let $\rho$ be a separable state on a bipartite space $\hilbertspaceone\otimes\hilbertspacetwo$. Thus, $\rho$ can be written as
\begin{equation}\label{eqdecompositionofrho}
\rho=\sum_{i=1}^l\rho_i\otimes\xi_i
\end{equation}
for some $l$ and some positive operators $\rho_i$, $\xi_i$ on $\hilbertspaceone$, $\hilbertspacetwo$, resp. The \textbf{length of $\bf{\rho}$} is the minimum number $l$ in a decomposition of the form \eqref{eqdecompositionofrho}.
\end{definition}
A generalization to a multipartite setting is straightforward, but we shall not discuss on this point here. Given a density matrix $\rho$ on a bipartite space, it is in general very difficult to tell whether and how it decomposes into a convex sum of products of positive operators. Successful attempts in this direction can be made in the situation described in \cite{HLVC2000}, but in general, no exact way to find an optimal decomposition of the form \eqref{eqdecompositionofrho} is known. In particular, it is difficult to determine the length $l$ of $\rho$. On the other hand, it is fairly simple to find a minimal decomposition of $\rho$ of the form
\begin{equation}\label{eqdecHermitian}
\rho=\sum_{i=1}^rF_i\otimes G_i,
\end{equation} 
where $F_i$, $G_i$ are Hermitian, but not required to be positive. The minimal number $r$ in the decomposition \eqref{eqdecHermitian} will be called the \textit{Schmidt rank} of $\rho$. The name originates from the well-known Schmidt decomposition of vectors, for which \eqref{eqdecHermitian} is an analogue. 

Intuitively, the length and the Schmidt rank of $\rho$ do not look entirely independent. Indeed, in \cite{S2010} we showed that separable states of small lengths $\leqslant 3$ necessarily have their Schmidt rank equal to their length. In the present section, we give an example of a state of Schmidt rank $3$ and length $4$. This shows that the mentioned result for lengths $\leqslant 3$ cannot be further generalized. Such conclusion should be expected from the beginning, but the concrete example is rather illustrative.

\begin{example}\label{exp.l4r3}
 Consider the following $4\times 4$ diagonal matrices
\begin{eqnarray}\label{defmatrE}
\delta_1=\diag{1,0,1,0},&\delta_2=\diag{0,1,0,1},\\
\delta_3=\diag{1,1,0,0},&\delta_4=\diag{0,0,1,1}.\nonumber
\end{eqnarray}
Let $\rho$ be a density matrix on $\setC^4\otimes\setC^4$ of the form
\begin{equation}\label{fml.exmplsep.2}
\rho:=\frac{1}{16}\sum_{i=1}^4\delta_i\otimes \delta_i. 
\end{equation}
This bipartite state is separable, has length $l=4$ and Schmidt rank $r=3$.
\begin{proof} 
 Obviously, $\rho$ is separable. For further convenience, let us denote the length of $\rho$ with $l$ and its Schmidt rank with $r$. We first prove that the Schmidt rank of $\rho$ is 3, which is equivalent to proving that the Schmidt rank of $\tilde\rho:=16\rho$ is 3. For that purpose, we observe that the operators $\delta_i$ in \eqref{defmatrE} are linearly dependent. For example, we can write $\delta_4$ as a linear combination of $\delta_1$, $\delta_2$ and $\delta_3$,
\begin{equation}\label{fml.lindepEi.1}
 \delta_4=\delta_1+\delta_2-\delta_3.
\end{equation}
We can put \fml{lindepEi.1} in \fml{exmplsep.2} and use distributivity of the tensor product to get
\begin{equation}\label{fml.exmplsep.4}
\tilde\rho=\delta_1\otimes\left(2\delta_1+\delta_2-\delta_3\right)+\delta_2\otimes\left(\delta_1+2\delta_2-\delta_3\right)+\delta_3\otimes\left(2\delta_3-\delta_1-\delta_2\right),
\end{equation}
From \fml{exmplsep.4}, we definitely see that $\tilde\rho$ has Schmidt rank lower than four. But the matrices $2\delta_1+\delta_2-\delta_3$, $\delta_1+2\delta_2-\delta_3$ and $\delta_1+\delta_2-2\delta_3$ are linearly independent\footnote{the matrix $\left[\begin{array}{ccc}2&1&-1\\1&2&-1\\1&1&-2 \end{array}\right]$ has a nonzero determinant}, just as the matrices $\delta_1$, $\delta_2$ and $\delta_3$ are. This implies that the number of product terms in \fml{exmplsep.4} cannot be reduced any further. Consequently, the Schmidt rank of $\tilde\rho$ and hence of $\rho$ is 3, $r=3$.

Of course, the length of $\rho$ is not lower than $r$, so we have $l\geqslant 3$. On the other hand, \fml{exmplsep.2} is an expression for $\rho$ as a sum of four products of positive operators $\delta_i$. Therefore $l$ cannot be higher than 4 and the only possibilities left are $l=3$ and $l=4$. In the following we show that $l=3$ is excluded. Put it in a different way, $\rho$ cannot be written as
\begin{equation}\label{fml.decofC2.1}
 \rho_1\otimes \xi_1+\rho_2\otimes \xi_2+\rho_3\otimes \xi_3,
\end{equation}
with $\rho_i$ and $\xi_i$ positive for $i=1,2,3$. It will be more convenient to show that $\tilde\rho$ cannot be written in the form \eqref{fml.decofC2.1} with all $\xi_i$, $\rho_i$ positive. To prove this, let us assume that a decomposition of $\rho$ of the form \fml{decofC2.1} exists. We should stress that \fml{exmplsep.4} is not an example of such a decomposition because $2\delta_3-\delta_1-\delta_2$ is not positive. The operators $\rho_i$ and $\xi_i$ are Hermitian, so we can write them as $\rho_i=\sum_{j=1}^{16}\alpha_i^jH_j$ and $\xi_i=\sum_{j=1}^{16}\beta_i^jH_j$, where $\alpha_i^j,\beta_i^j\in\setR\,\forall_{i,j}$, and \SEQ{H_j}{j=1}{16} is a basis of the $\setR$-linear space of Hermitian operators on $\setC^4$ such that
\begin{eqnarray}\label{fml.defmatrH.1}
 H_1=\diag{1,0,0,0},&H_2=\diag{0,1,0,0},\\
H_3=\diag{0,0,1,0},&H_4=\diag{0,0,0,1}.\nonumber
\end{eqnarray}
and $H_j$'s for $j\geqslant 5$ have only off-diagonal elements nonzero. Because of the form \eqref{defmatrE} of the operators $\delta_i$, $\tilde\rho$ does not have any off-diagonal elements and the decomposition of $\tilde\rho$ in the basis \SEQ{H_k\otimes H_l}{k,l=1}{16} of all Hermitian operators on $\setC^4\otimes\setC^4$ does not include any terms with $k\geqslant 5$ nor with $l\geqslant 5$. If there are any terms including $H_k$ with $k\geqslant 5$ in $\rho_i$ or $\xi_i$, they must eventually cancel out in the tensor product \fml{decofC2.1}. Therefore we may use $\widetilde \rho_i:=\sum_{j=1}^4\alpha_i^jH_j$ and $\widetilde \xi_i:=\sum_{j=1}^4\beta_i^jH_j$ instead of $\rho_i$ and $\xi_i$. The relation \fml{decofC2.1} still holds when $\rho_i$ is replaced with $\widetilde \rho_i$ and $\xi_i$ with $\widetilde \xi_i$. Positivity of $\widetilde \rho_i$ and $\widetilde \xi_i$ follows from the fact that they are diagonal parts of positive operators. We see that $\sum_{i=1}^4\widetilde \rho_i\otimes\widetilde \xi_i$ equals $\tilde\rho$, and it is also a sum of products of positive operators. Consequently, if there exists a decomposition of $\tilde\rho$ of the form \fml{decofC2.1} with $\rho_i$ and $\xi_i$ positive, another decomposition with diagonal and positive $\rho_i$ and $\xi_i$ must also exist. Therefore we can restrict our discussion to decompositions of the form
\begin{equation}\label{fml.decofC2.2}
 \tilde\rho=\sum_{i=1}^3\sum_{j,k=1}^4\alpha_i^jH_j\otimes\beta_i^kH_k=\sum_{i=1}^3\sum_{j,k=1}^4\alpha_i^j\beta_j^kH_j\otimes H_k,
\end{equation}
with $\alpha_j^j\geqslant 0$ and $\beta_i^k\geqslant 0$. Based on the definition \eqref{fml.exmplsep.2}, it can be easily checked that
\begin{equation}\label{fml.decofC2.3}
 \tilde\rho=\sum_{j,k=1}^4A^{jk}H_j\otimes H_k,
\end{equation}
with $A^{11}=A^{22}=A^{33}=A^{44}=2$, $A^{14}=A^{41}=A^{23}=A^{32}=0$ and $A^{ij}=1$ for the remaining eight coefficient pairs $\left(i,j\right)$. In order for equation \fml{decofC2.2} to be fulfilled, we must have
\begin{equation}\label{fml.coeffsofC2.1}
\sum_{i=1}^4\alpha_i^j\beta_i^k=A^{jk}\,\forall_{j,k\in\left\{1,2,3,4\right\}}.
\end{equation}
To see the consequences of \fml{coeffsofC2.1}, let us introduce vectors $\alpha^j\in\setR^3$ and $\beta^k\in\setR^3$ with coordinates \SEQ{\alpha_i^j}{i=1}{3} and \SEQ{\beta_i^k}{i=1}{3}, respectively. The conditions \fml{coeffsofC2.1} can be written as
\begin{eqnarray}
 \alpha^1\cdot\beta^1=\alpha^2\cdot\beta^2=\alpha^3\cdot\beta^3=\alpha^4\cdot\beta^4=2\label{fml.alphabetaprodC2.1},\\
\alpha^1\cdot\beta^4=\alpha^4\cdot\beta^1=\alpha^2\cdot\beta^3=\alpha^3\cdot\beta^2=0\label{fml.alphabetaprodC2.2},\\
\alpha^1\cdot\beta^2=\alpha^1\cdot\beta^3=\alpha^4\cdot\beta^2=\alpha^4\cdot\beta^3=1\label{fml.alphabetaprodC2.3},\\
\alpha^2\cdot\beta^1=\alpha^2\cdot\beta^4=\alpha^3\cdot\beta^1=\alpha^3\cdot\beta^4=1\label{fml.alphabetaprodC2.4}.
\end{eqnarray}
Keeping in mind nonnegativity of $\alpha_i^j$'s and $\beta_i^k$'s, we can draw some further conclusions about these numbers. First of all, we should notice that two real vectors with nonnegative coordinates are orthogonal if and only if a nonvanishing coordinate of one of the vectors corresponds to a vanishing coordinate of the other vector and vice versa.  As a consequence of this and \fml{alphabetaprodC2.2}, each of the vectors $\alpha^i$ and $\beta^i$ must have a vanishing coordinate. On the other hand, because of the formula \fml{alphabetaprodC2.1} neither of the vectors can be zero. In other words, each of them must have a nonvanishing coordinate. We are left with $\alpha^i$'s and $\beta^j$'s which have either one or two nonzero coordinates. Let us consider first a situation in which one of the vectors has two nonzero coordinates. Without any loss of generality we assume the vector to be $\alpha^1$ and we put $\alpha_1^1=0$, $\alpha^1_2>0$, $\alpha^1_3>0$. Because of \fml{alphabetaprodC2.2}, $\beta_1^4>0$, $\beta_2^4=0$, $\beta_3^4=0$. This in turn implies $\alpha_1^2>0$, $\alpha_1^3>0$ and $\alpha_1^4>0$ as a consequence of \fml{alphabetaprodC2.1}, \fml{alphabetaprodC2.3} and \fml{alphabetaprodC2.4}. Therefore $\beta_1^3=0$, $\beta_1^2=0$ and $\beta_1^1=0$. If $\alpha_2^2=\alpha_3^2=0$, the equality $\alpha^2\cdot\beta^1=1$ cannot hold. One of the coordinates $\alpha_2^2$, $\alpha_3^2$ must be nonzero. We may assume $\alpha_3^2>0$, so that we have $\alpha_1^2>0$, $\alpha_2^2=0$, $\alpha_3^2>0$. From \fml{alphabetaprodC2.2} it follows that $\beta_1^3=0$, $\beta_2^3>0$, $\beta_3^3=0$. Using \fml{alphabetaprodC2.1} we get $\alpha_2^3>0$ while \fml{alphabetaprodC2.3} yields $\alpha_2^4>0$. We have obtained $\alpha_1^3>0$ and $\alpha_2^3>0$, which implies $\alpha_3^3=0$. But now \fml{alphabetaprodC2.2} gives us $\beta_1^2=0$, $\beta_2^2=0$, $\beta_3^2>0$ and from $\alpha^4\cdot\beta^2=1$ we get $\alpha_3^4>0$.

In the successive steps above we obtained $\alpha_1^4>0$, $\alpha_2^4>0$ and finally the inequality $\alpha_3^4>0$. This is in contradiction with \fml{alphabetaprodC2.2}, so our initial assumption about the existence of a vector $\alpha^i$ (or $\beta^i$) with two nonzero coordinates, cannot be true for solutions of the equations \fml{alphabetaprodC2.1}-\fml{alphabetaprodC2.4}. None of the vectors $\alpha^i$, $\beta^i$ can have two nonvanishing coordinates. The only possibility we have not excluded yet is that of all the vectors $\alpha^i$, $\beta^i$ having precisely one nonzero coordinate each. Let us assume that this is the case and concentrate on $\alpha^i$'s. Because of the fact that $\alpha^i$'s are of dimension three, there must exist a pair of indices $i\neq j$ such that $\alpha^i$ is proportional to $\alpha^j$. Without loss of generality we may assume that either $\alpha^1=\alpha^2$ or $\alpha^1=\alpha^4$ holds. The first possibility is excluded because of the equalities $\alpha^1\cdot\beta^4=0$ and $\alpha^2\cdot\beta^4=1$.  The second is in contradiction with $\alpha^1\cdot\beta^4=0$ and $\alpha^1\cdot\alpha^1=2$. Thus we have excluded the only remaining possibility for $\alpha^i$'s and we conclude that \fml{coeffsofC2.1} has no solutions of the desired properties $\alpha_i^j,\beta_i^k\geqslant 0\,\forall_{i,j,k}$. Consequently, $\tilde\rho$ cannot be written in the form \fml{decofC2.1} with $\rho_i$'s and $\xi_i$'s positive. The same holds for $\rho$. Hence $l>3$, which in turn implies $l=4$ because $l\leqslant 4$. This proves our assertions about $\rho$.
\end{proof}
\end{example}

\chapter{Algebraic problems solved by using Groebner bases}\,\label{chGroebnerapplied}

\section{Compression equations -- a special case}\label{seccompeq}\,
In Section \ref{sec3x3matrix}, we defined the notion of a code carrier of an operator and outlined how it is related to the problem of finding solutions of generalized eigenvalue problems. Notably, the paper \cite{ref.CKZ06} contains an almost complete description of code carriers for Hermitian operators. Therefore, following similar steps to those described in Section \ref{sec3x3matrix}, one may attempt to find a general solution to a compression equation $P_kMP_k=\lambda P_k$ by first splitting $M$ into its Hermitian $M_H:=\left(M+\conj{M}\right)/2$ and anti-Hermitian part $M_A:=\left(M-\conj{M}\right)/2i$ and solving the respective compression equations 
\begin{equation}\label{eqcompeq1}
\begin{cases}P_kM_HP_k=\xi M_H\\P_kM_AP_k=\zeta M_A
\end{cases}
\end{equation}
\emph{separately}. Next, the sets of possible projectors $\Pi_k\left(M_A\right)$ and $\Pi_k\left(M_H\right)$ may be intersected to yield $\Pi_k\left(M\right)$. For a more detailed description of the method and for the definition of $\Pi_k$, cf. Section \ref{sec3x3matrix}. Finally, if we start with a system of equations of the form $P_kM^{\left(i\right)}P_k=\lambda_iP_k$ for $i=1,2,\ldots$ instead of a single one, we may first determine $\Pi_k\left(M^{\left(i\right)}\right)$ following the steps described above and then find the solutions to the initial set of equations by intersecting $\Pi\left(M^{\left(i\right)}\right)$ for $i=1,2,\ldots$ . 

The described procedure turns out to be rather difficult to implement in practice. However, in the present section we present a very simplified example where the method works. Let $k=2$. Let us also take $H_1$ and $H_2$ to be two Hermitian operators on $\setC^4$. We assume that $H_1$ and $H_2$ commute, so that they have a common eigenbasis $\left\{v_1,v_2,v_3,v_4\right\}$. Moreover, let the respective eigenvalues for $H_1$ fulfill $\lambda_1<\lambda_2<\lambda_3<\lambda_4$, while for $H_2$ we have $\chi_1<\chi_2<\chi_3<\chi_4$. In such case, it is relatively easy to find all possible solutions of the set of equations
\begin{equation}\label{eqH1H2}
P_2H_1P_2=\xi P_2\,\land\,P_2H_2P_2=\zeta P_2.
\end{equation}
This can be done using the technique of Groebner bases discussed in Part \ref{partII} of the thesis. 

In the setting described above, the general characterization of code carriers included in \cite[Section 4]{ref.CKZ06} reduces to 
\begin{multline}\label{eqprojH1}
\Pi\left(H_1\right)=\left\{\linspan\left\{\frac{v_1}{\sqrt{\lambda-\lambda_1}}+e^{i\phi}\left(\frac{av_3}{\sqrt{\lambda_3-\lambda}}-\frac{\bar bv_4}{\sqrt{\lambda_4-\lambda}}\right),\frac{v_2}{\sqrt{\lambda-\lambda_2}}+\right.\right.\\
\left.\left.+e^{i\phi}\left(\frac{bv_3}{\sqrt{\lambda_3-\lambda}}+\frac{\bar av_4}{\sqrt{\lambda_4-\lambda}}\right)\right\}\vline\lambda\in\left(\lambda_2,\lambda_3\right),\phi\in\setR,\left|a\right|^2+\left|b\right|^2=1\right\}.
\end{multline}
Similarly for $H_2$,
\begin{multline}\label{eqprojH2}
\Pi\left(H_2\right)=\left\{\linspan\left\{\frac{v_1}{\sqrt{\chi-\chi_1}}+e^{i\psi}\left(\frac{cv_3}{\sqrt{\chi_3-\chi}}-\frac{\bar dv_4}{\sqrt{\chi_4-\chi}}\right),\frac{v_2}{\sqrt{\chi-\chi_2}}+\right.\right.\\
\left.\left.+e^{i\psi}\left(\frac{dv_3}{\sqrt{\chi_3-\chi}}+\frac{\bar cv_4}{\sqrt{\chi_4-\chi}}\right)\right\}\vline\chi\in\left(\chi_2,\chi_3\right),\psi\in\setR,\left|c\right|^2+\left|d\right|^2=1\right\}.
\end{multline}
The question whether there exists a $P_2$ that satisfies the set of equations \eqref{eqH1H2} is equivalent to the existence of an identical pair of subspaces in the sets $\Pi_2\left(H_1\right)$ and $\Pi_2\left(H_2\right)$, given by the equations \eqref{eqprojH1} and \eqref{eqprojH2}. Fortunately, the existence can easily be checked. Due to the specific form of the subspaces in formulas \eqref{eqprojH1} and \eqref{eqprojH2}, the intersection of $\Pi_2\left(H_1\right)$ and $\Pi_2\left(H_2\right)$ is nonempty if and only if the following equations are satisfied for some admissible values of $a$, $b$, $c$, $d$, $\phi$, $\psi$, $\lambda$ and $\chi$.
\begin{eqnarray}
e^{i\phi}a\sqrt{\frac{\lambda-\lambda_1}{\lambda_3-\lambda}}=e^{i\psi}c\sqrt{\frac{\chi-\chi_1}{\chi_3-\chi}},&&e^{i\phi}\bar b\sqrt{\frac{\lambda-\lambda_1}{\lambda_4-\lambda}}=e^{i\psi}\bar d\sqrt{\frac{\chi-\chi_1}{\chi_4-\chi}},\label{eqcompression1}\\
e^{i\phi}b\sqrt{\frac{\lambda-\lambda_2}{\lambda_3-\lambda}}=e^{i\psi}d\sqrt{\frac{\chi-\chi_2}{\chi_3-\chi}},&&e^{i\phi}\bar a\sqrt{\frac{\lambda-\lambda_2}{\lambda_4-\lambda}}=e^{i\psi}\bar c\sqrt{\frac{\chi-\chi_2}{\chi_4-\chi}}.\label{eqcompression2}
\end{eqnarray}
The formulas above imply a weaker set of equations  
\begin{eqnarray}
\left|a\right|^2\frac{\lambda-\lambda_1}{\lambda_3-\lambda}=\left|c\right|^2\frac{\chi-\chi_1}{\chi_3-\chi},&&\left|b\right|^2\frac{\lambda-\lambda_1}{\lambda_4-\lambda}=\left|d\right|^2\frac{\chi-\chi_1}{\chi_4-\chi},\\
\left|b\right|^2\frac{\lambda-\lambda_2}{\lambda_3-\lambda}=\left|d\right|^2\frac{\chi-\chi_2}{\chi_3-\chi},&&\left|a\right|^2\frac{\lambda-\lambda_2}{\lambda_4-\lambda}=\left| c\right|^2\frac{\chi-\chi_2}{\chi_4-\chi}.
\end{eqnarray}
which can be rewritten in the form
\begin{equation}\label{eqset1}
\alpha h_3 l_1=\gamma h_1 l_3,\,\beta h_4 l_1 =\delta h_1 l_4,\,\beta h_3 l_2=\delta h_2 l_3,\,\alpha h_4 l_2=\gamma h_2 l_4.
\end{equation}
In the above expression, we the following notation was used: $\alpha:=\left|a\right|^2$, $\beta:=\left|b\right|^2$, $\gamma=\left|c\right|^2$, $\delta=\left|d\right|^2$, $l_1:=\lambda-\lambda_1$, $l_2:=\lambda-\lambda_2$, $l_3:=\lambda_3-\lambda$, $l_4:=\lambda_4-\lambda$. The newly introduced variables $\alpha$, $\beta$, $\gamma$, $\delta$ and $l_i$, $h_i$ for $i=1,2,3,4$ must be nonnegative and fulfill the additional conditions
\begin{eqnarray}
\alpha+\beta=1,\,\gamma+\delta=1,\label{eqset2}\\
l_1-l_2=\lambda_2-\lambda_1,\,l_2+l_3=\lambda_3-\lambda_2,\,l_3-l_4=\lambda_3-\lambda_4,\label{eqset3}\\
h_1-h_2=\chi_2-\chi_1,\,h_2+h_3=\chi_3-\chi_2,\,h_3-h_4=\chi_3-\chi_4.\label{eqset4}
\end{eqnarray}
The approach we take in the following is to solve \eqref{eqset1} together with \eqref{eqset2}--\eqref{eqset4} as if $\alpha$, $\beta$, $\gamma$, $\delta$ and $l_i$, $h_i$ for $i=1,2,3,4$ were allowed to take arbitrary values in $\setC$. Next, we look for real, nonnegative solutions. Note that the equalities \eqref{eqset1}, as well as \eqref{eqset2}--\eqref{eqset4}, can be rewritten as polynomial equations in the variables $\alpha$, $\beta$, $\gamma$, $\delta$ and $l_i$, $h_i$. Therefore, for fixed values of $\lambda_i$ and $\chi_i$, $i=1,2,3,4$, we can try to solve the equations \eqref{eqset1}, \eqref{eqset2}--\eqref{eqset4} using the Groebner basis approach described in Chapter~\ref{chVarIdGroeb}. As an example, let us consider $\lambda_1=1$, $\lambda_2=2$, $\lambda_3=3$, $\lambda_4=4$ and $\chi_1=1$, $\chi_2=4$, $\chi_3=9$, $\chi_4=16$.  Then, a Groebner basis calculation in $\setC\left[l_1,\ldots,l_4,h_1,\ldots,h_4,\alpha,\beta,\gamma,\delta\right]$ for the equations \eqref{eqset1} and \eqref{eqset2}--\eqref{eqset4} gives the following result,
\begin{multline}\label{eqGroebBascompression}
\left\{2560 \delta-5184 \delta^2+112 \delta^3+1704 \delta^4-224 \delta^5-102 \delta^6+9 \delta^7,-1+\gamma+\delta,\right.\\940800 \beta-11503040 \delta+2125136 \delta^2+3603128 \delta^3-630848 \delta^4-223842 \delta^5+20691 \delta^6,\\
-940800+940800 \alpha+11503040 \delta-2125136 \delta^2-3603128 \delta^3+630848 \delta^4+223842 \delta^5+\\-20691 \delta^6,-302400+236720 \delta+108512 \delta^2-45044 \delta^3-16046 \delta^4+141 \delta^5+117 \delta^6+\\+33600 h_4,
-67200+236720 \delta+108512 \delta^2-45044 \delta^3-16046 \delta^4+141 \delta^5+117 \delta^6+\\+33600 h_3,-100800-236720 \delta-108512 \delta^2+45044 \delta^3+16046 \delta^4-141 \delta^5-117 \delta^6+\\+33600 h_2,
-201600-236720 \delta-108512 \delta^2+45044 \delta^3+16046 \delta^4-141 \delta^5-117 \delta^6+\\+33600 h_1,-4032000-7312320 \delta+3861328 \delta^2+2882424 \delta^3-887584 \delta^4-221586 \delta^5+\\+22563 \delta^6+2688000 l_4,
-1344000-7312320 \delta+3861328 \delta^2+2882424 \delta^3-887584 \delta^4+\\-221586 \delta^5+22563 \delta^6+2688000 l_3,-1344000+7312320 \delta-3861328 \delta^2-2882424 \delta^3+\\+887584 \delta^4+221586 \delta^5-22563 \delta^6+2688000 l_2,
-4032000+7312320 \delta+\\\left.-3861328 \delta^2-2882424 \delta^3+887584 \delta^4+221586 \delta^5-22563 \delta^6+2688000 l_1\right\}.
\end{multline}
According to what we learned in Chapter~\ref{chVarIdGroeb}, we get a set of equations equivalent to \eqref{eqset1} and \eqref{eqset2}--\eqref{eqset4} by equating the above polynomials to zero. As expected, the first polynomial in \eqref{eqGroebBascompression} only involves the variable $\delta$. Moreover, its seven roots can be explicitly found. They are equal to $-4$, $-2$, $0$, $2$, $8/3$, $1/3\left(19-\sqrt{301}\right)$, $1/3\left(19+\sqrt{301}\right)$. The structure of the remaining equations resulting from the Groebner basis \eqref{eqGroebBascompression} is such that after we find $\delta$, the admissible values of the other variables can be determined by simple substitution. In this way we get the following solutions $\left(l_1,\ldots,l_4,h_1,\ldots,h_4,\alpha,\beta,\gamma,\delta\right)$
\begin{enumerate}[1)]
\item $\left(3,2,-1,0,15,12,-7,0,-0.714286,1.71429,-1,2\right)$,
\item $\left(2, 1, 0, 1, 8, 5, 0, 7, 2.14286, -1.14286, 3, -2\right)$,
\item $\left(1.5, 0.5, 0.5, 1.5, 6, 3, 2, 9, 1, 0, 1, 0\right)$,
\item $\left(1, 0, 1, 2, 3, 0, 5, 12, 3, -2, 5, -4\right)$,
\item $\left(0, -1, 2, 3, 0, -3, 8, 15, -1, 2, -1.66667, 2.66667\right)$,
\item $\left(2.31747, 1.31747, -0.317468, 0.682532, 1.91266, -1.08734,6.08734,\right.$\\ $\left. 
13.0873, 0.478479, 0.521521, -11.1165, 12.1165\right)$,
\item $\left(0.582532, -0.417468, 1.41747, 2.41747, 10.5873, 7.58734,-2.58734,\right.$\\ $\left. 
4.41266, -4.47848, 5.47848, 0.449784, 0.550216\right)$.
\end{enumerate}
The numerical values for the solutions were calculated using exact algebraic expressions. As we can see, only  solution number $3$ has all its coordinates nonnegative. Thus, if there exists a solution to equations \eqref{eqcompression1} and \eqref{eqcompression2}, the respective values of $a$, $b$, $c$, $d$, $\lambda$ and $\chi$ must be such that $\left|a\right|^2=\alpha=1$, $\left|b\right|^2=\beta=0$, $\left|c\right|^2=\gamma=1$, $\left|d\right|^2=\delta=0$, as well as $\lambda-\lambda_1=\lambda-1=l_1=3/2$, $\chi-\chi_1=\chi-1=h_1=6$. Hence $\lambda=5/2$, $\chi=7$. The formulas \eqref{eqcompression1} and \eqref{eqcompression2} take the form 
\begin{equation}\label{eqcompressionreduced1}
e^{i\phi}e^{i\mu}\sqrt{\frac{3/2}{1/2}}=e^{i\psi}e^{i\nu}\sqrt{\frac{6}{2}},\quad e^{i\phi}e^{-i\mu}\sqrt{\frac{1/2}{3/2}}=e^{i\psi}e^{-i\nu}\sqrt{\frac{3}{9}},
\end{equation}
where we introduced $e^{i\mu}:=a$ and $e^{i\nu}:=c$.
There are only two equations left, since the second one in \eqref{eqcompression1} and the first one in \eqref{eqcompression2} are trivially fulfilled for $b=d=0$. Clearly, the equalities in \eqref{eqcompressionreduced1} are equivalent to $e^{i\left(\phi+\mu-\psi-\nu\right)}=1$ and $e^{i\left(\phi-\mu-\psi+\nu\right)}=1$, or  $\phi+\mu-\psi-\nu=0\textnormal{ mod }2\pi$ and $\phi-\mu-\psi+\nu=0\textnormal{ mod }2\pi$, respectively. From the last two formulas, we get $\phi-\psi=0\textnormal{ mod }2\pi$ and $\mu-\nu=0\textnormal{ mod }2\pi$, which means that $\phi=\psi+n\pi$ and $\mu=\nu+m\pi$ for some $m,n\in\setZ$. Moreover, $\phi+\mu-\psi-\nu=0\textnormal{ mod }2\pi$ implies that $m=n\textnormal{ mod }2$. In conclusion, the full set of solutions are parametrized by the two angles $\psi$ and $\nu$. The solutions are of the form
\begin{equation}
\linspan\left\{
\sqrt{\frac{2}{3}}v_1+e^{i\left(\psi+\nu\right)}\sqrt{2}v_3,\sqrt{2}v_2+e^{i\left(\psi-\nu\right)}\sqrt{\frac{2}{3}}v_3
\right\}
\end{equation}
The corresponding compression value $\xi$ for $H_1$ is $5/2$, while for $H_2$, we get $\zeta=7$. Further investigation of equations \eqref{eqcompression1} and \eqref{eqcompression2} in Mathematica suggests that for any choice of the eigenvalues of $H_1$ and $H_2$, such that $\lambda_1<\lambda_2<\lambda_3<\lambda_4$ and $\chi_1<\chi_2<\chi_3<\chi_4$, there exists a single family of solutions to equations \eqref{eqcompression1} and \eqref{eqcompression2}, either with $a=c=0$ or with $b=d=0$. By choosing the eigenvalues from the set of rational numbers, we seem always to obtain polynomial equations that are exactly solvable.   
\section{Completely Entangled Subspaces}\,\label{secces}
Linear subspaces without a product vector are called \textbf{Completely Entangled Subspaces} or CES for short. In the present section, we shall discuss the question how to check whether a given subspace is a CES or not. In particular, we shall give an example of a one-parameter family of subspaces of $\setC^3\otimes\setC^4$ and characterize the values of the parameter for which the subspace and its orthogonal complement are completely entangled.

Let us start with the general question about the existence of a product vector in a linear subspace. Both the set of product states and a linear subspace are projective varieties and it should be possible to determine their intersection using the techniques described above. An approach we successfully used was very straightforward. The general algorithm we applied is shown in Figure \ref{figprodinlin}. The main idea is to write a set of polynomial equations, corresponding to $\left[a_1,\ldots a_n\right]\otimes\left[b_1,\ldots,b_m\right]\in V$ for a subspace $V$ and then generate the corresponding Groebner basis. The answer can often be read from the output. According to Proposition \ref{propexistenceofsolutions}, a necessary and sufficient condition for a set of polynomial equations to have a solution (over $\mathbbm{C}^n$) is that the corresponding reduced Groebner basis be different from $\left\{1\right\}$.  
\begin{figure}[h!]
\begin{verbatim}
symbols1={a1,a2,...,an}
symbols2={b1,b2,...,bm}
symbols12=Union(symbols1,symbols2)
productvector=KroneckerProduct(symbols1,symbols2)
subspace={{v11,v12,...,v1nm},{v21,v22,...,v2nm},...
...,{vd1,vd2,...,vdnm}}
positivematrix=Transpose(subspace).subspace
Diagonalize positivematrix
Choose eigenvectors= 
       ={{w11,w12,...,w1nm},{w21,w22,...,w2nm},...,
       {w(mn-d)1,w(mn-d)2,...,w(mn-d)nm}} 
       corresponding to eigenvalue 0
Calculate polynomialequations=eigenvectors.productvector
For i=1 to n do
    For j=1 to m do
        Calculate GroebnerBasis[{polynomialequations,ai-1,bj-1}]         
\end{verbatim}
\caption{An algorithm for testing whether a given linear subspace admits product vectors in the bipartite case.}\label{figprodinlin}
\end{figure}

 As a careful reader would notice, more than a single Groebner basis is actually calculated, and each of them has some additional polynomials. This is so because they are  different dehomogenizations of the set of equations $w_i\cdot\left(a\otimes b\right)=0$, $i=1,\ldots,nm - d$, which corresponds to $a\otimes b\in V$. We dehomogenize the equations in order to eliminate trivial solutions, corresponding to a zero ``product'' vector. Moreover, after dehomogenization product vectors that are a multiple of each other appear as a single solution, which is a desirable feature. For example, if we dehomogenize by adding the polynomials $a_1-1$ and $b_1-1$, we capture all product vectors $a\otimes b$ with the first coordinate in $a$ and $b$ nonvanishing. The method can be generalized in an obvious way to the multipartite case.
 
 In the sequel, we give details of the procedure for the particular case of product vectors in a family $V\left(z\right)$ of six-dimensional subspaces of $\mathbbm{C}^3\otimes\mathbbm{C}^4$. In this case, we can avoid considering $3\times 4=12$ different dehomogenizations and we get away with only four, three of which are different from those we would normally have used with the algorithm in Figure \ref{figprodinlin}.  The elements of the family $V\left(z\right)$ we consider are subspaces spanned by the vectors
\begin{multline}\label{vectorsv}
\left\{v_1\left(z\right),\ldots,v_6\left(z\right)\right\}=\\=\left\{e_1\otimes e_1+e_2\otimes e_2,e_2\otimes e_1+z e_3\otimes e_2,e_3\otimes e_1+z^2 e_1\otimes e_3\right.\\
\left.e_1\otimes e_2+z^3e_3\otimes e_4,e_2\otimes e_3+z^4e_1\otimes e_4,e_3\otimes e_3+z^5e_2\otimes e_4\right\},
\end{multline}
where $z\in\mathbbm{C}\setminus\left\{0\right\}$. As it can be easily checked, the orthogonal complement $V\left(z\right)^{\bot}$ is spanned by the vectors
\begin{multline}\label{vectorsw}
\left\{w_1\left(z\right),\ldots,w_6\left(z\right)\right\}=\\=\left\{e_1\otimes e_1-e_2\otimes e_2,\bar z e_2\otimes e_1- e_3\otimes e_2,\bar z^2e_3\otimes e_1- e_1\otimes e_3\right.\\
\left.\bar z^3e_1\otimes e_2-e_3\otimes e_4,\bar z^4e_2\otimes e_3-e_1\otimes e_4,\bar z^5e_3\otimes e_3-e_2\otimes e_4\right\}.
\end{multline}
Consider product vectors of the form $p=\left[a_1,a_2,a_3\right]\otimes\left[b_1,b_2,b_3,b_4\right]$. The condition $p\in V\left(z\right)$ is equivalent to $p\cdot w_i=0$ $\forall_{i=1,\ldots,6}$, which is a set of homogeneous polynomial equations. We would like to find their solutions with $\left[a_1,a_2,a_3\right]\neq\left[0,0,0\right]$ and $\left[b_1,b_2,b_3,b_4\right]\neq\left[0,0,0,0\right]$. A possible way to achieve this goal is to: i) add the polynomials $a_1-1$ and $b_1-1$ or equivalently, to substitute $a_1\rightarrow 1$, $b_1\rightarrow 1$. This gives us a dehomogenized set of polynomial equations, which capture all the nontrivial solutions of $p\cdot w_i=0$ $\forall_{i=1,\ldots,6}$, apart from those with $a_1=0$ or $b_1=0$. In order to account for the possible deficit, one needs to consider other  dehomogenizations. One way to do it is to proceed as in Figure \ref{figprodinlin} and dehomogenize in 12 different ways. However, in the case we consider it is easier to do the following substitutions: ii) $a_1\rightarrow 0$ and $b_1\rightarrow 1$, iii) $a_1\rightarrow 1$ and $b_1\rightarrow 0$ and iv) $a_1\rightarrow 0$, $b_1\rightarrow 0$. Equivalently, one adds ii) $a_1$ and $b_1-1$, iii) $a_1-1$ and $b_1$, iv) $a_1$ and $b_1$ to the ideal generated by the equations $p\cdot w_i=0$ for $i=1,\ldots,6$. The set of polynomials $p\cdot w_i$ reads 
\begin{multline}\label{equationspw}
\left\{b_1 a_1-b_2 a_2,-b_2 a_3+b_1 a_2 \bar{z},-a_1 b_3+b_1 a_3 \bar{z}^2,-a_3 b_4+a_1 b_2 \bar{z}^3,\right.\\\left.-a_1 b_4+a_2 b_3 \bar{z}^4,-a_2 b_4+b_3 a_3 \bar{z}^5,-1+b_1,-1+a_1\right\}.
\end{multline}
After dehomogenization i) and calculation of the corresponding Groebner basis in the ring $\mathbbm{C}\left[a_1,\ldots,a_n,b_1,\ldots,b_m,\bar{z}\right]$, we get
\begin{multline}\label{dehomoi}
\left\{-\bar{z}^3+\bar{z}^5,-b_4+b_4 \bar{z}^2,b_4^2-\bar{z}^3,-b_3+b_3 \bar{z}^2,b_3^3-\bar{z}^4,
-b_3 b_4+b_2 \bar{z}^3,\right.\\-b_3+b_2 b_4,b_2 b_3-b_3^2 b_4 \bar{z},-1+b_1,-b_3+a_3 \bar{z}^2,-b_3 b_4+a_3 b_4,\\
-b_3^2+b_3 a_3,b_2^2 a_3-\bar{z},-b_3^2 \bar{z}+a_3^2 \bar{z},a_3^3-\bar{z}^4,-b_2 a_3+a_2 \bar{z},\\\left.
a_2 b_4-b_3^2 \bar{z},a_2 b_3-b_4,-1+b_2 a_2,-1+a_1\right\}.
\end{multline}
We clearly see that after a substitution of a particular value of $z$, the first element of the basis is a nonzero constant in $\mathbbm{C}$ unless the substituted value is a solution of the equation $-z^3+z^5=0$. This implies that $1$ is in the ideal generated by $\left\{p\cdot w_1,\ldots p\cdot w_6,a_1-1,b_1-1\right\}$ unless $z=0$ or $z=\pm 1$. This implies that there is no solution to the corresponding equations for almost all choices of $z$. Equivalently, there is no product vector $a\otimes b$ with the first coordinate of $a$ and $b$ nonvanishing in $V\left(z\right)$ unless $z=0$ or $z=\pm 1$. Obviously, there exist product vectors in $V\left(z\right)$ when $z=0$, because the vectors $v_i\left(0\right)$ are of a product form. Thus we have already excluded $z=0$ in the definition of $V\left(z\right)$ given above. For $z= 1$, we get the following Groebner basis in the ring $\mathbbm{C}\left[a_1,\ldots,a_n,b_1,\ldots,b_m\right]$
\begin{equation}\label{casez1}
\left\{-1+b_4^2,-1+b_3^3,b_2-b_3 b_4,-1+b_1,-b_3+a_3,a_2-b_3^2 b_4,-1+a_1\right\}.
\end{equation}
It is easy to see that the above equations have six solutions, corresponding to the choices of $b_4=\pm 1$ and $b_3=e^{\frac{2\pi i}{3}n}$, $n=1,2,3$. Thus there are six product vectors $a\otimes b$ with nonvanishing first coordinates of $a$ and $b$ in $V\left(z\right)$. Similarly for $z=-1$, we get the following Groebner basis
\begin{equation}\label{casezm1}
\left\{1+b_4^2,-1+b_3^3,b_2+b_3 b_4,-1+b_1,-b_3+a_3,a_2-b_3^2 b_4,-1+a_1\right\}.
\end{equation}
Again, there are six product vectors $a\otimes b$ with nonvanishing first coordinates of $a$ and $b$ in $V\left(-1\right)$.

We still need to consider the dehomogenizations ii)-iv) for a general $V\left(z\right)$. In the case ii), we get the following Groebner basis 
\begin{equation}\label{dehomoi2}
\left\{-1+b_1,a_3 \bar{z}^2,a_3 b_4,b_2^2 a_3,b_1 a_3^2 \bar{z},-b_2 a_3+a_2 \bar{z},a_2 b_4,b_2 a_2,a_1\right\}.
\end{equation}
A solution for $z\neq 0$ must necessarily have $a_3=0$, which implies that $-b_2 a_3+a_2 \bar{z}=a_2 z$. Therefore also $a_2=0$. Thus $V\left(z\right)$ admits no product vector $a\otimes b$  with nonvanishing first coordinate in $b$ and vanishing first coordinate in $a$. In the case iii), we get the following Groebner basis in $\mathbbm{C}\left[a_1,\ldots,a_n,b_1,\ldots,b_m,\bar{z}\right]$
\begin{equation}\label{dehomoi3}
\left\{b_4, b_3, b_2 \bar{z}^3, b_1, b_2 a_3, b_2 a_2, -1 + a_1\right\}.
\end{equation}
One immediately sees that for $z\neq 0$, the above polynomials vanish only if $b_1=b_2=b_3=b_4=0$, which again gives a zero product vector. Therefore, there are no product vectors $a\otimes b$ with vanishing first coordinate of $b$ and nonvanishing first coordinate of $a$ in $V\left(z\right)$ for $z\neq 0$. We only need to consider the last case, number iv), when the first coordinates of both $a$ and $b$ vanish. The corresponding Groebner basis reads
\begin{equation}\label{dehomoi4}
\left\{b_1,a_3 b_4,b_3^2 a_3 \bar{z}^9,b_2 a_3,b_3 a_3^2 \bar{z}^5,a_2 b_4-b_3 a_3 \bar{z}^5,a_2 b_3 \bar{z}^4,b_2 a_2,a_1\right\}.
\end{equation}   
If $a_3\neq 0$, we see from the first four polynomials that $b_1=b_2=b_3=b_4=0$. Therefore we must have $a_3=0$ in order to obtain a nonzero vector $a\otimes b$. However, a substitution of $a_3=0$ to \eqref{dehomoi4} yields $\left\{b_1,a_2b_4,a_2b_3z^4,b_2a_2,a_1\right\}$. We see that these polynomials vanish simultaneously only if $a_2=0$ or $b_1=b_2=b_3=b_4=0$. In either case, $a\otimes b$ vanishes. Thus, there are no nonzero product vectors $a\otimes b$ with vanishing first coordinates of $a$ and $b$ in $V\left(z\right)$ for $z\neq 0$.

We can summarize our results by saying that $V\left(z\right)$ is a CES for all $z\not\in\left\{-1,0,1\right\}$. We can also easily repeat the above described procedure for the subspace $V\left(z\right)^{\bot}$ and obtain an analogous result. In this case $\left(V\left(z\right)^{\bot}\right)^{\bot}=V\left(z\right)$, so the r{\^o}le of the vectors $w_i\left(z\right)$ is played by the vectors $v_i\left(z\right)$. Otherwise, the calculation is almost the same. We obtain the following four Groebner Bases.
\begin{eqnarray}\label{groebnerbasesitoiv}
i)&\left\{-z^3+z^5,-b_4+b_4 z^2,b_4^2-z^3,-b_3+b_3 z^2,b_3^3-z^4,\right.\\
&-b_3 b_4+b_2 z^3,-b_3+b_2 b_4,b_2 b_3-b_3^2 b_4 z,-1+b_1,-b_3+a_3 z^2,\\&-b_3 b_4+a_3 b_4,
-b_3^2+b_3 a_3,b_2^2 a_3-z,-b_3^2 z+a_3^2 z,a_3^3-z^4,\\&\left.-b_2 a_3+a_2 z,
 a_2 b_4-b_3^2 z, a_2 b_3-b_4,-1+b_2 a_2,-1+a_1\right\},\\
ii)&\left\{-1+b_1,a_3 z^2,a_3 b_4,b_2^2 a_3,b_2 a_3^2 z,-b_2 a_3+a_2 z,a_2 b_4,b_2 a_2,a_1\right\},\\
iii)&\left\{b_4,b_3,b_2 z^3,b_1,b_2 a_3,b_2 a_2,-1+a_1\right\},\\
iv)&\left\{b_1,a_3 b_4,b_3^2 a_3 z^9,b_2 a_3,b_3 a_3^2 z^5,a_2 b_4-b_3 a_3 z^5,a_2 b_3 z^4,b_2 a_2,a_1\right\}.
\end{eqnarray}
with the notation i)-iv) referring to dehomogenizations of types i)-iv), as described above. An argument very similar to the one given above shows that there are no product vectors in $V\left(z\right)^{\bot}$, as long as $z\not\in\left\{-1,0,1\right\}$. The case $z=0$ is excluded by assumption, whereas for $z=\pm 1$ it can again be checked that there are six product vectors in the subspace in question, which this time is $V\left(z\right)^{\bot}$.

The results of the present section can be summarized by saying that, concerning the $3\times4$ CES problem considered above, the family of subspaces $V\left(z\right)$, $z\in\mathbbm{C}\setminus\left\{0\right\}$, spanned by the vectors \eqref{vectorsv}, \textbf{consists of CES}, with the exception of $z\in\left\{-1,1\right\}$. Moreover, the \textbf{orthogonal complement} $V\left(z\right)^{\bot}$ is also completely entangled for $z\not\in\left\{-1,1\right\}$. 
 
\section{Maximally entangled states in linear subspaces}\,\label{secmaxent}
In the previous section, we discussed the existence of product vectors in linear subspaces. It is natural to ask somewhat opposite question, under which conditions a linear subspace admits maximally entangled vectors, i.e. vectors of the form $\sum_{i=1}^ne_i\otimes f_i$, where the summation goes from $1$ to the dimension of the subsystems and $\left\{e_i\right\}_{i=1}^n$ and $\left\{f_i\right\}_{i=1}^n$ are orthonormal bases for the first and the second subsystem, respectively. By solving two examples, we will show that the problem can be tackled using the techniques of Groebner bases. 

Let us start with a subspace orthogonal to an Unextendible Product Basis in $\setC^3\otimes\setC^3$, i.e. to a set of orthogonal product vectors such that no other product vector in $\setC^3\otimes\setC^3$ is orthogonal to all of them. We shall discuss Unextendible Product Bases in more detail in Chapter \ref{chPPT3x3} and here we restrict our attention to the question whether there exist maximally entangled vectors in the orthogonal complement of a particular UPB, given by 
\begin{equation}\label{eqsymmetricUPB}
\left\{v_0\otimes v_2,v_1\otimes v_0,v_2\otimes v_3,v_3\otimes v_1,v_4\otimes v_4\right\},
\end{equation}
where
\begin{equation}\label{eqvvectors}
v_0=\left[\begin{array}{c}1\\0\\0\end{array}\right],v_1=\left[\begin{array}{c}\frac{1}{\sqrt{2}}\\0\\\frac{1}{\sqrt{2}}\end{array}\right],v_2=\left[\begin{array}{c}0\\\frac{1}{\sqrt{2}}\\\frac{1}{\sqrt{2}}\end{array}\right],v_3=\left[\begin{array}{c}0\\1\\0\end{array}\right],v_4=\left[\begin{array}{c}\frac{1}{\sqrt{3}}\\\frac{1}{\sqrt{3}}\\-\frac{1}{\sqrt{3}}\end{array}\right].
\end{equation} 
It will be clear from the following discussion that the methods we use can be applied in a much more general setting.

One can easily see that a vector $\sum_{i,j}A_{ij}e_i\otimes e_j\in\setC^3\otimes\setC^3$ is maximally entangled if and only if the matrix $A=\left[A_{ij}\right]$ is unitary. This yields a set of polynomial equations $\sum_{j=1}^3A_{ij}\bar A_{kj}=\delta_{ik}$ on the matrix elements $A_{ij}$ and their complex conjugates $\bar A_{ij}$. Another set of equations comes from the orthogonality conditions to the UPB given in \eqref{eqsymmetricUPB} and \eqref{eqvvectors}. The equations are linear and can be solved explicitly, which we leave as a simple exercise to the reader. The answer is
\begin{equation}\label{eqsolutionAij1}
\left[A_{ij}\right]=\left[
\begin{array}{ccc}
 a & b & -b \\
 d & e & -d \\
 -a & -e & -2 (a+b+d+e)
\end{array}
\right],
\end{equation} 
where $a,b,d,e$ are arbitrary complex parameters. The conditions $\sum_{j=1}^3A_{ij}\bar A_{kj}=\delta_{ik}$ for $i=k$ imply
\begin{eqnarray}
a \bar a+2 b \bar b-2 d \bar d-e \bar e=0,\label{equnitarity1}\\
2 d \bar d+e \bar e-a \bar a-e \bar e-2\left(a+b+d+e\right)\overline{\left(a+b+d+e\right)}=0.\label{equnitarity3}
\end{eqnarray}
For $i<k$, we have the following equations $\sum_{j=1}^3A_{ij}\bar A_{kj}=0$,
\begin{eqnarray}
a\bar d+b\overline{\left(e+d\right)}=0,\label{equnitarity4}\\
a\bar a+b\bar e-2b\overline{\left(a+b+d+e\right)}=0,\label{equnitarity5}\\
d\bar a+e\bar e-2d\overline{\left(a+b+d+e\right)}=0.\label{equnitarity6}
\end{eqnarray}
The crucial observation now is that the complex conjugates of \eqref{equnitarity4}-\eqref{equnitarity5} consist an independent set of equations if $a,b,d,e$ and $\bar a,\bar b,\bar d,\bar e$ are perceived as $8$ independent complex variables. This is the approach we are going to take in the following. The complex conjugates of  \eqref{equnitarity4}-\eqref{equnitarity5} read
\begin{eqnarray}
\bar a d+\bar b{\left(e+d\right)}=0\label{equnitarity44},\\
\bar a a+\bar b e-2\bar b{\left(a+b+d+e\right)}=0\label{equnitarity55},\\
\bar d a+\bar e e-2\bar d{\left(a+b+d+e\right)}=0\label{equnitarity66}.
\end{eqnarray}
A Groebner basis calculation in $\setC\left[a,\bar a,b,\bar b,d,\bar d,e,\bar e\right]$ for the nine polynomials in \eqref{equnitarity1}-\eqref{equnitarity66} yields the following basis
\begin{multline}
\label{eqGroebnerMax1}
\left\{e {\bar e}^3,e^2 {\bar e}^2,e^3 {\bar e},{\bar d} e {\bar e}^2,{\bar d} e^2 {\bar e},{\bar d} e^3,{\bar d}^2 e {\bar e},{\bar d}^2 e^2,
{\bar d}^3 e,d {\bar e}^3,d e {\bar e}^2,d e^2 {\bar e},\right.\\
d {\bar d} {\bar e}^2,d {\bar d} e {\bar e},d {\bar d} e^2,d {\bar d}^2 {\bar e},d {\bar d}^2 e,d {\bar d}^3,d^2 {\bar e}^2,d^2 e {\bar e},d^2 {\bar d} {\bar e},d^2 {\bar d} e,d^2 {\bar d}^2,d^3 {\bar e},d^3 {\bar d},{\bar b} e {\bar e}^2,{\bar b} e^2 {\bar e},\\
{\bar b} e^3,-2 d {\bar d}^2+5 {\bar b} {\bar d} e-4 d {\bar d} {\bar e}-2 {\bar b} e {\bar e}+3 {\bar d} e {\bar e},{\bar b} d+2 d {\bar d}-{\bar b} e+2 d {\bar e}-e {\bar e},{\bar b}^2 e {\bar e},{\bar b}^2 e^2,{\bar b}^3 e,\\
394 d {\bar d}^2-270 {\bar b}^2 e-80 {\bar d}^2 e-172 d {\bar d} {\bar e}-531 {\bar b} e {\bar e}+324 {\bar d} e {\bar e}+90 b {\bar e}^2-120 d {\bar e}^2+320 e {\bar e}^2,\\
b e^2 {\bar e},b {\bar d}+2 d {\bar d}+2 {\bar d} e-b {\bar e}-e {\bar e},-2 d^2 {\bar d}-4 d {\bar d} e+5 b d {\bar e}-2 b e {\bar e}+3 d e {\bar e},\\
-256 d {\bar d}^2+270 {\bar b}^2 e+20 {\bar d}^2 e+180 b {\bar b} {\bar e}+88 d {\bar d} {\bar e}+459 {\bar b} e {\bar e}-441 {\bar d} e {\bar e}+120 d {\bar e}^2-\\
260 e {\bar e}^2,46 d^2 {\bar d}+60 b {\bar b} e-28 d {\bar d} e+30 {\bar b} e^2-20 d^2 {\bar e}-24 b e {\bar e}-39 d e {\bar e}+20 e^2 {\bar e},\\
180 b {\bar b}^2+464 d {\bar d}^2+90 {\bar b}^2 e+20 {\bar d}^2 e+448 d {\bar d} {\bar e}-171 {\bar b} e {\bar e}-81 {\bar d} e {\bar e}+120 d {\bar e}^2-80 e {\bar e}^2,\\
-394 d^2 {\bar d}+172 d {\bar d} e-90 {\bar b} e^2+120 {\bar d} e^2+270 b^2 {\bar e}+80 d^2 {\bar e}+531 b e {\bar e}-324 d e {\bar e}-320 e^2 {\bar e},\\
540 b^2 {\bar b}+1786 d^2 {\bar d}+1172 d {\bar d} e+90 {\bar b} e^2+240 {\bar d} e^2-20 d^2 {\bar e}-1044 b e {\bar e}+81 d e {\bar e}+80 e^2 {\bar e},\\
96 d {\bar d}^2-90 {\bar b}^2 e-20 {\bar d}^2 e-48 d {\bar d} {\bar e}+20 {\bar a} e {\bar e}-149 {\bar b} e {\bar e}+161 {\bar d} e {\bar e}-40 d {\bar e}^2+100 e {\bar e}^2,\\
718 d^2 {\bar d}+716 d {\bar d} e+120 {\bar a} e^2+330 {\bar b} e^2+360 {\bar d} e^2-260 d^2 {\bar e}-612 b e {\bar e}-387 d e {\bar e}+410 e^2 {\bar e},\\
108 d {\bar d}^2-90 {\bar b}^2 e+40 {\bar a} {\bar d} e+20 {\bar d}^2 e-64 d {\bar d} {\bar e}-157 {\bar b} e {\bar e}+163 {\bar d} e {\bar e}-40 d {\bar e}^2+80 e {\bar e}^2,\\
{\bar a} d-2 d {\bar d}+2 {\bar b} e-2 d {\bar e}+e {\bar e},-16 d {\bar d}^2+4 {\bar a} {\bar b} e+26 {\bar b}^2 e+4 {\bar d}^2 e+16 d {\bar d} {\bar e}+\\
45 {\bar b} e {\bar e}-37 {\bar d} e {\bar e}+8 d {\bar e}^2-20 e {\bar e}^2,2 {\bar a} b+4 b {\bar b}-6 d {\bar d}-4 {\bar d} e+3 b {\bar e}+e {\bar e},\\
-247 d {\bar d}^2+20 {\bar a}^2 e+190 {\bar b}^2 e+20 {\bar d}^2 e+86 d {\bar d} {\bar e}+333 {\bar b} e {\bar e}-357 {\bar d} e {\bar e}+80 d {\bar e}^2-215 e {\bar e}^2,\\
-22 b {\bar b}+36 d {\bar d}+4 {\bar a} e-6 {\bar b} e+12 {\bar d} e+4 a {\bar e}-6 b {\bar e}+12 d {\bar e}+5 e {\bar e},a {\bar d}-2 d {\bar d}-2 {\bar d} e+2 b {\bar e}+e {\bar e},\\\left.
2 a {\bar b}+4 b {\bar b}-6 d {\bar d}+3 {\bar b} e-4 d {\bar e}+e {\bar e},a {\bar a}+2 b {\bar b}-2 d {\bar d}-e {\bar e}\right\}.
\end{multline}
Although the above formulas look very complicated, some of the polynomials in the ideal generated by \eqref{equnitarity1}-\eqref{equnitarity66} are of a very simple form. In particular, we obtain the corresponding equations $e\bar e=0$ and $d^3\bar d=0$ which clearly imply $e=0$ and $d=0$ if we recall the interpretation of $\bar e$ and $\bar d$ as complex conjugates of $e$ and $d$, resp. A substitution of $\left\{e\rightarrow 0,\bar e\rightarrow 0,d\rightarrow 0,\bar d\rightarrow 0\right\}$ in \eqref{eqGroebnerMax1} yields
\begin{equation}
\label{eqGroebnerMax2}
\left\{180 b {\bar b}^2,540 b^2 {\bar b},2 {\bar a} b+4 b {\bar b},-22 b {\bar b},2 a {\bar b}+4 b {\bar b},a {\bar a}+2 b {\bar b}\right\},
\end{equation}
where we removed all the zero polynomials. Again, because of the appearance of the polynomial $180 b {\bar b}^2$, a solution must have $b=0$ and $\bar b =0$. When this is substituted to \eqref{eqGroebnerMax2}, we obtain a single nonzero polynomial $a\bar a$, which in turn applies $a=0$. In summary, the only solution to the initial set of equations satisfying the constraint that $a,b,d,e$ and $\bar a,\bar b,\bar d,\bar e$ are complex conjugate is the zero matrix. Since it is clearly not unitary, we conclude that there exist no unitary matrices of the form \eqref{eqsolutionAij1}. This is equivalent to say that there are \textbf{no maximally entangled states} in the orthogonal complement of the UPB given by the formulas  \eqref{eqsymmetricUPB} and \eqref{eqvvectors}.

The example discussed above, although rather elegant mathematically, may seem unsatisfactory from the point of view of quantum information science. A natural question to ask is whether there exist Unextendible Product Bases in the $3\times 3$ case which admit a maximally entangled vector in their orthogonal complement. It turns out that the method presented above is powerful enough to give a affirmative answer to the question.Consider the following one-parameter family of Unextendible  Product Bases in $\setC^3\otimes\setC^3$.
\begin{equation}\label{eqUPBz}
\left[\begin{array}{ccccc}
\phi_1&\phi_2&\phi_3&\phi_4&\phi_5\\
\hline
\psi_1&\psi_2&\psi_3&\psi_4&\psi_5
\end{array}
\right]=
\left[
\begin{array}{ccccc}
1&0&z&1&0\\
0&1&0&1&1\\
0&0&1&-\bar z&1\\
\hline
0&1&1&1&0\\
1&0&0&1&1\\
1&1&0&-1&0
\end{array}
\right],
\end{equation} 
where $z\in\setC\setminus\left\{0\right\}$ is arbitrary  and we used the notation $\frac{\phi}{\psi}$ for a product vector $\phi\otimes\psi$, which is practical here. Note that the vectors are not normalized.

Our aim in the following is to decide whether the orthogonal complement to the UPB in \eqref{eqUPBz} contains a maximally entangled state for some  $z\in\setC\setminus\left\{0\right\}$ or not. Orthogonality conditions to the subspace spanned by $\phi_1\otimes\psi_1,\ldots,\phi_5\otimes\psi_5$ are a set of linear equations and can be solved explicitly. The result is
\begin{equation}\label{eqorthosolutions}
\left[A_{ij}\right]=\left[
\begin{array}{ccc}
 a & b & -b \\
 d & e & -d \\
 -a z & -e z & -\frac{b}{\bar z}-\frac{d}{\bar z}-\frac{a+b+d+e+a z \bar z+e z \bar z}{\bar z}
\end{array}
\right],\end{equation}
where $A_{ij}$ denote the coordinates of a vector $\sum_{i,j}A_{ij}e_i\otimes e_j\in\setC^3\otimes\setC^3$.

Taking the conjugate transpose of \eqref{eqorthosolutions} and multiplying by the matrix \eqref{eqorthosolutions} itself, we get the conditions for $\left[A_{ij}\right]$ to be unitary, or $\sum_{i,j}A_{ij}e_i\otimes e_j$ to be maximally entangled, in the form
\begin{eqnarray}
 d \bar d+a (\bar a+\bar a z \bar z)=1, \nonumber\\
  \bar d e+\bar a (b+e z \bar z)=0,\nonumber\\
   -d \bar d+a (\bar a+\bar a z \bar z)+\bar a (b+2 d+e+e z \bar z)=0, \nonumber\\
    d \bar e+a (\bar b+\bar e z \bar z)=0,\nonumber\\
     b \bar b+e \bar e (1+z \bar z)=1 \nonumber\\
     -b (\bar b-2 \bar e)+\bar e (a+d+e+a z \bar z+e z \bar z)=0, \nonumber\\
     -d \bar d+a (\bar a+\bar b+2 \bar d+\bar e+\bar a z \bar z+\bar e z \bar z)=0,\nonumber\\
      -b \bar b+e (\bar a+2 \bar b+\bar d+\bar e+\bar a z \bar z+\bar e z \bar z)=0,\nonumber\\
       b \bar b+d \bar d+\frac{(a+2 b+2 d+e+a z \bar z+e z \bar z) (\bar a+2 \bar b+2 \bar d+\bar e+\bar a z \bar z+\bar e z \bar z)}{z \bar z}=1.\nonumber
\end{eqnarray}
The last expression is not a polynomial in  $a,\bar a,b,\bar b,d,\bar d,e,\bar e,z,\bar z$, but can be easily transformed to
\begin{equation}
 b \bar b{z \bar z}+d \bar d{z \bar z}+{(a+2 b+2 d+e+a z \bar z+e z \bar z) (\bar a+2 \bar b+2 \bar d+\bar e+\bar a z \bar z+\bar e z \bar z)}={z \bar z},\nonumber
\end{equation}
if we remember that by assumption $z\neq 0$. Thus we get a set of nine polynomial equations in  the variables $a,\bar a,b,\bar b,d,\bar d,e,\bar e,z,\bar z$, equivalent to the condition that $\sum_{i,j}A_{ij}e_i\otimes e_j$ be maximally entangled. Next, we calculate the corresponding Groebner basis in $\setC\left[a,\bar a,b,\bar b,d,\bar d,e,\bar e,z,\bar z\right]$, where we take $a,\bar a,b,\bar b,d,\bar d,e,\bar e,z,\bar z$ as a set of \emph{independent} variables. In other words, we forget that numbers like $a$ and $\bar a$ are conjugate, and try to impose this condition only after a calculation of the Groebner basis. The basis reads
\begin{multline}\label{eqGrobnerBasisMaxEnt2}
\left\{-12 x-5 x^2+2 x^3,-27+27 e \bar e-12 x+18 e \bar e x+4 x^2,\right.\\
-14580-5832 e \bar e+729 e^2 \bar e^2+13122 e^3 \bar e^3+6561 e^4 \bar e^4-11688 x+3872 x^2,\\
7776 \bar d+8424 \bar e-405 e \bar e^2-4374 e^2 \bar e^3-3645 e^3 \bar e^4+3792 \bar e x-496 \bar e x^2,\\
7776 d+8424 e-405 e^2 \bar e-4374 e^3 \bar e^2-3645 e^4 \bar e^3+3792 e x-496 e x^2,\\
3888 \bar b-3078 \bar e-567 e \bar e^2+2187 e^2 \bar e^3+1458 e^3 \bar e^4+ 978 \bar e x+436 \bar e x^2,\\
3888 b-3078 e-567 e^2 \bar e+2187 e^3 \bar e^2+1458 e^4 \bar e^3+978 e x+436 e x^2,\\
1296 \bar a+4536 \bar e-1053 e \bar e^2-1458 e^2 \bar e^3-729 e^3 \bar e^4+1968 \bar e x-848 \bar e x^2,\\ 
\left.1296 a+4536 e-1053 e^2 \bar e-1458 e^3 \bar e^2-729 e^4 \bar e^3+1968 e x-848 e x^2\right\},
\end{multline}
where we introduced the notation $x:=z\bar z$. From the first polynomial we see that a solution can exist only if $x\in\left\{-\frac{3}{2},0,4\right\}$. But $x=-\frac{3}{2}$ is impossible according to our definition of $x$, and $x=0$ is excluded by the assumption $z\neq 0$. Therefore maximally entangled vectors in the orthogonal complement to the UPB in \eqref{eqUPBz} can exist only if $x=\left|z\right|^2=4$, thus if $\left|z\right|=2$. If we substitute $x=4$ in \eqref{eqGrobnerBasisMaxEnt2} and calculate the Groebner basis of the resulting polynomials in $\setC\left[a,\bar a,b,\bar b,d,\bar d,e,\bar e\right]$, we get
\begin{equation}
\{-1+9 e \bar e,\bar d+2 \bar e,d+2 e,\bar b+2 \bar e,b+2 e,\bar a-\bar e,a-e\}.
\end{equation}
Clearly, all the polynomials can be made zero by choosing $e=\frac{e^{i\phi}}{3}$, $d=-2e$, $b=-2e$ and $a=e$. Therefore there exist a single, up to an overall phase factor, maximally entangled state in the orthogonal complement of the UPB in \eqref{eqUPBz}. It has the following coordinate matrix
\begin{equation}\label{eqmaxentinlin}
\left[A_{ij}\right]=\frac{1}{3}
 \left[
\begin{array}{ccc}
 {1} & -{2} & {2} \\
 -{2} & {1} & {2} \\
 -{2} & -{2} & -{1}
\end{array}
\right].
\end{equation}
In summary, we have shown that the UBP given in equation \eqref{eqUPBz} does not admit a maximally entangled vector in its complement, with the \textbf{only exception of $\bf\left|z\right|=2$}. When $\left|z\right|=2$, there is a maximally entangled vector in the orthogonal complement of the UPB \eqref{eqUPBz}, which has a coordinate matrix of the form \eqref{eqmaxentinlin}.

\section{Mutually Unbiased Bases}\label{secMUBs}
As we already explained in Section \ref{seccrypto}, a generalization of quantum cryptography protocols such as BB84 to multidimensional quantum systems \cite{CBKG2001} relies on the notion of \textit{mutually unbiased bases}, MUBs for short. Two orthonormal bases $\left\{\psi_i\right\}_{i=1}^d$, $\left\{\phi_j\right\}_{j=1}^d$ of $\setC^d$ are said to be (mutually) unbiased if and only if
\begin{equation}\label{eqdefunbiased}
\left|\innerpr{\psi_i}{\phi_j}\right|^2=\frac{1}{d}
\end{equation}
holds for all $i,j\in\left\{1,2,\ldots,d\right\}$. The importance of the above relation for quantum state determination has been first pointed out by Ivanović \cite{Ivanovic81}, who also proved the existence of $d+1$ mutually unbiased bases in $\setC^d$ when $d$ is a prime number. Later, Wootters and Fields \cite{WF89} showed that there are at most $d+1$ mutually unbiased bases in $\setC^d$ and gave examples of full sets of MUBs when $d$ is a prime power. Moreover, they demonstrated that quantum state determination using a full set of MUBs is optimal in the sense of giving minimum statistical errors. A broader view of the known constructions of MUBs was then provided in \cite{BBRV02}, where the authors related MUBs to classes of pairwise orthogonal and commuting unitary matrices. The main efforts in the field concentrated on proving or disproving the existence of maximal sets of MUBs in non-prime power dimensions \cite{Grassl04,A05,BH07,BBELTZ07,BW08,BW09}, which still remains an open problem. However, on the basis of the extensive searches presented in \cite{BW09}, the existence of four, let alone seven, MUBs in $\setC^6$ is almost certainly excluded.
   
In the present section, we briefly describe how the authors of \cite{BW09} used the technique of Groebner bases to provide a large number of examples where a set of two MUBs in $\setC^6$ cannot be extended to a set of four. We first need to introduce the notion of \textit{complex Hadamard matrices} (cf. e.g. \cite{TZ06}). Such matrices are by definition unitaries $H$ with the property that $\left|H_{ij}\right|=1/\sqrt{d}$ for all matrix elements $H_{ij}$ of $H$. It is easy to notice that for any Hadamard matrix $H$, the canonical basis $\left\{e_1,\ldots,e_d\right\}$ and the columns of $H$, $\left\{H^1,\ldots,H^d\right\}$, constitute a pair of MUBs. Any other basis mutually unbiased with respect to these two, must also consist of columns of some complex Hadamard matrix. Now, the strategy applied in \cite{BW09} was the following: 
\begin{enumerate}[1)]
\item Select a known Hadamard matrix $H$ in $\setC^d$, the vast majority of which can be found in the online catalogue \cite{catalogue},
\item Parametrize a general, up to a phase, vector in $\setC^d$ mutually unbiased with respect to $\left\{e_1,\ldots,e_d\right\}$, as 
\begin{equation}\label{eqparam6}
v=\frac{1}{\sqrt{d}}\left[\begin{array}{ccccc}1&x_1+i y_1&x_2+i y_2&\cdots&x_{d-1}+i y_{d-1}\end{array}\right]^T,
\end{equation}
with $T$ denoting matrix transposition, $x_i,y_i\in\setR$ and $x_i^2+y_i^2=1$,
\item Multiply $v$ from the left by $\conj{H}$ and equate the squared moduluses of the coordinates of the resulting vector to $1/d$. This gives a set of polynomial equations in $x_i$ and $y_i$, equivalent to the unbiasedness condition
\begin{equation}\label{eqHunbiased}
\left|\innerpr{H^k}{v}\right|^2=\frac{1}{d}
\end{equation} 
for $k=1,2,\ldots,d$,
\item Solve the resulting equations, together with $x_i^2+y_i^2=1$, $i=1,2,\ldots,d-1$, for $x_i,y_i\in\setR$. In this way, the set of all vectors in $\setC^d$ unbiased with respect to $\left\{e_1,\ldots,e_d\right\}$ and $\left\{H^1,\ldots,H^d\right\}$ is obtained,
\item Check whether it is possible to arrange the resulting vectors in $d$-tuples that consist MUBs, and how many such MUBs can be obtained altogether, including $\left\{e_1,\ldots,e_d\right\}$ and $\left\{H^1,\ldots,H^d\right\}$.
\end{enumerate}

The authors of \cite{BW09} worked mainly with the case $d=6$, but the above steps can be followed also when the MUB problem in dimension different from $6$ is considered. For purely expository purposes, in order not to resort to numerical solutions necessary in $\setC^6$, we shall now explain how the above method yields a complete set of MUBs in $\setC^3$, which is well-known to exist \cite{Ivanovic81}. This is in contrast with the main findings of \cite{BW09} in dimension $6$, where the authors conclude that for no single one of the nearly 6000 Hadamard matrices $H$ they studied, there exists more than three mutually unbiased bases including $\left\{e_1,\ldots,e_6\right\}$ and $\left\{H^1,\ldots,H^6\right\}$.

Up to some simple invariances (for more details, cf. e.g. \cite{TZ06}), there only exists one Hadamard matrix when $d=3$, which is the Fourier matrix
\begin{equation}
F_3=\frac{1}{\sqrt{3}}\left[\begin{array}{ccc}1&1&1\\1&\omega&\omega^2\\1&\omega^2&\omega\end{array}\right],
\end{equation}
where $\omega=e^{2\pi i/3}$. The corresponding unbiasedness conditions $\left|\innerpr{F^j_3}{v}\right|^2=1/3$ read
\begin{equation}\left\{
\begin{array}{r}
-2+2 x_1+x_1^2+2 x_2+2 x_1 x_2+x_2^2+y_1^2+2 y_1 y_2+y_2^2=0,\\
-2-x_1+x_1^2-x_2-x_1 x_2+x_2^2+\sqrt{3} y_1-\sqrt{3} x_2 y_1+y_1^2+\\-\sqrt{3} y_2+\sqrt{3} x_1 y_2-y_1 y_2+y_2^2=0,\\
-2-x_1+x_1^2-x_2-x_1 x_2+x_2^2-\sqrt{3} y_1+\sqrt{3} x_2 y_1+y_1^2+\\+\sqrt{3} y_2-\sqrt{3} x_1 y_2-y_1 y_2+y_2^2=0.
\end{array}\right.
\end{equation}
If we take into account the relations $x_1^2+y_1^2=1$ and $x_2^2+y_2^2=1$, the above equations take the form
\begin{equation}\left\{
\begin{array}{r}
x_1+x_2+x_1x_2+y_1y_2=0,\\
x_1+x_2+x_1x_2+y_1y_2-\sqrt{3}\left(y_1-x_2y_1-y_2+x_1y_2\right)=0,\\
x_1+x_2+x_1x_2+y_1y_2+\sqrt{3}\left(y_1-x_2y_1-y_2+x_1y_2\right)=0.
\end{array}\right.
\end{equation}
which are clearly equivalent to the following system of equations
\begin{equation}\left\{
\begin{array}{r}
x_1+x_2+x_1x_2+y_1y_2=0,\\
y_1-x_2y_1-y_2+x_1y_2=0.
\end{array}\right.
\end{equation}
Taking the above equalities together with $x_1^2+y_1^2=1$ and $x_2^2+y_2^2=1$, we get the following set of polynomial equations
\begin{equation}\label{equnbias0000}
\left\{
\begin{array}{r}
x_1^2+y_1^2=0,\\ x_2^2+y_2^2=0,\\
x_1+x_2+x_1x_2+y_1y_2=0,\\
y_1-x_2y_1-y_2+x_1y_2=0.
\end{array}\right.
\end{equation}
As we know from Chapter \ref{chVarIdGroeb}, a possible approach to solving equations like \eqref{equnbias0000} is by the calculation of the corresponding Groebner basis, preferably with respect to the lexicographic order. The result is
\begin{equation}
\left\{-3 y_2+4 y_2^3,-1+x_2+2 y_2^2,-3+4 y_1^2-4 y_1 y_2+4 y_2^2,1+2 x_1+4 y_1 y_2-4 y_2^2\right\}.
\end{equation}
By equating the above polynomials to zero, we get a system of equations equivalent to \eqref{equnbias0000}, which can be readily solved by backward substitution. The corresponding solutions $\left(x_1,y_1,x_2,y_2\right)$ are the elements of the following set
\begin{multline}
\left\{\left(-\frac{1}{2},-\frac{\sqrt{3}}{2}, 1, 0\right), \left(-\frac{1}{2}, 
  \frac{\sqrt{3}}{2}, 1, 0\right), \left(1, 0, -\frac{1}{2},
   -\frac{\sqrt{3}}{2}\right),\right.\\\left.\left(-\frac{1}{2},-\frac{\sqrt{3}}{2}, 
  -\frac{1}{2}, -\frac{\sqrt{3}}{2}\right), \left(1, 0, -\frac{1}{2}, 
   \frac{\sqrt{3}}{2}\right), \left(-\frac{1}{2}, \frac{\sqrt{3}}{2}, -\frac{1}{2}, 
  \frac{\sqrt{3}}{2}\right)\right\}.
\end{multline}
Hence, we get six vectors in total that are unbiased with respect to $\left\{e_1,e_2,e_3\right\}$ and $\left\{F_3^1,F_3^2,F_3^3\right\}$. Explicitly, we have the following vectors
\begin{eqnarray}
v_1=\frac{1}{\sqrt{3}}\left[\begin{array}{c}1\\\omega^2\\1\end{array}\right],&v_2=\frac{1}{\sqrt{3}}\left[\begin{array}{c}1\\\omega\\ 1\end{array}\right],&v_3=\frac{1}{\sqrt{3}}\left[\begin{array}{c}1\\1\\\omega^2\end{array}\right],\\
v_4=\frac{1}{\sqrt{3}}\left[\begin{array}{c}1\\\omega^2\\\omega^2\end{array}\right],&v_5=\frac{1}{\sqrt{3}}\left[\begin{array}{c}1\\1\\\omega\end{array}\right],&v_6=\frac{1}{\sqrt{3}}\left[\begin{array}{c}1\\\omega\\\omega\end{array}\right].
\end{eqnarray}
By examining the inner products $\innerpr{v_i}{v_j}$ for $i,j\in\left\{1,\ldots,6\right\}$, we get to the conclusion that $\left\{v_1,v_3,v_6\right\}$ and $\left\{v_2,v_4,v_5\right\}$ are two orthonormal bases, mutually unbiased with respect to each other. Consequently, all the four bases $\left\{e_1,e_2,e_3\right\}$, $\left\{F_3^1,F_3^2,F_3^3\right\}$, $\left\{v_1,v_3,v_6\right\}$ and $\left\{v_2,v_4,v_5\right\}$ together constitute a full set of MUBs in $\setC^3$.

The authors of \cite{BW09} followed the same path of reasoning as in the example described above, however they worked with $d=6$ and needed to resort to numerical methods in order to obtain the solutions of the respective polynomial equations. In their case, it turned out not to be possible to find four mutually unbiased bases, starting from $\left\{e_1,\ldots,e_6\right\}$ and $\left\{H^1,\ldots,H^6\right\}$ for any $6\times 6$ Hadamard matrix $H$ they examined. 
\section{Symmetric Informationally Complete vectors}\label{secSICs}
When discussing the applications of polynomial equations in quantum information science, it seems impossible to neglect the prominent role they play in the research on so-called \textit{Symmetric Informationally Complete Positive Operator Valued Measures}, or SIC-POVMs for short. A SIC-POVM in $\setC^d$ is by definition a set of normalized vectors $\left\{\psi_i\right\}_{i=1}^{d^2}$ with the property
\begin{equation}\label{eqdefSICs}
\left|\innerpr{\psi_i}{\psi_j}\right|^2=\frac{1}{d+1}
\end{equation}
for all $i,j\in\left\{1,\ldots,d\right\}$, $i\neq j$. The first generally recognized work on SIC POVMs, although it uses a different name for the same object, is by Zauner \cite{Zauner}, who famously states a (stronger) version of the following conjecture
\begin{conjecture}[Zauner]\label{conjZauner}
For every dimension $d\geqslant 2$ there exists a SIC-POVM whose elements are the orbit of a vector $\psi_0$ under the Heisenberg group, which consists of elements $\omega^aX^bZ^c$, where $a,b,c\in\left\{0,1,\ldots,d-1\right\}$, $\omega=e^{2\pi i/d}$ and
\begin{equation}
X=\left[\begin{array}{ccccc}
0&0&\cdots&0&1\\
1&0&\cdots&0&0\\
0&1&\cdots&0&0\\
\vdots&\vdots&\ddots&\vdots&\vdots\\
0&0&\cdots&1&0
\end{array}\right],\quad Z=\left[\begin{array}{ccccc}1&0&\cdots&0&0\\
0&\omega&\cdots&0&0\\
\vdots&\vdots&\ddots&\vdots&\vdots\\
0&0&\cdots&\omega^{d-1}&0\\
0&0&\cdots&0&\omega^d\end{array}\right]
\end{equation}
\qed
\end{conjecture}
The term SIC-POVM was coined by the authors of \cite{RBSC04}, and SIC-POVMs became popular as a consequence of the usefulness for quantum state tomography \cite{Scott06} and the rich mathematical structure they have \cite{Appleby05,Zhu2010,Zhu2010b}. In the following, we outline how they relate to polynomial equations and we solve a very simplified example where it is possible to find explicit algebraic expressions for vectors constituting a SIC-POVM. Note, however, that the example we solve is only a subcase of the general solution for $d=3$, provided in \cite{RBSC04}. 

An approach to searching SIC-POVM vectors successfully applied in papers like \cite{Grassl04} and \cite{ScottGrassl} starts from writing \eqref{eqdefSICs} as a set of polynomial equations for the real and imaginary parts of the coefficients of the vectors $\psi_i$. Such equations may contain a reasonably small number of variables only if the vectors $\psi_i$ are not assumed to be independent. The standard way to follow consists in assuming that the requested SIC-POVM satisfies the Zauner conjecture, therefore all the $\psi_i$, $i=1,2,\ldots,d$, are determined by a single vector $\psi_0$, called the \textit{fiducial}. In this way, the number of real variables in the polynomial equations is reduced to $2d-1$, where the factor $-1$ comes from the fact that we can take the first coefficient of $\psi_0$ to be real without affecting the whole SIC-POVM construction as described by Conjecture \ref{conjZauner}. Further simplifications also follow from the full statement of the Zauner conjecture, which involves elements of the Clifford group, cf. e.g. \cite{Appleby05}. 

In the following, we show how to find an exemplary SIC-POVM in dimension $3$ by solving a set of polynomial equations, based on the ideas sketched above. Since a general form of SIC-POVMs in $\setC^3$ is known \cite{RBSC04}, our discussion should be perceived as a purely expository one, aimed at giving a rough picture of what happens in real science applications. 

In our very simplified example, we are looking for $9$ normalized vectors $\left\{\psi_1,\ldots,\psi_9\right\}\subset\setC^3$ that would satisfy $\left|\innerpr{\psi_i}{\psi_j}\right|^2=1/4$ for all $i\neq j$. As explained above, the related polynomial equations become much easier to tackle if a form of Conjecture \ref{conjZauner} is assumed to hold. Hence, instead of looking for general sets of nine vectors $\psi_i\in\setC^3$, we assume that $\left\{\psi_1,\ldots,\psi_9\right\}$ is equal to the set $\left\{X^nZ^m\psi_0\,\vline\, m,n\in\left\{0,1,2\right\}\right\}$, where $\psi_0=\left[\begin{array}{ccc}a&x+iy&z+it\end{array}\right]^T$ is a normalized fiducial vector in $\setC^3$, $a,x,y,z,t\in\setR$, and
\begin{equation}\label{HWdim3}
X=\left[\begin{array}{ccc}0&0&1\\1&0&0\\0&1&0\end{array}\right],\quad Z=\left[\begin{array}{ccc}1&0&0\\0&-\frac{1-\sqrt{3}i}{2}&0\\0&0&-\frac{1+\sqrt{3}i}{2}\end{array}\right].
\end{equation}
Under the above assumption, the equations $\left|\innerpr{\psi_i}{\psi_j}\right|^2=1/4$ become equivalent to $\left|\innerpr{\psi_0}{X^nZ^m\psi_0}\right|^2=1/4$ for all such that $n\neq 0$ or $m\neq 0$. The latter imply another set of equalities, $\left|\innerpr{\psi_0}{X^nZ^m\psi_0}\right|^2=\left|\innerpr{\psi_0}{X^{n'}Z^{m'}\psi_0}\right|^2$, where $n\neq 0$ or $m\neq 0$ and $n'\neq 0$ or $m'\neq 0$. In our case, the last set of equations take the explicit form  
\begin{align*}\scriptstyle
\left(t x+a (-t+y)-y z\right)^2-\frac{3}{4} \left(t^2-x^2-y^2+z^2\right)^2-\frac{1}{4} \left(-2 a^2+t^2+x^2+y^2+z^2\right)^2+\left(t y+x z+a (x+z)\right)^2&\scriptstyle=0,\\\scriptstyle
3 \left(t^2-x^2-y^2+z^2\right)^2+\left(-2 a^2+t^2+x^2+y^2+z^2\right)^2-\left(t \left(-\sqrt{3} x+y\right)+\left(x+\sqrt{3} y\right) z+a \left(-\sqrt{3} t-2 x+z\right)\right)^2+&\\\scriptstyle
\left.-\left(-t \left(x+\sqrt{3} y\right)+\left(-\sqrt{3} x+y\right) z+a \left(t+2 y+\sqrt{3} z\right)\right)^2\right)&\scriptstyle=0,\\\scriptstyle
2 \sqrt{3} a \left(-t^2 y+y z (-2 x+z)+t \left(x^2-y^2-2 x z\right)+a (t x+y z)\right)&\scriptstyle=0,\\\scriptstyle
\left(t x+\sqrt{3} t y+a \left(2 t-\sqrt{3} x+y\right)+\sqrt{3} x z-y z\right)^2+&\\\scriptstyle
\left(t \left(-\sqrt{3} x+y\right)+a \left(x+\sqrt{3} y-2 z\right)+\left(x+\sqrt{3} y\right) z\right)^2-3 \left(t^2-x^2-y^2+z^2\right)^2-\left(-2 a^2+t^2+x^2+y^2+z^2\right)^2&\scriptstyle=0,\\\scriptstyle
3 \left(t^2-x^2-y^2+z^2\right)^2+\left(-2 a^2+t^2+x^2+y^2+z^2\right)^2-\left(t \left(\sqrt{3} x+y\right)+\left(x-\sqrt{3} y\right) z+a \left(\sqrt{3} t-2 x+z\right)\right)^2&\\\scriptstyle
-\left(-t x+\sqrt{3} t y+\sqrt{3} x z+y z+a \left(t+2 y-\sqrt{3} z\right)\right)^2&\scriptstyle=0,\\\scriptstyle
-2 \sqrt{3} a \left(-t^2 y+y z (-2 x+z)+t \left(x^2-y^2-2 x z\right)+a (t x+y z)\right)&\scriptstyle=0.
\end{align*}
In order to find exemplary SIC-POVMs in $\setC^3$, we add the normalization condition $a^2+x^2+y^2+z^2+t^2=1$ for $\psi_0$ to the above equations, and then we try the substitution $a\rightarrow\sqrt{2/3}$. Note that the value $\sqrt{2/3}$ has not been selected at random, and the specific choice of $a$ makes the subsequent calculations rather straightforward. However, any other number of modulus $<1$ can be tried as well, and would typically lead to a few fiducial vectors or to the conclusion that no suitable fiducials exist. For $a$ not an algebraic number, numerical methods would be required to find the solutions or to show they are non-existent.

Once we substituted $\sqrt{2/3}$ for $a$, we are left with a set of seven polynomial equations for $x,y,z,t$, some of which are redundant. Calculation of the corresponding Groebner basis with respect to the lexicographic order gives the following result,
\begin{multline}\label{eqGroebnerSIC}
\left\{t^2-24 t^4+192 t^6-512 t^8,3 \sqrt{2} t^2-32 \sqrt{2} t^4+64 \sqrt{2} t^6-8 \sqrt{3} t^2 z+64 \sqrt{3} t^4 z,\right.\\
-2 \sqrt{3}+93 \sqrt{3} t^2-1008 \sqrt{3} t^4+2880 \sqrt{3} t^6-12 \sqrt{2} z+144 \sqrt{2} t^2 z-12 \sqrt{3} z^2,\\
6 \sqrt{3} t-167 \sqrt{3} t^3+1296 \sqrt{3} t^5-3264 \sqrt{3} t^7+6 \sqrt{3} y-105 \sqrt{3} t^2 y+1008 \sqrt{3} t^4 y+\\
-2880 \sqrt{3} t^6 y-16 \sqrt{3} y^3+12 \sqrt{2} t z-48 \sqrt{2} t^3 z+12 \sqrt{2} y z-144 \sqrt{2} t^2 y z,\\
8 \sqrt{3}-105 \sqrt{3} t^2+1008 \sqrt{3} t^4-2880 \sqrt{3} t^6+12 \sqrt{2} x+900 \sqrt{3} t^3 y-15552 \sqrt{3} t^5 y+\\
+48384 \sqrt{3} t^7 y-1260 \sqrt{3} t^2 y^2+12096 \sqrt{3} t^4 y^2-34560 \sqrt{3} t^6 y^2+12 \sqrt{2} z-144 \sqrt{2} t^2 z+\\
\left.-216 \sqrt{2} t y z+3456 \sqrt{2} t^3 y z+144 \sqrt{2} y^2 z-1728 \sqrt{2} t^2 y^2 z\right\}.
\end{multline}
By equating the above polynomials to $0$, we get a system of equations that can readily be solved by backward substitution, provided that one can find solutions to the equation $t^2-24 t^4+192 t^6-512 t^8=0$. Fortunately, this problem can easily be solved explicitly, as $t^2-24 t^4+192 t^6-512 t^8=-t^2(8t^2-1)^3$. Thus, we have $t=0$, $t=-1/2\sqrt{2}$ and $t=1/2\sqrt{2}$ as the possible values for the $t$ coordinate. Substitution of any of these values to \eqref{eqGroebnerSIC} gives us a set of polynomials in $x,y,z$ of maximum degree $2$, whose common zeros are easy to find. Altogether, there are nine solutions $\left(x,y,z,t\right)$ to the equations $\left|\innerpr{\psi_0}{X^nZ^m\psi_0}\right|^2=\left|\innerpr{\psi_0}{X^{n'}Z^{m'}\psi_0}\right|^2$ and $\left|\psi_0\right|^2=1$, corresponding to nine fiducials. We give a list in Table \ref{tabelaSIC}. Note that a vector $\psi_0$ is a fiducial if and only if $\conj{\psi_0}$ also has this property. This is a general fact, which can be confirmed with Table \ref{tabelaSIC}. Thus we have completed the task of finding a set of three-dimensional SIC-POVM vectors with help of the Groebner basis method.

\begin{table}
\renewcommand{\arraystretch}{1.35}
\centering
\begin{tabular}{|c|c|c|c|}
\hline
$t$&$z$&$y$&$x$\\
\hline
$0$&$-\frac{1}{\sqrt{6}}$&$0$&$-\frac{1}{\sqrt{6}}$\\
$0$&$-\frac{1}{\sqrt{6}}$&$-\frac{1}{2\sqrt{2}}$&$\frac{1}{2\sqrt{6}}$\\
$0$&$-\frac{1}{\sqrt{6}}$&$\frac{1}{2\sqrt{2}}$&$\frac{1}{2\sqrt{6}}$\\
$-\frac{1}{2\sqrt{2}}$&$\frac{1}{2\sqrt{6}}$&$0$&$-\frac{1}{\sqrt{6}}$\\
$-\frac{1}{2\sqrt{2}}$&$\frac{1}{2\sqrt{6}}$&$-\frac{1}{2\sqrt{2}}$&$\frac{1}{2\sqrt{6}}$\\
$-\frac{1}{2\sqrt{2}}$&$\frac{1}{2\sqrt{6}}$&$-\frac{1}{2\sqrt{2}}$&$\frac{1}{2\sqrt{6}}$\\
$\frac{1}{2\sqrt{2}}$&$\frac{1}{2\sqrt{6}}$&$0$&$-\frac{1}{\sqrt{6}}$\\
$\frac{1}{2\sqrt{2}}$&$\frac{1}{2\sqrt{6}}$&$-\frac{1}{2\sqrt{2}}$&$\frac{1}{2\sqrt{6}}$\\
$\frac{1}{2\sqrt{2}}$&$\frac{1}{2\sqrt{6}}$&$\frac{1}{2\sqrt{2}}$&$\frac{1}{2\sqrt{6}}$\\
\hline
\end{tabular}
\caption{Solutions of the Heisenberg group-invariant SIC-POVM equations in case of $\setC^3$.\label{tabelaSIC}}
\end{table}

\chapter{A structure theorem for PPT bound entangled states of lowest rank}\label{chPPT3x3}

The aim of the present section is to present the main result of the thesis, concerning positive-partial-transpose non-separable states of rank $4$ in $3\times 3$ systems. As indicated in \cite{LS2010}, they all seem to be possible to locally transform to projections onto the orthogonal complement to a subspace spanned by an orthogonal Unextendible Product Basis \cite{Bennett99}. Thus, there is strong numerical evidence that they are all locally equivalent to bound entangled states of the form discussed in \cite{Bennett99}. In the following, we analytically prove that this is actually the case. Note that according to the results of \cite{HLVC2000}, four is the minimal rank for an entangled PPT state. Therefore it is correct to say that our theorem concerns non-separable PPT states of lowest rank. Very shortly after our paper \cite{S2011} was available as a preprint on arXiv, Chen and {\DJ}oković \cite{CD2011} presented an alternative proof of the theorem. The research reported here was conducted independently of \cite{CD2011}, and the author had no prior knowledge about the manuscript by the other authors. An important related work by Chen and {\DJ}oković is also \cite{CD2011distill}. 

Before we start with the proof, it will be useful to introduce the concept of general Unextendible Product Bases, discussed in more detail elsewhere \cite{S2011}.

\section{General Unextendible Product Bases}\label{secgUPB}
The most common definition of an Unextendible Product Basis (UPB), in accordance with \cite{Bennett99}, has already been phrased in Section \ref{secexamplesbound}.
Here, we start with a definition of a \emph{general} UPB.
\begin{definition}\label{defnonortUPB} Take $n,m\in\mathbbm{N}$. By a {\bf general Unextendible Product Basis}, or a gUPB for short, we mean a set  $\left\{\wektor{\phi_i\otimes\psi_i}\right\}_{i=1}^k$ of product vectors in $\mathbbm{C}^n\otimes\mathbbm{C}^m$, $0<k<mn$, such that there is no product vector in $\linspan{\left\{\wektor{\phi_i\otimes\psi_i}\right\}_{i=1}^k}^{\bot}$, the orthogonal complement to the linear span of $\left\{\wektor{\phi_i\otimes\psi_i}\right\}_{i=1}^k$.
\end{definition}
In other words, a gUPB is a set of product vectors $\left\{\wektor{\phi_i\otimes\psi_i}\right\}_{i=1}^k$ such that there is no product vector orthogonal to all of them. Note that we do not require the vectors to be linearly independent, and this choice is somewhat arbitrary. Yet another way of phrasing the above definition is that the orthogonal complement to a gUPB is a Completely Entangled Subspace \cite{Partha04,Bhat04}, or CES for short, cf. Section \ref{secces}.

\begin{remark}The definition of a gUPB can be trivially extended to a multipartite setting.\end{remark}

We know that gUPBs do exist. Any UPB consisting of orthogonal vectors is an example (concrete UPBs can be found e.g. in  \cite{DiVicenzo04}). We also know that for some spaces, no UPB consisting of orthogonal vectors can exist. For example, it has been noticed as early as in \cite{Bennett99} that $2\times n$ systems do not admit an orthogonal UPB, and a more general discussion of existence questions for orthogonal UPBs has been included in \cite{AL2001}. In the following, we show that gUPBs are much more common than the usual UPBs, and give a characterization of gUPBs of minimal number of elements. 

First, let us answer a question about the minimum number of elements in a gUPB in $\mathbbm{C}^n\otimes\mathbbm{C}^m$. 

\begin{proposition}\label{propgUPBnonexistence}A set of vectors $\left\{\wektor{\phi_i\otimes\psi_i}\right\}_{i=1}^k\subset\mathbbm{C}^n\otimes\mathbbm{C}^m$ consisting of $k<m+n-1$ elements is not a generalized UPB.
\begin{proof} There exists a vector $\wektor{f}\in\mathbbm{C}^n$ orthogonal to all the vectors $\wektor{\phi_i}$ with $i=1,\ldots,n-1$. Moreover, there exists a $\wektor{g}\in\mathbbm{C}^m$ orthogonal to the vectors $\wektor{\psi_j}$ with $j=n,\ldots,k$ (because $k-n<m$). The product vector $\wektor{f\otimes g}$ is orthogonal to all $\wektor{\phi_i\otimes\psi_i}$ for $i=1,\ldots, k$.
\end{proof}
\end{proposition}

\begin{proposition}\label{propgUPBsuffnecessary}
A set of vectors $\left\{\wektor{\phi_i\otimes\psi_i}\right\}_{i=1}^{m+n-1}\subset\mathbbm{C}^n\otimes\mathbbm{C}^m$ is a gUPB if and only if any $n$-tuple of vectors in $\left\{\wektor{\phi_i}\right\}_{i=1}^{m+n-1}$ consists of linearly independent vectors, the same as any $m$-tuple of vectors in $\left\{\wektor{\psi_j}\right\}_{j=1}^{m+n-1}$.
\begin{proof}
In order to prove necessity, assume that an $n$-tuple of vectors $\left\{\wektor{\phi_{i_l}}\right\}_{l=1}^n$ is linearly dependent. Therefore there exists a vector $\wektor{f}\in\mathbbm{C}^n$ orthogonal to all of them. Vectors of the form $\wektor{f\otimes g}\in\mathbbm{C}^n\otimes\mathbbm{C}^m$ with an arbitrary $\wektor{g}\in\mathbbm{C}^m$ are orthogonal to all the vectors $\left\{\wektor{\phi_{i_l}\otimes\psi_{i_l}}\right\}_{l=1}^n\subset\mathbbm{C}^n\otimes\mathbbm{C}^m$. Obviously, $\wektor{g}$ can be chosen in such a way that $\wektor{f\otimes g}$ is orthogonal to the remaining $m-1$ elements of $\mathbbm{C}^m$ (because $m-1<m=\dim\mathbbm{C}^m$). For a sufficiency proof, assume that $\wektor{f\otimes g}$ is orthogonal to $\wektor{\phi_i\otimes\psi_i}$ for $i=1,\ldots,m+n-1$. The vector $\wektor{f}$ can be orthogonal to at most $n-1$ of $\wektor{\phi_i}$'s, whereas $\wektor{g}$ cannot be orthogonal to more than $m-1$ $\wektor{\psi_i}$'s (remember the linear independence of $n$-tuples and $m$-tuples, respectively). This gives a maximum of $\left(n-1\right)+\left(m-1\right)$ vectors in $\left\{\wektor{\phi_i\otimes\psi_i}\right\}_{i=1}^{m+n-1}$ orthogonal to $\wektor{f\otimes g}$. Therefore $\wektor{f\otimes g}$ cannot be orthogonal to all the $\wektor{\phi_i\otimes\psi_i}$'s, the set $\left\{\wektor{\phi_i\otimes\psi_i}\right\}_{i=1}^{m+n-1}$ is a gUPB.
\end{proof}
\end{proposition}

It is natural to ask for a generalization of Proposition \ref{propgUPBsuffnecessary} for sets of product vectors consisting of more than $m+n-1$ elements. We have the following

\begin{proposition}\label{progUPBnecsuffnonminimal}
A set of vectors $\left\{\wektor{\phi_i\otimes\psi_i}\right\}_{i=1}^N\subset\mathbbm{C}^n\otimes\mathbbm{C}^m$ with $N\geqslant m+n-1$ is a gUBP if and only if for any $\mathcal{N},\mathcal{M}\subset\mathbbm{N}$ such that $\mathcal{N}\cap\mathcal{M}=\emptyset$ and $\mathcal{N}\cup\mathcal{M}=\left\{1,2,\ldots,N\right\}$, at least one of the sets of vectors $\left\{\phi_i\right\}_{i\in\mathcal{N}}$ and $\left\{\psi_j\right\}_{j\in\mathcal{M}}$ spans the entire corresponding vector space (\/$\mathbbm{C}^n$ or $\mathbbm{C}^m$, resp.).
\begin{proof}Let us first prove necessity. Assume that the vectors $\left\{\phi_i\otimes\psi_i\right\}_{i=1}^N$ constitute a gUPB and choose some $\mathcal{N},\mathcal{M}\subset\mathbbm{N}$ as in the statement of the proposition. If neither of the sets $\left\{\phi_i\right\}_{i\in\mathcal{N}}$ and $\left\{\psi_j\right\}_{j\in\mathcal{M}}$ spans the respective vector space, there exist $\wektor{f}\in\mathbbm{C}^n$ and $\wektor{g}\in\mathbbm{C}^m$ such that $\innerpr{f}{\phi_i}=0$ and $\innerpr{g}{\psi_j}=0$ for all $i\in\mathcal{N}$ and $j\in\mathcal{M}$. Because of the condition $\mathcal{N}\cup\mathcal{M}=\left\{1,2,\ldots,N\right\}$, we clearly have $\innerpr{f\otimes g}{\phi_i\otimes\psi_i}=0$ for $i=1,2,\ldots,N$. This contradicts the fact that the vectors $\phi_i\otimes\psi_i$ constitute a gUPB. In order to show sufficiency, assume that $f\otimes g\in\mathbbm{C}^n\otimes\mathbbm{C}^m$ is such that $\innerpr{f\otimes g}{\phi_i\otimes\psi_i}=0$ for all $i=1,2,\ldots,N$. Define the set of indices $\mathcal{N}_f:=\left\{i\,\vline\innerpr{f}{\phi_i}=0\right\}$ and $\mathcal{M}_g:=\left\{j\,\vline\innerpr{g}{\psi_j}=0\right\}$. Clearly, we must have $\mathcal{N}_f\cup\mathcal{M}_g=\left\{1,2,\ldots,N\right\}$. Thus it is possible to choose $\mathcal{N}\subset\mathcal{N}_f$ and $\mathcal{M}\subset\mathcal{M}_g$ such that $\mathcal{N}\cup\mathcal{M}=\left\{1,2,\ldots,N\right\}$ and $\mathcal{N}\cap\mathcal{M}=\emptyset$. By the very definition of $\mathcal{N}_f$ and $\mathcal{M}_g$, we have $\innerpr{f}{\phi_i}=0$ for all $i\in\mathcal{N}$ and $\innerpr{g}{\psi_j}$ for  $j\in\mathcal{M}$. But according to the assumptions of the theorem, this is only possible if $f$ or $g$ is equal to zero. Thus $\left\{\phi_i\otimes\psi_i\right\}_{i=1}^N$ is a gUPB.
\end{proof}
\end{proposition}

Certain characterizations of gUPBs were earlier obtained in \cite{Pittenger03}, but the above results, rather surprisingly, seem to appear for the first time in our work \cite{S2011}. They can also easily be generalized to a multipartite setting.

\begin{proposition}\label{progUPBnecsuffmulti}
A set of vectors $\left\{\wektor{\phi^1_i\otimes\phi^2_i\otimes\ldots\otimes\phi^l_i}\right\}_{i=1}^N\subset\mathbbm{C}^{n_1}\otimes\mathbbm{C}^{n_2}\otimes\ldots\mathbbm\otimes{C}^{n_l}$ with $N\geqslant\sum_{i=1}^ln_i+l-1$ is a gUBP if and only if for any $\mathcal{N}_1,\ldots,\mathcal{N}_l\subset\mathbbm{N}$ such that $\mathcal{N}_i\cap\mathcal{N}_j=\emptyset$ for all $i\neq j$ and $\bigcup_{i=1}^l\mathcal{N}_i=\left\{1,2,\ldots,N\right\}$, at least one of the sets of vectors $\left\{\phi^i_j\right\}_{j\in\mathcal{N}_i}$, $i=1,2,\ldots,l$, spans the entire corresponding vector space $\mathbbm{C}^{n_i}$.
\begin{proof}
Follows the same lines as the proof of Proposition \ref{progUPBnecsuffnonminimal} and will be omitted here.
\end{proof}
\end{proposition}

\section{The concept of local equivalence}\label{secequivalence}

Before we present the proof of the main result of the thesis (Theorem \ref{maintheorem}), we also need to introduce the concept of local equivalence. Numerous questions of physical or mathematical origin need the proper identification of a symmetry group relevant to the problem in order to simplify the solution, or even to find it at all. The same is the case for the result we are going to obtain below. For PPT states, a natural group of symmetries should be of a product form, $\rho\mapsto\left(A\otimes B\right)^{\ast}\rho\left(A\otimes B\right)$, because all such transformations preserve the property of being PPT. In physical terms, they preserve the splitting of a composite system into subsystems, which is a highly desirable property. The remaining question is, what group should $A$ and $B$ belong to. When the amount of entanglement between the two subsystems is in question, a natural choice is $A$ and $B$ in the Unitary or Special Unitary group. Such transformations cannot change any measure of entanglement. However, if the aim is to classify PPT states with respect to the property of being extreme, being an edge state \cite{LKHC2001}, or the number and dimensionalities spanned by the product vectors in their kernels or ranges, $A$ and $B$ should most naturally belong to the General Linear or Special Linear group. There is no essential difference between the two latter choices. Since we are not interested in positive scaling factors in front of the states, we choose to work with the Special Linear group. This was also the approach so successfully used by the authors of \cite{LMO2006,LS2010,HHMS2011}. We should remark that, while a PPT state is transformed according to $\rho\mapsto\left(A\otimes B\right)^{\ast}\rho\left(A\otimes B\right)$, the product vectors in its kernel and its range undergo the following transformation, $\phi\otimes\psi\mapsto\left(A^{-1}\otimes B^{-1}\right)\phi\otimes\psi$. Conversely, a transformation $\phi\otimes\psi\mapsto\left(A\otimes B\right)\phi\otimes\psi$ forces a change of $\rho$ into $\left(A^{-1}\otimes B^{-1}\right)^{\ast}\rho\left(A^{-1}\otimes B^{-1}\right)$. It is these kind of transformations we will have in mind when we talk about ``local equivalence'', ``local $\SL$ equivalence'' or ``$\SLtt$ equivalence'' in the following sections.

Any similar terms, even not listed here, will also refer to precisely the same situation. Nevertheless, when product vectors in the kernel of a PPT state $\rho$ are in question, it is more convenient to look at them as rays, points in the projective space. In such case, it is also more accurate to refer to the projectivisation of the group $\SLtt$, namely to $\PSLtt$, with $\textnormal{PSL}$ referring to the Projective Special Linear group.  In simple words, we may multiply vectors $\left\{\phi_1\otimes\psi_1,\phi_2\otimes\psi_2,\ldots\right\}\subset\kernel{\rho}$ by arbitrary individual factors, and they will remain elements of the kernel of $\rho$. We may also transform them by a $\SLtt$ transformation. 
All in all, we have a group of transformations that is most properly described as $\PSLtt$. Note that the use of this term is motivated mainly by the possibility to avoid excessive comments about constant factors in front of the product vectors in $\kernel{\rho}$. We are legitimate to use the previously introduced name ``local equivalence'' also for the $\PSLtt$ transformations we just described because constant multiplicative factors in front of vectors in $\kernel{\rho}$ are completely irrelevant to $\rho$ itself.

The ultimate reason for using equivalences of the form described above will be the simplicity of our main result, a characterization theorem that we are going to obtain in Section \ref{secmainresult}. The equivalence classes under $\SLtt$ of non-separable PPT states of rank $4$ in $3\times 3$ systems turn out to be parametrized by just four real, positive numbers. Moreover, each class has a representative which is a projection onto a Completely Entangled Subspace complementary to a $3\times 3$ orthogonal UBP. This is quite a striking result, for which strong numerical evidence was provided by Leinaas \textit{et al.} in \cite{LS2010} and later supported by certain analytical results of \cite{HHMS2011}.

\section{Outline of the proof}\label{secoutline}
The proof of our main result is not excessively complicated, but it needs a considerable amount of work. It also consists of a number of steps which do not seem easy to merge. In order to simplify the reading, we start with a list of building blocks. We will elaborate on each of them in the following sections.
\begin{enumerate}
\item The kernel of a rank four PPT state $\rho$ must intersect the Segre variety $\Sigma_{2,2}$ in a transverse way. In particular, according to the Bezout's Theorem, the intersection must consist of exactly six points.
\item The product vectors in the kernel of a rank $4$ PPT state in the $3\times 3$ case span the kernel. As a result, they must be a generalized UPB. There cannot exist a product vector orthogonal to all of them.
\item A generalized UPB in the $3\times 3$ case is locally equivalent to an orthogonal one if and only if certain invariants $s_1,\ldots,s_4$, introduced by Leinaas \textit{et al.} in \cite{LS2010}, are all positive, possibly after the vectors are permuted.
\item A generalized UPB in a $3\times 3$ system is contained in a kernel of some rank four PPT state if and only if the corresponding values of $s_1,\ldots,s_4$ are positive, possibly after the vectors are permuted. Moreover, in such case the PPT state in question is uniquely determined.
\end{enumerate}
The final conclusion from the facts mentioned in items $1.-4.$ is that the only non-separable PPT states of rank $4$ in $3\times 3$ systems are local transforms of projections onto orthogonal complements of orthogonal pentagram-type Unextendible Product Bases \cite{Bennett99,DiVicenzo04}.

\section{Product vectors in the kernel of a PPT state}\label{secprodvecPPT}
The present section elaborates on item $1.$ in the list given above and on related topics. Let us start with an elementary fact.
\begin{lemma}\label{lemmaconj}
A product vector $\wektor{\phi\otimes\psi}$ is in the kernel of a PPT state $\rho$ if and only if the partially conjugated states $\wektor{\phi^{\ast}\otimes\psi}$ and $\wektor{\phi\otimes\psi^{\ast}}$ are in the kernels of $\rho^{T_1}$ and $\rho^{T_2}$, respectively.
\begin{proof}
It follows from the equality between the expressions $\innerpr{\phi\otimes\psi}{\rho\left(\phi\otimes\psi\right)}$, $\innerpr{\phi\otimes\psi^{\ast}}{\rho^{T_2}\left(\phi\otimes\psi^{\ast}\right)}$ and $\innerpr{\phi^{\ast}\otimes\psi}{\rho^{T_1}\left(\phi^{\ast}\otimes\psi\right)}$, by the positivity of $\rho$, $\rho^{T_1}$ and $\rho^{T_2}$.
\end{proof}
\end{lemma}
In the above lemma, we did not assume anything about the dimensionality of the system. Neither we do it in the following.
\begin{lemma}\label{lemmaprodPPT}
Assume that a product vector $\wektor{\phi\otimes\psi}$ is in the kernel of a PPT state $\rho$. In such case
\begin{equation}\label{condPPTproduct}
\innerpr{\phi'\otimes\psi}{\rho\left(\phi\otimes\psi'\right)}=\innerpr{\phi\otimes\psi'}{\rho\left(\phi'\otimes\psi\right)}=0\quad\forall_{\phi',\psi'}.
\end{equation}
\begin{proof} Since $\rho\left(\wektor{\phi\otimes\psi}\right)=0$, we know from Lemma \ref{lemmaconj} that $\rho^{T_1}\left(\wektor{\phi^{\ast}\otimes\psi}\right)=0$, which obviously implies $\innerpr{\phi'^{\ast}\otimes\psi'}{\rho^{T_1}\left(\phi^{\ast}\otimes\psi\right)}=\innerpr{\phi\otimes\psi'}{\rho\left(\phi'\otimes\psi\right)}=0$. This is the first equality in \eqref{condPPTproduct}. The second one can be obtained in a similar way.
\end{proof}
\end{lemma}
Let us denote by $r\left(\rho\right)$, $R\left(\rho\right)$ and $\ker\rho$ the rank, the range and the kernel of $\rho$. Our next lemma applies specifically to the $3\times n$ case and concerns so-called \textit{edge states}. For more information about this topic, consult \cite{LKHC2001}. In short, edge PPT states are PPT states $\rho$ that {do \emph{not} admit a product vector $\phi\otimes\xi\in\ker\rho$ such that $\conj{\phi}\otimes\xi\in\ker\rho^{T_1}$}.  
\begin{lemma}\label{lemma3x3twoproducts}
Assume that both $\phi\otimes\psi$ and $\wektor{\phi'\otimes\psi}$, with $\phi$, $\phi'$ in $\mathbbm{C}^3$ and $\psi$ in $\mathbbm{C}^n$, $\phi\neq\phi'$, belong to the kernel of a PPT state $\rho$, acting on $\mathbbm{C}^3\otimes\mathbbm{C}^3$. The state $\rho$ is either supported on a $3\times\left(n-1\right)$ or smaller subspace, or it can be written as $\rho=\rho'+\lambda\proj{\phi''\otimes\xi}
$ for some $\lambda>0$, $\xi\in\mathbbm{C}^n$, $\phi''\in\mathbbm{C}^3$ linearly independent of $\phi$ and $\phi'$, and a PPT state $\rho'$, supported on a $3\times\left(n-1\right)$ or smaller subspace. Moreover, the rank $\rank{\rho'}=\rank{\rho}-1$ and $\rank{\left(\rho'\right)^{T_1}}=\rank{\rho^{T_1}}-1$.
In a situation when the reduction is possible, the state $\rho$ is not an edge PPT state. In particular, $\rho$ is not an extreme and non-separable PPT state. 
\begin{proof}
Let us assume that the product states $\phi'\otimes\psi$ and $\wektor{\phi_2\otimes\psi}$ belong to the kernel of $\rho$. Let $A$ be an $\SLt$ transformation that brings $e_1,e_2\subset\mathbbm{C}^3$ to $\phi_1$ and $\phi_2$. A little inspection shows that Lemmas~1 and 2 of \cite{HLVC2000} can be applied to $\tilde\rho:=\left(A\otimes\mathbbm{1}\right)^{\ast}\rho\left(A\otimes\mathbbm{1}\right)$. Consequently, we see that either $\rho$ is supported on a $3\times\left(n-1\right)$ or smaller space, or the assertion of Lemma~2 of \cite{HLVC2000} tells us that $\tilde\rho=\rho_1+\lambda\proj{e_3\otimes\xi}
$ for some $\xi\in\mathbbm{C}^n$, and moreover, $\rho_1$ is a PPT state supported on a $3\times\left(n-1\right)$ or smaller subspace, with $\rank{\rho_1}=\rank{\rho}-1$ and $\rank{\rho_1^{T_1}}=\rank{\rho^{T_1}}-1$. We have $\rho=\left(A^{-1}\otimes\mathbbm{1}\right)^{\ast}\tilde\rho\left(A^{-1}\otimes\mathbbm{1}\right)=\rho'+\lambda\proj{\phi'''\otimes\xi}
$, where  $\phi'''=A^{-1}e_3$ and $\rho'=\left(A^{-1}\otimes\mathbbm{1}\right)^{\ast}\rho_1\left(A^{-1}\otimes\mathbbm{1}\right)$. The states $\rho'$ and $\left(\rho'\right)^{T_1}$ still have their ranks reduced by one with respect to the ranks of $\rho$ and $\rho^{T_1}$, respectively. The subspaces on which they are supported are of the same type as for $\rho_1$, hence $3\times\left(n-1\right)$ or smaller. The statement that $\rho$ is not an edge state simply follows because $\proj{\phi'''\otimes\xi}
$ is in $\range{\rho}$ while its partial conjugation is in $\range{\rho^{T_1}}$.
\end{proof}
\end{lemma}
The following result reduces a more general case to the situation considered above. However, this time we assume $n=3$.
\begin{lemma}\label{lemmaSR2}
Let $\wektor{\phi\otimes\psi}\in\mathbbm{C}^3\otimes\mathbbm{C}^3$ be an element of a PPT state $\rho$, acting on $\mathbbm{C}^3\otimes\mathbbm{C}^3$. There cannot exist a nonzero vector $\wektor{\phi\otimes\psi'}+\wektor{\phi'\otimes\psi}$, with $\phi'\neq\phi$ or $\psi'\neq\psi$, in the kernel of $\rho$, unless one of the following is true: i) $\rho=\rho'+\lambda\proj{\zeta\otimes\xi}
$ for $\lambda>0$, $\xi,\zeta\in\mathbbm{C}^3$ and $\rho'$ a PPT state supported on a $2\times 3$ or smaller subspace with $\rank{\rho'}=\rank{\rho}-1$ and  $\rank{\left(\rho'\right)^{T_1}}=\rank{\left(\rho\right)^{T_1}}-1$ or ii) $\rho$ is supported on a $2\times 3$ or smaller subspace itself.
\begin{proof}
Assume that there is a state of the form $\wektor{\phi\otimes\psi'}+\wektor{\phi'\otimes\psi}$ in the kernel of $\rho$. This is equivalent to saying that $\innerpr{\phi\otimes\psi'+\phi'\otimes\psi}{\rho\left(\phi\otimes\psi'+\phi'\otimes\psi\right)}=0$. The inner product factorizes as
\begin{equation}\label{factorinnerpr}
\innerpr{\phi\otimes\psi'}{\rho\left(\phi\otimes\psi'\right)}+\innerpr{\phi\otimes\psi'}{\rho\left(\phi'\otimes\psi\right)}+\innerpr{\phi'\otimes\psi}{\rho\left(\phi\otimes\psi'\right)}+\innerpr{\phi'\otimes\psi}{\rho\left(\phi'\otimes\psi\right)}\nonumber
\end{equation}
The two factors in the middle vanish according to Lemma \ref{lemmaprodPPT}, while the two remaining factors are nonnegative as a consequence of positivity of $\rho$. Therefore, the only possibility for the above expression to vanish is when $\innerpr{\phi\otimes\psi'}{\rho\left(\phi\otimes\psi'\right)}=0$ and $\innerpr{\phi'\otimes\psi}{\rho\left(\phi'\otimes\psi\right)}=0$. This in turn means that $\rho\left(\wektor{\phi'\otimes\psi}\right)=0$ and $\rho\left(\wektor{\phi\otimes\psi'}\right)=0$. According to our assumptions, at least one of these equalities is nontrivial (i.e. $\phi'\neq 0$ or $\psi'\neq 0$). Lemma \ref{lemma3x3twoproducts} can be applied, and Lemma \ref{lemmaSR2} follows directly.
\end{proof}
\end{lemma}
The importance of Lemma \ref{lemmaSR2} is evident if we realize that the tangent space to the Segre variety, or to the set of product states at a point $\wektor{\phi\otimes\psi}$, consists precisely of the vectors  of the form considered above. We have
\begin{lemma}\label{lemmatangentSegre}
Elements of the tangent space to the Segre variety, or to the set of product vectors at a point $\wektor{\phi\otimes\psi}$, are of the form
\begin{equation}\label{formtangent}
\wektor{\phi\otimes\psi'}+\wektor{\phi'\otimes\psi},
\end{equation} 
with $\psi'$ and $\phi'$ arbitrary.
\begin{proof} A heuristic proof may consist in writing $\wektor{\left(\phi+\delta\phi\right)\otimes\left(\psi+\delta\psi\right)}\approx\wektor{\phi\otimes\psi}+\wektor{\delta\phi\otimes\psi}+\wektor{\phi\otimes\delta\psi}$, where the approximate equality holds to the first order. A more rigorous proof can be found in Example \ref{exSegretangent} of Section \ref{sectangent}, as well as in Example 14.16 of the textbook by Harris \cite{Harris}.
\end{proof}
\end{lemma}
Next, we specify the rank of $\rho$ to be $4$ and keep the assumption that $\rho$ acts on $\mathbbm{C}^3\otimes\mathbbm{C}^3$. Thus the kernel of $\rho$ is of dimension $5$, which is the smallest number $d$ such that a $d$-dimensional linear subspace must intersect the set of product vectors in $\mathbbm{C}^3\otimes\mathbbm{C}^3$, cf. e.g. \cite{Partha04}. Following Lemmas \ref{lemmaSR2} and \ref{lemmatangentSegre}, we can show that the nonempty intersection is generic in the sense of Bezout's theorem \cite[Theorem 18.3]{Harris} and thus it consists of exactly six points.
\begin{lemma}\label{lemmageneralintersection}
Let $\rho$ be a non-separable PPT state of rank $4$ acting on $\mathbbm{C}^3\otimes\mathbbm{C}^3$. The intersection between the respective Segre variety and the five-dimensional kernel of $\rho$ is transverse at every point. There are exactly six product vectors in the kernel of $\rho$.
\begin{proof}
Let us take $\wektor{\phi\otimes\psi}\in\kernel{\rho}$. As we mentioned above, such a vector exists \cite{Partha04,Cubitt07} by a dimensionality argument  for projective varieties. We easily see from Lemma \ref{lemmatangentSegre} that the dimension of the tangent space $\mathbbm{T}_{\wektor{\phi\otimes\psi}}\left(\Sigma_{2,2}\right)$ to the Segre variety at $\wektor{\phi\otimes\psi}$ is $5$, and thus the projective dimension is $4$. Being more explicit, any vector of the form $\wektor{\phi\otimes\psi'}+\wektor{\phi'\otimes\psi}$ can be written in the form $\lambda\phi\otimes\psi+\sum_{i=1}^2\xi_i\phi\otimes\psi_i+\sum_{j=1}^2\zeta_j\phi_j\otimes\psi$, where $\left\{\phi,\phi_1,\phi_2\right\}$ and $\left\{\psi,\psi_1,\psi_2\right\}$ are two sets of  three linearly independent vectors in $\mathbbm{C}^3$ and $x_i$, $\zeta_j$ are arbitrary complex coefficients. From Lemmas \ref{lemmaSR2} and \ref{lemmatangentSegre} we know that the only vector in the intersection of $\kernel{\rho}$ and $\mathbbm{T}_{\wektor{\phi\otimes\psi}}\left(\Sigma_{2,2}\right)$ is $\wektor{\phi\otimes\psi}$ itself. It must be so, because otherwise we could reduce the rank of $\rho$ by subtracting a projection onto a product state. After the reduction, we would be left with a PPT state of rank $3$. However, all such PPT states are separable according to \cite{HLVC2000}, and $\rho$ would have to be separable as well. The other option is that $\rho$ could be supported on $\mathbbm{C}^2\otimes\mathbbm{C}^3$ or even a less dimensional space itself. But then it is well-known that $\rho$ is separable as as consequence of being PPT \cite{HHH96}. In either case, we get a contradiction with the assumption that $\rho$ is non-separable. Therefore, $\wektor{\phi\otimes\psi}$ must be, up to a scalar factor, the only element of the intersection between $\kernel{\rho}$ and $\mathbbm{T}_{\wektor{\phi\otimes\psi}}\left(\Sigma_{2,2}\right)$. Consequently, the dimension of $\kernel{\rho}+\mathbbm{T}_{\wektor{\phi\otimes\psi}}\left(\Sigma_{2,2}\right)$ equals $5+5-1=9$, while its projective dimension is $9-1=8$. This equals the projective dimension of $\mathbbm{T}_{\wektor{\phi\otimes\psi}}\left(\mathbbm{P}^8\right)$, or simpler, the dimension of the complex projective space $\mathbbm{P}^8$. In other words, $\kernel{\rho}$ and $\mathbbm{T}_{\wektor{\phi\otimes\psi}}\left(\Sigma_{2,2}\right)$ span $\mathbbm{T}_{\wektor{\phi\otimes\psi}}\left(\mathbbm{P}^8\right)$, which is equivalent to saying that the intersection between $\kernel{\rho}$ and the Segre variety is transverse at $\wektor{\phi\otimes\psi}$. Since we did not make any additional assumptions about $\wektor{\phi\otimes\psi}$ apart from that it belongs to the intersection, we see that the intersection is transverse at every point. Therefore Bezout's theorem applies. The fact that there are exactly six points in the intersection follows because the degree of the Segre variety $\Sigma_{2,2}$ is six \cite[Example 18.15]{Harris}. 
\end{proof}
\end{lemma}
In summary, in the present section we have shown that a non-separable rank $4$ PPT state in a $3\times 3$ system must have exactly six vectors in its kernel. This is in full agreement with an assertion of \cite{LS2010n}. It should be noticed that, as a part of the proof of the above lemma, we have shown that non-separable PPT states of rank $4$ in $3\times 3$ systems  are edge states.  Thus, Lemmas \ref{lemma3x3twoproducts} and \ref{lemmaSR2} can be directly applied. We will frequently use them in the following section. 

\section{Product vectors in the kernel must be a gUPB}\label{secgUPBkernel}
We already know that the number of product vectors in the kernel of a rank $4$ non-separable PPT state of a $3\times 3$ system is six. In the following, we discuss more specific properties of the set of six product vectors. Let us denote them with $\phi_i\otimes\psi_i$, $i=1,2,\ldots,6$. It turns out that, up to local equivalence, five of them can always be brought to a special form, which has only four real parameters, the numbers $s_1,\ldots,s_4$ introduced in \cite{LS2010}. It then follows that the vectors $\phi_i\otimes\psi_i$, if they belong to the kernel of a rank $4$ PPT state, must span a five-dimensional subspace. Thus they span the kernel.

In order to prove our assertion, first observe that $\phi_i\neq\phi_j$ for $i\neq j$ (cf. Lemmas  \ref{lemmatangentSegre} \& \ref{lemmageneralintersection}), and thus they must span at least a two-dimensional subspace of $\mathbbm{C}^3$. Similarly for the $\psi$'s. Let us try to assume first that one of the sets $\left\{\phi_i\right\}_{i=1}^6$ and $\left\{\psi_j\right\}_{j=1}^6$ spans a two-dimensional subspace. We may, for example, try to assume this about $\left\{\phi_i\right\}_{i=1}^6$. Up to $\PSLt$ transformations, we have
\begin{equation}\label{twodimphi}
\left[\begin{array}{cccccc}\phi_1 & \phi_2 & \phi_3 & \phi_4 & \phi_5 & \phi_6\end{array}\right]=\left[\begin{array}{cccccc}
1&0&1&1&1&1\\
0&1&1&p&q&r\\
0&0&0&0&0&0
\end{array}
\right],
\end{equation}
where $p$, $q$, $r$ are all different and different from $0$ and $1$. When writing \eqref{twodimphi}, we used the fact that there is no pair of identical vectors in $\left\{\phi_i\right\}_{i=1}^6$.  Up to local transformations, we have $\psi_1=e_1$ and $\psi_2=e_2$. As for the other vectors $\psi$, we use the following notation, $\psi_i=\left[\begin{array}{ccc}\psi_{1i}&\psi_{2i}&\psi_{3i}\end{array}\right]$, $i=3,4,5,6$. We also introduce coordinates $\omega^{ij}$ for general vectors $\wektor{\omega}=\sum_{i,j}\omega^{ij}e_i\otimes e_j$ in $\mathbbm{C}^3\otimes\mathbbm{C}^3$. Our aim is to show that there exists a linear combination of the vectors $\phi_i\otimes\psi_i$ of the form  $\wektor{\phi\otimes\psi'}+\wektor{\phi'\otimes\psi}$ from Lemma \ref{lemmatangentSegre}. This will lead us to a contradiction and show that $\phi_i$'s cannot be as in \eqref{twodimphi}, and must span $\mathbbm{C}^3$. An analogous conclusion for $\psi$'s will be immediate.

Let us first observe that $\psi_{3i}\neq 0$ for all $i\in\left\{3,4,5,6\right\}$. Otherwise, we would have three product vectors supported on $\linspan{\left\{e_1,e_2\right\}}\otimes\linspan{\left\{e_1,e_2\right\}}$. Up to local equivalence, they would be of the form $e_1\otimes e_1$, $e_2\otimes e_2$ and $\left(e_1+e_2\right)\otimes\left(e_1+e_2\right)$. In such case, $e_1\otimes e_2+e_2\otimes e_1=\left(e_1+e_2\right)\otimes\left(e_1+e_2\right)-e_1\otimes e_1-e_2\otimes e_2$ would be in the kernel of $\rho$, which contradicts Lemma \ref{lemmaSR2}. Therefore we must have $\psi_{3i}\neq 0$ for all $i$. Let us choose $\alpha$ and $\beta$ so that $\alpha\psi_{33}+\beta p\psi_{34}=0$. The vector $\alpha\phi_3\otimes\psi_3+\beta\phi_4\otimes\phi_4$ has a vanishing coordinate $\omega^{23}=\alpha\psi_{33}+\beta p\psi_{34}$ and a non-vanishing coordinate $\omega^{13}=\alpha\psi_{33}+\beta\psi_{34}$ (remember that $p\neq 1$). By subtracting $e_2\otimes e_2$ times $\alpha\psi_{23}+\beta p\psi_{24}$, we can cancel the $\omega^{22}$ coordinate, and similarly cancel $\omega^{11}$ by subtracting $\alpha\psi_{13}+\beta \psi_{14}$ times $e_1\otimes e_1$. In the end, we see that a vector of the form $\omega^{21} e_2\otimes e_1+\omega^{12} e_1\otimes e_2+\omega^{13}e_1\otimes e_3$ with $\omega^{13}\neq 0$ is in the kernel of $\rho$. But this contradicts Lemma \ref{lemmaSR2}. In summary, the vectors $\phi_i$ cannot be brought to the form \eqref{twodimphi}, or in other words, they span $\mathbbm{C}^3$. Obviously, the same is true for the set $\left\{\psi_i\right\}_{i=1}^6$. A more careful analysis of the above argument leads to even stronger conclusions. Firstly, an assumption that there exist three vectors $\phi_i\otimes\psi_i$ supported on a $2\times 2$ dimensional subspace lead us to a contradiction. Therefore we have the following
\begin{lemma}\label{lemmanotwotwo}
Let $\left\{\phi_i\otimes\psi_i\right\}_{i=1}^6$ be the six product vectors in the kernel of a non-separable PPT state of rank $4$ in the $3\times 3$ case. For any triple $\left\{\phi_{i_j}\otimes\psi_{i_j}\right\}_{j=1}^3\subset\left\{\phi_i\otimes\psi_i\right\}_{i=1}^6$, at least one of the sets of vectors $\left\{\phi_{i_j}\right\}_{j=1}^3$ or $\left\{\psi_{i_j}\right\}_{j=1}^3$ spans $\mathbbm{C}^3$.
\end{lemma}
Moreover, we only needed four product vectors with $\phi$'s as in \eqref{twodimphi} to arrive at a contradiction with Lemma \ref{lemmaSR2}. As a consequence, we have
\begin{lemma}\label{lemmaquadruples}
For any quadruple $\left\{\phi_{i_j}\otimes\psi_{i_j}\right\}_{j=1}^4\subset\left\{\phi_i\otimes\psi_i\right\}_{i=1}^6$, both the sets of vectors $\left\{\phi_{i_j}\right\}_{j=1}^4$ and $\left\{\psi_{i_j}\right\}_{j=1}^4$ span $\mathbbm{C}^3$.
\end{lemma}
As an immediate consequence of Lemma \ref{lemmanotwotwo}, there exists a set of three linearly independent vectors in $\left\{\phi_i\right\}_{i=1}^6$. With no loss of generality, we may assume that $\left\{\phi_1,\phi_2,\phi_6\right\}$ is a linearly independent set. After a $\PSLt$ transformation, $\phi_1=e_1$, $\phi_2=e_2$ and $\phi_6=e_3$. There are in principle two possibilities concerning the remaining vectors $\phi_3$, $\phi_4$ and $\phi_5$. Either one of them is of the form $\left[\begin{array}{ccc}x&y&z\end{array}\right]$ with $xyz\neq 0$, or all of them have exactly one coordinate equal to zero. Two vanishing coordinates in a single vector cannot occur because there is no pair of identical vectors among $\phi_1,\ldots,\phi_6$. Moreover, according to Lemma \ref{lemmaquadruples}, the zeros must occur in different places in $\phi_3$, $\phi_4$ and $\phi_5$. Up to $\PSLt$ transformations and permuting the vectors, we may assume that $\phi_3=\left[\begin{array}{ccc}x&0&1\end{array}\right]$,  $\phi_4=\left[\begin{array}{ccc}0&1&z\end{array}\right]$, $\phi_5=\left[\begin{array}{ccc}1&y&0\end{array}\right]$ with $x$, $y$, $z$ all different from $0$. But then, write the coordinate matrix for $\left\{\phi_1,\phi_2,\phi_3,\phi_4\right\}$,
\begin{equation}\label{coordmatrix2345}
\left[\begin{array}{cccc}\phi_1 & \phi_2 & \phi_3 & \phi_4\end{array}\right]=\left[\begin{array}{cccc}
1&0&x&0\\
0&1&0&1\\
0&0&1&z
\end{array}
\right].
\end{equation} 
It is easy to check that all the $3\times 3$ minors in \eqref{coordmatrix2345} are non-vanishing. In other words, any triple of vectors in $\left\{\phi_1,\phi_2,\phi_3,\phi_4\right\}$ spans $\mathbbm{C}^3$. The corresponding vectors $\psi_1$, $\psi_2$, $\psi_3$ and $\psi_4$ may or may not have all triples linearly independent. It is not difficult to show that if all the triples span $\mathbbm{C}^3$, we can simultaneously, by using a $\PSLtt$ transformation, bring $\left\{\phi_1,\phi_2,\phi_3,\phi_4\right\}$ and $\left\{\psi_1,\psi_2,\psi_3,\psi_4\right\}$ to the form
\begin{equation}\label{formtwoindeptriples}
\left[\begin{array}{cccc}\phi_1 & \phi_2 & \phi_3 & \phi_4\end{array}\right]=\left[\begin{array}{cccc}\psi_1 & \psi_2 & \psi_3 & \psi_4\end{array}\right]=\left[\begin{array}{cccc}
1&0&0&1\\
0&1&0&1\\
0&0&1&1
\end{array}
\right].
\end{equation} 
By adding a fifth product vector, say $\phi_5\otimes\psi_5$, we get, up to local transformation and relabelling the vectors $\phi_i\otimes\psi_i$,
\begin{equation}\label{formtwoindeptriples2}
\left[\begin{array}{ccccc}\phi_1&\phi_2&\phi_3&\phi_4&\phi_5\\
\hline
\psi_1&\psi_2&\psi_3&\psi_4&\psi_5
\end{array}
\right]=\left[\begin{array}{ccccc}
1&0&0&1&1\\
0&1&0&1&p\\
0&0&1&1&q\\
\hline
1&0&0&1&1\\
0&1&0&1&r\\
0&0&1&1&s
\end{array}
\right],
\end{equation}
where $p,q,r,s$ are some complex numbers. We should remark that the possibility to have $1$ in the first coordinate of $\phi_5$ and $\psi_5$ follows because there must exist $i\in\left\{1,2,3\right\}$ such that $\phi_{i5}\psi_{i5}\neq 0$, where $\phi_{i5}$ and $\psi_{i5}$ denote the $i$-th coordinate of $\phi$ and $\psi$, respectively. Otherwise, $\phi_5$ or $\psi_5$ would have to be proportional to $e_i$ for some $i\in\left\{1,2,3\right\}$.
 
If not all triples in $\left\{\psi_1,\psi_2,\psi_3,\psi_4\right\}$ are linearly independent, it is still possible, according to Lemma~\ref{lemmaquadruples}, to find a linearly independent triple among them. Without loss of generality, we may assume that the triple is $\left\{\psi_1,\psi_2,\psi_3\right\}$. By an identical argument as for the $\phi$'s, we know that there is a vector $\psi_i$, $i\in\left\{5,6\right\}$ such that $\left\{\psi_1,\psi_2,\psi_3,\psi_i\right\}$ have all triples linearly independent. Without loss of generality, we may assume that $\phi_i=\phi_5$. This time, a local transformation and possible relabelling brings the product vectors $\phi_i\otimes\psi_i$ with $i=1,2,\ldots,5$ to the form
\begin{equation}\label{formtwoindeptriples3}
\left[\begin{array}{ccccc}\phi_1&\phi_2&\phi_3&\phi_4&\phi_5\\
\hline
\psi_1&\psi_2&\psi_3&\psi_4&\psi_5
\end{array}
\right]=\left[\begin{array}{ccccc}
1&0&0&1&1\\
0&1&0&1&p\\
0&0&1&1&q\\
\hline
1&0&0&1&1\\
0&1&0&r&1\\
0&0&1&s&1
\end{array}
\right].
\end{equation}

To make a final touch to this section, we need to show that product vectors of the form \eqref{formtwoindeptriples2} or \eqref{formtwoindeptriples3} are linearly independent if no two of them coincide, and thus they span the five-dimensional kernel of $\rho$. We will also show that they constitute a minimal gUPB, and that the parameters $p,q,r,s$ have to be real when the vectors are in the kernel of a PPT state.

Let us use $\left[\begin{array}{ccccccccc}\omega^{11}&\omega^{12}&\omega^{13}&\omega^{21}&\omega^{22}&\omega^{23}&\omega^{31}&\omega^{32}&\omega^{33}\end{array}\right]$ to denote vectors $\omega=\sum_{i,j}\omega^{ij}e_i\otimes e_j$ in $\mathbbm{C}^3\otimes\mathbbm{C}^3$. In the case \eqref{formtwoindeptriples2}, we have
\begin{equation}\label{sit1coordinates}
\left[\begin{array}{c}
\phi_1\otimes\psi_1\\
\phi_2\otimes\psi_2\\
\phi_3\otimes\psi_3\\
\phi_4\otimes\psi_4\\
\phi_5\otimes\psi_5\\
\end{array}
\right]=
\left[\begin{array}{ccccccccc}
1&0&0&0&0&0&0&0&0\\
0&0&0&0&1&0&0&0&0\\
0&0&0&0&0&0&0&0&1\\
1&1&1&1&1&1&1&1&1\\
1&r&s&p&pr&ps&q&qr&qs\end{array}
\right].
\end{equation}
In the case \eqref{formtwoindeptriples3}, the coordinates of the product vectors are the following,
\begin{equation}\label{sit2coordinates}
\left[\begin{array}{c}
\phi_1\otimes\psi_1\\
\phi_2\otimes\psi_2\\
\phi_3\otimes\psi_3\\
\phi_4\otimes\psi_4\\
\phi_5\otimes\psi_5\\
\end{array}
\right]=
\left[\begin{array}{ccccccccc}
1&0&0&0&0&0&0&0&0\\
0&0&0&0&1&0&0&0&0\\
0&0&0&0&0&0&0&0&1\\
1&r&s&1&r&s&1&r&s\\
1&1&1&p&p&p&q&q&q\end{array}
\right].
\end{equation}
It is an elementary exercise to check that the matrices on the right-hand side of \eqref{sit1coordinates} and \eqref{sit2coordinates} are of rank $5$ for all choices of $p,q,r,s$, with the only exception of $p=q=r=s=1$. But the last possibility is excluded because it implies $\phi_4\otimes\psi_4=\phi_5\otimes\psi_5$.

Next, we can show that the vectors $\phi_i\otimes\psi_i$ with $i=1,2,\ldots,5$, chosen as above, constitute a general Unextendible Product Basis.  In order to prove it, let us first show that the rank of $\rho^{T_1}$ has to be $4$. 
\begin{proposition}\label{propequalranks}
Let $\rho$ be a non-separable PPT state of rank $4$ acting on $\mathbbm{C}^3\otimes\mathbbm{C}^3$. The rank of the partially transposed state $\rho^{T_1}$ is also $4$.
\begin{proof}
If $\rho$ is non-separable, we know by the above argument that the product vectors $\left\{\phi_i\otimes\psi_i\right\}_{i=1}^6$ in the kernel of $\rho$ span a five-dimensional subspace, which is the kernel itself. Moreover, five of them are, up to local transformations, of the form \eqref{formtwoindeptriples2} or \eqref{formtwoindeptriples3}. But this implies that the corresponding product vectors in the kernel of $\rho^{T_1}$, which are $\phi_i^{\ast}\otimes\psi_i$ according to Lemma \ref{lemmaconj}, can also be brought to the form \eqref{formtwoindeptriples2} or \eqref{formtwoindeptriples3}. To be more explicit, if a local transformation $A\otimes B$ brings the vectors $\phi_i\otimes\psi_i$ with $i=1,2,\ldots,5$ to the form \eqref{formtwoindeptriples2} or \eqref{formtwoindeptriples3}, $A^{\ast}\otimes B$ does the same to the partial conjugations $\phi_i^{\ast}\otimes\psi_i$. The only difference is that $p$ and $q$ change into $p^{\ast}$ and $q^{\ast}$ in \eqref{formtwoindeptriples2} or \eqref{formtwoindeptriples3}. But this does not change the conclusion about the dimensionality of the subspace spanned by vectors of the form \eqref{formtwoindeptriples2} or \eqref{formtwoindeptriples3}. As a consequence, the product vectors in the kernel of $\rho^{T_1}$ span at least a five-dimensional subspace. Thus the kernel of $\rho^{T_1}$ is at least five-dimensional. If it had higher dimension, the rank of $\rho^{T_1}$ would be lower or equal $3$, which is, according to \cite{HLVC2000}, impossible for non-separable $\rho$. Therefore, the dimension of the kernel equals $5$, and the rank of $\rho^{T_1}$ is $4$.
\end{proof}
\end{proposition}
There exist  separable states $\rho$ of rank $4$ in $3\times 3$ systems that have the rank of $\rho^{T_1}$ different from $4$. However, our next proposition shows that if $\rho$ is supported on $\mathbbm{C}^3\otimes\mathbbm{C}^3$ and it cannot be written as $\rho'+\lambda\proj{\zeta\otimes\xi}
$ with $\lambda>0$ and $\rho'$ supported on a $2\times 2$ subspace, the rank of $\rho^{T_1}$ is also $4$ (cf. Figure 4 in \cite{LS2010n}, which we reproduce here as Table \ref{tab3N}).

\begin{table}[h!]
\centering
\scalebox{0.85}{%
\begin{tabular}{|cccccc|}
\hline
\multicolumn{1}{|c|}{\textbf{$\mathbf{(m,n)}$}} & \multicolumn{1}{c|}{$\mathbf{m^2 + n^2 - N^2}$} & \multicolumn{1}{c|}{\textbf{dim}\,{\boldmath$\cal F$}} & \multicolumn{1}{c|}{$\mathbf{(r_A,r_B)}$} & \multicolumn{1}{c|}{\textbf{\#pv [Im\,$\rho$]}} & \multicolumn{1}{c|}{\textbf{\#pv  [Ker\,$\rho$]}} 
\\
\hline\hline
(9,9) & 81 & 81 & (3,3) & $\infty$/9 & 0 \\ \hline
(9,8) & 64 & 64 & (3,3) & $\infty$/9 & 0 \\ \hline
(9,7) & 49 & 49 & (3,3) & $\infty$/9 & 0 \\ \hline
(8,8) & 47 & 47 & (3,3) & $\infty$/8 & 0 \\ \hline
(9,6) & 36 & 36 & (3,3) & $\infty$/9 & 0 \\ \hline 
(8,7) & 32 & 32 & (3,3) & $\infty$/8 & 0 \\ \hline
(8,6) &  19 & 19 & (3,3) & $\infty$/8 & 0 \\ \hline
(7,7) &  17 & 17 & (3,3) & $\infty$/7 & 0 \\ \hline
(8,5) &    8 &    8 & (3,3) & $\infty$/8 & 0 \\ \hline
(7,6) &    4 &    4 & (3,3) & $\infty$/7 & 0 \\ \hline
(7,5) &   -7 &    1 & (3,3) & $\infty$/7 & 0 \\ \hline
(6,6) &   -9 &    1 & (3,3) & $\infty$/6 & 0 \\ \hline
(6,5) & -20 &    1 & (3,3) & $\infty$/6 & 0 \\ \hline 
(5,5) & -31 &    1 & (3,3) & 6/5 & 0 \\ \hline
(4,4) & -49 &    1 & (3,3) & 0 & 6/5 \\ \hline
(3,3) & -63 &    3 & (3,3) & 3/3 & $\infty$/6 \\ \hline
(2,2) & -73 &    2 & (2,2) & 2/2 & $\infty$/7 \\ \hline
(1,1) & -79 &    1 & (1,1) & 1/1 & $\infty$/8 \\ \hline
\end{tabular}}
\caption{Numerical results for $3\times 3$ PPT states $\rho$. The numbers $m$ and $n$ denote the ranks of $\rho^{T_1}$ and $\rho^{T_2}$, resp. The number $N=9$ is the dimension of the space on which the states act. The number dim\,$\cal F$ is the dimension of the face of the cone of PPT states on which the given state lives. The symbols $r_A$ and $r_B$ denote the ranks of the partially traced states $\Tr_B\rho$ and $\Tr_A\rho$. The fifth and the sixth column list the number of product vectors in the image and the kernel of $\rho$.} \label{tab3N}
\end{table}

\begin{proposition}\label{propranksseparable}
Let $\rho$ be a separable state of rank $4$ supported on $\mathbbm{C}^3\otimes\mathbbm{C}^3$, which cannot be written as $\rho'+\lambda\proj{\zeta\otimes\xi}
$ with $\lambda>0$ and $\rho'$ supported on a $2\times 2$ subspace of $\mathbbm{C}^3\otimes\mathbbm{C}^3$. The rank of $\rho^{T_1}$ is also $4$.
\begin{proof} 
First, we should remark that $\rank{\rho^{T_1}}=\rank{\rho^{T_2}}$. This fact will be important for some parts of the proof, although never explicitly referred to. The main idea that we are going to use is that the argument preceding formulas \eqref{formtwoindeptriples2} and \eqref{formtwoindeptriples3} works for separable states as well, provided that they cannot be reduced according to Lemma \ref{lemmaSR2}. In other words, the argument works when the kernel of a PPT state in question does intersect the Segre variety in a transverse way, irrespectively of the state being entangled or not. Thus, if a reduction according to Lemma \ref{lemmaSR2} is not possible for a separable state $\rho$, we have vectors of the form \eqref{formtwoindeptriples2} or \eqref{formtwoindeptriples3} in $\kernel{\rho}$, and they span a five-dimensional space. This is also the dimensionality of the subspace spanned by their partial conjugates, which are in $\kernel{\rho^{T_1}}$. Therefore, the rank of $\rho^{T_1}$ is not bigger than $4$. If it was less than four, the intersection between $\kernel{\rho^{T_1}}$ and the Segre variety $\Sigma_{2,2}$ would be more than zero-dimensional, according to the Projective Dimension Theorem \cite[Theorem 7.2]{Hartshorne}. But this contradicts the fact that there are only a finite number of product vectors in $\kernel{\rho^{T_1}}$ (equal to $\phi^{\ast}\otimes\psi$ for all $\phi\otimes\psi\in\kernel{\rho}$). In summary, the rank of $\rho^{T_1}$ has to be $4$ when $\kernel{\rho}$ intersects the Segre variety transversely. If not, we know from Lemmas and \ref{lemmaSR2} and \ref{lemmatangentSegre}  that there are two options:
\begin{enumerate}[i)]
 \item it is possible to write $\rho$ as $\rho'+\lambda\proj{\zeta\otimes\xi}
$, where $\lambda$ and $\rho'$ is a rank $3$ PPT state supported on a $2\times 3$ or smaller subspace of $\mathbbm{C}^3\otimes\mathbbm{C}^3$, with $\rank{\rho'}=3$ and $\rank{\left(\rho'\right)^{T_1}}=d-1$, 
\item $\rho$ is supported on a $2\times 3$ or smaller subspace itself.
\end{enumerate}
 Option ii) is excluded because of the assumption of $\rho$ supported on $\mathbbm{C}^3\otimes\mathbbm{C}^3$. Our aim in the following will be to show that $\rank{\rho'}=\rank{\left(\rho'\right)^{T_1}}$ unless $\rho'$ is supported on a $2\times 2$ subspace, which is precisely the second possibility we allow in the proposition. First, observe that if $\rho'$ is supported on a $2\times 3$ subspace, we can use an analogue of Lemma \ref{lemmaSR2}. Either we have $\rho'=\rho''+\lambda'\proj{\zeta'\otimes\xi'}
$ where $\lambda'>0$ and $\rho''$ is supported on a $2\times 2$, $1\times 3$ or $1\times 2$ subspace, $\rank{\rho''}=2$ and $\rank{\left(\rho''\right)^{T_1}}=\rank{\left(\rho'\right)^{T_1}}-1$, or $\kernel{\rho'}$ intersects the respective Segre variety $\Sigma_{1,2}$ transversely.  In the latter case, by Bezout's Theorem the $3$-dimensional kernel of $\rho'$ has precisely three product vectors in it.  Actually, we can repeat the argument preceding Lemmas \ref{lemmanotwotwo} and \ref{lemmaquadruples} to conclude that the product vectors in $\kernel{\rho'}$ have to be locally equivalent to 
\begin{equation}\label{threeproductvectors}
\left[\begin{array}{ccc}
\phi_1&\phi_2&\phi_3\\
\hline
\psi_1&\psi_2&\psi_3
\end{array}
\right]=
\left[\begin{array}{ccc}
1&0&1\\
0&1&1\\
\hline
1&0&0\\
0&1&0\\
0&0&1
\end{array}
\right].
\end{equation}
Obviously, these vectors span the kernel. We see that there are, within the $2\times 3$ subspace, only three product vectors in $\range{\rho'}=\left(\kernel{\rho'}\right)^{\bot}$. They are locally equivalent to
\begin{equation}\label{anotherthree}
\left[\begin{array}{ccc}
\zeta_1&\zeta_2&\zeta_3\\
\hline
\xi_1&\xi_2&\xi_3
\end{array}
\right]=
\left[\begin{array}{ccc}
0&1&1\\
1&0&-1\\
\hline
1&0&0\\
0&1&0\\
0&0&1
\end{array}
\right].
\end{equation}
Since $\rho'$ is separable and of rank $3$, it must be locally equivalent to a convex sum of projections onto  the vectors $\zeta_i\otimes\xi_i$ in \eqref{anotherthree}, which implies that $\tilde\rho^{T_1}$ is an analogous sum of projections onto  $\zeta_i^{\ast}\otimes\xi_i$. But $\zeta_i^{\ast}\otimes\xi_i=\zeta_i\otimes\xi_i$ if the product vectors are as in \eqref{anotherthree}. Therefore  $\rank{\rho'}=\rank{\left(\rho'\right)^{T_1}}$, which implies $\rank{\rho}=\rank{\rho^{T_1}}$, as expected. This proves our assertion for $\rho'$ supported on a $2\times 3$ subspace with $\kernel{\rho'}$ that intersects the corresponding Segre variety $\Sigma_{1,2}$ transversely. For the other nontrivial cases, we can have $\rho''$ separable and of rank $2$, supported on a $2\times 2$ subspace. There is also the trivial case of $\rho''$ supported on a $1\times 2$ or $1\times 3$ subspace, in which the equality $\rank{\rho''}=\rank{\left(\rho''\right)^{T_1}}$ clearly holds, and it implies equality of ranks of $\rho$ and $\rho^{T_1}$.

In the case of $\rho''$ supported on a $2\times 2$ subspace, we can repeat the argument with transverse intersections. Either $\rho''$ can be reduced once again, in which case it turns out to be equal to $\lambda'''\proj{\zeta'''\otimes\xi'''}
+\lambda''\proj{\zeta''\otimes\xi''}
$ with $\lambda''>0$, $\lambda'''>0$ and $\zeta'''\otimes\xi'''$ not proportional to $\zeta''\otimes\xi''$, or $\kernel{\rho''}$ must intersect the respective Segre variety $\Sigma_{1,1}$ in a transverse way. The first possibility clearly gives us $\rank{\rho''}=2=\rank{\left(\rho''\right)^{T_1}}$. The latter implies, by Bezout's Theorem, that there are exactly two product vectors in $\kernel{\rho''}$. Similarly as for \eqref{threeproductvectors}, we can prove that the two product vectors must be locally equivalent to $e_1\otimes e_1$ and $e_2\otimes e_2$. Clearly, they span the kernel of $\rho''$ and there are only two product vectors, locally equivalent to $e_1\otimes e_2$ and $e_2\otimes e_1$, in $\range{\rho''}$. But $\rho''$ is separable and of rank $2$. Therefore it must be locally equivalent to a convex sum of projections onto these two vectors. Accordingly, $\left(\rho''\right)^{T_1}$ is locally equivalent to a sum of two projections onto product vectors, which are $e_1^{\ast}\otimes e_2$ and $e_2^{\ast}\otimes e_1$, actually equal to $e_1\otimes e_2$ and  $e_2\otimes e_1$. This implies $\rank{\rho''}=\rank{\left(\rho''\right)^{T_1}}$ and the equality between the ranks of $\rho$ and $\rho^{T_1}$ follows.
\end{proof}
\end{proposition}

\begin{remark}\label{remarkone}
The two propositions above explain why PPT states of ranks $\left(4,n\right)$, $n\neq 4$ should not be expected to appear in the upper part of \textrm{Table II} in \cite{LS2010n}, which we reproduced above as Table \ref{tab3N}. They do exist, but they are always separable and of a rather special form. 
\end{remark}

It is useful to formulate the following
\begin{corollary}\label{coredge}
All rank $4$ non-separable PPT states $\rho$ in $3\times 3$ systems are edge states.
\begin{proof}
If some non-separable $\rho$ of rank $4$ had a product vector $\phi\otimes\psi$ in its range, and the partial conjugated vector $\phi^{\ast}\otimes\psi$ was in the range of $\rho^{T_1}$, we could diminish the rank of $\rho$ or $\rho^{T_1}$ by subtracting $\lambda\proj{\phi\otimes\psi}
$, where
\begin{equation}\label{defLambda} \lambda=\min\left\{\innerpr{\phi\otimes\psi}{\rho^{-1}\left(\phi\otimes\psi\right)}^{-1},\innerpr{\phi\otimes\psi}{\left(\rho^{T_1}\right)^{-1}\left(\phi\otimes\psi\right)}^{-1}\right\},
\end{equation}  
cf. \cite{LKHC2001}. In such case, $\rho$ could be written as $\rho=\rho'+\lambda\proj{\phi\otimes\psi}
$ with $\rho'$ PPT and of rank $3$ or with $\rho^{T_1}$ of rank $3$. But this implies, by \cite{HLVC2000}, that $\rho'$ would have to be separable. This further implies separability of $\rho$, which is a contradiction. 
\end{proof}
\end{corollary}
At this point, we can easily prove that the vectors $\phi_i\otimes\psi_i$ in the kernel of a non-separable $\rho$ of rank $4$, chosen as in \eqref{formtwoindeptriples2} or \eqref{formtwoindeptriples3}, constitute a generalized Unextendible Product Basis. If there was a product vector $\phi\otimes\psi$ orthogonal to all of them, it would be an element of the range of $\rho$. From the proof of Proposition \ref{propequalranks} we know that the partially conjugated vectors $\phi_i^{\ast}\otimes\psi_i$ span the kernel of $\rho^{T_1}$. Since $\innerpr{\phi\otimes\psi}{\phi_i\otimes\psi_i}=0=\innerpr{\phi^{\ast}\otimes\psi}{\phi_i^{\ast}\otimes\psi_i}$ for all $i$, we see that $\phi^{\ast}\otimes\psi$ is in the range of $\rho^{T_1}$, $\left(\kernel\rho^{T_1}\right)^{\bot}$. Therefore we have a product vector $\phi\otimes\psi$ in the range of $\rho$ such that its partial conjugation is in the range $\rho^{T_1}$. In other words, $\rho$ is not an edge state. But this contradicts Corollary \ref{coredge} and therefore cannot happen. In this way, we have proved the following.
\begin{proposition}\label{propgUPB}
Let $\rho$ be a rank $4$ non-separable PPT state in a $3\times 3$ system. The six vectors in the kernel of $\rho$ constitute a generalized UPB. There is a subset of five of them that constitutes a minimal gUPB in the sense of Proposition \ref{propgUPBsuffnecessary}.
\begin{proof}
Most of the proof has already been provided above. We only need to comment on the fact that five of the product vectors constitute a minimal gUPB. It must be so because the five vectors we brought to the form \eqref{formtwoindeptriples2} or \eqref{formtwoindeptriples3} span the kernel of $\rho$, and  the orthogonal complement to the kernel has no product vector in it. Thus, the five vectors are a gUPB of $\kernel{\rho}$, which is minimal according to Proposition \ref{propgUPBsuffnecessary}, because $m+n-1=5$ for $m=n=3$. 
\end{proof}
\end{proposition}
By Proposition \ref{propgUPBsuffnecessary} we know that a minimal gUPB $\left\{\phi_i\otimes\psi_i\right\}_{i=1}^6$ has the property that all triples in $\left\{\phi_i\right\}_{i=1}^6$ and in $\left\{\psi_i\right\}_{i=1}^6$ are linearly independent. In such case, the forms \eqref{formtwoindeptriples2} and \eqref{formtwoindeptriples3} are locally equivalent, and we may choose to work with only one of them. In the sequel, we prefer to assume the form  \eqref{formtwoindeptriples2} of the product vectors, which is in agreement with the convention used in \cite{HHMS2011}. Our next step is to prove that the parameters $p$, $q$, $r$ and $s$ in \eqref{formtwoindeptriples2} must be real if the corresponding product vectors belong to the kernel of a rank $4$ PPT state in the $3\times 3$ case. This is not of much use here, but will prove to be important in Section \ref{secdetermination}.

We know from Lemma \ref{lemmageneralintersection} that there are exactly six product vectors in the kernel of $\rho$, while we have only five of them in \eqref{formtwoindeptriples2}, and we know that they span the kernel. Consequently, the sixth vector is a linear combination of the other five ones,
\begin{equation}\label{lincombsixth}
\phi_6\otimes\psi_6=\sum_{i=1}^5\lambda_i\phi_i\otimes\psi_i
\end{equation}
Note that explicit formulas for the sixth vector can be found in \cite[Section 5.2]{HHMS2011}. Interestingly, since $\phi_6\otimes\psi_6\in\kernel{\rho}$, we know from Lemma \ref{lemmaprodPPT} that $\phi_6^{\ast}\otimes\psi_6\in\kernel{\rho}$ is in the kernel of $\rho^{T_1}$. However, the vectors $\phi_i^{\ast}\otimes\psi_i$ with $i=1,2,\ldots,5$ are also there and moreover, since they are, up to local equivalence, of the form \eqref{formtwoindeptriples2} with $p$ and $q$ complex conjugated, we already know that they span $\kernel{\rho^{T_1}}$. Thus the sixth partially conjugated vector must be a linear combination of the former five,
\begin{equation}\label{lincombsixthconj}
\phi_6^{\ast}\otimes\psi_6=\sum_{i=1}^5\xi_i\phi_i^{\ast}\otimes\psi_i,
\end{equation}
where the coefficients $\xi_i$ are in principle not related to the $\lambda_i$'s in \eqref{lincombsixth}. However, we can already see at this point that it may be very difficult to simultaneously satisfy equations \eqref{lincombsixth} and \eqref{lincombsixthconj}, if we do not assume that $\phi_i=\phi_i^{\ast}$ for all $i$. In the latter case, one can obviously choose $\xi_i=\lambda_i$. Our aim in the following will be to show that $\xi_i=\lambda_i$ is the only possible choice. By projecting \eqref{lincombsixth} onto the first, the second and the third coordinate in the first subsystem, we get
\begin{eqnarray}\label{lambdaeqs1}
\lambda_1\psi_1+\lambda_4\psi_4+\lambda_5\psi_5&=&\phi_{16}\,\psi_6,\\
\lambda_2\psi_2+\lambda_4\psi_4+p\lambda_5\psi_5&=&\phi_{26}\,\psi_6,\label{lambdaeqs2}\\
\lambda_3\psi_3+\lambda_4\psi_4+q\lambda_5\psi_5&=&\phi_{36}\,\psi_6,\label{lambdaeqs3}
\end{eqnarray}
where $\left\{\phi_{i6}\right\}_{i=1}^3$ are coordinates of $\phi_6$. Similarly, from \eqref{lincombsixthconj} we get
\begin{eqnarray}\label{xieqs1}
\xi_1\psi_1+\xi_4\psi_4+\xi_5\psi_5&=&\phi_{16}^{\ast}\,\psi_6,\\
\xi_2\psi_2+\xi_4\psi_4+p^{\ast}\xi_5\psi_5&=&\phi_{26}^{\ast}\,\psi_6,\label{xieqs2}\\
\xi_3\psi_3+\xi_4\psi_4+q^{\ast}\xi_5\psi_5&=&\phi_{36}^{\ast}\,\psi_6.\label{xieqs3}
\end{eqnarray}
Let us note that the triples $\left\{\psi_1,\psi_4,\psi_5\right\}$, $\left\{\psi_2,\psi_4,\psi_5\right\}$, $\left\{\psi_3,\psi_4,\psi_5\right\}$ all consist of linearly independent vectors, according to Proposition \ref{propgUPB}. This implies that each of the formulas \eqref{lambdaeqs1}--\eqref{xieqs3} gives exactly one solution for the coefficients $\lambda_i$ or $\xi_i$ which it contains. For one of the consequences, all the coefficients $\psi_{i6}$ must be non-vanishing. Two of them cannot vanish, because $\psi_6$ proportional to any of $\psi_i$ with $i=1,2,3$ would contradict $\phi_6\otimes\psi_6\neq\phi_i\otimes\psi_i$ or Lemma \ref{lemma3x3twoproducts}. To see this, let us assume that one of them vanishes, e.g. $\phi_{36}=0$. In such case, equation \eqref{lambdaeqs3} implies $\lambda_3=\lambda_4=\lambda_5=0$, where we used the fact that $q\neq 0$. Hence \eqref{lambdaeqs1} and \eqref{lambdaeqs2} reduce to $\phi_{16}\psi_6=\lambda_1\psi_1$ and $\phi_{26}\psi_6=\lambda_2\psi_2$. But neither of these equalities can hold, since $\phi_{16}\neq 0$ and $\phi_{26}\neq 0$, while $\psi_6$ proportional to $\psi_1$ or $\psi_2$ contradicts Lemma~\ref{lemma3x3twoproducts}. Thus our assumption $\phi_{36}=0$ must have been false. By repeating the same argument for $\phi_{16}$ and $\phi_{26}$, we arrive at $\phi_{16}\phi_{26}\phi_{36}\neq 0$. Let us also notice that necessarily $\lambda_4\neq 0$ and $\xi_4\neq 0$. We cannot have, for example $\xi_1\psi_1+\xi_5\psi_5=\phi_{16}^{\ast}\psi_6$ and $\xi_2\psi_2+p^{\ast}\xi_5\psi_5=\phi_{26}^{\ast}\psi_6$ since the only vector in the intersection of $\linspan{\left\{\psi_1,\psi_5\right\}}$ and $\linspan{\left\{\psi_2,\psi_5\right\}}$ is $\psi_5$, and we know that $\psi_6\neq\psi_5$ by Lemma \ref{lemma3x3twoproducts}. In a similar way, one obtains $\lambda_5\neq 0$ and $\xi_5\neq 0$. With such amount of knowledge, we can prove the expected result.
\begin{proposition}\label{proprealparams}
Let $\phi_i\otimes\psi_i$ for $i=1,2,\ldots,5$ be product vectors of the form \eqref{formtwoindeptriples2} in the kernel of a non-separable PPT state of rank four, acting on $\mathbbm{C}^3\otimes\mathbbm{C}^3$. The parameters $p$, $q$, $r$ and $s$ must necessarily be real.
\begin{proof}
By dividing \eqref{lambdaeqs1} by $\phi_{16}$ and \eqref{xieqs1} by  $\phi_{16}^{\ast}$, which is possible according to $\phi_{16}\neq 0$, we get
\begin{equation}\label{eqpsisixeq}
\frac{\lambda_1}{\phi_{16}}\psi_1+\frac{\lambda_4}{\phi_{16}}\psi_4+\frac{\lambda_5}{\phi_{16}}\psi_5=\psi_6=\frac{\xi_1}{\phi_{16}^{\ast}}\psi_1+\frac{\xi_4}{\phi_{16}^{\ast}}\psi_4+\frac{\xi_5}{\phi_{16}^{\ast}}\psi_5.
\end{equation}
Since $\left\{\psi_1,\psi_4,\psi_5\right\}$ is a linearly independent triple, the above equality implies $\lambda_1/\phi_{16}=\xi_1/\phi_{16}^{\ast}$, $\lambda_4/\phi_{16}=\xi_4/\phi_{16}^{\ast}$ and $\lambda_5/\phi_{16}=\xi_5/\phi_{16}^{\ast}$. In a similar way, from \eqref{lambdaeqs2} and \eqref{xieqs2} we can get $\lambda_2/\phi_{26}=\xi_2/\phi_{26}^{\ast}$, $\lambda_4/\phi_{26}=\xi_4/\phi_{26}^{\ast}$ and $p\lambda_5/\phi_{26}=p^{\ast}\xi_5/\phi_{26}^{\ast}$, whereas \eqref{lambdaeqs3} and \eqref{xieqs3} give us $\lambda_2/\phi_{36}=\xi_2/\phi_{36}^{\ast}$, $\lambda_4/\phi_{36}=\xi_4/\phi_{36}^{\ast}$ and $q\lambda_5/\phi_{36}=q^{\ast}\xi_5/\phi_{36}^{\ast}$. From the equalities involving $\lambda_4$ and $\xi_4$, we get
\begin{equation}\label{xifourlambdafour}
\frac{\phi_{16}}{\phi_{16}^{\ast}}=\frac{\phi_{26}}{\phi_{26}^{\ast}}=\frac{\phi_{36}}{\phi_{36}^{\ast}}.
\end{equation}
Together with $\lambda_5/\phi_{16}=\xi_5/\phi_{16}^{\ast}$, the above equations give us $\lambda_5/\phi_{26}=\xi_5/\phi_{26}^{\ast}$ and $\lambda_5/\phi_{36}=\xi_5/\phi_{36}^{\ast}$. But 
\begin{equation}\label{eqimplication}
\left(\frac{\lambda_5}{\phi_{26}}=\frac{\xi_5}{\phi_{26}^{\ast}}\,\land\,\frac{p\lambda_5}{\phi_{26}}=\frac{p^{\ast}\xi_5}{\phi_{26}^{\ast}}\right)\,\Rightarrow\,p=p^{\ast}.
\end{equation}
In a similar way, from $\lambda_5/\phi_{36}=\xi_5/\phi_{36}^{\ast}$ and $q\lambda_5/\phi_{36}=q^{\ast}\xi_5/\phi_{36}^{\ast}$ we can get $q=q^{\ast}$.
\end{proof}
\end{proposition}

\section[An equivalence between generalized and orthonormal Unextendible Product Bases]{An equivalence between generalized and\\orthonormal Unextendible Product Bases}\label{secequiv}

In the following, we discuss item $4.$ of the list given in Section \ref{secoutline}. Let us start with a set of five vectors in $\mathbbm{C}^3$,
\begin{equation}\label{fivevectors}
\left[\begin{array}{ccccc}
\phi_1&\phi_2&\phi_3&\phi_4&\phi_5
\end{array}\right]=\left[\begin{array}{ccccc}
\phi_{11}&\phi_{12}&\phi_{13}&\phi_{14}&\phi_{15}\\
\phi_{21}&\phi_{22}&\phi_{23}&\phi_{24}&\phi_{25}\\
\phi_{31}&\phi_{32}&\phi_{33}&\phi_{34}&\phi_{35}
\end{array}\right],
\end{equation}
and assume that any three of them are linearly independent, as in Proposition \ref{propgUPB}. For the moment, we do not require the vectors in \eqref{fivevectors} to be equal to $\phi_1,\ldots,\phi_5$ in \eqref{formtwoindeptriples2}, but our ultimate goal is to apply the results we are going to obtain to \eqref{formtwoindeptriples2}. $\textnormal{PSL}\left(3,\mathbbm{C}\right)$ transformations of the above set correspond to the multiplication of the $3\times 5$ matrix in \eqref{fivevectors} from the left by an element of $\textnormal{SL}\left(3,\mathbbm{C}\right)$ and to the multiplication of the columns of \eqref{fivevectors} by arbitrary non-zero scalar factors. It is clear that we can transform \eqref{fivevectors} by a $\textnormal{PSL}\left(3,\mathbbm{C}\right)$ transformation to the following form,
\begin{equation}\label{fivevectors2}
\left[\begin{array}{ccccc}
1&0&\phi'_{13}&\phi'_{14}&\phi'_{15}\\
0&1&\phi'_{23}&\phi'_{24}&\phi'_{25}\\
0&0&\phi'_{33}&\phi'_{34}&\phi'_{35}
\end{array}\right].
\end{equation}
By another $\textnormal{PSL}\left(3,\mathbbm{C}\right)$ transformation, we get
\begin{equation}\label{fivevectors3}\left[\begin{array}{ccc}1&0&0\\0&1&-\frac{\phi'_{23}}{\phi'_{33}}\\0&0&1\end{array}\right]\left[\begin{array}{ccccc}
1&0&\phi'_{13}&\phi'_{14}&\phi'_{15}\\
0&1&\phi'_{23}&\phi'_{24}&\phi'_{25}\\
0&0&\phi'_{33}&\phi'_{34}&\phi'_{35}
\end{array}\right]=
\left[\begin{array}{ccccc}
1&0&\phi'_{13}&\phi''_{14}&\phi''_{15}\\
0&1&0&\phi''_{24}&\phi''_{25}\\
0&0&\phi'_{33}&\phi''_{34}&\phi''_{35}
\end{array}\right].
\end{equation}
We should remark that the matrix we multiply with from the left is well-defined, since $\phi'_{33}\neq 0$ according to the assumption about linear independence of triples. Let us transform once again, in the following way, 
\begin{equation}\label{fivevectors4}\left[\begin{array}{ccc}1&0&-\frac{\phi''_{15}}{\phi''_{35}}\\0&1&0\\0&0&1\end{array}\right]\left[\begin{array}{ccccc}
1&0&\phi'_{13}&\phi''_{14}&\phi''_{15}\\
0&1&0&\phi''_{24}&\phi''_{25}\\
0&0&\phi'_{33}&\phi''_{34}&\phi''_{35}
\end{array}\right]=
\left[\begin{array}{ccccc}
1&0&\phi'_{13}&\phi'''_{14}&0\\
0&1&0&\phi'''_{24}&\phi''_{25}\\
0&0&\phi'_{33}&\phi'''_{34}&\phi''_{35}
\end{array}\right].
\end{equation}
This is again possible because $\phi''_{35}\neq 0$ according to our assumptions. 

In a similar way as before, we see that $\phi''_{24}\neq 0$ and $\phi''_{35}\neq 0$. If we multiply the fourth column by ${1}/{\phi'''_{24}}$ and the fifth by ${1}/{\phi''_{35}}$, the above transforms to
\begin{equation}\label{fivevectors5}
\left[\begin{array}{ccccc}
1&0&x&y&0\\
0&1&0&1&z\\
0&0&t&u&1
\end{array}\right],
\end{equation}
where we introduced the notation $x:=\phi'_{13}$, $t:=\phi'_{33}$, $y:={\phi'''_{14}}/{\phi''_{24}}$, $u:={\phi'''_{34}}/{\phi''_{24}}$, $z:={\phi''_{25}}/{\phi''_{35}}$. It is quite straightforward to see that all the coefficients $x,y,z,t,u$ have to be different from zero according to the independent triples assumption.

Now, introduce the following invariants \cite{LS2010},
\begin{equation}\label{invariants1}
s_1=-\,\frac{\left|\begin{array}{ccc}\phi_{1}&\phi_{2}&\phi_{4}\end{array}\right|\cdot\left|\begin{array}{ccc}\phi_{1}&\phi_{3}&\phi_{5}\end{array}\right|}{\left|\begin{array}{ccc}\phi_{1}&\phi_{2}&\phi_{5}\end{array}\right|\cdot\left|\begin{array}{ccc}\phi_{1}&\phi_{3}&\phi_{4}\end{array}\right|},
\end{equation}
\begin{equation}\label{invariants2}
s_2=-\,\frac{\left|\begin{array}{ccc}\phi_{1}&\phi_{2}&\phi_{3}\end{array}\right|\cdot\left|\begin{array}{ccc}\phi_{2}&\phi_{4}&\phi_{5}\end{array}\right|}{\left|\begin{array}{ccc}\phi_{1}&\phi_{2}&\phi_{4}\end{array}\right|\cdot\left|\begin{array}{ccc}\phi_{2}&\phi_{3}&\phi_{5}\end{array}\right|}.
\end{equation}
The numbers $s_1$, $s_2$ are indeed invariant. They do not change under the family of $\textnormal{PSL}\left(3,\mathbbm{C}\right)$ transformations we were using in the consecutive steps \eqref{fivevectors}--\eqref{fivevectors5}. Thus we can substitute
\begin{equation}\label{subst}
\left[\begin{array}{ccccc}
\phi_{11}&\phi_{12}&\phi_{13}&\phi_{14}&\phi_{15}\\
\phi_{21}&\phi_{22}&\phi_{23}&\phi_{24}&\phi_{25}\\
\phi_{31}&\phi_{32}&\phi_{33}&\phi_{34}&\phi_{35}
\end{array}\right]\rightarrow\left[\begin{array}{ccccc}
1&0&x&y&0\\
0&1&0&1&z\\
0&0&t&u&1
\end{array}\right].
\end{equation}
in the above formulas for $s_1$ and $s_2$. In this way, we can quickly calculate the values of the invariants,
\begin{equation}\label{valuesofinvariants}
s_1=-uz\quad\textrm{and}\quad s_2=-\frac{ty}{ux}.
\end{equation}
Now, impose the conditions $s_1>0$ and $s_2>0$. From the first one, we clearly get $u=-rz^{\ast}$, where $r$ is a positive real number. Thus, we have the vectors
\begin{equation}\label{fivevectors6}
\left[\begin{array}{ccccc}
1&0&x&y&0\\
0&1&0&1&z\\
0&0&t&-rz^{\ast}&1
\end{array}\right].
\end{equation}
Next, let us multiply from the left by a diagonal matrix $\textnormal{diag}\left(1,\sqrt{\sqrt{r'}},{1}/{\sqrt{\sqrt{r'}}}\right)$, as well as multiply the second column by ${1}/{\sqrt{\sqrt{r'}}}$, the fourth by ${1}/{\sqrt{\sqrt{r'}}}$ and the fifth by $\sqrt{\sqrt{r'}}$, where $r':=r{z^{\ast}}/{z}$ and $\sqrt{\zeta}$ stands for the square root of $\zeta\in\mathbbm{C}$ with the argument in $\left[0,\pi\right)$. Under such $\textnormal{PSL}\left(3,\mathbbm{C}\right)$ transformation the vectors \eqref{fivevectors6} change into 
\begin{equation}\label{fivevectors7}
\left[\begin{array}{ccccc}
1&0&x'&y'&0\\
0&1&0&1&z'\\
0&0&t'&-z'&1
\end{array}\right],
\end{equation}
where $z'$ is real and positive, and all the other parameters $x',y',t'$ are non-zero. Moreover, the conditon $s_2>0$ transforms to
\begin{equation}\label{conds2}
s_2=-\frac{ty}{ux}=\frac{t'y'}{z'x'}>0\quad\Leftrightarrow\quad\frac{t'y'}{x'}>0,
\end{equation}
simply by formula \eqref{valuesofinvariants} and the invariance of $s_2$. The last equivalence holds by strict positivity of $z'$. In our next step, we we are going to multiply \eqref{fivevectors7} from the left by a diagonal matrix $\textnormal{diag}\left(\zeta_1,\zeta_2,\zeta_3\right)$, with $\zeta_1,\zeta_2,\zeta_3\in\mathbbm{C}$ and $\zeta_1\zeta_2\zeta_3=1$, and also multiply the consecutive columns, beginning with the first, by ${1}/{\zeta_1}$, ${1}/{\zeta_2}$, $\zeta_4$, $\zeta_5$ and $\zeta_6$, where $\zeta_4\zeta_5\zeta_6\neq 0$. Our aim is to choose the numbers $\zeta_1,\ldots,\zeta_6$ in such a way that $\eqref{fivevectors7}$ transforms to a set of vectors with orthogonality relations given by a pentagon graph (that is, any two consecutive ones are orthogonal, and these are the only orthogonality relations). We would like to have
\begin{equation}\label{fivevectors8}
\left[\begin{array}{ccccc}
1&0&a&b&0\\
0&1&0&1&a\\
0&0&b&-a&1
\end{array}\right],
\end{equation}
where $a=z'$ and $b$ is a positive real number in place of \eqref{fivevectors7}. Let us write the numbers $\zeta_j$ as $r_je^{i\alpha_j}$, where $r_j$ is a positive real number and $\alpha_j\in\mathbbm{R}$. In order to obtain \eqref{fivevectors8} with $a$ and $b$ real and positive, certain phase matching conditions have to be fulfilled. Let us consider them first. If $\alpha_{y'}$, $\alpha_{t'}$, $\alpha_{x'}$ are such that $y'=r_{y'}e^{i\alpha_{y'}}$, $t'=r_{t'}e^{i\alpha_{t'}}$ and $x'=r_{x'}e^{i\alpha_{x'}}$ with $r_{x'}$, $r_{t'}$ and $r_{x'}$ real and positive, complex phases match correctly if and only if the following set of equations hold
\begin{eqnarray}
\alpha_2+\alpha_5&=&0\quad\textnormal{ mod }2\pi,\label{eq1}\\
\alpha_3+\alpha_6&=&0\quad\textnormal{ mod }2\pi,\label{eq2}\\
\alpha_2+\alpha_6&=&0\quad\textnormal{ mod }2\pi,\label{eq3}\\
\alpha_3+\alpha_5&=&0\quad\textnormal{ mod }2\pi,\label{eq4}\\
\alpha_5+\alpha_1+\alpha_{y'}&=&0\quad\textnormal{ mod }2\pi,\label{eq5}\\
\alpha_3+\alpha_4+\alpha_{t'}&=&0\quad\textnormal{ mod }2\pi,\label{eq6}\\
\alpha_4+\alpha_1+\alpha_{x'}&=&0\quad\textnormal{ mod }2\pi.\label{eq7}
\end{eqnarray}
The requirement that $\zeta_1\zeta_2\zeta_3=1$ adds a condition $\alpha_1+\alpha_2+\alpha_3=0\textnormal{ mod }2\pi$ to equations \eqref{eq1}--\eqref{eq2}. However, a substitution of the form
\begin{equation}\nonumber
\left(\alpha_1,\alpha_2,\alpha_3\right)\rightarrow\left(\alpha_1+\beta,\alpha_2+\beta,\alpha_3+\beta\right)\quad\left(\alpha_4,\alpha_5,\alpha_6\right)\rightarrow\left(\alpha_4-\beta,\alpha_5-\beta,\alpha_6-\beta\right)
\end{equation}
with an appropriately chosen $\beta$ can always bring $\alpha_1+\alpha_2+\alpha_3$ to zero and it has no effect on \eqref{eq1}--\eqref{eq7}. Therefore, as long as existence of solutions is in question, we may neglect the additional condition. It is easy to see that the relations \eqref{eq1}--\eqref{eq4} are fulfilled if and only if $\alpha_2=\alpha_3=-\alpha_5=-\alpha_6=-\alpha\textnormal{ mod }2\pi$ for some $\alpha\in\mathbbm{R}$. Thus the set of equations \eqref{eq1}--\eqref{eq7} are reduced to 
\begin{equation}\label{matrixeq}
\left[\begin{array}{ccc}1&1&0\\-1&0&1\\0&1&1\end{array}\right]\left[\begin{array}{c}\alpha\\\alpha_1\\\alpha_4\end{array}\right]=\left[\begin{array}{c}-\alpha_{y'}\\-\alpha_{t'}\\-\alpha_{x'}\end{array}\right]\textnormal{ mod }2\pi.
\end{equation}
Interestingly, the $3\times 3$ matrix in equation \eqref{matrixeq} has rank $2$. A solution $\left(\alpha,\alpha_1,\alpha_4\right)$ exists if and only if 
\begin{equation}\label{phasematchingfinal}
\alpha_{y'}+\alpha_{t'}-\alpha_{x'}=0\textnormal{ mod }2\pi.
\end{equation} 
But this is exactly the positivity condition \eqref{conds2} for the invariant $s_2$. Thus, if $s_2>0$ in addition to $s_1>0$, we can cancel the complex phases, as in \eqref{fivevectors8}. The only remaining thing to do  is to match the modules, which gives us the following set of equations,
\begin{equation}\label{eqsmodulematch}
r_2r_5=1,\quad r_3r_6=1,\quad r_2r_6=r_3r_5,\quad r_4r_1r_{x'}=a,\quad r_5r_1r_{y'}=r_3r_4r_{t'}.
\end{equation}
There is also an equation $r_1r_2r_3=1$, following from the requirement that $\zeta_1\zeta_2\zeta_3=1$. As we see, there are five equations in \eqref{eqsmodulematch}, and the variables $r_1,\ldots,r_6$ are six in number. Therefore, one can expect a solution to exist. It can easily be checked that the following, with $r\in\mathbbm{R}$, is a one-parameter family of solutions,
\begin{equation}\label{solutionmodulesmatch}
r_1=\sqrt{\frac{ar_{t'}}{r_{x'}r_{y'}}}r,\quad r_2=r,\quad r_3=r, \quad r_4=\sqrt{\frac{ar_{t'}}{r_{x'}r_{y'}}}\frac{1}{r}, \quad r_5=\frac{1}{r},\quad r_6=\frac{1}{r}.
\end{equation}
By choosing $r=1/\sqrt[6]{{ar_{t'}}/{r_{x'}r_{y'}}}$ we can satisfy the additional condition $r_1r_2r_3=1$. Thus we have proved that the positivity of the invariants $s_1$, $s_2$ guarantees that the family of five vectors \eqref{fivevectors} can be transformed by a $\textnormal{PSL}\left(3,\mathbbm{C}\right)$ transformation, without permuting them, to the form \eqref{fivevectors8}. Obviously, a converse statement is also true, since the values of $s_1$ and $s_2$ calculated from \eqref{fivevectors8} are $a^2$ and $b^2/a^2$, respectively. In this way we arrive at the following
\begin{proposition}\label{pentagon}
A set of five vectors $\left\{\alpha_i\right\}_{i=1}^5\subset\mathbbm{C}^3$ with the property that any triple of them is linearly independent, can be transformed by a $\textnormal{PSL}\left(3,\mathbbm{C}\right)$ transformation, without permuting them, to the form \eqref{fivevectors8} with $a$ and $b$ real and positive, if and only if the invariants $s_1$ and $s_2$, defined in \eqref{invariants1}, are positive.
\end{proposition}
Let us note that any set of five vectors $\left\{v_1,\ldots,v_5\right\}\subset\mathbbm{C}^3$ with orthogonality relations $\innerpr{v_i}{v_{\left(i+1\right)\textnormal{ mod }5}}=0$ can be transformed by $\PSLt$ transformations to the form \eqref{fivevectors8}. A simple argument shows that they can be transformed to
\begin{equation}\label{fivevectors9}
\left[\begin{array}{ccccc}v_1&v_2&v_3&v_4&v_5\end{array}\right]=\left[\begin{array}{ccccc}
1&0&x&y^{\ast}&0\\
0&1&0&1&x\\
0&0&y&-x^{\ast}&1
\end{array}\right],
\end{equation}
with $x$ and $y$ complex. But since $s_1=\left|x\right|^2>0$ and $s_2=\left|y/x\right|^2>0$ in the above case, the argument following equation \eqref{fivevectors7} tells us that a $\PSLt$ transformation brings \eqref{fivevectors9} to the form \eqref{fivevectors8}. As a consequence, Proposition \ref{pentagon} is a necessary and sufficient criterion for a set of five vectors $\phi_1,\ldots,\phi_5$ to be $\SLt$ equivalent, without permuting them, to a set of vectors  $v_1,\ldots,v_5$ with orthogonality relations $\innerpr{v_i}{v_{i\textnormal{ mod }5+1}}=0$.

From \cite{DiVicenzo04} we know that orthogonal UPBs in the $3\times 3$ case always have five elements, and they are, up to permutations, precisely the sets of product vectors $\left\{v_i\otimes w_i\right\}_{i=1}^5$ with orthogonality relations $\innerpr{v_i}{v_{i\textnormal{ mod }5+1}}=0$ and $\innerpr{w_j}{w_{\left(j+1\right)\textnormal{ mod }5+1}}=0$. Consider the question, whether an arbitrary set of five vectors $\left\{\phi_i\otimes\psi_i\right\}_{i=1}^5\subset\mathbbm{C}^3\otimes\mathbbm{C}^3$ with linearly independent triples can be brought by $\PSLtt$ transformations to such $\left\{v_i\otimes w_i\right\}_{i=1}^5$, without permuting the vectors. In other words, what are the necessary and sufficient conditions for $\phi_i\otimes\psi_i$'s to be convertible into $v_i\otimes w_i$'s with the orthogonality conditions given above. By using Proposition \ref{pentagon}, we can already deal with the question about $\phi_i$'s being convertible into $v_i$'s. Namely, an $\PSLt$ transformation on the first subsystem can bring the vectors $\left\{\phi_i\right\}_{i=1}^5$, without permuting them, to $\left\{v_i\right\}_{i=1}^5$ with $\innerpr{v_i}{v_{i\textnormal{ mod }5+1}}=0$ if and only if the corresponding values of the invariants $s_1$ and $s_2$ are positive. We are only missing a similar criterion for $\psi_i$'s and $w_i$'s. However, it is not difficult to check that a permutation $\sigma=\left(\begin{array}{ccccc}1&2&3&4&5\\1&3&5&2&4\end{array}\right)$ brings any $\left\{w_i\right\}_{i=1}^5$ with $\innerpr{w_j}{w_{\left(j+1\right)\textnormal{ mod }5+1}}=0$ to $\left\{w'_i\right\}_{i=1}^5=\left\{w_{\sigma\left(i\right)}\right\}_{i=1}^5$ with $\innerpr{w'_i}{w'_{i\textnormal{ mod }5+1}}=0$. Therefore, it is sufficient to calculate the invariants \eqref{invariants1} and \eqref{invariants2} corresponding to the permuted vectors $\psi'_i:=\psi_{\sigma\left(i\right)}$ and check their positivity in order to tell whether the vectors $\psi_i$ are convertible into some $\left\{w_i\right\}_{i=1}^5$ with the desired orthogonality relations. Following the definitions \eqref{invariants1} and \eqref{invariants2}, let us introduce additional invariants
\begin{multline}\label{invariant3}
s_3=
-\,\frac{\left|\begin{array}{ccc}\psi_{1}&\psi_{3}&\psi_{2}\end{array}\right|\cdot\left|\begin{array}{ccc}\psi_{1}&\psi_{5}&\psi_{4}\end{array}\right|}{\left|\begin{array}{ccc}\psi_{1}&\psi_{3}&\psi_{4}\end{array}\right|\cdot\left|\begin{array}{ccc}\psi_{1}&\psi_{5}&\psi_{2}\end{array}\right|}=\\=
-\,\frac{\left|\begin{array}{ccc}\psi_{\sigma\left(1\right)}&\psi_{\sigma\left(2\right)}&\psi_{\sigma\left(4\right)}\end{array}\right|\cdot\left|\begin{array}{ccc}\psi_{\sigma\left(1\right)}&\psi_{\sigma\left(3\right)}&\psi_{\sigma\left(5\right)}\end{array}\right|}{\left|\begin{array}{ccc}\psi_{\sigma\left(1\right)}&\psi_{\sigma\left(2\right)}&\psi_{\sigma\left(5\right)}\end{array}\right|\cdot\left|\begin{array}{ccc}\psi_{\sigma\left(1\right)}&\psi_{\sigma\left(3\right)}&\psi_{\sigma\left(4\right)}\end{array}\right|}
\end{multline}
and
\begin{multline}\label{invariant4}
s_4=-\,\frac{\left|\begin{array}{ccc}\psi_{1}&\psi_{3}&\psi_{5}\end{array}\right|\cdot\left|\begin{array}{ccc}\psi_{3}&\psi_{2}&\psi_{4}\end{array}\right|}{\left|\begin{array}{ccc}\psi_{1}&\psi_{3}&\psi_{2}\end{array}\right|\cdot\left|\begin{array}{ccc}\psi_{3}&\psi_{5}&\psi_{4}\end{array}\right|}=\\=
-\,\frac{\left|\begin{array}{ccc}\psi_{\sigma\left(1\right)}&\psi_{\sigma\left(2\right)}&\psi_{\sigma\left(3\right)}\end{array}\right|\cdot\left|\begin{array}{ccc}\psi_{\sigma\left(2\right)}&\psi_{\sigma\left(4\right)}&\psi_{\sigma\left(5\right)}\end{array}\right|}{\left|\begin{array}{ccc}\psi_{\sigma\left(1\right)}&\psi_{\sigma\left(2\right)}&\psi_{\sigma\left(4\right)}\end{array}\right|\cdot\left|\begin{array}{ccc}\psi_{\sigma\left(2\right)}&\psi_{\sigma\left(3\right)}&\psi_{\sigma\left(5\right)}\end{array}\right|},
\end{multline}
in accordance with \cite{LS2010}. From the discussion above it follows that arbitrary five vectors $\psi_1,\ldots,\psi_5$ in $\mathbbm{C}^3$ can be transformed, without permuting them, to $\left\{w_i\right\}_{i=1}^5$ with orthogonality relations $\innerpr{w_j}{w_{\left(j+1\right)\textnormal{ mod }5+1}}=0$ if and only if the above invariants $s_3$ and $s_4$ are positive. Together with the previously obtained convertibility result between $\phi_1,\ldots,\phi_5$ and $v_1,\ldots,v_5$, the last result gives us the following.
\begin{proposition}\label{propinvariantspositiveortho}
A set of product vectors $\left\{\phi_i\otimes\psi_i\right\}_{i=1}^5\subset\mathbbm{C}^3\otimes\mathbbm{C}^3$ can be transformed by a $\PSLtt$ transformation to an orthogonal UPB $\left\{v_i\otimes w_i\right\}_{i=1}^5$ with orthogonality relations $\innerpr{v_i}{v_{i\textnormal{ mod }5+1}}=0$ and $\innerpr{w_j}{w_{\left(j+1\right)\textnormal{ mod }5+1}}=0$, without permuting the $\phi_i\otimes\psi_i$'s, if and only if the invariants $s_1$, $s_2$, $s_3$ and $s_4$, defined in \eqref{invariants1}, \eqref{invariants2}, \eqref{invariant3} and \eqref{invariant4}, are positive.
\begin{proof}
Most of the proof has already been included above. Let $\left\{v_i\otimes w_i\right\}_{i=1}^5$ denote an orthogonal UPB  with the orthogonality relations $\innerpr{v_i}{v_{i\textnormal{ mod }5+1}}=0$ and $\innerpr{w_j}{w_{\left(j+1\right)\textnormal{ mod }5+1}}=0$ for all $i,j\in\left\{1,2,3,4,5\right\}$. The possibility to convert
\begin{equation}\label{posconvert}
\left[\begin{array}{ccccc}\phi_1&\phi_2&\phi_3&\phi_4&\phi_5\\
\hline
\psi_1&\psi_2&\psi_3&\psi_4&\psi_5
\end{array}
\right]\rightarrow\left[\begin{array}{ccccc}
v_1&v_2&v_3&v_4&v_5\\
\hline
w_1&w_2&w_3&w_4&w_5
\end{array}
\right]
\end{equation}
by $\PSLtt$ transformations, or by local equivalence in our usual terms, is the same as the possibility to separately convert $\left\{\phi_i\right\}_{i=1}^5$ into $\left\{v_i\right\}_{i=1}^5$ and $\left\{\psi_j\right\}_{j=1}^5$ into $\left\{w_j\right\}_{j=1}^5$ by some $\PSLt$ transformations. However, we know that the first conversion is possible if and only if $s_1$ and $s_2$ are positive, while the second needs positivity of $s_3$ and $s_4$. Altogether, positivity of all the invariants $s_i$, $i=1,2,3,4$ is a necessary and sufficient criterion for the transformation \eqref{posconvert} to be possible.  
\end{proof}
\end{proposition}
In the context of product vectors in the kernel of a PPT state, as well as elements of an orthogonal UPB, permutations are obviously possible. Therefore we would like to have a version of Proposition \ref{propinvariantspositiveortho} with no restriction on the ordering of the vectors $\left\{\phi_i\otimes\psi_i\right\}_{i=1}^5$. 
\begin{proposition}\label{propinvariantspositiveortho2}
A set of product vectors $\left\{\phi_i\otimes\psi_i\right\}_{i=1}^5\subset\mathbbm{C}^3\otimes\mathbbm{C}^3$ can be transformed by a $\PSLtt$ transformation to an orthogonal UPB, if and only if for some permutation $\kappa$ the invariants $s_1$, $s_2$, $s_3$ and $s_4$, calculated with the permuted vectors $\phi_{\kappa\left(i\right)}$ and $\psi_{\kappa\left(i\right)}$ substituted for $\phi_i$ and $\psi_i$, respectively, are all positive.
\begin{proof}
Immediate given the fact \cite{DiVicenzo04} that an orthogonal UPB in a $3\times 3$ system can always be brought by a permutation to a $\left\{v_i\otimes w_i\right\}_{i=1}^5$ with the orthogonality relations as in Proposition \ref{propinvariantspositiveortho}.
\end{proof}
\end{proposition} 
Let us also note that, in accordance with \cite{HHMS2011}, not every single permutation of the five product vectors needs to be considered if we want to check whether they can be transformed into an orthogonal UPB or not.
\begin{remark}\label{remarkpermutation}
Only $12$ permutations, given in 
Table \ref{tblpentagram}, have to be checked  in order to obtain a decisive answer to the question raised in Proposition \ref{propinvariantspositiveortho2}.
\begin{proof}
An explanation is included in \cite{LS2010} and \cite{HHMS2011}, but we repeat it quickly here for completeness. Let us denote by $S_5$ the symmetric group of $\left\{1,2,\ldots,5\right\}$. The permutations given in 
Table \ref{tblpentagram} are representatives of equivalence classes in $S_5$ of the regular pentagon subgroup $G$, generated by the cycle $\left(1\,2\,3\,4\,5\right)$ and the inversion $\left(\begin{array}{ccccc}1&2&3&4&5\\5&4&3&2&1\end{array}\right)$. The regular pentagon symmetry subgroup has the expected property that it does not change signs of $s_1$, $s_2$, $s_3$ and $s_4$, just as it does not change orthogonality relations between the vectors $\left\{v_i\right\}_{i=1}^5$ and $\left\{w_j\right\}_{j=1}^5$. Therefore, we may divide $S_5$ by $G$ when we check positivity of the invariants in Proposition \ref{propinvariantspositiveortho2}. The number of invariance classes is $12$ because $\#S_5=5!=120$ and $\#G=10$.
\end{proof}
\end{remark}

\begin{table}\centering
\begin{tabular}{|c|c|c|c|}
\hline
$\sigma_1:$&$\left(\begin{array}{ccccc}1&2&3&4&5\\1&2&3&4&5\end{array}\right)$&$\sigma_2:$&$\left(\begin{array}{ccccc}1&2&3&4&5\\1&3&2&4&5\end{array}\right)$\\

$\sigma_3:$&$\left(\begin{array}{ccccc}1&2&3&4&5\\2&1&3&4&5\end{array}\right)$&$\sigma_4:$&$\left(\begin{array}{ccccc}1&2&3&4&5\\2&3&1&4&5\end{array}\right)$\\

$\sigma_5:$&$\left(\begin{array}{ccccc}1&2&3&4&5\\3&1&2&4&5\end{array}\right)$&$\sigma_6:$&$\left(\begin{array}{ccccc}1&2&3&4&5\\3&2&1&4&5\end{array}\right)$\\

$\sigma_7:$&$\left(\begin{array}{ccccc}1&2&3&4&5\\1&2&4&3&5\end{array}\right)$&$\sigma_8:$&$\left(\begin{array}{ccccc}1&2&3&4&5\\1&4&2&3&5\end{array}\right)$\\

$\sigma_9:$&$\left(\begin{array}{ccccc}1&2&3&4&5\\2&1&4&3&5\end{array}\right)$&$\sigma_{10}:$&$\left(\begin{array}{ccccc}1&2&3&4&5\\2&4&1&3&5\end{array}\right)$\\
$\sigma_{11}:$&$\left(\begin{array}{ccccc}1&2&3&4&5\\1&3&4&2&5\end{array}\right)$&$\sigma_{12}:$&$\left(\begin{array}{ccccc}1&2&3&4&5\\1&4&3&2&5\end{array}\right)$\\
\hline
\end{tabular}
\caption{A list of representatives of the $12$ equivalence classes of the symmetric group $S_5$ under left multiplication by the regular pentagram group.\label{tblpentagram}}
\end{table}

\section{Determination of a PPT state by product vectors in its kernel}\label{secdetermination}
In the last part of the proof of our main result, concerning PPT states of rank four in two qutrit systems, we recall a number of surprising facts that were earlier reported in \cite[Section 5]{HHMS2011} without a complete explanation. Here we fill in that little gap, and we collect a sufficient amount of information to quickly explain the findings of Leinaas \textit{et al.}, concerning the relation of extreme PPT states to Unextendible Product Bases \cite{LS2010}.

Note that, given a set of product vectors in $\kernel{\rho}$, the conditions in Lemma \ref{lemmaprodPPT} are a set of linear equations for $\rho$. An idea, earlier presented in \cite{HHMS2011}, is to try to solve these equations assuming a specific form of the product vectors, namely \eqref{formtwoindeptriples2}. Let us repeat formula \eqref{formtwoindeptriples2} here for the convenience of the reader.
\begin{equation}\label{formtwoindeptriples4}
\left[\begin{array}{ccccc}\phi_1&\phi_2&\phi_3&\phi_4&\phi_5\\
\hline
\psi_1&\psi_2&\psi_3&\psi_4&\psi_5
\end{array}
\right]=\left[\begin{array}{ccccc}
1&0&0&1&1\\
0&1&0&1&p\\
0&0&1&1&q\\
\hline
1&0&0&1&1\\
0&1&0&1&r\\
0&0&1&1&s
\end{array}
\right],
\end{equation}
We actually know from Proposition \ref{propgUPB} that there always exists a local $\SLtt$ transformation $A\otimes B$ that brings five vectors in the kernel of a non-separable PPT state of rank $4$, possibly multiplied by some scalar factors, into the form \eqref{formtwoindeptriples4} with \emph{all triples linearly independent}. Moreover, Proposition \ref{proprealparams} tells us that the parameters $p$, $q$, $r$ and $s$ are necessarily real numbers. By solving the linear conditions on a PPT state following from Lemma \ref{lemmaprodPPT} with  $\phi_i\otimes\psi_i$, $i=1,2,\ldots,5$ as in \eqref{formtwoindeptriples4} substituted for $\phi\otimes\psi$, we will actually be solving a set of constraints on $\left(A^{-1}\otimes B^{-1}\right)^{\ast}\rho\left(A^{-1}\otimes B^{-1}\right)$. However, according to the discussion in Section \ref{secequivalence}, such local transformations are irrelevant to all the questions considered in this paper. Therefore we may simply assume that a PPT state $\rho$ in question has the product vectors \eqref{formtwoindeptriples4} in its kernel and check the consequences. As previously reported by the authors of \cite{HHMS2011}, the conditions $\innerpr{\phi_i\otimes\psi_j}{\rho\left(\phi_k\otimes\psi_i\right)}=0$ for $i,j,k\in\left\{1,2,3\right\}$ together with $\rho\left(\phi_4\otimes\psi_4\right)=0$ and $\innerpr{\phi_1\otimes\psi_4}{\rho\left(\phi_4\otimes\psi_2\right)}=0$ imply the following form of $\rho$,
\begin{equation}
\rho=\left[\begin{array}{ccc|ccc|ccc}
0&0&0&0&0&0&0&0&0\\
0&a_1&b_1&0&0&0&0&b_2&0\\
0&b_1&a_2&0&0&b_3&0&0&0\\
\hline
0&0&0&a_3&0&b_4&b_5&0&0\\
0&0&0&0&0&0&0&0&0\\
0&0&b_3&b_4&0&a_4&0&0&0\\
\hline
0&0&0&b_5&0&0&a_5&b_6&0\\
0&b_2&0&0&0&0&b_6&a_6&0\\
0&0&0&0&0&0&0&0&0
\end{array}\right],
\label{eqtabelaab}
\end{equation}
with $a_i$ and $b_j$ \emph{real} for all $i,j\in\left\{1,2,\ldots,6\right\}$ and such that
\begin{eqnarray}\label{eqslinearab1}
a_1+b_1+b_2=0,&\,b_1+a_2+b_3=0,&\,a_3+b_4+a_4=0,\\
b_3+b_4+a_4=0,&\,b_5+a_5+b_6=0,&\,b_2+b_6+a_6=0,\label{eqslinearab2}\\
&a_1+b_1+b_2=0.&\label{eqslinearab3}
\end{eqnarray}
Derivation of the equations \eqref{eqtabelaab} and \eqref{eqslinearab1}--\eqref{eqslinearab3} is left as a simple exercise for the reader. It may be useful to consult Section 5.4 of \cite{HHMS2011} in order to solve it.

We still have not used the condition $\rho\left(\phi_5\otimes\psi_5\right)=0$, which gives us additional six linear equations on $a_1,\ldots,a_6$ and $b_1,\ldots,b_6$,
\begin{eqnarray}\label{eqlinearab4}
-r\left(b_1+ b_2\right)+q r b_2 + s b_1 = 0,&\ r b_1 - s\left(b_1+b_3\right) +p s b_3 = 0,\\
-p\left(b_4 + b_5\right)+ q b_5  +p s b_4  = 0,&\  p b_4  + s b_3  - p s\left(b_3 + b_4\right) = 0,\label{eqlinearab5}\\ 
 p b_5  - q\left(b_5 + b_6\right)  + q r b_6  = 0,&\ q b_6  + r b_2 - q r\left(b_2 + b_6\right)  = 0.\label{eqlinearab6}
\end{eqnarray}
Under the assumption of $\left\{\phi_i\otimes\psi_i\right\}_{i=1}^5$ of the form \eqref{formtwoindeptriples4} being a gUPB, there exists, up to scaling by arbitrary real factors, exactly one solution to the equations \eqref{eqslinearab1}--\eqref{eqlinearab6}. We know from Proposition \ref{propgUPB} that the assumption is true for vectors $\phi_i\otimes\psi_i$ in the kernel of a non-separable rank $4$ PPT state in $3\times 3$ systems. It is most important for us that there exist, up to scaling by arbitrary \emph{positive} factors, exactly two solutions
\begin{equation}\label{eqsolutionpm}\scalebox{1.5}{$\pm$}\scalebox{0.85}{
$\left[
\begin{array}{ccccccccc}
 0 & 0 & 0 & 0 & 0 & 0 & 0 & 0 & 0 \\
 0 & \frac{q r-s}{r\left(q-1\right)} & 1 & 0 & 0 & 0 & 0 & \frac{r-s}{r\left(1-q\right) } & 0 \\
 0 & 1 & \frac{r-p s}{s\left(1-p\right) } & 0 & 0 & \frac{r-s}{s\left(p-1\right)} & 0 & 0 & 0 \\
 0 & 0 & 0 & \frac{\left(r-s\right) \left(p s-q\right)}{p \left(p-q\right) \left(s-1\right)} & 0 & \frac{r-s}{p \left(1-s\right)} & \frac{r-s}{q-p} & 0 & 0 \\
 0 & 0 & 0 & 0 & 0 & 0 & 0 & 0 & 0 \\
 0 & 0 & \frac{r-s}{s\left(p-1\right)} & \frac{r-s}{p \left(1-s\right)} & 0 & \frac{\left(p-s\right)\left(r-s\right)}{p\left(p-1\right)  s\left(s-1\right) } & 0 & 0 & 0 \\
 0 & 0 & 0 & \frac{r-s}{q-p} & 0 & 0 & \frac{\left(q r-p\right)\left(r-s\right)}{q \left(1-q\right) \left(r-1\right)} & \frac{r-s}{q\left(q-1\right)} & 0 \\
 0 & \frac{r-s}{r\left(1-q\right) } & 0 & 0 & 0 & 0 & \frac{r-s}{q\left(q-1\right)} & \frac{\left(q-r\right)\left(r-s\right)}{q\left(1-q\right) r \left(r-1\right) } & 0 \\
 0 & 0 & 0 & 0 & 0 & 0 & 0 & 0 & 0
\end{array}
\right].$
}
\end{equation}
The above matrix is well-defined since all the numbers $p$, $q$, $r$, $s$, $p-1$, $q-1$, $r-1$, $s-1$, $p-q$ and $r-s$ are nonzero as a consequence of all triples of vectors in \eqref{formtwoindeptriples4} being linearly independent.

Note that, for both choices of sign, \eqref{eqsolutionpm} is a symmetric matrix. Moreover, it is \emph{symmetric with respect to the partial transpose}. Therefore $\rho$ is PPT iff it is positive definite. A necessary condition for \eqref{eqsolutionpm} to be positive definite is that all the nonzero elements on its diagonal, as well as all nontrivial $2\times 2$ minors of the form $\left|\begin{array}{cc}\rho_{ii}&\rho_{ij}\\\rho_{ji}&\rho_{jj}\end{array}\right|$ are positive. Altogether, we have six nonzero elements on the diagonal
\begin{multline}\label{eqelondiag}
\pm\left\{\frac{q r - s}{r\left( q-1\right)},-\frac{r - p s}{s\left(p - 
 1\right)}, \frac{\left(r - s\right)\left(p s-q\right)}{p \left(p - q\right)\left( s-1\right)},\right.\\\left. \frac{\left(p - s\right)\left(r - 
   s\right)}{p \left(p-1\right)s\left(s-1\right)}, -\frac{\left( q r-p\right) (r - s)}{q\left (p-q\right)\left(r-1\right)}, \frac{\left(r-q\right)\left(r - s\right)}{q\left(q-1\right)r\left(r-1\right)}\right\},
\end{multline}
and six nontrivial minors
\begin{multline}\label{eqminors}
\left\{-\frac{\left(r - s\right)\left(q r - p s\right)}{r\left( p-1\right)s\left(q-1\right)}, -\frac{\left(q - s\right)\left(r - 
   s\right)}{q\left( q-1\right) r\left(r-1\right) },\right.\\
   \frac{\left(p - r\right)\left(r - s\right)}{p\left( p-1\right) s\left( 
   s-1\right)},
   \frac{\left(q - s\right) \left(r - s\right)^2}{p\left(p-1\right) \left(p - q\right)s\left(s-1\right) },\\\left. \frac{\left(r - 
   s\right)^2 \left(q r - p s\right)}{p\left(p - q\right) q\left(r-1\right)\left( 
   s-1\right)},
   -\frac{\left(p - r\right)\left(r - s\right)^2}{\left(p - q\right)q\left(q-1\right) r \left(r-1\right) }\right\}.
\end{multline}

The $\pm$ sign in \eqref{eqelondiag} corresponds to the choice we make in \eqref{eqsolutionpm}. We see that all the expressions in \eqref{eqminors} and \eqref{eqelondiag} are quotients and products of the following nineteen numbers
\begin{eqnarray}
&p,\,q,\,r,\,s,\,p-1,\,q-1,\,r-1,\,s-1,\,p-q,\,r-s,\label{numerkilista1}\\
&p-r,\,q-s,\,p-s,\,r-q,\,ps-q,\,qr-p,\,r-ps,\,qr-s,\,qr-ps.\label{numerkilista2}
\end{eqnarray}
Concerning the list \eqref{numerkilista1}, we already know that all its elements have to be nonzero. This follows from the condition of $\left\{\phi_i\otimes\psi_i\right\}_{i=1}^5$ being a gUPB. It turns out that the same holds for the elements of \eqref{numerkilista2}. The number $qr-ps$  must be nonzero, because otherwise the vector 
\begin{equation}
\phi_5\otimes\psi_5-qr\phi_4\otimes\psi_4-q\left(s-r\right)\phi_3\otimes\psi_3-r\left(p-q\right)\phi_2\otimes\psi_2
\end{equation}
would be of the form $\phi_1\otimes\psi'+\phi'\otimes\psi_1$, thus contradicting Lemma \ref{lemmaSR2} and Corollary \ref{coredge}. In a similar way, one can show that $p-r\neq 0$ and $q-s\neq 0$. Let us now assume that $ps-q=0$. In such case, we have the following submatrix in \eqref{eqsolutionpm}
\begin{equation}\label{submatrix}
\pm\left[\begin{array}{cc}\frac{\left(r-s\right)\left(ps-q\right)}{p\left(p-q\right)\left(s-1\right)}&\frac{r-s}{p\left(1-s\right)}\\
\frac{r-s}{p\left(1-s\right)}&\frac{\left(p-s\right)\left(r-s\right)}{p\left(p-1\right)s\left(s-1\right)}
\end{array}\right]=\pm\left[\begin{array}{cc}0&\frac{r-s}{p\left(1-s\right)}\\
\frac{r-s}{p\left(1-s\right)}&\frac{\left(p-s\right)\left(r-s\right)}{p\left(p-1\right)s\left(s-1\right)}
\end{array}\right].
\end{equation}
In order for \eqref{submatrix} to be positive definite for some choice of the sign $\pm$, we need to have $r-s=0$, which we know is impossible. Thus we have proved that $ps-q\neq 0$ for $\rho$ positive definite. Finally, the fact that $qr-p$, $r-ps$ and $qr-s$ must also be nonvanishing for $\rho$ positive definite follows by a suitable modification of the above argument. Different submatrices need to be chosen, but otherwise the proof is identical. 

Our task in the following will be to relate positivity of all the numbers in \eqref{eqelondiag} and \eqref{eqminors} to the fact that all the invariants $s_1,\ldots,s_4$, given in Section \ref{secequiv}, are positive, possibly after we suitably permute the vectors $\phi_i\otimes\psi_i$. Note that we already know that only the $12$ permutations listed in 
Table \ref{tblpentagram} need to be considered. An explanation is included in the proof related to Remark \ref{remarkpermutation}. Not to much surprise, the formulas for the invariants $s_1,\ldots,s_4$ for permuted vectors of the form \eqref{formtwoindeptriples4} are always expressed as products and quotients including only the numbers listed in \eqref{numerkilista1}. Explicit formulas can be found in Table \ref{tabinvariants}. To explain the notation we used in the table, it is sufficient to say, for example, that by using $\sigma_6$ from 
Table \ref{tblpentagram} to permute the product vectors \eqref{formtwoindeptriples4}, we obtain $s_1=-p$, $s_2={\left(1-q\right)}/{q}$, $s_3=r-1$ and $s_4={r}/{\left(s-r\right)}$ as the expressions for the invariants. 
\begin{table}
\begin{tabular}{|l|l|l|l|}
\hline
$\sigma_1:$&$-\frac{p}{q}, q-1, \frac{r-s}{s}, \frac{r}{1-r}$&$\sigma_2:$&$-\frac{q}{p}, 
  p-1, \frac{s-r}{r}, \frac{s}{1-s}$\\
$\sigma_3:$&$-\frac{1}{q}, \frac{q-p}{p}, \frac{1-s}{s}, \frac{1}{r-1}$&$\sigma_4:$&$-q, \frac{1-p}{p},s-1,\frac{s}{r-s}$\\
$\sigma_5:$&$-\frac{1}{p}, \frac{p-q}{q}, \frac{1-r}{r}, \frac{1}{s-1}$&$\sigma_6:$&$-p,\frac{1-q}{q}, 
  r-1, \frac{r}{s-r}$\\
$\sigma_7:$&$\frac{p-q}{q}, \frac{1}{q-1}, -\frac{r}{s}, \frac{s-r}{r-1}$&$\sigma_8:$&$\frac{q}{p-q}, \frac{1-p}{q-1}, \frac{r}{s-r}, -s$\\
$\sigma_9:$&$-\frac{q-1}{q}, \frac{p}{q-p}, -\frac{1}{s}, \frac{1-s}{r-1}$&$\sigma_{10}:$&$\frac{q}{1-q}, \frac{p-1}{q-p}, \frac{1}{s-1}, -\frac{s}{r}$\\
$\sigma_{11}:$&$\frac{q-p}{p}, \frac{1}{p-1}, -\frac{s}{r}, \frac{r-s}{s-1}$&$\sigma_{12}:$&$\frac{p}{q-p}, \frac{1-q}{p-1}, \frac{s}{r-s}, -r$\\
\hline
\end{tabular}\centering
\vskip 2 pt
\caption{Formulas for the invariants $s_1,\ldots,s_4$, calculated for vectors of the form \eqref{formtwoindeptriples4} permuted by the $12$ inequivalent permutations 
in Table \ref{tblpentagram}.}
\label{tabinvariants}
\end{table}

It turns out that the values of $s_1,\ldots,s_4$ corresponding to one of the permutations $\sigma_i$ have to be all positive to assure that $\rho$, given in \eqref{eqsolutionpm}, is a positive matrix for some choice of the sign $\pm$. Our computer-aided proof of this fact consisted in simply checking all admissible sign choices for the numbers listed in \eqref{numerkilista1} and \eqref{numerkilista2}. We already know that neither of those numbers can be zero, and thus it seems that we have $2^{19}$ cases to check. However, some further constraints apply, which reduce this number considerably. First of all, the requirement that $\pm{\left(p - s\right)\left(r - 
   s\right)}/{\left(p \left(p-1\right)s\left(s-1\right)\right)}$ of the list \eqref{eqelondiag} and a very similar element ${\left(p - r\right)\left(r - s\right)}/{\left(p\left( p-1\right) s\left( 
   s-1\right)\right)}$ of \eqref{eqminors} have the same sign implies that $p-r=\pm\left(p-s\right)$, with the $\pm$ sign depending on the choice we made in \eqref{eqsolutionpm}. Along the same lines, by comparing the last element of \eqref{eqelondiag} with the second element of \eqref{eqminors}, one can prove that $r-q=\mp\left(q-s\right)$. More importantly, the signs of the numbers listed in \eqref{numerkilista1} and \eqref{numerkilista2} are not all independent. Various relations have to hold between them. For example, $p-1>0$ clearly implies $p>0$, and we cannot have a plus sign for $p-1$ and a minus sign for $p$. More sophisticated relations like
\begin{equation}\label{sophistic}
\left(r<0\land q-1<0\land r-ps>0\right)\Rightarrow qr-ps>0.
\end{equation}
have to hold as well. Alternatively, the above formula can be written as
\begin{equation}\label{sophistic2}
\neg\left(r<0\land q-1<0\land r-ps>0\land qr-ps<0\right).
\end{equation}
 We provide a more or less exhaustive list, consisting of 76 elements, in Tables \ref{tabsigns3} and \ref{tabsigns2} on pages \pageref{tabsigns3} and \pageref{tabsigns2}. 
For example, the relation \ref{sophistic2} corresponds to the following row in Table \ref{tabsigns3},
\begin{equation}\nonumber
\scalebox{0.80}{
\begin{tabular}{|c|c|c|c|c|c|c|c|c|c|c|c|c|c|c|c|c|c|c|}
\hline
p&q&r&s&pp&qq&rr&ss&pq&rs&pr&ps&rq&qs&qrp&qrs&psq&rps&qrps\\
\hline
&&--&&&--&&&&&&&&&&&&+&--\\
\hline
\end{tabular}
}
\end{equation}
which should explain the notation we used\footnote{To better explain the symbols in the header of Tables \ref{tabsigns3}, \ref{tabsigns2} and \ref{tabsignchoices}, let us add that $pp$, $qq$, $rr$ and $ss$ denote $p-1$, $q-1$, $r-1$ and $s-1$, respectively, while $pq$, $rs$, $pr$, $rq$, $qs$, $qrp$, $qrs$, $psq$, $rps$ and $qrps$ stand for $p-q$, $r-s$, $p-r$, $r-q$, $qr-p$, $qr-s$, $ps-q$, $r-ps$ and $qr-ps$, respectively.}. While some further relations could still possibly exist, the use of those listed in the appendix allowed us to confirm the necessity result mentioned above. When all the constraints are imposed, a comparably small number of $761$ or $352$ out of the $2^{19}$ sign choices remain possible when ``$+$'' or ``$-$'' is fixed in \eqref{eqsolutionpm}, respectively. It then turns out that, by choosing an admissible sign configuration, all the numbers in the lists \eqref{eqelondiag} and \eqref{eqminors} can be made positive only if one of the quadruples listed in Table \ref{tabinvariants} consists solely of positive numbers. This is in full agreement with, and provides a rigorous, although not very insightful proof of the results reported in Section 5 of \cite{HHMS2011}. Actually, it turns out that there are precisely $12$ admissible sign configurations that correspond to a positive $\rho$ for some choice of the sign $\pm$ in \eqref{eqsolutionpm} and each of the quadruples in Table \ref{tabinvariants} is positive precisely for one of them. A complete list of the selected sign choices and the corresponding permutations is given in 
 Table \ref{tabsignchoices}. Interestingly, $10$ of them correspond to choosing the plus sign in \eqref{eqsolutionpm}, while only $2$ to the minus sign. This is rather an uneven partitioning of the total of $12$ configurations, which is somewhat puzzling.


\begin{table}[ht!]\centering
\scalebox{0.70}{
\begin{tabular}{|c|c|c|c|c|c|c|c|c|c|c|c|c|c|c|c|c|c|c|}
\hline
p&q&r&s&pp&qq&rr&ss&pq&rs&pr&ps&rq&qs&qrp&qrs&psq&rps&qrps\\
\hline
&&+&&&+&&&&+&&&&&&--&&&\\
&+&&&&&+&&&&&&&+&&--&&&\\
&&+&&&--&&&&--&&&&&&+&&&\\
&+&&&&&--&&&&&&&--&&+&&&\\
&&&+&+&&&&&&&&&--&&&--&&\\
+&&&&&&&+&+&&&&&&&&--&&\\
&&&+&--&&&&&&&&&+&&&+&&\\
+&&&&&&&--&--&&&&&&&&+&&\\
&+&&&&&+&&--&&&&&&--&&&&\\
&&+&&&+&&&&&--&&&&--&&&&\\
&+&&&&&--&&+&&&&&&+&&&&\\
&&+&&&--&&&&&+&&&&+&&&&\\
+&&&&&&&+&&&+&&&&&&&+&\\
&&&+&+&&&&&--&&&&&&&&+&\\
+&&&&&&&--&&&--&&&&&&&--&\\
&&&+&--&&&&&+&&&&&&&&--&\\
&&&+&--&&&&&&&&&&&+&&&--\\
&&&+&+&&&&&&&&&&&--&&&+\\
&+&&&&&+&&&&&&&&&&--&&--\\
&+&&&&&--&&&&&&&&&&+&&+\\
+&&&&&&&--&&&&&&&+&&&&--\\
+&&&&&&&+&&&&&&&--&&&&+\\
&&+&&&+&&&&&&&&&&&&+&--\\
&&+&&&--&&&&&&&&&&&&--&+\\
&&--&&&+&&&&--&&&&&&+&&&\\
&&--&&&--&&&&+&&&&&&--&&&\\
&--&&&&&--&&&&&&&+&&--&&&\\
&--&&&&&+&&&&&&&--&&+&&&\\
&&&--&--&&&&&&&&&--&&&--&&\\
--&&&&&&&--&+&&&&&&&&--&&\\
&&&--&+&&&&&&&&&+&&&+&&\\
--&&&&&&&+&--&&&&&&&&+&&\\
&--&&&&&--&&--&&&&&&--&&&&\\
&&--&&&--&&&&&--&&&&--&&&&\\
&--&&&&&+&&+&&&&&&+&&&&\\
&&--&&&+&&&&&+&&&&+&&&&\\
--&&&&&&&+&&&--&&&&&&&--&\\
&&&--&+&&&&&+&&&&&&&&--&\\
--&&&&&&&--&&&+&&&&&&&+&\\
&&&--&--&&&&&--&&&&&&&&+&\\
\hline
\end{tabular}
}
\vskip 2 pt
\caption{Non-admissible sign choices. Part I.}
\label{tabsigns3}
\end{table}
\begin{table}[ht!]\centering
\scalebox{0.7}{
\begin{tabular}{|c|c|c|c|c|c|c|c|c|c|c|c|c|c|c|c|c|c|c|}
\hline
p&q&r&s&pp&qq&rr&ss&pq&rs&pr&ps&rq&qs&qrp&qrs&psq&rps&qrps\\
\hline
&&&--&+&&&&&&&&&&&+&&&--\\
&&&--&--&&&&&&&&&&&--&&&+\\
&--&&&&&--&&&&&&&&&&--&&--\\
&--&&&&&+&&&&&&&&&&+&&+\\
--&&&&&&&+&&&&&&&+&&&&--\\
--&&&&&&&--&&&&&&&--&&&&+\\
&&--&&&--&&&&&&&&&&&&+&--\\
&&--&&&+&&&&&&&&&&&&--&+\\
--&&&&+&&&&&&&&&&&&&&\\
&--&&&&+&&&&&&&&&&&&&\\
&&--&&&&+&&&&&&&&&&&&\\
&&&--&&&&+&&&&&&&&&&&\\
--&+&&&&&&&+&&&&&&&&&&\\
+&--&&&&&&&--&&&&&&&&&&\\
&&+&--&&&&&&--&&&&&&&&&\\
&&--&+&&&&&&+&&&&&&&&&\\
&&&&--&+&&&+&&&&&&&&&&\\
&&&&+&--&&&--&&&&&&&&&&\\
&&&&&&--&+&&+&&&&&&&&&\\
&&&&&&+&--&&--&&&&&&&&&\\
%
&&&&&--&&+&&&&&&+&&&&&\\
&--&&+&&&&&&&&&&+&&&&&\\
&+&&--&&&&&&&&&&--&&&&&\\
&&&&&+&&--&&&&&&--&&&&&\\
--&&+&&&&&&&&+&&&&&&&&\\
&&&&--&&+&&&&+&&&&&&&&\\
+&&--&&&&&&&&--&&&&&&&&\\
&&&&+&&--&&&&--&&&&&&&&\\
&+&--&&&&&&&&&&+&&&&&&\\
&&&&&+&--&&&&&&+&&&&&&\\
&--&+&&&&&&&&&&--&&&&&&\\
&&&&&--&+&&&&&&--&&&&&&\\
--&&&+&&&&&&&&+&&&&&&&\\
&&&&--&&&+&&&&+&&&&&&&\\
+&&&--&&&&&&&&--&&&&&&&\\
&&&&+&&&--&&&&--&&&&&&&\\
\hline
\end{tabular}
}
\vskip 2 pt
\caption{Non-admissible sign choices. Part~II.}
\label{tabsigns2}
\end{table}


\begin{table}[ht]
\scalebox{0.73}{
\begin{tabular}{|>{$}c<{$}|c|c|c|c|c|c|c|c|c|c|c|c|c|c|c|c|c|c|c|}
\cline{2-20}
\multicolumn{1}{c|}{}&p&q&r&s&pp&qq&rr&ss&pq&rs&pr&ps&rq&qs&qrp&qrs&psq&rps&qrps\\
\hline
\sigma_1&--&+&+&+&--&+&--&--&--&+&--&--&--&+&+&+&--&+&+\\
\sigma_2&+&--&+&+&+&--&--&--&+&--&+&+&+&--&--&--&+&--&--\\
\sigma_3&--&--&+&+&--&--&+&--&+&+&--&--&+&--&--&--&+&+&--\\
\sigma_4&+&--&+&+&--&--&+&+&+&+&--&--&+&--&--&--&+&+&--\\
\sigma_5&--&--&+&+&--&--&--&+&--&--&--&--&+&--&+&--&--&+&+\\
\sigma_6&--&+&+&+&--&--&+&+&--&--&--&--&+&--&+&--&--&+&+\\
\sigma_7&+&+&+&--&+&+&--&--&+&+&+&+&--&+&--&+&--&+&+\\
\sigma_8&+&+&--&--&+&--&--&--&+&+&+&+&--&+&--&+&--&+&+\\
\sigma_9&+&+&+&--&--&--&+&--&--&+&--&+&+&+&+&+&--&+&+\\
\sigma_{10}&+&+&--&+&--&--&--&+&+&--&+&--&--&--&--&--&+&--&--\\
\sigma_{11}&+&+&--&+&+&+&--&--&--&--&+&+&+&+&--&--&--&--&--\\
\sigma_{12}&+&+&--&--&--&+&--&--&--&--&+&+&--&+&--&--&--&--&--\\
\hline
\end{tabular}
}
\vskip 2 pt
\caption{Sign choices that yield a positive $\rho$ and obey all the constraints of Tables \ref{tabsigns3} and \ref{tabsigns2} on pages \pageref{tabsigns3} and \pageref{tabsigns2}.}
\label{tabsignchoices}
\end{table} 

To summarize, the computer-aided proof we carried out allows us to state the following.
\begin{proposition}\label{propdeterminationbygUPB}
A necessary and sufficient criterion for a generalized Unextendible Product Basis $\left\{\phi_i\otimes\psi_i\right\}_{i=1}^5\subset\mathbbm{C}^3\otimes\mathbbm{C}^3$ to belong to the kernel of a rank $4$ PPT state $\rho$ is that there exists a permutation of the vectors $\phi_i\otimes\psi_i$ that it yields all the values of the invariants $s_1$, $s_2$, $s_3$ and $s_4$, defined as in equations \eqref{invariants1}, \eqref{invariants2}, \eqref{invariant3} and \eqref{invariant4}, positive.  When checking positivity of $s_i$, it is possible to consider only the $12$ permutations, listed in 
Table \ref{tblpentagram}, and the corresponding expressions for the invariants, given in Table \ref{tabinvariants}. 
\begin{proof}
First of all, let us note that a separable state $\rho$ cannot have a gUPB in its kernel, since it must have a product state in its range. Thus in the following we may always assume that $\rho$ is entangled.
Let us prove sufficiency first. If the invariants are positive for the permuted vectors $\phi'_i\otimes\psi'_i:=\phi_{\sigma\left(i\right)}\otimes\psi_{\sigma\left(i\right)}$, we know from Proposition~\ref{propinvariantspositiveortho2} that there exists a $\SLtt$ transformation $A\otimes B$ such that the transformed vectors $\left(A\otimes B\right)\phi'_{i}\otimes\psi'_{i}=\left(A\otimes B\right)\phi_{\sigma\left(i\right)}\otimes\psi_{\sigma\left(i\right)}$ are elements of an orthogonal UPB $\left\{v_i\otimes w_i\right\}_{i=1}^5$. With no loss of generality, we may assume that the vectors $v_i\otimes w_i$ are normalized to unity. In such case the projection
\begin{equation}\label{eqprojector}
\rho':=\mathbbm{1}-\sum_{i=1}^5\proj{v_i\otimes w_i}
\end{equation}
has all the vectors $v_i\otimes w_i$ in its kernel and it is a PPT entangled state \cite{Bennett99}. The locally transformed PPT state $\rho=\left(A\otimes B\right)^{\ast}\rho'\left(A\otimes B\right)$ has all the vectors $\phi_i\otimes\psi_i$ in its kernel.

In order to prove necessity, note that from the discussion above we know that positivity of $s_1,\ldots,s_4$, possibly after a permutation, is a necessary condition for a PPT entangled state $\rho'$ with vectors $\phi_i\otimes\psi_i$ in its kernel to exist, provided that the vectors are as in equation \eqref{formtwoindeptriples4}. But any gUPB $\left\{\phi_i\otimes\psi_i\right\}_{i=1}^5$ can be brought to the form \eqref{formtwoindeptriples4} by a local transform, say $C\otimes D$. If we assume that a PPT state $\rho$ has $\left\{\phi_i\otimes\psi_i\right\}_{i=1}^5$ in its kernel, then the locally transformed $\rho'':=\left(C^{-1}\otimes D^{-1}\right)^{\ast}\rho\left(C^{-1}\otimes D^{-1}\right)$ has $\left(C\otimes D\right)\phi_i\otimes\psi_i$ in its kernel. But $\left(C\otimes D\right)\phi_i\otimes\psi_i$ are of the form \eqref{formtwoindeptriples4}. From the above discussion, $\rho''$ is PPT if and only if the invariants $s_1,\ldots,s_4$ are positive, possibly after we permute the vectors $\left(C\otimes D\right)\phi_i\otimes\psi_i$. But $C\otimes D$ does not change the value of the invariants, and thus $\phi_i\otimes\psi_i$, permuted in the same way as the $\left(C\otimes D\right)\phi_i\otimes\psi_i$, must also have all of them positive.
\end{proof}
\end{proposition}
Let us also state the following result, which should be expected from the discussion above.
\begin{proposition}\label{propuniqueness}
Let $\left\{\phi_i\otimes\psi_i\right\}_{i=1}^5\subset\mathbbm{C}^3\otimes\mathbbm{C}^3$ be a gUPB that yields, after a suitable permutation of the product vectors, positive values of all the invariants $s_1,\ldots,s_4$. The PPT state $\rho$ with $\left\{\phi_i\otimes\psi_i\right\}_{i=1}^5$ in its kernel is uniquely determined, up to scaling by a constant positive factor.
\begin{proof}
We already know that the assertion of the proposition holds for gUPBs of the form \eqref{formtwoindeptriples4}. We also know that any gUPB $\left\{\phi_i\otimes\psi_i\right\}_{i=1}^5$ can be locally transformed so that it looks like in \eqref{formtwoindeptriples4}. Let us denote the transformation which does it by $C\otimes D$. There cannot exist two PPT states $\rho_1$ and $\rho_2$ with $\left\{\phi_i\otimes\psi_i\right\}_{i=1}^5$ in their kernels, because in such case the PPT states $\left(C^{-1}\otimes D^{-1}\right)^{\ast}\rho_1\left(C^{-1}\otimes D^{-1}\right)$ and $\left(C^{-1}\otimes D^{-1}\right)^{\ast}\rho_2\left(C^{-1}\otimes D^{-1}\right)$ would both have the same gUPB of the form \eqref{formtwoindeptriples4} in their kernel, which we know is not possible.
\end{proof}
\end{proposition}
\section{The main result}\label{secmainresult}
Using the knowledge from the previous sections, we can now easily prove our main result.
\begin{theorem}\label{maintheorem}
Positive-partial-transpose states of rank $4$ in $3\times 3$ systems are either separable or they are of the form
\begin{equation}\label{eqlocaltransform}
\rho=\left(A\otimes B\right)^{\ast}\left(\mathbbm{1}-\sum_{i=1}^5\proj{v_i\otimes w_i}
\right)\left(A\otimes B\right)
\end{equation}
with $A,B\in\SLt$ and $\left\{v_i\otimes w_i\right\}_{i=1}^5$ an orthonormal Unextendible Product Basis. In the latter case, they are entangled, and extreme in the set of PPT states. The rank of the partial transpose of the state is  $4$ in case of nonseparable states. 
\begin{proof}
In case of separable states, there is nothing to prove.
Let $\rho$ be a non-separable PPT state of rank $4$ in a $3\times 3$ system. We know from Proposition \ref{propgUPB} that there is a generalized UPB, say $\left\{\phi_i\otimes\psi_i\right\}_{i=1}^5$, in the kernel of $\rho$. From Proposition \ref{propdeterminationbygUPB} we know that the corresponding values of the invariants $s_1,\ldots,s_4$ must be all positive after we suitably permute the vectors $\phi_i\otimes\psi_i$. Next, Proposition \ref{propinvariantspositiveortho} tells us that there exists a $\SLtt$ transformation $A\otimes B$ that brings $\left\{\phi_i\otimes\psi_i\right\}_{i=1}^5$ to an orthogonal UPB $\left\{v_i\otimes w_i\right\}_{i=1}^5$. With no loss of generality, we may assume that the vectors $v_i\otimes w_i$ are normalized. From Proposition \ref{propuniqueness} we know that there exists, up to scaling, exactly one PPT state which has $\left\{v_i\otimes w_i\right\}_{i=1}^5$ in its kernel. It must be $\mathbbm{1}-\sum_{i=1}^5\proj{v_i\otimes w_i}
$. The state given by the formula \eqref{eqlocaltransform} clearly is PPT, and it has all the vectors $\phi_i\otimes\psi_i$ in its kernel. By using Proposition~\ref{propuniqueness} again, we see that it must be equal to the $\rho$ we started with. The fact that the rank of the partial transpose is $4$ for non-separable states, is simply the assertion of Proposition \ref{propequalranks}.
\end{proof}
\end{theorem}  
In this way, we have obtained a full characterization of bound entangled states of minimal rank. Let us also mention a special property they have, which can be loosely described as saying that it is not enough for an entanglement witness to be indecomposable in order to detect them.
\begin{remark}\label{remarkatomicity}
According to \cite[Lemma 3]{SBL2001}, all PPT states of rank $4$ in $3\times 3$ systems can be written as a sum of four projections onto vectors of Schmidt rank $2$. By Theorem \ref{maintheorem}, or Proposition \ref{propequalranks}, their partial transposes are also of rank $4$ and thus can be decomposed in an analogous way. Using the notation of \cite{ref.SSZ09}, we can write that all such PPT states are elements of the cone $\mathcal{S}_{2,2}$. The dual cone $\mathcal{S}_{2,2}^{\circ}=\mathcal{D}_{2,2}$ consists of Jamiołkowski-Choi transforms of convex sums of $2$-positive and $2$-co-positive maps. Consequently, any entanglement witness that detects a PPT state of rank $4$ in a $3\times 3$ system is {\bf atomic} \cite{Ha98}. This applies in particular to the witness discussed in Example 1 of \cite{Terhal2001} and the Choi map, in relation to the PPT state discussed in Section 4 of \cite{HaKyePark2003}.
\end{remark}

\chapter*{Conclusion}\label{chconclusion}\addcontentsline{toc}{chapter}{Conclusion}
Computational advances in the field of algebraic geometry have not become well-known among the quantum information community, despite a number of problems that are, at the very bottom, systems of polynomial equations. In the present thesis, I tried to outline a few possible applications of Groebner basis methods in quantum information and quantum entanglement science, including:
\begin{itemize}
\item Compression equations for Quantum Error Correction (QEC), Section \ref{seccompeq}
\item Completely Entangled Subspaces (CES), Section \ref{secces}
\item Maximally entangled states, Section \ref{secmaxent}
\item Mutually Unbiased Bases (MUBs) and Symmetric Informationally Complete vectors (SICs), Sections \ref{secMUBs} and \ref{secSICs}
\end{itemize}
The main result, which is a characterization of rank four  entangled states of two qutrits with positive-partial-transpose (PPT), was presented in Chapter~\ref{chPPT3x3}. Its proof uses a tool from algebraic geometry, but this time it is the theorem of Bezout, a basic result in intersection theory. In the thesis, I also included a few problems that I solved during my PhD studies using simple algebra tricks. They can be found in Chapter~\ref{chhand}. Moreover, I felt it was appropriate to present a characterization result for certain cones of positive maps, included in Chapter~\ref{chmappingcones}. 

The central idea of the thesis was that the problems solved should be algebraic in nature. Obviously, I also required them to be of interest for the quantum information community. I did not presume the readers to be experts neither in mathematics, nor in foundational or practical questions relating to quantum mechanics. Hence, I included introduction to both the mathematical apparatus I used and to certain aspects of quantum theory. I hope the thesis may contribute to a better understanding of some tools of algebraic geometry among the quantum information community and hence lead to their new applications in areas such as the classification of PPT states or Completely Entangled Subspaces, solving QEC equations or the investigation of MUBs and SICs,
and hopefully a few more. One of big questions that remains open is how to understand all the numerical findings on PPT states included in the work by Leinaas, Myrheim and Sollid \cite{LS2010n}. I believe algebraic geometry, which turned out to be so useful in the three-by-three, rank four case, could still be used to explain properties observed for higher rank and/or higher dimensional cases. However, there does not seem to exist a direct generalization of the results of Chapter~\ref{chPPT3x3} to these cases.

\chapter*{A list of papers published}\addcontentsline{toc}{chapter}{A list of papers published}\,
The following papers were published by the author as a part of the PhD project reported in this thesis (inverse chronological order):
\begin{enumerate}
\item Ł. Skowronek, E. St{\o}rmer, \textit{Choi matrices, norms and entanglement associated with positive maps on matrix algebras}, J. Func. Analysis 262 (2012), 639--647
\item Ł. Skowronek, \textit{Three-by-three bound entanglement with general unextendible product bases},  J. Math. Phys. 52 (2011), 122202
\item Ł. Skowronek, \textit{Cones with a mapping cone symmetry in the finite-dimensional case}, Lin. Alg. Appl. 435 (2011), 361--370
\item Z. Puchała, P. Gawron, J. A. Miszczak, Ł. Skowronek, M.-D. Choi, K. Życzkowski, \textit{Product numerical range in a space with tensor product structure}, Lin. Alg. Appl. 434 (2011), 327--342 
\item P. Gawron, Z. Puchała, J. A. Miszczak, Ł. Skowronek, K. Życzkowski, \textit{Restricted numerical range: a versatile tool in the theory of quantum information}, J. Math. Phys. 51 (2010), 102204 
\item Ł. Skowronek, \textit{Dualities and positivity in the study of quantum entanglement},  Int. J. Quantum Inf. Vol. 8, No. 5 (2010), 721--754
\item Ł. Skowronek, K. Życzkowski, \textit{Positive maps, positive polynomials and entanglement witnesses},  J. Phys. A: Math. Theor. 42 (2009), 325302
\item Ł. Skowronek, E. St{\o}rmer, K. Życzkowski, \textit{Cones of positive maps and their duality relations},  J. Math. Phys. 50 (2009), 062106 
\end{enumerate}


\tableofcontents\newpage
\bibliographystyle{unsrt}\addcontentsline{toc}{chapter}{Bibliography}
\bibliography{PhDThesis}

\end{document}